\documentclass{elsart}
\usepackage{color,setspace,times}
\usepackage{amssymb,amsmath,graphicx,color,rotating,subfigure,url}
\usepackage{lineno}
\usepackage{times}
\usepackage[square,sort&compress,comma]{natbib}
\usepackage{url}
\newcommand{\be}{\begin{equation}}
\newcommand{\ee}{\end{equation}}
\newcommand{\bea}{\begin{eqnarray}}
\newcommand{\eea}{\end{eqnarray}}

\journal{Physica A} 

\begin{document}


\runauthor{Zhou and Sornette} \markboth{A}{B}
\begin{frontmatter}
\title{A case study of speculative financial bubbles in the South African stock market 2003-2006}
\author[BS,SS,RCE,RCSE]{\small{Wei-Xing Zhou}},
\author[ETH]{\small{Didier Sornette}\thanksref{EM}}
\address[BS]{School of Business, East China University of Science and Technology, Shanghai 200237, China}
\address[SS]{School of Science, East China University of Science and Technology, Shanghai 200237, China}
\address[RCE]{Research Center for Econophysics, East China University of Science and Technology, Shanghai 200237, China}
\address[RCSE]{Research Center of Systems Engineering, East China University of Science and Technology, Shanghai 200237, China}
\address[ETH]{Chair of Entrepreneurial Risks, Department of Management, Technology, and Economics, ETH Zurich, CH-8032 Zurich, Switzerland}
\thanks[EM]{Corresponding author. {\it E-mail address:}\/
dsornette@ethz.ch (D. Sornette)\\
http://www.er.ethz.ch/}

\begin{abstract}
We tested 45 indices and common stocks in the South African stock
market for the possible existence of a bubble over the period from
January 2003 to May 2006. A bubble is defined by a
faster-than-exponential acceleration with significant log-periodic
oscillations. These two traits are analyzed using different methods.
Sensitivity tests shows that the estimated parameters are robust.
With the insight of 6 additional month of data since the analysis
was performed, we observe that many of the stocks on the South
Africa market experienced an abrupt drop mid-June 2006, which is
compatible with the predicted $t_c$ for several of the stocks, but
not all. This suggests that the mini-crash that occurred around
mid-June of 2006 was only a partial correction, which has resumed
into a renewed bubbly acceleration bound to end some times in 2007,
similarly to what happened on the US market from October 1997 to
August 1998.
\end{abstract}


\end{frontmatter}

\section{Introduction}
\label{sec:intro}

One of the most robust characteristics of humans, which has arguably
the most visible imprint in our social affairs, is imitation and
herding. Imitation has been documented in psychology and in
neuro-sciences as one of the most evolved cognitive process,
requiring a developed cortex and sophisticated processing abilities.
In short, we learn our basics and how to adapt mostly by imitation
all along our life. It seems that imitation has evolved as an
evolutionary advantageous trait, and may even have promoted the
development of our anomalously large brain (compared with other
mammals) \cite{Dunbar-1998-EA}. It is actually ``rational'' to
imitate when lacking sufficient time, energy and information to take
a decision based only on private information and processing, that
is..., most of the time. Imitation, in obvious or subtle forms, is a
pervasive activity of humans. In the modern business, economic and
financial worlds, the tendency for humans to imitate leads in its
strongest form to herding and to crowd effects \cite{Sornette-2003}.

Models of cooperative herding and imitation have been built on the
notion that imitation leads to positive feedbacks, that is, an
action leads to consequences which themselves reinforce the action
and so on, leading to virtuous or vicious circles. We have
formalized these ideas in the mathematical theory of rational
expectation bubbles in the presence of noisy imitative traders.
The main idea is to take into account positive feedbacks,
due for instance to derivative hedging, portfolio insurance and foremost to
imitative trading, as an essential cause for the appearance
of non-sustainable bubble regimes. Specifically, the positive feedbacks
give rise to power law (i.e., faster than exponential) acceleration of prices.
Previous works by us and our co-workers as well as a few other groups
suggest that a robust additional feature characterizes the faster-than-exponential
growth of prices during bubbles: the existence of accelerating ups and downs,
roughly organized according to a geometrically convergence series of characteristic
time scales decorating the power law acceleration. Such patterns have
been coined ``log-periodic power law'' (LPPL).

Several mechanisms are known to generate LPPL, suggesting
several complementary non-necessarily exclusive explanations.
A dynamical explanation \cite{Ide-Sornette-2002-PA,Sornette-Ide-2003-IJMPC}
consists in taking into account the
competition between positive feedback (self-fulfilling sentiment),
negative feedbacks (contrariant behavior and fundamental value
analysis) and inertia (everything takes time to adjust). The
competition between nonlinear trend followers and nonlinear value
investors together with inertia between investor decisions and their
market impact may lead to nonlinear oscillations
approximating log-periodicity
\cite{Ide-Sornette-2002-PA,Sornette-Ide-2003-IJMPC}.
In this case, log-periodicity is nothing but the observable signature of the
developing discrete hierarchy of alternating positive and negative
feedbacks culminating in the final ``rupture,'' which is the end of
the bubble often associated with a crash.
Another explanation is based on the natural hierarchical structure
of human groups \cite{Zhou-Sornette-Hill-Dunbar-2005-PRSB,Hill-Dunbar08},
applied to the network structure of traders \cite{Johansen-Ledoit-Sornette-2000-IJTAF,Johansen-Sornette-2001-IJTAF}.
In this class of models, investors in the stock market form a complex connected
hierarchical network and interact with each other ``locally'' through transfers of
information, leading to what we refer broadly as ``imitation.''
Local interactions propagate spontaneously into global cooperation
leading to herding behaviors, which result in bubbles. These ingredients, together with the fact that prices
reflect the aggregate decisions of investors, is formalized by a
rational expectation model of bubbles with imitation between the
noise traders
\cite{Johansen-Sornette-Ledoit-1999-JR,Johansen-Ledoit-Sornette-2000-IJTAF,Johansen-Sornette-2001-IJTAF}.
The main consequence of these models is that the dynamics may evolve
towards a critical point at a critical time
$t_c$ corresponding to the most probable end of the bubble. We refer to the book by
Sornette \cite{Sornette-2003} for a general introduction, a
synthesis of the models and examples of empirical tests and applications.

Mathematically, in its simplest version, the LPPL model is represented by the equation
(\ref{Eq:LPPL}) giving
the anticipated expected trajectory of the log-price $I(t) \equiv {\rm E}[\ln p(t)]$
of a given asset as a function of time,
expressed in terms of the distance $\tau=t_c-t$ to the
critical time $t_c$ for bubbles (respectively for antibubbles):
\begin{equation}
I\left( t\right) = A + B \tau ^{m} +C \tau^m \cos\left[ \omega \ln
\left( \tau \right) - \phi \right]~. \label{Eq:LPPL}
\end{equation}
$A$ is the expected log-price at $t_c$ (since $\tau =0$ at $t=t_c$, all the
other terms are vanishing at $t=t_c$), $B$ (respectively) controls
the amplitude of the power law acceleration (respectively the
log-periodic component) of the log-price. The exponent $m$ encodes
the structural shape of the acceleration. It is usually found
between $0$ and $1$, which ensures a finite price at $t_c$ together
with an asymptotic infinite rate of change close to $t_c$. The
parameter $\omega$ is the log-periodic angular frequency of the
log-periodic oscillations. It should be stressed that $\omega$ is
not the inverse of a time scale, but rather it is proportional to
the inverse of the logarithm of a scale factor $\lambda$, where
$\lambda$ is roughly speaking the ratio of the distances between
successive peaks of the log-periodic oscillations. Finally, the
phase $\phi$ contains two ingredients: the information on the
mechanism of interactions between investors and a rescaling of time
\cite{Johansen-2003-QF,Sornette-Zhou-2003-QF}.

In a nutshell, the LPPL model  (\ref{Eq:LPPL}) describes two
phenomena: (i) a faster-than-exponential growth (for $0<m<1$ and
$B<0$) of the expected log-price culminating in a singularity of its
slope at the critical time $t_c$ and (ii) an accelerating sequence
of local rallies and corrections decorating the overall power law
acceleration (the ``log-periodicity''). The literature cited above
contains several derivations of this equation (\ref{Eq:LPPL}) and we
refer the reader to them. Basically, the power law part $A + B \tau
^{m}$ embodies the effect of collective imitation leading to global
herding and a critical behavior characterized mathematically by the
singularity at $t_c$ (in a way reminiscent to phase transitions in
physics, but here in the time domain). The log-periodic component
reflects at least two possible effects as mentioned above: (1) an
inherent discrete hierarchical structure in the social network of
investors and (2) a nonlinear mean reversal behavior of fundamental
investing styles. See Ref.~\cite{Sornette-1998-PR} for a general
review on the symmetry of ``discrete scale invariance'' at the basis
of log-periodicity.

The organization of the paper is as follows. The next section
\ref{s1:Class} tests for the existence of faster-than-exponential
acceleration in the price of the 45 stocks used in this study (their ticker code
used in the South African stock exchange and the
corresponding numbering we use are given in Table \ref{TB:StockNames}). It
also presents preliminary screening tests on the existence of
log-periodicity. Combining these evidences, section \ref{s1:Class}
finally qualifies 5 assets out of the 45 as exhibiting a significant
bubble regime. Then, section \ref{jgnroefnqv} presents a detailed
analysis of the log-periodic characteristics of these 5 assets,
using four different techniques. The estimations of the angular
log-frequency $\omega$'s are found consistent and robust across the
four methods. Section \ref{jgnbaa} presents a sensitivity analysis
of the critical time $t_c$ and of the angular log-frequency $\omega$
by varying both the starting time as well as the ending time of the
interval over which the fits are performed. Section 5 summarizes and
concludes with respect to the quality of the forecasts using $t_c$
as the most probable time of the corrections associated with the end
of the bubbles, using the insight obtained from waiting an
additional 6 months since the end of the study in May 2006.

\begin{table}
\begin{center}
\caption{The ticker codes and corresponding numbering of 45 South African
stocks (http://www.jse.co.za/) analyzed in the present work. }\label{TB:StockNames}
\medskip
\begin{tabular}{cccccccccccccccccccccccccccccccccccccccccccccccccccccc}
  \hline\hline
 Num  &    1  &    2  &    3  &    4  &    5  &    6  &    7  &    8  &    9  \\
 Code &  J203 &  J210 &  J257 &  J580 &   ABL &   AGL &   AMS &   ANG &   APN \\
 Num  &   10  &   11  &   12  &   13  &   14  &   15  &   16  &   17  &   18  \\
 Code &   ASA &   BAW &   BIL &   BVT &   ECO &   FSR &   GFI &   HAR &   IMP \\
 Num  &   19  &   20  &   21  &   22  &   23  &   24  &   25  &   26  &   27  \\
 Code &   INL &   INP &   IPL &   JDG &   KMB &   LBT &   LGL &   MLA &   MTN \\
 Num  &   28  &   29  &   30  &   31  &   32  &   33  &   34  &   35  &   36  \\
 Code &   NED &   NPN &   NTC &   OML &   PIK &   PPC &   RCH &   REM &   RMH \\
 Num  &   37  &   38  &   39  &   40  &   41  &   42  &   43  &   44  &   45  \\
 Code &   SAB &   SAP &   SBK &   SHF &   SLM &   SOL &   TBS &   TKG &   WHL \\
\hline\hline
\end{tabular}
\end{center}
\end{table}

\section{Classification of potential speculative behavior among 45 representative
stocks on the Johannesburg Stock Exchange} \label{s1:Class}

One of the most advanced and productive economies in Africa, South
Africa is characterized by a developed first world economic
infrastructure and an emerging market economy. Its financial market
is organized by the Johannesburg Stock Exchange (JSE)
(http://www.jse.co.za/), which is the only securities exchange in
South Africa. It was officially established on 8 November, 1887, and
has now more than 400 listed companies \cite{VanZyl-2006}.

Our study is performed on a subset of 45 companies among the largest
companies listed on the JSE, whose ticker codes are given in table \ref{TB:StockNames}
(for detailed information, see the extended version only available
online at http://arXiv.org/abs/physics/0701171). The selection of
these 45 companies is representative of the diversification
performed by major investment banks and brokerage houses in South
Africa, as communicated to us by one of the major brokerage house in
South Africa. The daily price series run from the first trading day
of January 2003 till May 2006.

\subsection{Is there an acceleration?}
\label{s2:Class:Acceleration}

Our first goal is to perform a robust analysis of the the 45
financial time series, to identify those potential candidates for a
bubble behavior. There is a large literature on the empirical issue
of identifying financial bubbles (see
\cite{Hamilton-Whiteman-1985-JMonE,Evans-1986-AER,Meese-1986-JPE,Woo-1987-JMCB,Camerer-1989-JES,Evans-1991-AER,Adam-Szafarz-1992-OX,Sornette-2003,Gurkaynak-2005,Johansen-Sornette-2006-BER}
and references therein). A key problem is that bubbles are usually
defined as exponentially explosive growth phases, but how can one
then distinguish them from the growth of a fundamental valuation
process which is also generically expected to follow an exponential
growth path? We address this problem by defining a bubble as a
faster-than-exponential accelerating price
\cite{Sornette-2003,Johansen-Sornette-2006-BER}, which we refer to
as ``super-exponential.'' Being faster than exponential, i.e., the
growth rate is itself growing, it is necessarily unsustainable if we
assume a standard geometric growth for the underlying economy. A
super-exponential path can only be transient, reflecting various
positive feedback processes that lead to reinforced growth.

Our goal is thus to contrast
a standard exponential growth from a super-exponential growth.
The former is characterized by the logarithm
of the price of a given asset being linear in time as
\begin{equation}
\label{Eq:Class:Linear}
    \ln[p(t)] = a + b t + \epsilon_1~,
\end{equation}
where $\epsilon_1$ is a short-hand notation for a random walk
component.
The parameter $b$ is then the rate of return for continuous
compounding of the interests. Expression (\ref{Eq:Class:Linear})
is nothing but the integrated form of the standard geometrical
random walk model with drift.
The arguably simplest extension (\ref{Eq:Class:Linear}) which
gives super-exponential growth is obtained as
\begin{equation}
\label{Eq:Class:Quadratic}
    \ln[p(t)] = a + b t + c t^2 + \epsilon_2~.
\end{equation}

\begin{figure}[htb]
\centering
\includegraphics[width=9cm]{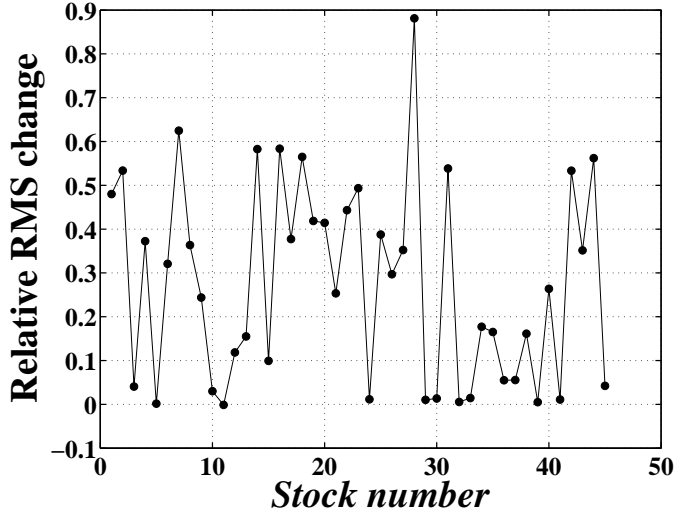}
\caption{Relative difference of the improvement of the fits resulting from the additional
quadratic term in model (\ref{Eq:Class:Quadratic}) in the left-hand-side
of (\ref{clondakkfq}), as defined by (\ref{Eq:LQ:D}).}
\label{Fig:SouthAfrica:LQ}
\end{figure}

The null hypothesis is $c=0$. If it is rejected with $c>0$ at a
sufficiently large significance level, we would conclude that the
price process is growing super-exponentially, qualifying a bubble
regime. Since the residual is not Gaussian and not stationary, we cannot use directly
the Wilks' likelihood ratio test or t-test of significance for the
coefficient $c$. We rather use an approach in the spirit of the
Akaike Information Criterion (AIC), which amounts to test the two
models by comparing $[RMS_1(i)]^2$ for model (\ref{Eq:Class:Linear})
with $[RMS_2(i)]^2 + 2 \sigma^2$ for model
(\ref{Eq:Class:Quadratic}) \cite{Akaike-1974-IEEEtac}. The terms
$RMS_1(i)$ and $RMS_2(i)$ are the root-mean-square of the residuals
of the fits to the price time series of asset $i$ with model
(\ref{Eq:Class:Linear}) and model (\ref{Eq:Class:Quadratic})
respectively. More precisely, the calibration of model (\ref{Eq:Class:Linear})
determines the best values of the parameters $a$ and $b$ for each stock $i$
and $RMS_1(i)$ then provides a measure of the average standard deviation
of the residual $\epsilon_1$. Similarly, the calibration of model (\ref{Eq:Class:Quadratic})
determines the best values of the parameters $a$, $b$ and $c$ for each stock $i$
and $RMS_2(i)$ then provides a measure of the average standard deviation
of the residual $\epsilon_2$.  The term $\sigma^2$ is the square of the standard deviation of the
noise of the price process. The term $2 \sigma^2$ is the cost
attributed to model (\ref{Eq:Class:Quadratic}) for having one more
adjustable parameter compared with model (\ref{Eq:Class:Linear}). In
other words, AIC not only rewards goodness of fit, but also includes
a penalty linear in the number of estimated parameters. Then, the
AIC qualifies model (\ref{Eq:Class:Quadratic}) if $[RMS_2(i)]^2 + 2
\sigma^2 < [RMS_1(i)]^2$. This is equivalent to
\be
{[RMS_1(i)]^2 - [RMS_2(i)]^2 \over [RMS_1(i)]^2} > {2 \sigma^2 \over [RMS_1(i)]^2}~.
\label{clondakkfq}
\ee
The ratio in the l.h.s. of inequality (\ref{clondakkfq}) quantifies the relative
goodness of fit of expressions (\ref{Eq:Class:Linear}) and (\ref{Eq:Class:Quadratic}).
Since in practice, $\sigma^2$ is not known to
us, we calculate the relative difference of the improvement of
the fits resulting from the additional quadratic term in model
(\ref{Eq:Class:Quadratic}) in the left-hand-side of
(\ref{clondakkfq}):
\begin{equation}
\label{Eq:LQ:D}
    D(i) = [RMS_1(i)-RMS_2(i)]/RMS_1(i) ~,
\end{equation}
for each asset. Then, the larger is the value of $D(i)$, the more
probable is the rejection of the null hypothesis and the relevance
of the quadratic term qualifying a bubble regime. Here, we put
aside the subtle issues \cite{GrangerNewbold74,Phillips-1986-JEm} stemming from possible spurious regressions
resulting from the non-stationarity of the residuals in models
(\ref{Eq:Class:Linear}) and (\ref{Eq:Class:Quadratic}), which
are left for another work.
Fig.~\ref{Fig:SouthAfrica:LQ} gives the 45 values of $D(i)$ for all
the 45 stocks.

Fig.~\ref{Fig:SouthAfrica:LQ} suggests the existence of two clusters, characterized
by values of the ``Relative RMS change'' above or below $\approx 25\%$.
We thus disqualify stocks as not being in a bubble regime if they obey
at least one of the following criteria:
\begin{enumerate}
\item $D(i) \leqslant 25\%$,
\item the quadratic term is positive ($c>0$),
\item the overall price has been increasing from the beginning to the end of the
period.
\end{enumerate}
This leads us to reject the following 27 stocks as not in a bubble
regime from Jan. 2003 till May 2006: 3, 5, 9, 10, 11, 12, 13, 14,
15, 21, 22, 24, 26, 27, 29, 30, 32, 33, 34, 35, 36, 37, 38, 39, 41,
44, 45. This selection is in line with visual inspection for all
except stock 30, for which the analysis comparing
(\ref{Eq:Class:Linear}) and (\ref{Eq:Class:Quadratic}) has not
enough power. Indeed, in this case of stock 30, the weak improvement
of model (\ref{Eq:Class:Quadratic}) may be attributed to the
combination of a large drawup followed by a large drawdown from
end-of-2003 to middle-of-2004 followed by an upward acceleration. We
thus keep this stock 30 for further analysis.

\subsection{Log-periodic oscillations}
\label{s2:Class:LP}

As we just mentioned, the analysis comparing (\ref{Eq:Class:Linear}) and
(\ref{Eq:Class:Quadratic}) is not claimed to have universal absolute power:
it may have not enough power to reject a stock when it should have been rejected
(error of type I or false positive), or it may reject a stock that should be
kept in the bubble class (error of type II or false negative) as we argue is
perhaps the case of stock 30.

It is thus useful to examine the results of another test, based on a
specification which is well-adapted to test for the presence of
super-exponential behavior, in the possible presence of oscillatory
intermittent fluctuations. This specification uses the so-called
log-periodic formula (\ref{Eq:LPPL}) introduced first in
\cite{Sornette-Johansen-Bouchaud-1996-JPIF,Feigenbaum-Freund-1996-IJMPB}
and expanded upon in several subsequent papers (see for instance
\cite{Sornette-2003,Sornette-2003-PR}. We have fitted the 45 stock
price series with the log-periodic power-law formula (\ref{Eq:LPPL})
and obtained the key parameters $t_c$, $m$, and $\omega$.

In the above characterization in terms of (\ref{Eq:Class:Linear})
versus (\ref{Eq:Class:Quadratic}), the non-accelerating log-price
was characterized essentially by the absence of a significant
positive quadratic term $c t^2$ in (\ref{Eq:Class:Quadratic}). In
the present log-periodic power law analysis, a non-accelerating
log-price should be qualified by an exponent $m$ close to $1$, while
an accelerating log-price corresponds to $m$ small or even negative.
We find that this correspondence holds to a large degree, as seen
from the following list: (3, 0.86), (5, 0.96), (9, 0.52), (10,
0.86), (11, 0.84), (12, 0.58), (13, 0.84), (14, 2), (15, 0.71), (21,
0.27), (22, 1.6), (24, 0.5), (26, 1.6), (27, 1.8), (29, 0.99), (32,
1), (33, 0.63), (34, 0.41), (35, 0.84), (36, 1.4), (37, 1.3),
(38,-2), (39, 0.99), (41, 0.64), (44, 1.4), (45, 0.55). Here, the
first number in each parenthesis stands for the stock number and the
second one is the value of the exponent $m$. For the stock prices
which have been qualified as non-accelerating by the method of the
previous section, we find values of $m$ typically larger than $0.8$, confirming the
classification. The following cases can be considered marginal:
\begin{enumerate}
\item (9, 0.52), which has a relative RMS change $D(9) \approx 0.24$
as defined by (\ref{Eq:LQ:D}),
\item (12, 0.58), which has a relative RMS change $D(12) \approx 0.12$,
\item (21, 0.27), which has a
relative RMS change $D(21) \approx 0.24$,
\item  (24, 0.5), which has a relative RMS change $D(24)$ of almost zero,
while presenting an intermediate $m$ value,
\item (34, 0.41), which has a relative RMS change
$D(34)=0.18$ and the acceleration is more an abrupt change of slope
or kink occuring in the first
quarter of 2005, and
\item (38, -2), which exhibits a log-price which has decreased over most
of the time period and which should be rejected.
\end{enumerate}

\begin{figure}[htb]
\centering
\includegraphics[width=9cm]{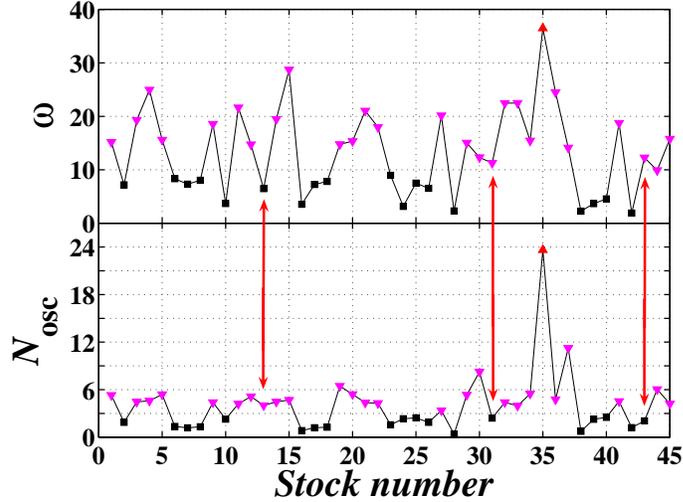}
\caption{(Color online). Classification of the 45 analyzed South
African stocks with respect to the properties of their log-periodic
oscillations. Upper panel: The angular logarithmic frequency
$\omega$ for each of the 45 stock indexed with its number $i$.
Stocks having $\omega\leqslant9$ are marked with solid squares,
stocks with $9<\omega\leqslant30$ are indicated by solid downward
triangles, and stocks with $\omega\geqslant30$ are shown with a
solid upward triangle. Lower panel: The number of oscillations
$N_{\rm{osc}}$ over which log-periodicity is found, as a function of
the stock number $i$. Stocks having $N_{\rm{osc}}\leqslant3$ are
marked with solid squares, those with $3<N_{\rm{osc}}\leqslant10$
are indicated by solid downward triangles, and the stocks with
$N_{\rm{osc}}\geqslant10$ are shown with a solid upward triangle.}
\label{Fig:SouthAfrica:LP}
\end{figure}

In addition to the value of the exponent $m$, the presence of
log-periodicity has been argued as a signature of a bubble regime
\cite{Sornette-Johansen-Bouchaud-1996-JPIF,Feigenbaum-Freund-1996-IJMPB}.
The upper panel of Fig.~\ref{Fig:SouthAfrica:LP} plots the value of
the fitted log-periodic angular frequency $\omega$ for the 45
analyzed stocks. Previous studies on a large number of bubbles
mostly on financial indices, bonds and currencies of many developed
and emerging countries have shown that the distribution of $\omega$
exhibits several peaks, the prominent one being on the so-called
fundamental log-periodic angular frequency $\omega_1 \approx 6.4 \pm
1.5$ \cite{Johansen-2003-PA}, with other peaks on its harmonics
$\omega_n = n \omega_1$. The importance of the high-order harmonics
is expected to decrease exponentially
\cite{Gluzman-Sornette-2002-PRE}, but large amplitudes for the
second-order and third-order harmonics $\omega_2$ and $\omega_3$
have been observed to be sometimes very significant
\cite{Zhou-Sornette-2002-PD,Zhou-Sornette-2003c-PA}. It thus seems
difficult to use here a filter based solely on $\omega$, in
particular for individual stocks which are necessarily more noisy
than aggregate indices.

Ref.~\cite{Zhou-Sornette-2002-IJMPC} has investigated all kinds of
scaling series (Bm, fBm, Levy model) to establish the statistical
significance level of periodicity and log-periodicity in noisy time
series. Ref.~\cite{Johansen-Ledoit-Sornette-2000-IJTAF} also reports
extensive tests with GARCH processes to assess the statistical
significance level of log-periodicity. Our systematic statistical
analysis of the significance level of periodic (as well as
log-periodic) signals performed in
Ref.~\cite{Zhou-Sornette-2002-IJMPC} shows that the hypothesis, that
the observed log-periodicity results from noise, can be rejected at
a confidence level higher than $95\%$ as soon as the the number
$N_{\rm{osc}}$ of oscillations is  $3$ or more, for most types of
noises. For instance, it has been shown that multiplicative noise on
a power law accelerating function leads naturally to stochastic
log-periodic oscillations with a most probable number equal to
$N_{\rm{osc}} \approx 1.5$
\cite{Huang-Johansen-Lee-Saleur-Sornette-2000-JGR}. The tests of
statistical significance performed in
\cite{Zhou-Sornette-2002-IJMPC} have shown that, for most types of
noise, three oscillations are in general sufficient to qualify a
genuine oscillatory component. Rigorously, by ``genuine'', we mean
that the probability that the observed oscillatory behavior results
from some random noise configuration rather than from some
informative signal is below a standard significance level, typically
$1-p$ with $p=95\%$ or $99\%$. We thus complement the determination
of the angular log-frequency $\omega$ by the measure $N_{\rm{osc}}$
of log-periodic oscillations in each of the 45 stock prices. Given a
LPPL fit and the obtained calibrated parameters $t_c$ and $\omega$,
the number of oscillations is determined by
\begin{equation}\label{Eq:Class:Nosc}
    N_{\rm{osc}} = \frac{\omega}{2\pi}\ln\left|\frac{t_c-t_{\rm{first}}}{t_c-t_{\rm{last}}}\right|~,
\end{equation}
where $[t_{\rm{first}}, t_{\rm{last}}]$ is the interval over which
the LPPL fitting is performed. The variable $N_{\rm{osc}}$ for each
of the 45 stocks is shown in the lower panel of
Fig.~\ref{Fig:SouthAfrica:LP}.

One can observe that the two measures $\omega$ and $N_{\rm{osc}}$
provide consistent results. As shown by the correspondence between
the symbols in the upper and lower panel of
Fig.~\ref{Fig:SouthAfrica:LP} for most of the stocks, large
$\omega$'s are associated with a large number of oscillations and
vice-versa. There are three exceptions to this observation: $i=13$
(BVT), $31$ (OML), and $43$ (TBS), outlined in the figure by the
arrows. Given the large range of observed $\omega$'s, we take the
number $N_{\rm{osc}}$ of LPPL oscillations as the more robust
indicator of genuine log-periodicity. This leads to qualify the
following stocks as exhibiting a significant log-periodicity: 1, 3,
4, 5, 9, 11, 12, 13, 14, 15, 19, 20, 21, 22, 27, 29, 30, 32, 33, 34,
36, 37, 41, 44, and 45.

Combining this with the results shown in
Sec.~\ref{s2:Class:Acceleration}, we are left with five stocks
which qualify as being in a bubble regime, defined as a significant
super-exponential acceleration with the presence of significant
log-periodicity: 1 (J203: an index), 4 (J580: an index), 19 (INL), 20 (INP), and 30 (NTC).

In the remaining of this paper, we analyze these five stocks extensively.

\section{Analyzing the log-periodic structure of stocks 1 (J203),
4 (J580), 19 (INL), 20 (INP), and 30 (NTC)  \label{jgnroefnqv}}

\subsection{First-order LPPL model}

\begin{figure}[h]
\begin{center}
\includegraphics[width=7cm]{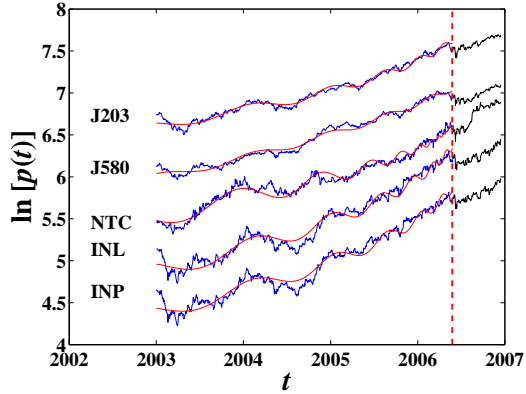}
\caption{(Color online). Fits of the first-order LPPL model
(\ref{Eq:LPPL}) to the logarithm of the five South African stock
prices 1 (J203),  4 (J580), 19 (INL), 20 (INP), and 30 (NTC)
from January 2003 to May 2006. The vertical dashed line
indicated the end of the fitting interval: May 2006.  The rough
curves are the historical raw data and extends till Dec. 2006
and the smooth curves are the
LPPL fits. The data from June to Dec. 2006 to the right
of the vertical dashed line has not been used in the fits.
The values of the fit
parameters are listed in Table \ref{TB:LPPL}. The curves have been
shifted vertically for clarity.} \label{Fig:LPPL}
\end{center}
\end{figure}

The fits of the logarithms of the prices of the five stocks 1
(J203),  4 (J580), 19 (INL), 20 (INP), and 30 (NTC) by the
log-periodic power law model (\ref{Eq:LPPL}) are shown in
Fig.~\ref{Fig:LPPL} and the parameters are given in Table
\ref{TB:LPPL}. The small value of the exponent $m$ (except for NTC)
confirms a clear super-exponential acceleration. The log-periodic
angular frequencies are found to be close to twice (J203, INL, INP,
NTC) or four times (J580) the value $\omega_1 \approx 6.4 \pm 1.5$
of the fundamental log-periodic angular frequency found in many
previous studies \cite{Johansen-2003-PA,Johansen-Sornette-2006-BER}.
The LPPL fits suggested at the time of the fits (end of May 2006)
that the bubbles would end either immediately (NTC) or during the
second part of the year. We discuss this prediction below.

\begin{table}[htb]
\begin{center}
\caption{\label{TB:LPPL} Parameters of the first-order
log-periodic power law model providing the best fits to the
the five South African stocks 1 (J203),  4 (J580), 19 (INL), 20 (INP), and 30 (NTC)
shown in Fig.~\ref{Fig:LPPL}}
\medskip
\begin{tabular}{ccccccccccc}
  \hline\hline
  Stock & $t_c$ & $m$ & $\omega$ & $\phi$ & $A$ & $B$ & $10^3C$ & $\chi$
  \\\hline
J203 (1) & 2006/11/01 & 0.27 & 15.3 & 5.41 & 11.3 & -0.336 &   8.73 &  0.036\\
J580 (4) & 2007/11/12 & 0.30 & 24.1 & 4.20 & 12.4 & -0.388 &   4.84 &  0.040\\
INL (19) & 2006/08/07 & 0.37 & 14.1 & 1.89 &  11.3 & -0.159 &   8.08 &  0.072\\
INP (20) & 2006/09/19 & 0.30 & 14.0 & 4.81 &  11.9 & -0.307 & -11.27 &  0.074\\
NTC (30) & 2006/05/17 & 0.60 & 12.8 & 2.21 &   6.9 & -0.014 &  -1.91 &  0.054\\
\hline\hline
\end{tabular}
\end{center}
\end{table}

The detection of log-periodic oscillations, if any, is conveniently
performed by removing the global trend of the price of a given
stock. One way is to subtract the power law trend from the price and
then to analyze the wobbles of the obtained residuals $s(t)$ by an
adequate spectral analysis \cite{Johansen-Sornette-Ledoit-1999-JR}.
We shall also use a non-parametric approach called the
$(H,q)$-analysis \cite{Zhou-Sornette-2002-PRE}. Since
log-periodicity corresponds to regular oscillations in the variable
$\ln (t-t_c)$, we use a Lomb periodogram analysis which is
well-adapted to the uneven sampling of the variable $\ln (t-t_c)$
\cite{Press-Teukolsky-Vetterling-Flannery-1996}. The Lomb periodogram analysis
is nothing but a spectral analysis based on a least squares fit of sinusoids to data samples,
similar to Fourier analysis. The Lomb analysis
also allows us to assess the statistical significance level of the
extracted log-periodicity
\cite{Press-Teukolsky-Vetterling-Flannery-1996,Zhou-Sornette-2002-IJMPC,Bothmer-Meister-2003-PA}.

\subsection{Parametric detrending approach \label{s3:paradet}}

\begin{figure}[h]
\begin{center}
\includegraphics[width=7cm]{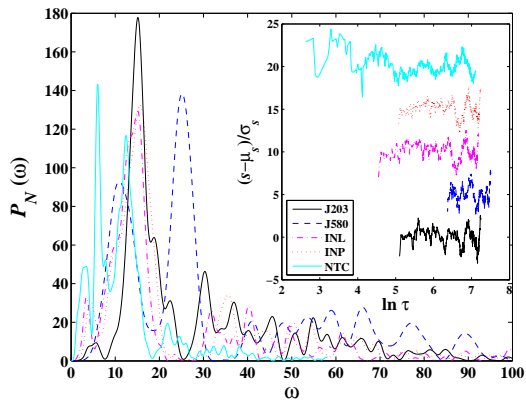}
\caption{(Color online). Lomb periodogram analysis of the five
standardized residuals $[s(t)-\mu_s]/\sigma_s$ shown in the inset,
where $s(t)$ is defined by (\ref{Eq:Res}), considered as a function
of $\ln(\tau)= \ln (t-t_c)$. The curves in the inset have been
translated vertically for clarity.} \label{Fig:Lomb}
\end{center}
\end{figure}

Following
\cite{Johansen-Sornette-Ledoit-1999-JR,Johansen-Ledoit-Sornette-2000-IJTAF},
the first method of analysis of the log-periodicity consists in
removing the power law trend and then testing for a possible pure
log-periodicity without acceleration.  We construct the residual
$s(t)$ in the following way
\begin{equation}
s(t) = [I(t)-A]/(t-t_c)^{m}~. \label{Eq:Res}
\end{equation}
This residual $s(t)$ has a nonzero mean $\mu_s$ associated with the
coefficient $B$ and a given variance $\sigma_s^2$ (in general
different from $1$). The inset of Fig.~\ref{Fig:Lomb} plots the
standardized residuals $[s(t)-\mu_s]/\sigma_s$ as a function of $\ln
\tau = \ln (t-t_c)$ for the five stocks 1 (J203), 4 (J580), 19
(INL), 20 (INP), and 30 (NTC). The five Lomb periodograms of these
five residuals are presented in Fig.~\ref{Fig:Lomb}. Since
$P_N(\omega)$ is a normalized Lomb power, $s(t)$ and
$[s(t)-\mu_s]/\sigma_s$ have identical Lomb periodogram. The angular
log-periodic frequencies associated with the highest Lomb peaks are
respectively 15.1, 25.1, 15.0, 15.6, and 12.5, which are close to
the values obtained with the parametric fit with expression
(\ref{Eq:LPPL}) listed in Table \ref{TB:LPPL}. Their statistical
significance is extremely high, much larger than 99\%, for all
possible noise processes, according to the benchmarks developed in
\cite{Zhou-Sornette-2002-IJMPC}.

\subsection{$(H,q)$-analysis \label{s3:HqA}}

We have also performed a generalized $q$-analysis, called
$(H,q)$-analysis
\cite{Zhou-Sornette-2002-PRE,Zhou-Sornette-2003-IJMPC}, on each of
the logarithm of the five price trajectories. The $(H,q)$-analysis
is a non-parametric method for characterizing self-similar
functions, which generalizes the $q$-analysis
\cite{Erzan-1997-PLA,Erzan-Eckmann-1997-PRL}. The later is a natural
tool for the description of discretely scale invariant fractals. The
$(H,q)$-derivative of a function $I(\tau)$ is defined as
\begin{equation}
D_q^H I(\tau) \stackrel{\triangle}{=} \frac
{I(\tau)-I(q\tau)}{[(1-q)\tau]^H}~. \label{Eq:HqD}
\end{equation}
The special case $H=1$ recovers the standard $q$-derivative, which
itself reduces to the standard derivative in the limit $q \to 1^-$.
There is no loss of generality by constraining $q$ in the open
interval $(0,1)$ \cite{Zhou-Sornette-2002-PRE}. The advantage of the
$(H,q)$-analysis is that there is no need for detrending, as trends
are automatically accounted for by the finite difference and the
normalization by the denominator upon a systematic sweeping of the
parameter $H$.

We apply the $(H,q)$-analysis to $I(x) = \ln p(t)$, where $p(t)$ is
the price of each of the five stocks 1 (J203), 4 (J580), 19 (INL),
20 (INP), and 30 (NTC), as an independent powerful test of
log-periodicity. The independent variable is taken to be $\ln \tau$
\cite{Zhou-Sornette-2002-PRE}. The same method has been applied to
test for log-periodicity in stock market bubbles and antibubbles
\cite{Sornette-Zhou-2002-QF,Zhou-Sornette-2003-IJMPC}, in the USA
foreign capital inflow bubble ending in early 2001
\cite{Sornette-Zhou-2004-PA}, in the UK real estate bubble
\cite{Zhou-Sornette-2003a-PA}, in the Chinese stock market
antibubble \cite{Zhou-Sornette-2004a-PA}, and in the US treasury
bond yield antibubble \cite{Zhou-Sornette-2004b-PA}.

\begin{figure}[h]
\begin{center}
\includegraphics[width=7cm]{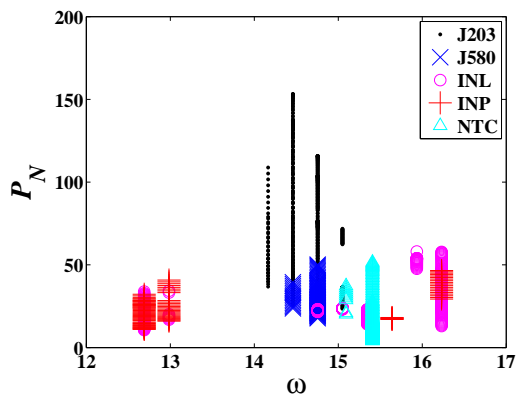}
\end{center}
\caption{(Color online). Numerically constructed discrete binned
bivariate distribution of pairs $(\omega,P_N)$, defined as the
angular log-frequencies and corresponding highest Lomb peaks of the
$(H,q)$-derivative of $\ln[p(t)]$, when scanning a $100 \times 50$
rectangular grid in the $(H,q)$ plane, ($H$ from $-0.99$ to $+0.99$
with increment $0.02$ and $q$ from $0.01$ to $0.99$ with increment
$0.02$). Each marker in the figure corresponds to the highest Lomb
peak and its associated angular log-frequency in the Lomb
periodogram of the $(H,q)$-derivative of $\ln[p(t)]$ for a given
pair $(H,q)$.} \label{Fig:HqA}
\end{figure}

We scan a $100 \times 50$ rectangular grid in the $(H,q)$ plane,
with $H = -0.99:0.02:0.99$ (from $-0.99$ to $+0.99$ with increment
$0.02$) and $q = 0.01:0.02:0.99$ (from $0.01$ to $0.99$ with
increment $0.02$). For each pair of $(H,q)$ values, we calculate the
$(H,q)$-derivative (\ref{Eq:HqD}), on which we perform a Lomb
analysis. The highest Lomb peak of the resulting periodogram has
height $P_N$ and abscissa $\omega$, both $P_N$ and $\omega$ being
functions of $H$ and $q$. Figure~\ref{Fig:HqA} shows the numerically
constructed discrete binned bivariate distribution of pairs
$(\omega,P_N)$ for $12\leqslant\omega\leqslant17$. For values
$\omega \leqslant 12$, only one cluster with small values
$\omega<4.75$ can be observed: such small values of $\omega$ are
associated with only one or at most two oscillations and correspond
most probably to the most probable oscillatory structure of
multiplicative noise of power law function
\cite{Huang-Johansen-Lee-Saleur-Sornette-2000-JGR}. Another possible
origin of this cluster at $\omega<4.75$ is a residual global trend
which has not been completely accounted for by the
$(H,q)$-derivative
\cite{Zhou-Sornette-2002-PRE,Zhou-Sornette-2003-IJMPC}.

Most of the Lomb periodograms associated with the points drawn in
Fig.~\ref{Fig:HqA} have a shape similar to that shown in
Fig.~\ref{Fig:Lomb}. Specifically, the average angular
log-frequencies determined from this $(H,q)$ analysis are
$14.7\pm0.3$ for stock ``J203'', $14.7\pm0.1$
for stock ``J580'', $15.0\pm1.3$ for stock ``INL'', $13.7\pm1.5$ for
stock ``INP'', and $15.4\pm0.1$ for stock ``NTC''. In summary, this
$(H,q)$-analysis provides even stronger evidence for the existence
of log-periodicity than the parametric detrending approach of the
previous section \ref{s3:paradet}.

\subsection{Second-order Weierstrass-type LPPL model}

As already pointed out above, the log-periodic angular frequencies
are found to be close to twice (J203, INL, INP, NTC) or four times
(J580) the value $\omega_1 \approx 6.4 \pm 1.5$ of the fundamental
log-periodic angular frequency found in many previous studies
\cite{Johansen-2003-PA,Johansen-Sornette-2006-BER}. This suggests
that the first-order LPPL formula (\ref{Eq:LPPL}) used until now
should be extended to include several harmonics. Indeed, the general
mathematical formulation of discrete scale invariance
\cite{Sornette-1998-PR,Sornette-2003-PR} shows that a log-periodic
function is expected in general to be represented by a systematic
series of log-periodic terms of the form
\cite{Gluzman-Sornette-2002-PRE}
\begin{equation}
I(t) = A + B {\tau}^{m} + {\Re{\left(\sum_{n=1}^N C_n
{\rm{e}}^{i\psi_n}{\tau}^{-s_n}\right)}}~, \label{Eq:Wei}
\end{equation}
where $\tau=t_c-t$ as before. The main advantage of the high-order
Weierstrass-type LPPL models is that they allow to identify the
fundamental log-frequency and its harmonics. This family of models
have been applied to the case study of many bubbles and antibubbles
\cite{Zhou-Sornette-2003c-PA}, such as the UK real estate bubble in
the last decade \cite{Zhou-Sornette-2003a-PA}, the 1975-2001 bubble
in the American foreign assets capital inflow
\cite{Sornette-Zhou-2004-PA}, the Chinese stock market antibubble
since 2001 \cite{Zhou-Sornette-2004a-PA}, and the USA treasury bond
yield antibubble since 2000 \cite{Zhou-Sornette-2004b-PA}.

\begin{figure}[h]
\begin{center}
\includegraphics[width=7cm]{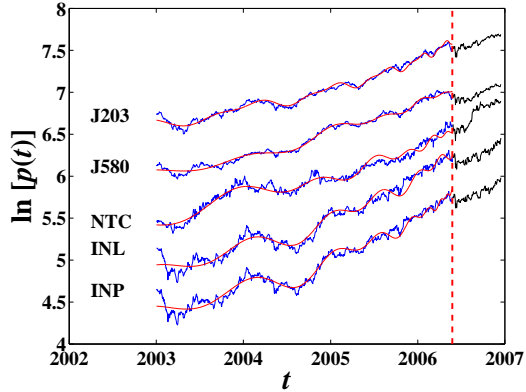}
\caption{(Color online). Best fits of the second-order
Weierstrass-type LPPL formula (\ref{Eq:Wei}) with $N = 2$ to the
logarithm of the five South African stock prices from January 2003
to May 2006. The rough curves are the historical raw
data and the smooth curves are the associated LPPL fits.
The values of the fit parameters are listed in Table
\ref{TB:Wei}. The curves have been shifted vertically for clarity.
}
\label{Fig:Wei}
\end{center}
\end{figure}

We adopt the second-order Weierstrass-type LPPL model (\ref{Eq:Wei})
with $N = 2$ to fit the logarithm of the five South African stock
prices. The time evolution of the five stocks and the corresponding fits are
drawn in Fig.~\ref{Fig:Wei}.

The parameters of the five fits using the second-order
Weierstrass-type LPPL model are presented in Table \ref{TB:Wei}. We
observe that, except from stock J203, the angular log-frequencies
lie in the range $6.5 \leq \omega \leq 8.1$, which is consistent
with previous results on the fundamental log-frequency $\omega_1=6.4
\pm 1.5$ \cite{Johansen-Sornette-2006-BER,Johansen-2003-PA}.
Specifically, this confirms the coexistence of this fundamental
log-frequency together with its harmonics, justifying the
interpretation of the large values reported above with the
first-order formula (\ref{Eq:LPPL}) as corresponding to the
harmonics of $\omega_1$. The fact that the angular log-frequency
$\omega=15.4$ for stock ``J203'' is close to the second-order
harmonic of $\omega_1$ is probably associated with a very strong
amplitude of the second harmonics, which may hide the existence of
$\omega_1$. It is also interesting to notice that the absolute
values of the linear parameters $C_1$ and $C_2$ are comparable. In
three cases $|C_2|>|C_1|$, indicating that the amplitudes of the
second-order harmonic oscillations are considerable, again
consistent with our previous interpretation of the results obtained
above.

Except for NTC, the exponents $m$ are larger in the second-order LPPL fit
than in the first-order case. The critical times $t_c$ predicted to be
the end of the bubbles are quite robust: they are essentially
unchanged for J203 and J580 while they are pushed towards the future
by roughly three months for the three other stocks, when going
from the first-order to the second-order formula.

\begin{table}[htb]
\begin{center}
\caption{Parameters of the fits with the second-order Weierstrass-type
LPPL model (\ref{Eq:Wei}) of the five South African stocks}\label{TB:Wei}
\medskip
\begin{tabular}{ccccccccccc}
  \hline\hline
  Stock & $t_c$ & $m$ & $\omega$ & $\phi_1$ & $\phi_2$ & $A$ & $B$ & $10^3C_1$ & $10^3C_2$ & $\chi$
  \\\hline
J203 & 2006/11/05 & 0.27 & 15.4 & 5.71 &   3.2 &  11.3 & -0.336 &   8.51 &   4.16 &  0.032\\
J580 & 2006/10/27 & 0.68 & 7.5 & 4.14 &   3.3 &  10.2 & -0.010 &   0.53 &   0.47 &  0.029\\
INL  & 2006/11/11 & 0.39 & 8.1 & 4.24 &   2.4 &  11.7 & -0.156 &  -4.11 &  -6.72 &  0.066\\
INP  & 2006/06/22 & 0.61 & 6.5 & 1.58 &   6.2 &  10.7 & -0.021 &   1.39 &   1.87 &  0.062\\
NTC  & 2006/08/26 & 0.51 & 7.2 & 5.20 &   5.3 &   7.2 & -0.036 &   1.98 &   2.44 &  0.050\\
\hline\hline
\end{tabular}
\end{center}
\end{table}

\subsection{Comparison of the different methods\label{s3:cpm}}

Let us now compare the estimated angular log-frequencies of the five stock
bubbles obtained with the different methods presented above:
(i) the fit with the first order LPPL function, (ii) the
parametric detrending approach, (iii) the $(H,q)$-analysis, and (iv) the fit with the
second-order Weierstrass-type function. The obtained
angular log-frequencies are listed in Table \ref{TB:omega}. The
results are self-consistent in the sense that all the bubbles have
the same fundamental angular log-frequency $\omega=7.6\pm1.9$ and
the large values are its higher-order harmonics.

\begin{table}[htb]
\begin{center}
\caption{Comparison of the angular log-frequencies of the five South
African stock bubbles estimated by the four different
methods.}\label{TB:omega}
\medskip
\begin{tabular}{cccccc}
  \hline\hline
  Stock & LPPL & Detrending & $(H,q)$-analysis & Weierstrass 2
  \\\hline
J203 & 15.3 & 15.1 & $14.7\pm0.3$ & $15.4$\\
J580 & 24.1 & 25.1 & $14.7\pm0.1$ & $~~7.5$\\
INL  & 14.1 & 15.0 & $15.0\pm1.3$ & $~~8.1$\\
INP  & 14.0 & 15.6 & $13.7\pm1.5$ & $~~6.5$\\
NTC  & 12.8 & 12.5 & $15.4\pm0.1$ & $~~7.2$\\
\hline\hline
\end{tabular}
\end{center}
\end{table}

For J203, the four methods give essentially the same value $\omega
\approx 15$, which can be interpreted as a very strong second
harmonic $2 \omega_1$ of the fundamental log-periodic angular
log-frequency $\omega_1 \approx 6.4 \pm 1.5$ found in many previous
studies \cite{Johansen-2003-PA,Johansen-Sornette-2006-BER}.

For J580, the second-order LPPL fit correctly identifies
the presence of $\omega_1$, and of course its second-order
harmonics. The $(H,q)$ analysis identifies the second harmonics
$2 \omega_1$ while the two other methods seem to be most
sensitive to the fourth-order harmonics.

For the three other stocks, the second-order LPPL fit correctly identifies
the presence of $\omega_1$, while the three other methods
extract its second harmonics $2 \omega_1$ as being the dominant contribution,
in agreement with the amplitude $C_2 > C_1$ determined in their
second-order LPPL fit.

We conclude that genuine log-periodicity exists the price trajectories of these five stocks
with high statistical significance, and
that the extracted values of the angular log-frequencies are
compatible with previous results on other bubbles
\cite{Johansen-2003-PA,Johansen-Sornette-2006-BER}. The main novelty
lies in the importance of the second-order and fourth-order
harmonics, which is larger than usual.

\section{Sensitivity analysis of the critical times $t_c$  \label{jgnbaa}}

The determination of the critical time $t_c$ is particularly
important since it gives the estimated termination time of the
bubbles, which can occur approximately two times out of three in the
form of a significant correction or a crash. It is noteworthy to
stress that a bubble does not end necessarily with a crash as there
is a finite probability for a bubble to end with a transition to
another regime such as slow deflation or correction
\cite{Johansen-Sornette-Ledoit-1999-JR,Johansen-Ledoit-Sornette-2000-IJTAF}.
The critical time $t_c$ is thus the end of the LPPL bubble and the
time at which the crash is most probable, if it ever occurs.

Given its high significance, we have performed a sensitivity
analysis of $t_c$ for the five stocks with respect to different
starting time $t_{\rm{first}}$ and ending time $t_{\rm{last}}$ of
the price time series used in the fitting procedure, following
previous works
\cite{Zhou-Sornette-2003a-PA,Zhou-Sornette-2005-PA,Zhou-Sornette-2006a-PA}.

\subsection{The impact of $t_{\rm{first}}$}

\begin{figure}[htp]
\begin{center}
\includegraphics[width=6.5cm]{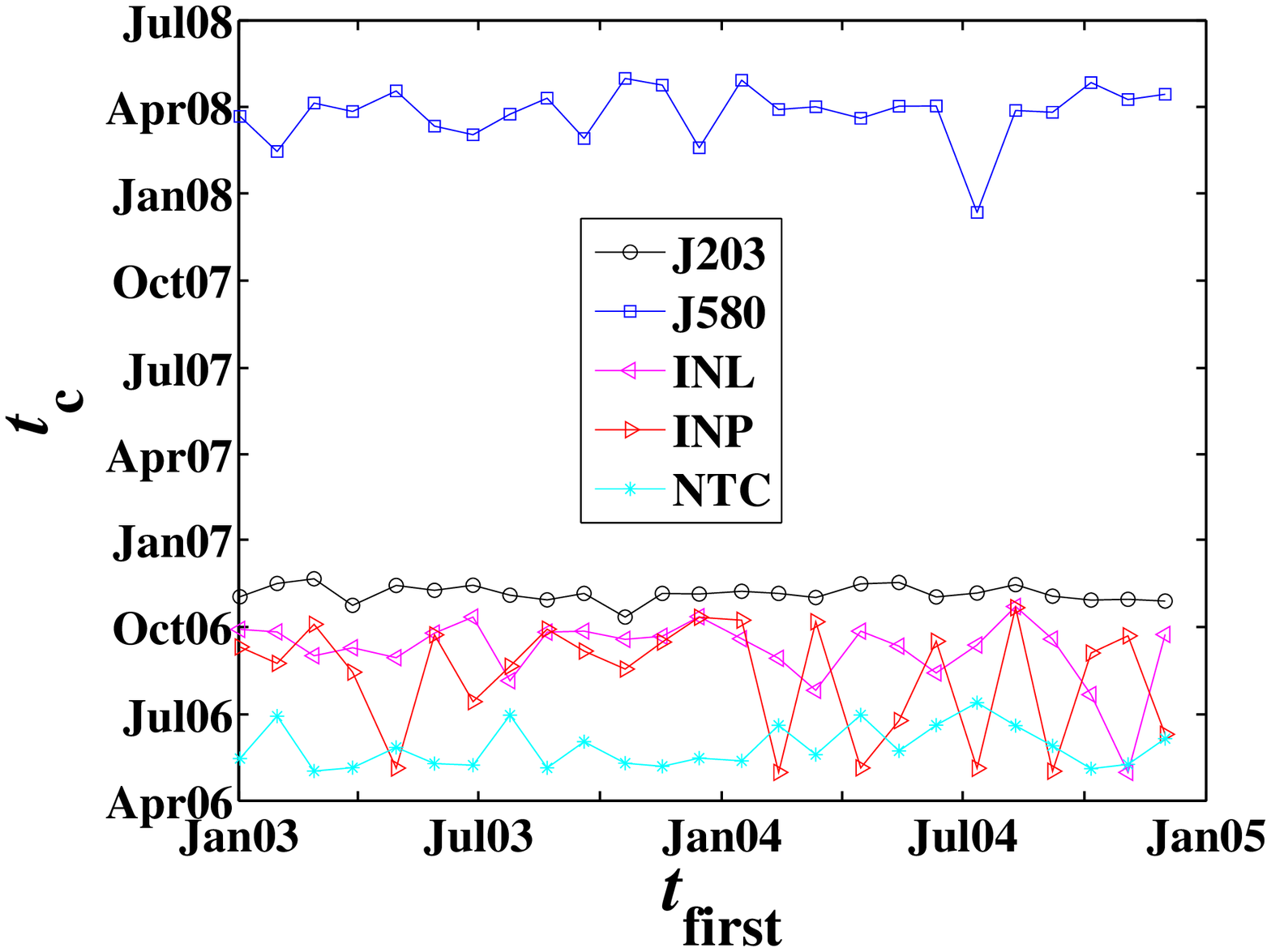}
\includegraphics[width=6.5cm]{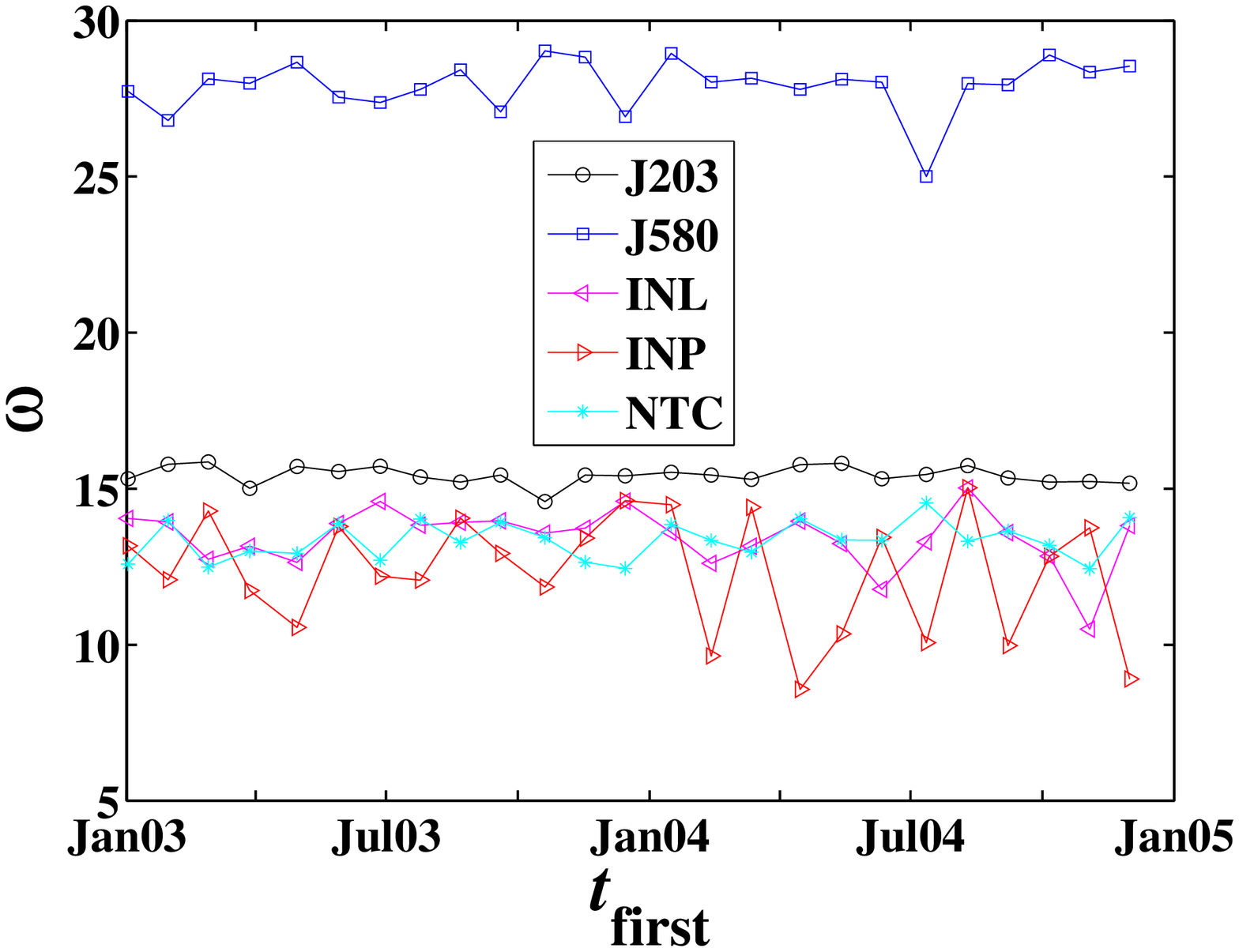}
\caption{(Color online). Sensitivity analysis of the estimated
critical time $t_c$ and the angular log-frequency $\omega$ for the
five stocks obtained by varying the last point $t_{\rm{first}}$ of
the time series up to which the fits using formula (\ref{Eq:LPPL})
are performed.} \label{Fig:Tfirst}
\end{center}
\end{figure}

\begin{figure}[htp]
\begin{center}
\includegraphics[width=6.5cm]{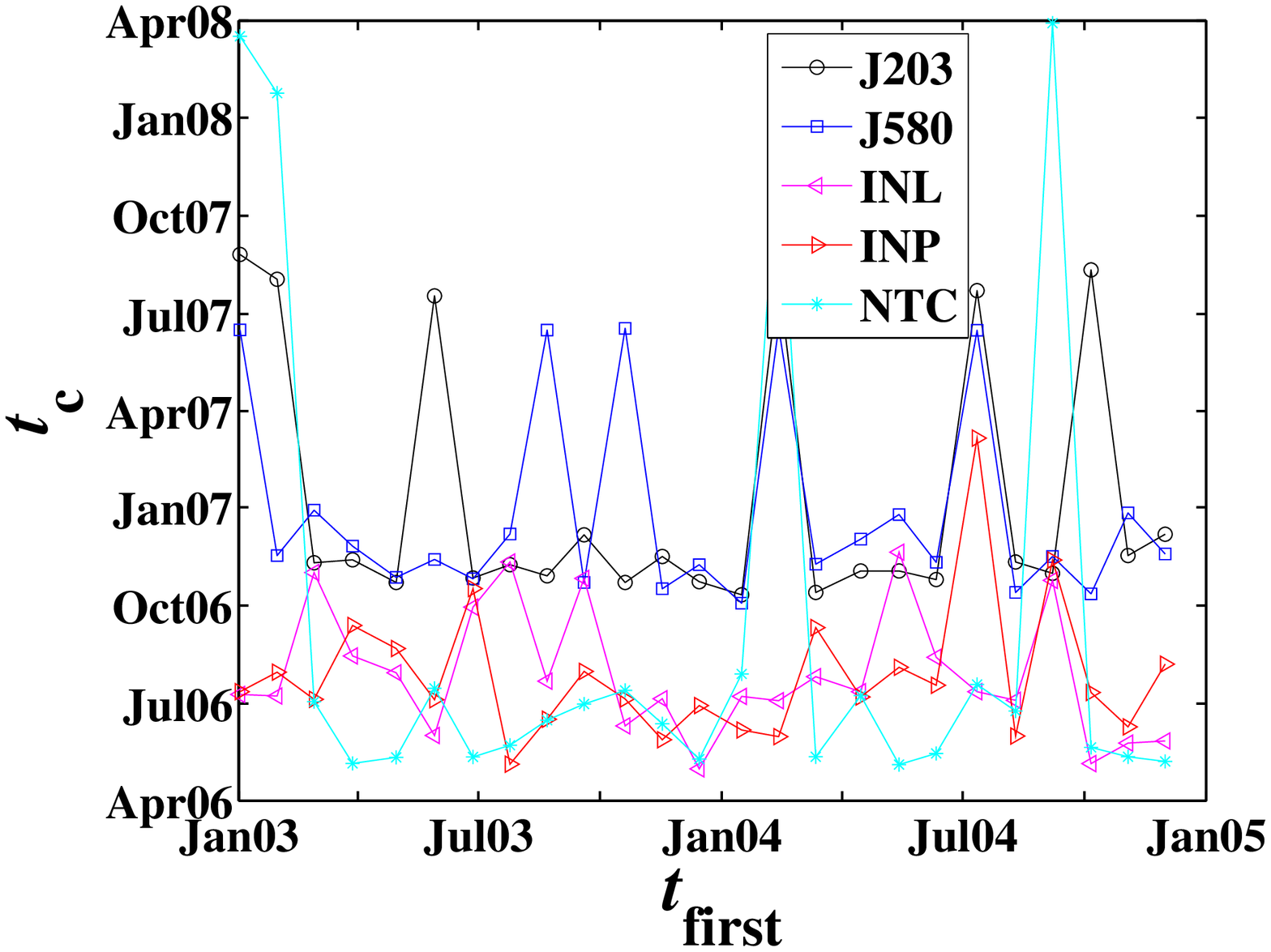}
\includegraphics[width=6.5cm]{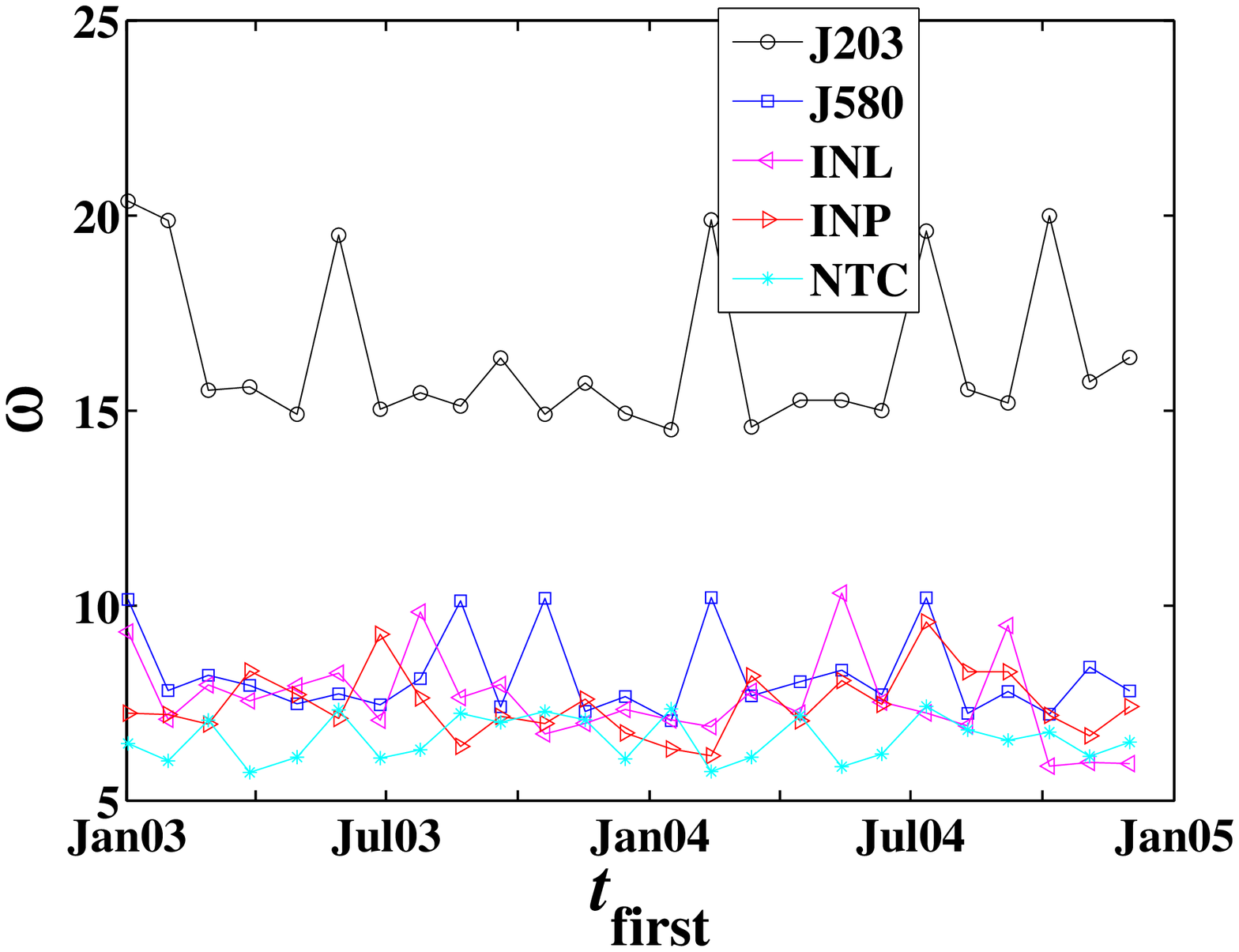}
\caption{(Color online). Sensitivity analysis of the estimated
critical time $t_c$ and the angular log-frequency $\omega$ for the
five stocks obtained by varying the last point $t_{\rm{first}}$ of
the time series up to which the fits using formula  (\ref{Eq:Wei})
with $N=2$ are performed.} \label{Fig:Tfirst2}
\end{center}
\end{figure}

We first study the impact of $t_{\rm{first}}$ on $t_c$ and $\omega$ to
check the stability of the estimated critical time and the
significance of the log-periodic pattern. For each stock, we use
the first-order LPPL formula (\ref{Eq:LPPL})
(respectively the second-order LPPL formula (\ref{Eq:Wei}))
to fit the price time series from
$t_{\rm{first}}$ to $\rm{2006/05/26}$, where $t_{\rm{first}}$ ranges
from $\rm{2003/01/02}$ to $\rm{2004/12/01}$ with a spacing of 20
trading days. The results are presented in Fig.~\ref{Fig:Tfirst}
(respectively Fig.~\ref{Fig:Tfirst2}).
The overall conclusion is that both $t_c$ and $\omega$ are very
robust with respect to the choice of the starting time $t_{\rm{first}}$
of the fitting interval.

\subsection{The impact of $t_{\rm{last}}$}

\begin{figure}[htp]
\begin{center}
\includegraphics[width=6.5cm]{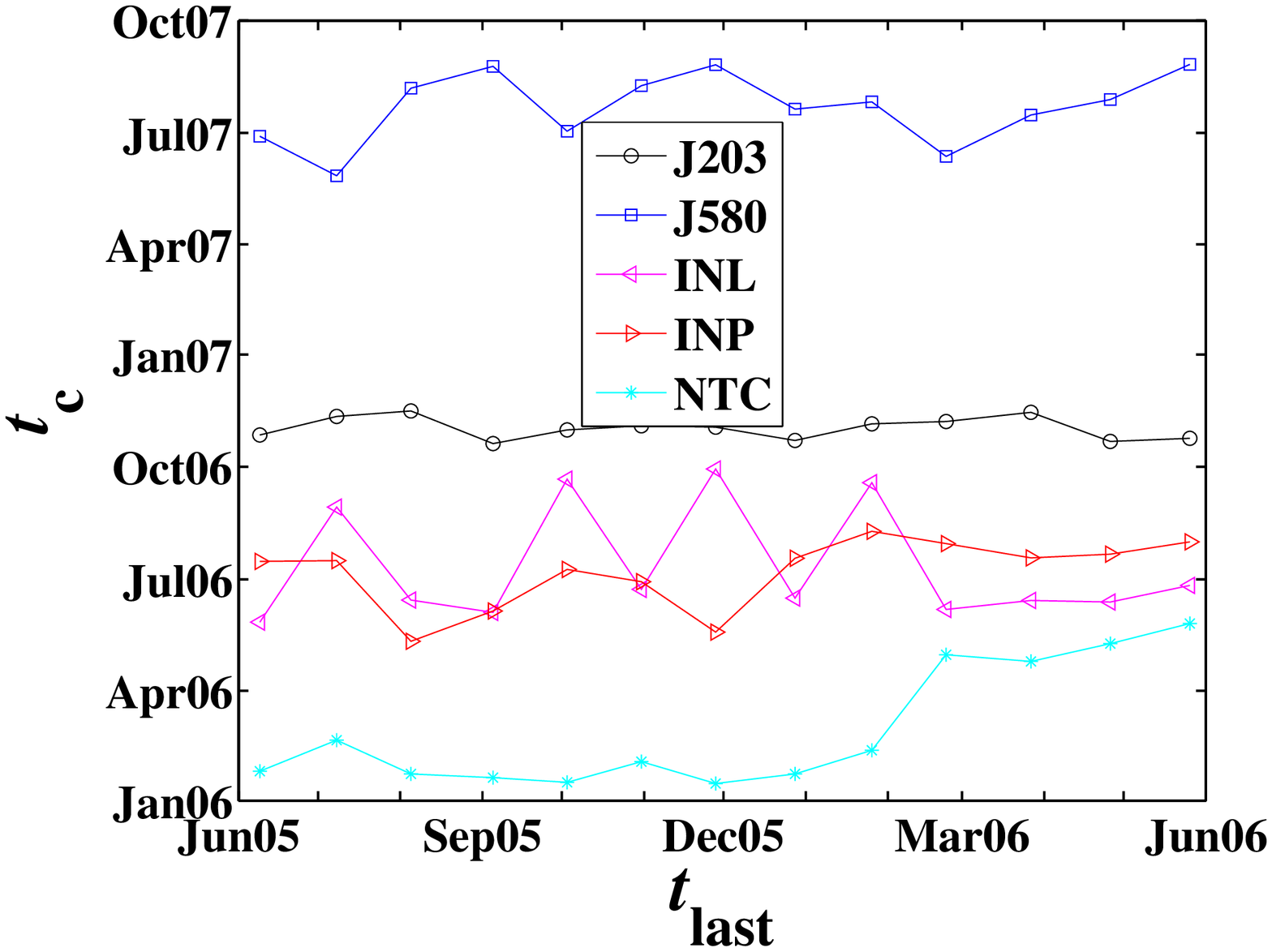}
\includegraphics[width=6.5cm]{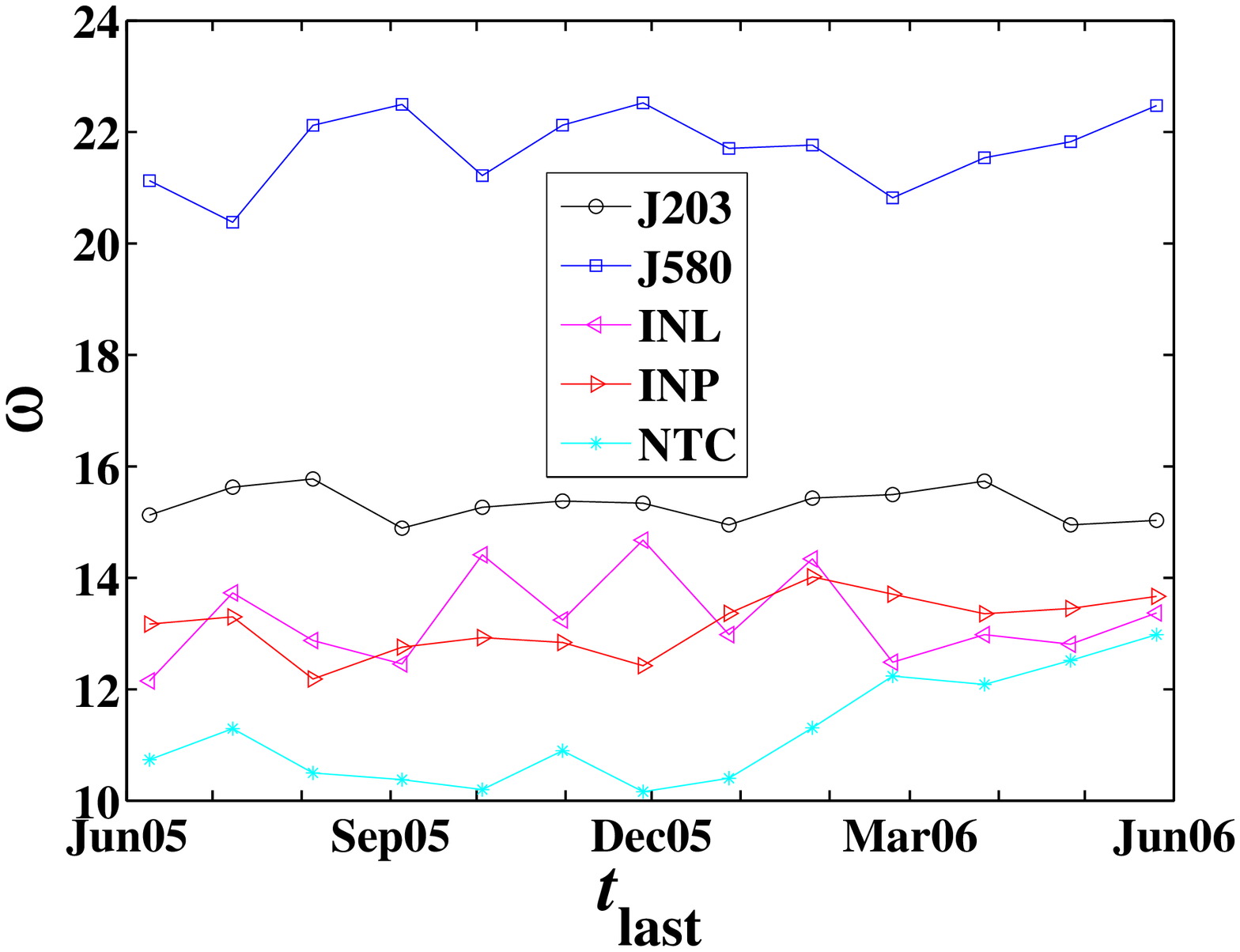}
\caption{(Color online). Sensitivity analysis of the estimated
critical time $t_c$ and the angular log-frequency $\omega$ for the
five stocks obtained by varying the last point $t_{\rm{last}}$ of
the time interval in which the fits using formula (\ref{Eq:LPPL})
are performed.} \label{Fig:Tlast}
\end{center}
\end{figure}

\begin{figure}[htp]
\begin{center}
\includegraphics[width=6.5cm]{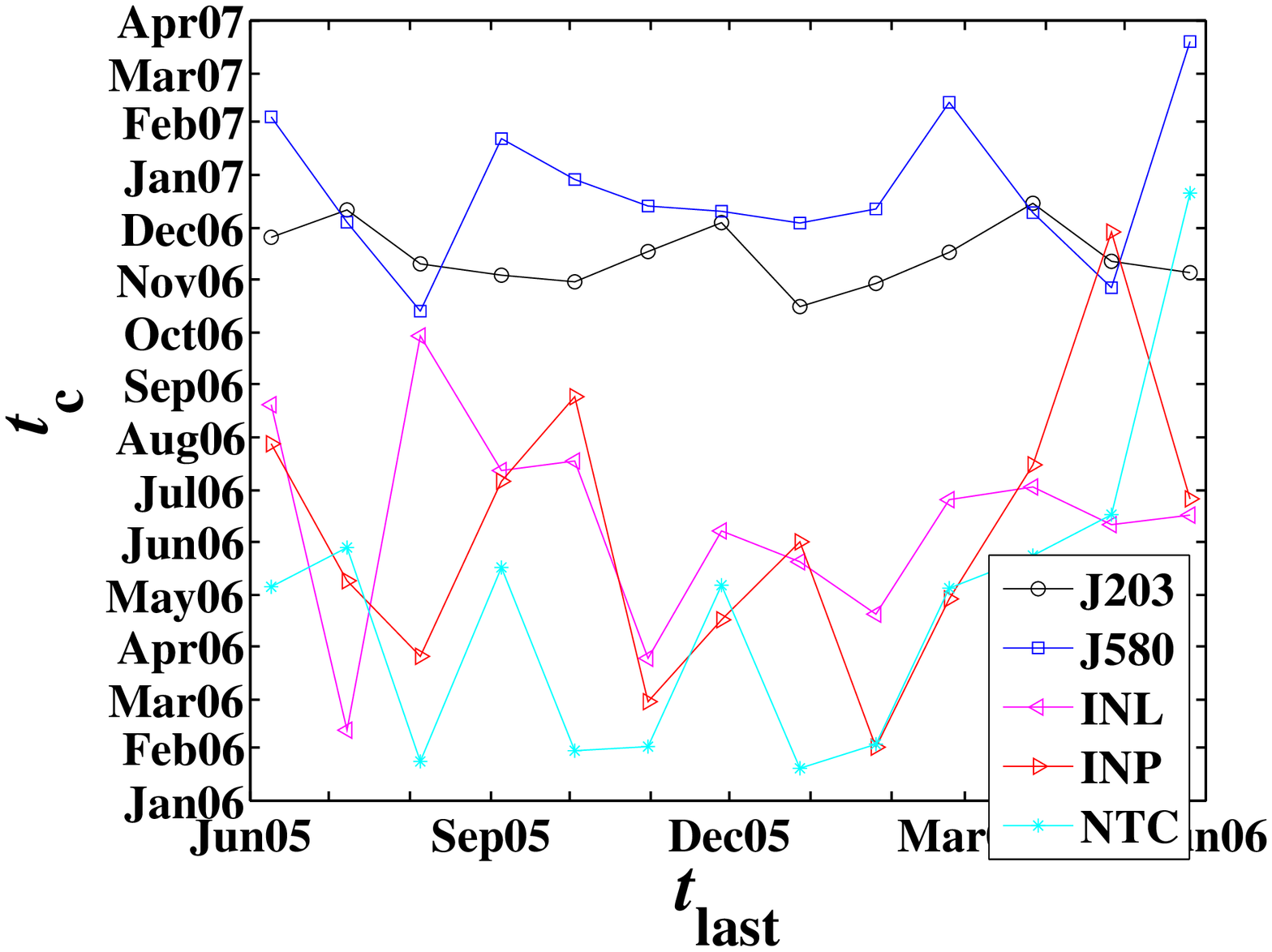}
\includegraphics[width=6.5cm]{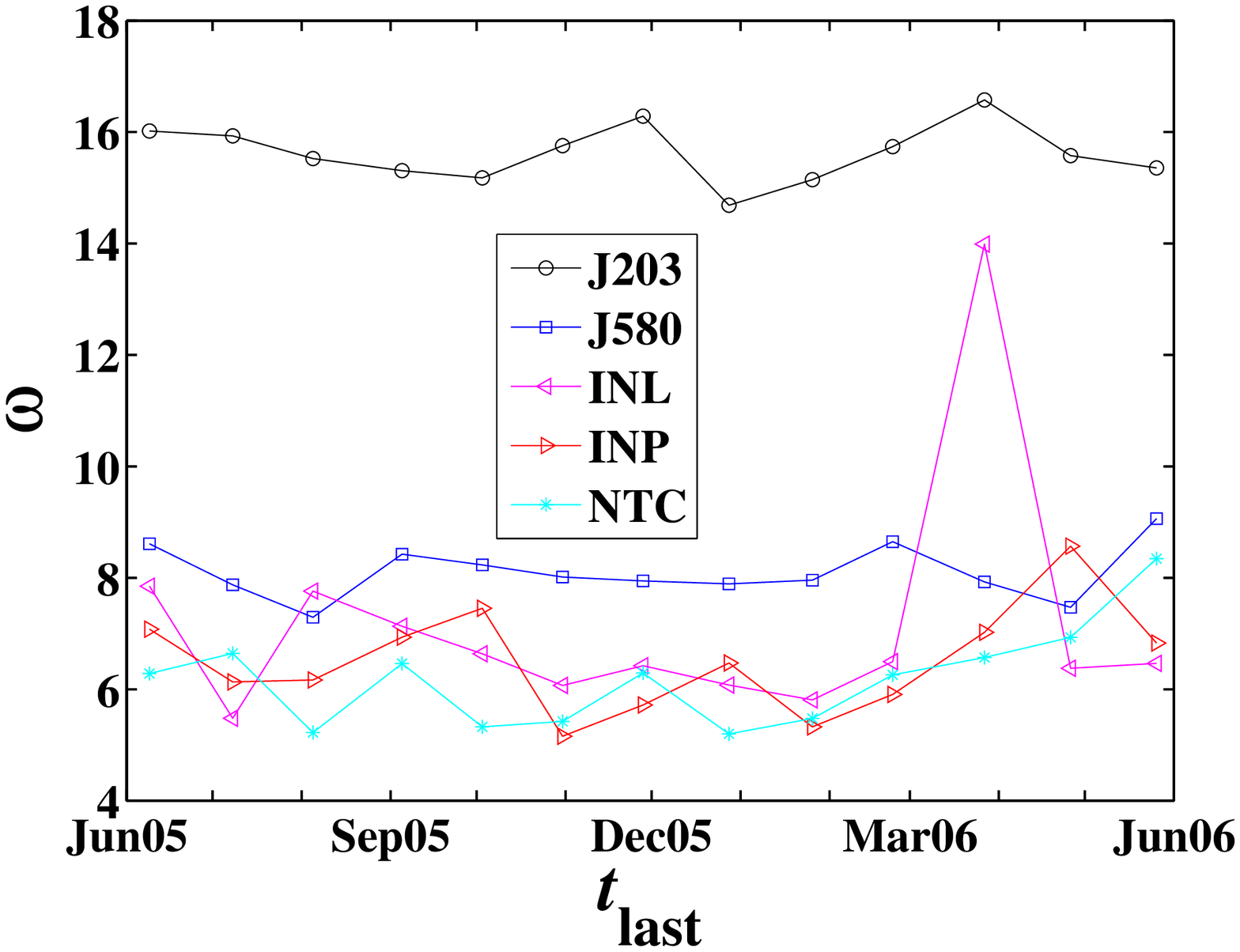}
\caption{(Color online). Sensitivity analysis of the estimated
critical time $t_c$ and the angular log-frequency $\omega$ for the
five stocks obtained by varying the last point $t_{\rm{last}}$ of
the time series up to which the fits using formula (\ref{Eq:Wei})
with $N=2$ are performed.} \label{Fig:Tlast2}
\end{center}
\end{figure}

We then study the impact of $t_{\rm{last}}$ on $t_c$ and $\omega$ to
check the stability of the estimated critical time and the
significance of the log-periodic pattern. For each stock, we use the
first-order LPPL formula (\ref{Eq:LPPL}) (respectively the
second-order LPPL formula (\ref{Eq:Wei})) to fit the price time
series from $\rm{2003/01/02}$  to $t_{\rm{last}}$, where
$t_{\rm{last}}$ ranges from $\rm{2003/01/02}$ to $\rm{2004/12/01}$
with a spacing of 20 trading days. The results are presented in
Fig.~\ref{Fig:Tlast} (respectively Fig.~\ref{Fig:Tlast2}). Overall,
the conclusion is similar than for the dependence on
$t_{\rm{first}}$, confirming the robustness of the fits and the
reliability of our conclusions. The stock NTC is the only one
exhibiting a change of regime in the first quarter of 2006, at which
the fitted $t_c$ jumps from Jan. 2006 to May 2006.

\section{Concluding remarks}

In summary, we have identified five stocks (1 (J203),  4 (J580), 19
(INL), 20 (INP), and 30 (NTC)) out of a representative sample of
forty five South African stocks, that we qualified as being in a
bubble regime defined as a super-exponential growth regime from Jan.
2003 to May 2006 with significant log-periodic oscillations. We
studied the log-periodic characteristics of these stocks using four
different techniques, the parametric fits with the first-order LPPL
formula, with the second-order Weierstrass-type model, the
parametric detrending method, and the $(H,q)$-analysis. The four
techniques give consistent estimations for the value of the
fundamental angular log-frequency $\omega_1$ in agreement with
previous works on many other bubbles in developed and emergent
markets, confirming with very high statistical confidence the
existence of genuine log-periodicity.

Sensitivity tests of the estimated critical times and of the angular
log-frequency by varying the first date and the last date of the stock
price time series over which the fits are performed confirm the
robustness of the estimated parameters.

This study was performed at the end of May 2006 and we waited
another six months before completing this paper to see what were the
subsequent evolutions of the five stocks. It turns out that the five
selected stocks on the South Africa market experienced an abrupt
drop in mid-June 2006, as can be seen from Fig.~\ref{Fig:LPPL}  in
which we have shown the subsequent price evolution after May 2006.
Quantitatively, the cumulative drawdown (defined as the
peak-to-valley relative price variation) measured from the highest
price value prior to the end of May 2006 to the price bottom
thereafter are respectively: 16.8\% for  1 (J203), 16.6\% for 4
(J580), 20.1\% for 19 (INL), 17.9\% for 20 (INP), and 20.1\% for 30
(NTC). These drawdowns occurred over a time period of less than one
month and their amplitudes belong to the $1$-percentile of the
distribution of drawdowns
\cite{Johansen-Sornette-1998-EPJB,Johansen-Sornette-2001-JR,Sornette-Johansen-2001-QF}.
The occurrence of these large market price corrections and their
timing (mid-June 2006) are compatible with the predicted $t_c$ for
INL, INP and NTC and to a lesser extend for J203. However, other
stocks including J580 give a much large $t_c$, some time during
2007, suggesting that the potential for growth in several of these
stocks is not exhausted. It is possible that the mini-crash that
occurred in mid-June 2006 was only a partial correction, similarly
to the Oct. 1997 8\% drop witnessed on the S\&P500 US market which,
after being followed by a plateau of three months, resumed in a
strong acceleration, to finally end with the real crash in
August-September 1998 (see discussion of this sequence in
\cite{Sornette-2003}). Finally, we refer to the extended version
only available online at http://arXiv.org/abs/physics/0701171, in
which the 45 price trajectories for the 45 assets and the fit of the
logarithm of their price with the linear model and with the
nonlinear model are shown.

\bigskip
{\textbf{Acknowledgments:}}

We are grateful to Franco Busetti for providing us the data. This
work was partly supported by the National Natural Science Foundation
of China (Grant No. 70501011), the Fok Ying Tong Education
Foundation (Grant No. 101086), the Program for New Century Excellent
Talents in University (Grant No. NCET-07-0288), and the Alfred
Kastler Foundation in France.

\bibliography{E:/papers/Auxiliary/Bibliography}

\begin{thebibliography}{10}
\expandafter\ifx\csname url\endcsname\relax
  \def\url#1{\texttt{#1}}\fi
\expandafter\ifx\csname urlprefix\endcsname\relax\def\urlprefix{URL }\fi

\bibitem{Dunbar-1998-EA}
R.~I.~M. Dunbar, {The social brain hypothesis}, Evol. Anthrop. 6 (1998)
  178--190.

\bibitem{Sornette-2003}
D.~Sornette, {Why Stock Markets Crash: Critical Events in Complex Financial
  Systems}, Princeton University Press, Princeton, 2003.

\bibitem{Ide-Sornette-2002-PA}
K.~Ide, D.~Sornette, {Oscillatory finite-time singularities in finance,
  population and rupture}, Physica A 307 (2002) 63--106.

\bibitem{Sornette-Ide-2003-IJMPC}
D.~Sornette, K.~Ide, {Theory of self-similar oscillatory finite-time
  singularities}, Int. J. Modern Phys. C 14 (2003) 267--275.

\bibitem{Zhou-Sornette-Hill-Dunbar-2005-PRSB}
W.-X. Zhou, D.~Sornette, R.~A. Hill, R.~I.~M. Dunbar, {Discrete hierarchical
  organization of social group sizes}, Proc. Royal Soc. B 272 (2005) 439--444.

\bibitem{Hill-Dunbar08}
R.~A. Hill, R.~A. Bentley, R.~Dunbar, {Network scaling reveals consistent
  fractal pattern in hierarchical mammalian societies}, Biology Letters (2008)
  doi:10.1098/rsbl.2008.0393.

\bibitem{Johansen-Ledoit-Sornette-2000-IJTAF}
A.~Johansen, O.~Ledoit, D.~Sornette, {Crashes as critical points}, Int. J.
  Theoret. Appl. Financ. 3 (2000) 219--255.

\bibitem{Johansen-Sornette-2001-IJTAF}
A.~Johansen, D.~Sornette, {Bubbles and anti-bubbles in Latin-American, Asian
  and western stock markets: An empirical study}, Int. J. Theoret. Appl.
  Financ. 4 (2001) 853--920.

\bibitem{Johansen-Sornette-Ledoit-1999-JR}
A.~Johansen, D.~Sornette, O.~Ledoit, {Predicting financial crashes using
  discrete scale invariance}, J. Risk 1 (1999) 5--32.

\bibitem{Johansen-2003-QF}
A.~Johansen, {An alternative view}, Quant. Financ. 3 (2003) C6--C7.

\bibitem{Sornette-Zhou-2003-QF}
D.~Sornette, W.-X. Zhou, {The US 2000-2003 market descent: Clarifications},
  Quant. Financ. 3 (2003) C39--C41.

\bibitem{Sornette-1998-PR}
D.~Sornette, {Discrete scale invariance and complex dimensions}, Phys. Rep. 297
  (1998) 239--270.

\bibitem{VanZyl-2006}
C.~Van~Zyl, {Understanding South African Financial Markets}, Schaik Publishers,
  South Africa, 2006.

\bibitem{Hamilton-Whiteman-1985-JMonE}
J.~Hamilton, C.~Whiteman, {The observable implications of self-fulfilling
  speculative price}, J. Monet. Econ. 16 (1985) 353--373.

\bibitem{Evans-1986-AER}
G.~W. Evans, {A test for speculative bubbles in the Sterling-Dollar exchange
  rate: 1981-84}, Am. Econ. Rev. 76 (1986) 621--636.

\bibitem{Meese-1986-JPE}
R.~A. Meese, {Testing for bubbles in exchange rates: A case of sparkling
  rates?}, J. Polit. Econ. 94 (1986) 345--373.

\bibitem{Woo-1987-JMCB}
W.~T. Woo, {Some evidence of speculative bubbles in the foreign exchange
  markets}, J. Money, Credit, and Banking 19 (1987) 499--514.

\bibitem{Camerer-1989-JES}
C.~Camerer, {Bubbles and fads in asset prices}, J. Econ. Surveys 3 (1989)
  3--41.

\bibitem{Evans-1991-AER}
G.~W. Evans, {Pitfalls in testing for explosive bubbles in asset prices}, Am.
  Econ. Rev. 81 (1991) 922--930.

\bibitem{Adam-Szafarz-1992-OX}
M.~C. Adam, A.~Szafarz, {Speculative bubbles and financial markets}, Oxford
  Econ. Papers 44 (1992) 626--640.

\bibitem{Gurkaynak-2005}
R.~S. Gurkaynak, Econometric tests of asset price bubbles: Taking stock, FEDS
  Working Paper (2005) 2005--04.

\bibitem{Johansen-Sornette-2006-BER}
A.~Johansen, D.~Sornette, {Shocks, crashes and bubbles in financial markets},
  Brussels Economic Review 49, (http://arXiv.org/abs/cond-mat/0210509).

\bibitem{Akaike-1974-IEEEtac}
H.~Akaike, {A new look at the statistical model identification}, IEEE Trans.
  Automat. Contr. 19 (1974) 716--723.

\bibitem{GrangerNewbold74}
C.~Granger, P.~Newbold, {Spurious regressions in economics,}, J. Econometrics 2
  (1974) 111--120.

\bibitem{Phillips-1986-JEm}
P.~C.~B. Phillips, {Understanding spurious regressions in econometrics}, J.
  Econometrics 33 (1986) 311--340.

\bibitem{Sornette-Johansen-Bouchaud-1996-JPIF}
D.~Sornette, A.~Johansen, J.-P. Bouchaud, {Stock market crashes, precursors and
  replicas}, J. Phys. I France 6 (1996) 167--175.

\bibitem{Feigenbaum-Freund-1996-IJMPB}
J.~A. Feigenbaum, P.~G.~O. Freund, {Discrete scale invariance in stock markets
  before crashes}, Int. J. Modern Phys. B 10 (1996) 3737--3745.

\bibitem{Sornette-2003-PR}
D.~Sornette, {Critical market crashes}, Phys. Rep. 378 (2003) 1--98.

\bibitem{Johansen-2003-PA}
A.~Johansen, {Characterization of large price variations in financial markets},
  Physica A 324 (2003) 157--166.

\bibitem{Gluzman-Sornette-2002-PRE}
S.~Gluzman, D.~Sornette, {Log-periodic route to fractal functions}, Phys. Rev.
  E 65 (2002) 036142.

\bibitem{Zhou-Sornette-2002-PD}
W.-X. Zhou, D.~Sornette, {Evidence of intermittent cascades from discrete
  hierarchical dissipation in turbulence}, Physica D 165 (2002) 94--125.

\bibitem{Zhou-Sornette-2003c-PA}
W.-X. Zhou, D.~Sornette, {Renormalization group analysis of the 2000-2002
  anti-bubble in the US S\&P 500 index: Explanation of the hierarchy of five
  crashes and prediction}, Physica A 330 (2003) 584--604.

\bibitem{Zhou-Sornette-2002-IJMPC}
W.-X. Zhou, D.~Sornette, {Statistical significance of periodicity and
  log-periodicity with heavy-tailed correlated noise}, Int. J. Modern Phys. C
  13 (2002) 137--170.

\bibitem{Huang-Johansen-Lee-Saleur-Sornette-2000-JGR}
Y.~Huang, A.~Johansen, M.~W. Lee, H.~Saleur, D.~Sornette, {Artifactual
  log-periodicity in finite-size data: Relevance for earthquake aftershocks},
  J. Geophys. Res. 105 (2000) 25451--25471.

\bibitem{Zhou-Sornette-2002-PRE}
W.-X. Zhou, D.~Sornette, {Generalized $q$-analysis of log-periodicity:
  Applications to critical ruptures}, Phys. Rev. E 66 (2002) 046111.

\bibitem{Press-Teukolsky-Vetterling-Flannery-1996}
W.~Press, S.~Teukolsky, W.~Vetterling, B.~Flannery, {Numerical Recipes in
  FORTRAN: The Art of Scientific Computing}, Cambridge University Press,
  Cambridge, 1996.

\bibitem{Bothmer-Meister-2003-PA}
H.-C.~G. van Bothmer, C.~Meister, {Predictingcritical crashes? A new
  restriction for the free variables}, Physica A 320 (2003) 539--547.

\bibitem{Zhou-Sornette-2003-IJMPC}
W.-X. Zhou, D.~Sornette, {Nonparametric analyses of log-periodic precursors to
  financial crashes}, Int. J. Modern Phys. C 14 (2003) 1107--1125.

\bibitem{Erzan-1997-PLA}
A.~Erzan, {Finite $q$-differences and the discrete renormalization group},
  Phys. Lett. A 225 (1997) 235--238.

\bibitem{Erzan-Eckmann-1997-PRL}
A.~Erzan, J.-P. Eckmann, {$q$-analysis of fractal sets}, Phys. Rev. Lett. 78
  (1997) 3245--3248.

\bibitem{Sornette-Zhou-2002-QF}
D.~Sornette, W.-X. Zhou, {The US 2000-2002 market descent: How much longer and
  deeper?}, Quant. Financ. 2 (2002) 468--481.

\bibitem{Sornette-Zhou-2004-PA}
D.~Sornette, W.-X. Zhou, {Evidence of fueling of the 2000 new economy bubble by
  foreign capital inflow: Implications for the future of the US economy and its
  stock market}, Physica A 332 (2004) 412--440.

\bibitem{Zhou-Sornette-2003a-PA}
W.-X. Zhou, D.~Sornette, {2000-2003 real estate bubble in the UK but not in the
  USA}, Physica A 329 (2003) 249--263.

\bibitem{Zhou-Sornette-2004a-PA}
W.-X. Zhou, D.~Sornette, {Antibubble and prediction of China's stock market and
  real-estate}, Physica A 337 (2004) 243--268.

\bibitem{Zhou-Sornette-2004b-PA}
W.-X. Zhou, D.~Sornette, {Causal slaving of the U.S. treasury bond yield
  antibubble by the stock market antibubble of August 2000}, Physica A 337
  (2004) 586--608.

\bibitem{Zhou-Sornette-2005-PA}
W.-X. Zhou, D.~Sornette, {Testing the stability of the 2000 US stock market
  ``antibubble''}, Physica A 348 (2005) 428--452.

\bibitem{Zhou-Sornette-2006a-PA}
W.-X. Zhou, D.~Sornette, {Fundamental factors versus herding in the 2000-2005
  US stock market and prediction}, Physica A 360 (2006) 459--482.

\bibitem{Johansen-Sornette-1998-EPJB}
A.~Johansen, D.~Sornette, {Stock market crashes are outliers}, Eur. Phys. J. B
  1 (1998) 141--143.

\bibitem{Johansen-Sornette-2001-JR}
A.~Johansen, D.~Sornette, {Large stock market price drawdowns are outliers}, J.
  Risk 4~(2) (2001) 69--110.

\bibitem{Sornette-Johansen-2001-QF}
D.~Sornette, A.~Johansen, {Significance of log-periodic precursors to financial
  crashes}, Quant. Financ. 1 (2001) 452--471.

\end{thebibliography}

\newpage

\begin{sidewaystable}
\begin{center}
\caption{The 45 financial time series analyzed. Rows in red are
indices.}\label{TB:com1}
\medskip
\begin{tabular}{llll}
  \hline\hline
  Ticker & Company name & Major Industry & Sub Industry
  \\\hline
{\textcolor[rgb]{1,0,0}{J203}} & {\textcolor[rgb]{1,0,0}{JH-OVER}} &\multicolumn{2}{l}{{\textcolor[rgb]{1,0,0}{FTSE/JSE Africa All Share}}} \\
{\textcolor[rgb]{1,0,0}{J210}} & {\textcolor[rgb]{1,0,0}{JSE-RESI}}&\multicolumn{2}{l}{{\textcolor[rgb]{1,0,0}{FTSE/JSE Africa Resource 20}}} \\
{\textcolor[rgb]{1,0,0}{J257}} & {\textcolor[rgb]{1,0,0}{JH-ASIN}} &\multicolumn{2}{l}{{\textcolor[rgb]{1,0,0}{FTSE/JSE Africa All Share Industrials}}} \\
{\textcolor[rgb]{1,0,0}{J580}} & {\textcolor[rgb]{1,0,0}{JSE-FINA}}&\multicolumn{2}{l}{{\textcolor[rgb]{1,0,0}{FTSE/JSE Africa All Share Financial}}}\\\hline%
ABL & African Bank Investments Ltd & Financial & Commercial banks \\
{\textcolor[rgb]{0,0,1}{AGL}} & Anglo American Plc & Metal producers \& Products manufacturers & Miscellaneous metal producers \\
AMS & Anglo Platinum Ltd & Metal producers \& Products manufacturers & Miscellaneous metal producers\\%
ANG & Anglogold Ashanti Ltd & Metal producers \& Products manufacturers & Gold producers\\
APN & Aspen Pharmacare Holdings Ltd & Drugs, cosmetics \& health care & Diversified drugs, cosmetics, \& health care \\%
ASA & ABSA Group Ltd & Financial & Commercial banks \\
BAW & Barloworld Limited & Diversified & General diversified\\
{\textcolor[rgb]{0,0,1}{BIL}} & BHP Billiton Plc & Metal producers \& Products manufacturers & Diversified metal producers\\
BVT & Bidvest Group Ltd & Diversified & General diversified\\
ECO & Edgars Consolidated Stores Ltd & Retailers & Apparel store chains \\%
FSR & FirstRand Ltd & Financial & Commercial banks \\
\hline\hline
\end{tabular}
\end{center}
\end{sidewaystable}

\begin{sidewaystable}
\begin{center}
\caption{The 45 financial time series analyzed
({\em{Continued}}).}\label{TB:com2}
\medskip
\begin{tabular}{llll}
  \hline\hline
  Ticker & Company name & Major Industry & Sub Industry
  \\\hline
GFI & Gold Fields Ltd & Metal producers \& Products manufacturers & Gold producers \\
HAR & Harmony Gold Mining Company Ltd & Metal producers \& Products manufacturers & Gold producers \\
IMP & Impala Platinum Holdings Ltd & Metal producers \& Products manufacturers & Miscellaneous metal producers \\%
INL & Investec Ltd & Financial & Commercial banks\\
{\textcolor[rgb]{0,0,1}{INP}} & Investec Ltd & Financial & Commercial banks \\
IPL & Imperial Holdings Ltd & Transportation & Other transportation\\
JDG & JD Group Ltd & Retailers & Miscellaneous retailers \\
{\textcolor[rgb]{0,0,1}{KMB}} & Kumba Resources Ltd & Metal producers \& Products manufacturers & Diversified metal producers\\
{\textcolor[rgb]{0,0,1}{LBT}} & Liberty International Plc & Financial & Land and real estate \\%
LGL & Liberty Group Ltd & Financial & Insurance companies\\
MLA & Mittal Steel South Africa Ltd & Metal producers \& Products manufacturers & Diversified metal producers \\
MTN & MTN Group Ltd & Utilities & Telecommunications \\
NED & Nedbank Group Ltd & Financial & Commercial banks\\
NPN & Naspers Ltd & Recreation & Radio \& TV broadcasts \\
NTC & Network Healthcare Holdings Ltd & Miscellaneous & Medical services\\
\hline\hline
\end{tabular}
\end{center}
\end{sidewaystable}

\begin{sidewaystable}
\begin{center}
\caption{The 45 financial time series analyzed
({\em{Continued}})}\label{TB:com3}
\medskip
\begin{tabular}{llll}
  \hline\hline
  Ticker & Company name & Major Industry & Sub Industry
  \\\hline
{\textcolor[rgb]{0,0,1}{OML}} & Old Mutual plc & Financial & Insurance companies \\
PIK & Pick n Pay Stores Ltd & Retailers & Miscellaneous retailers\\
PPC & Pretoria Portland Cement Company Ltd & Construction & Cement producers \\
{\textcolor[rgb]{0,0,1}{RCH}} & Richemont Securities Ag & Clothing \& Footware \\
REM & Remgro Ltd & Diversified & General diversified\\
RMH & RMB Holdings Ltd & Financial & Other financial services\\
{\textcolor[rgb]{0,0,1}{SAB}} & SABMiller plc & Food \& beverages & Brewers \\%
SAP & Sappi Ltd & Paper & Diversified paper\\
SBK & Standard Bank Group Ltd & Financial & Commercial banks  \\
SHF & Steinhoff International Holdings Ltd & Miscellaneous & Furnishings\\
SLM & Sanlam Ltd & Financial & Insurance companies \\
SOL & Sasol Ltd & Chemicals & Diversified chemical mfrs. \\
TBS & Tiger Brands Ltd & Diversified & General diversified \\
TKG & Telkom SA Ltd & Utilities & Telecommunications\\
WHL & Woolworths Holdings Ltd & Retailers & Miscellaneous retailers \\
\hline\hline
\end{tabular}
\end{center}
\end{sidewaystable}

\newpage

\begin{figure}[htb]
\begin{center}
  \subfigure[Stock No. 1: J203]{
  \label{Fig:SouthAfrica:LQ:J203}
  \begin{minipage}[b]{0.31\textwidth}
    \includegraphics[width=4.5cm,height=4.5cm]{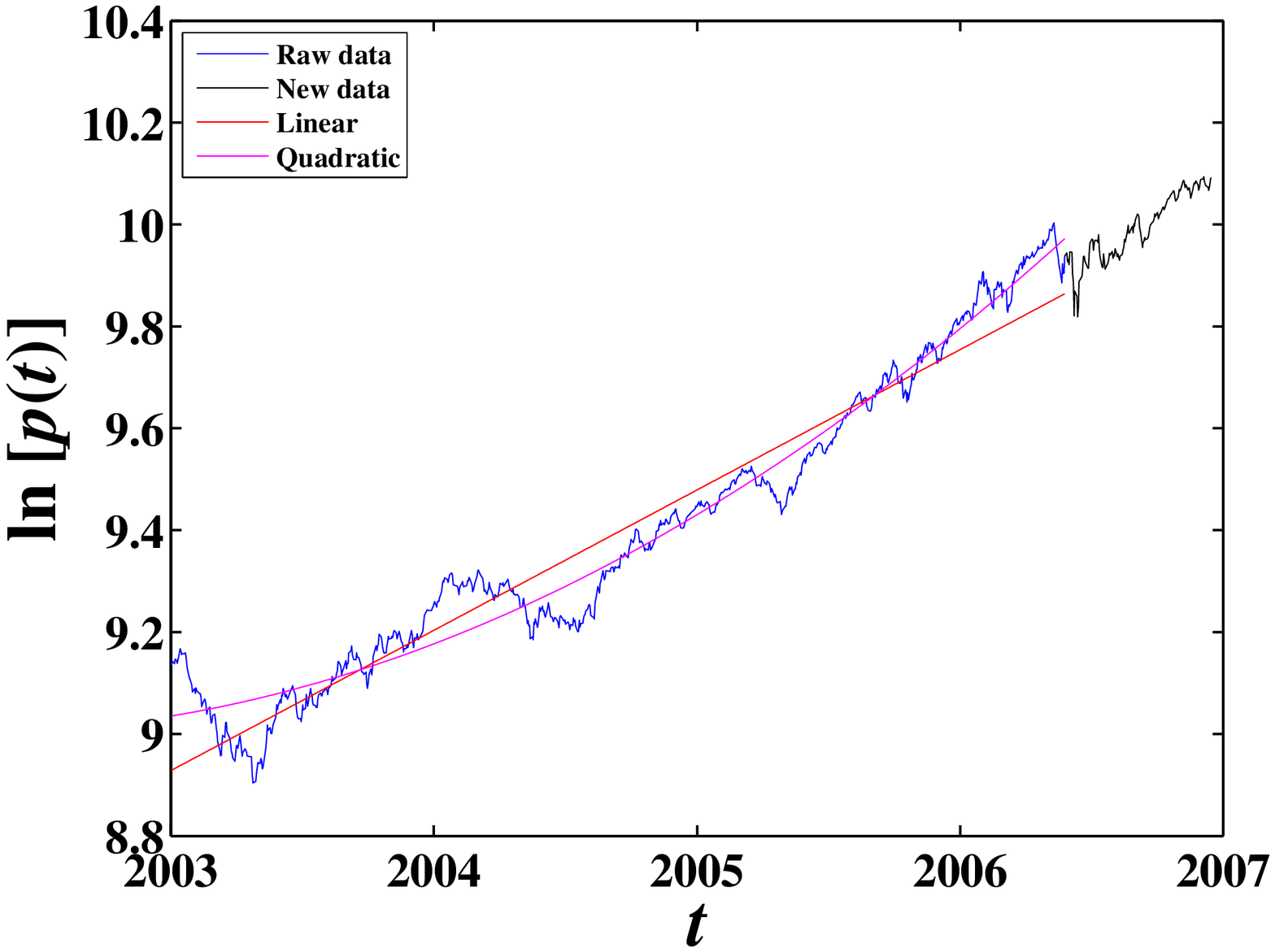}
  \end{minipage}}
  \hspace{0.1cm}
  \subfigure[Stock No. 2: J210]{
  \label{Fig:SouthAfrica:LQ:J210}
  \begin{minipage}[b]{0.31\textwidth}
    \includegraphics[width=4.5cm,height=4.5cm]{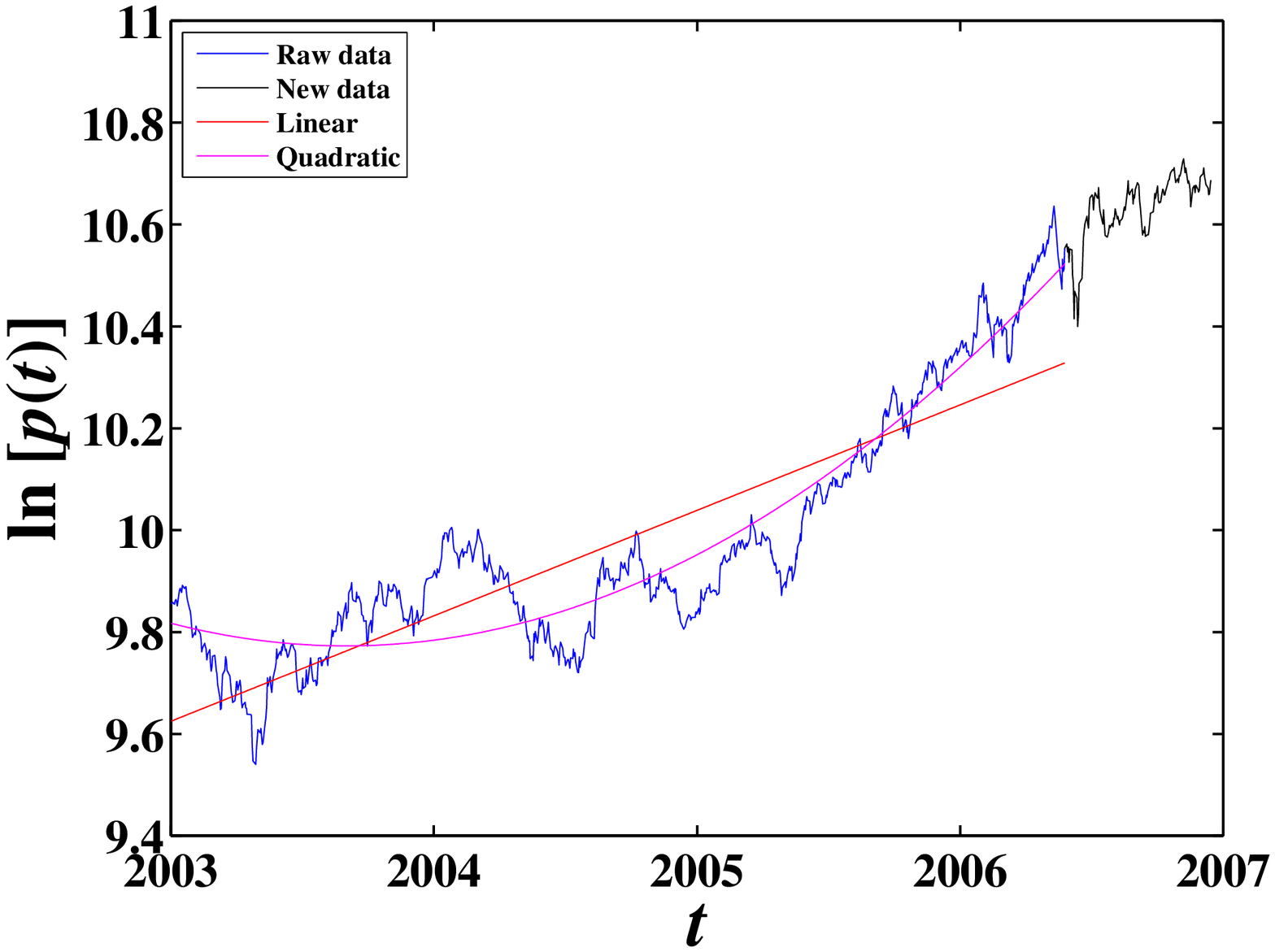}
  \end{minipage}}
  \hspace{0.1cm}
  \subfigure[Stock No. 3: J257]{
  \label{Fig:SouthAfrica:LQ:J257}
  \begin{minipage}[b]{0.31\textwidth}
    \includegraphics[width=4.5cm,height=4.5cm]{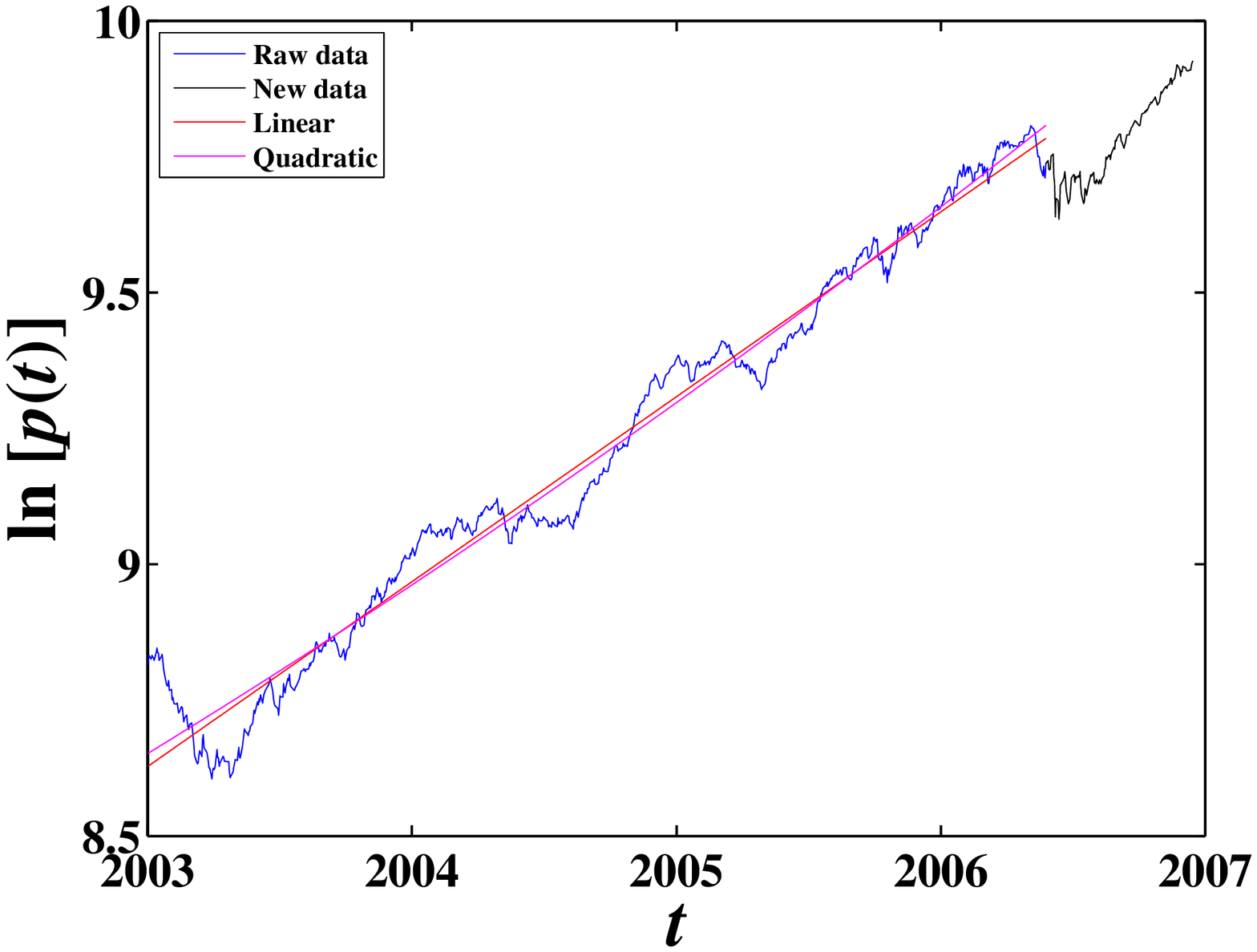}
  \end{minipage}}\\[10pt]
  \subfigure[Stock No. 4: J580]{
  \label{Fig:SouthAfrica:LQ:J580}
  \begin{minipage}[b]{0.31\textwidth}
    \includegraphics[width=4.5cm,height=4.5cm]{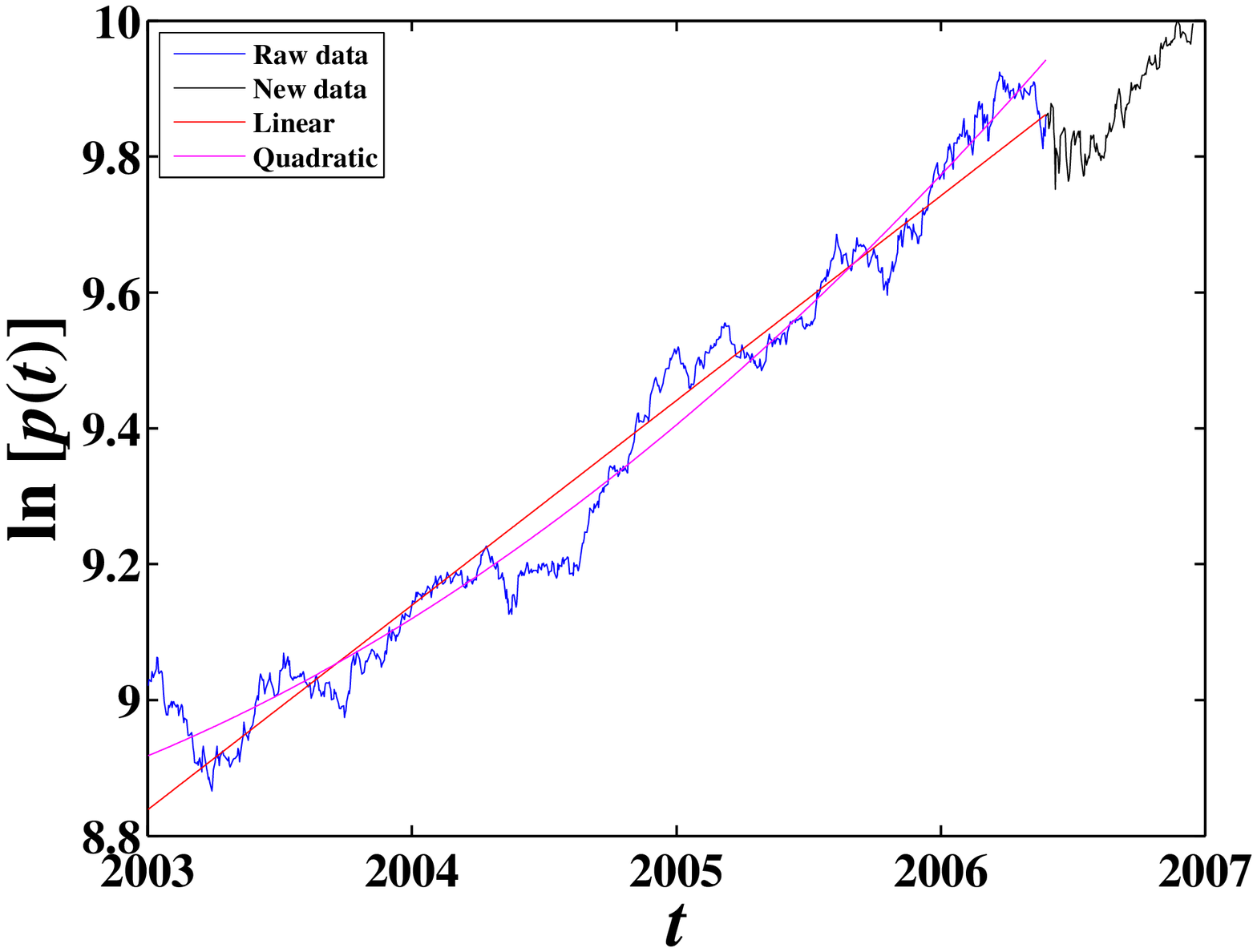}
  \end{minipage}}
  \hspace{0.1cm}
  \subfigure[Stock No. 5: ABL]{
  \label{Fig:SouthAfrica:LQ:ABL}
  \begin{minipage}[b]{0.31\textwidth}
    \includegraphics[width=4.5cm,height=4.5cm]{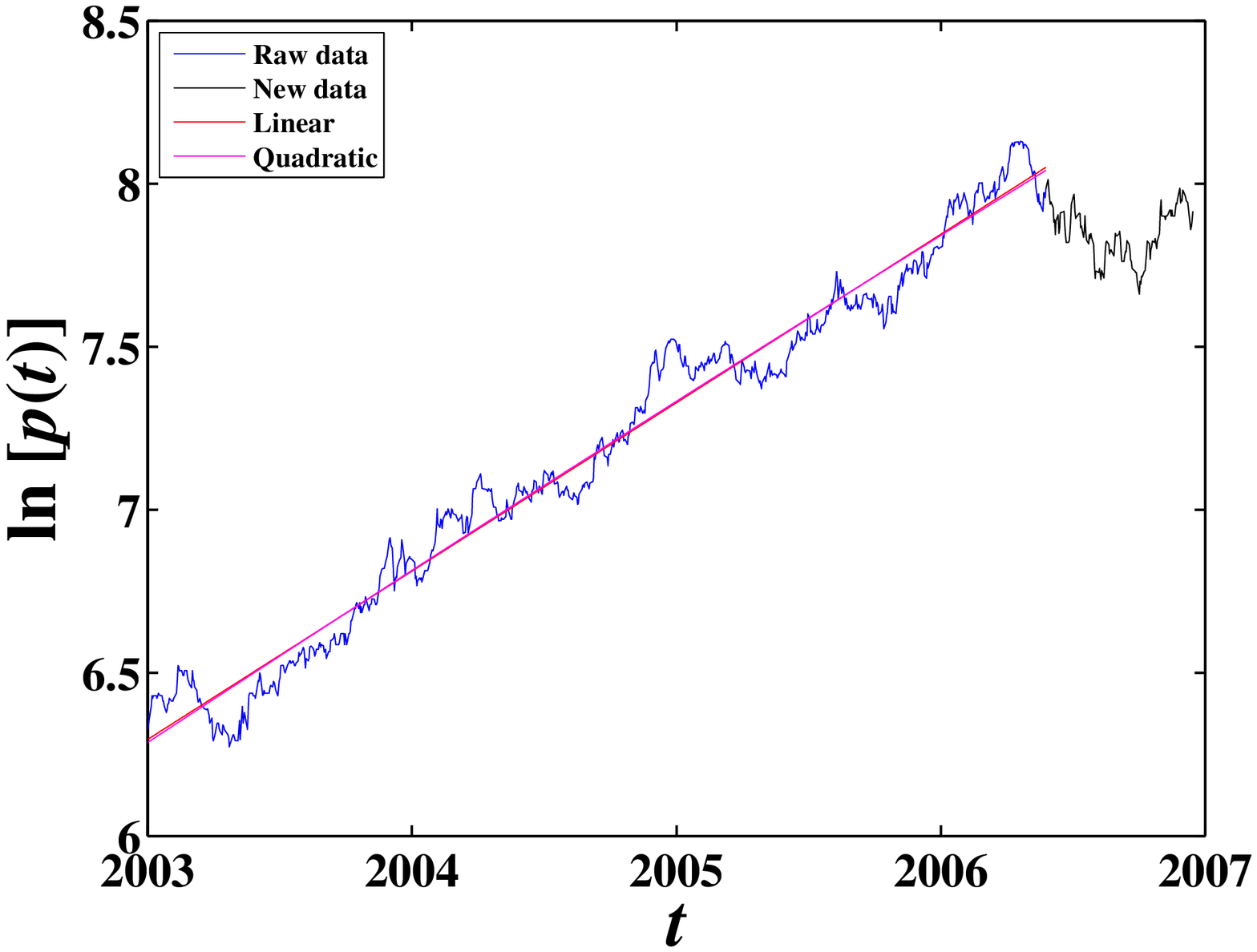}
  \end{minipage}}
  \hspace{0.1cm}
  \subfigure[Stock No. 6: AGL]{
  \label{Fig:SouthAfrica:LQ:AGL}
  \begin{minipage}[b]{0.31\textwidth}
    \includegraphics[width=4.5cm,height=4.5cm]{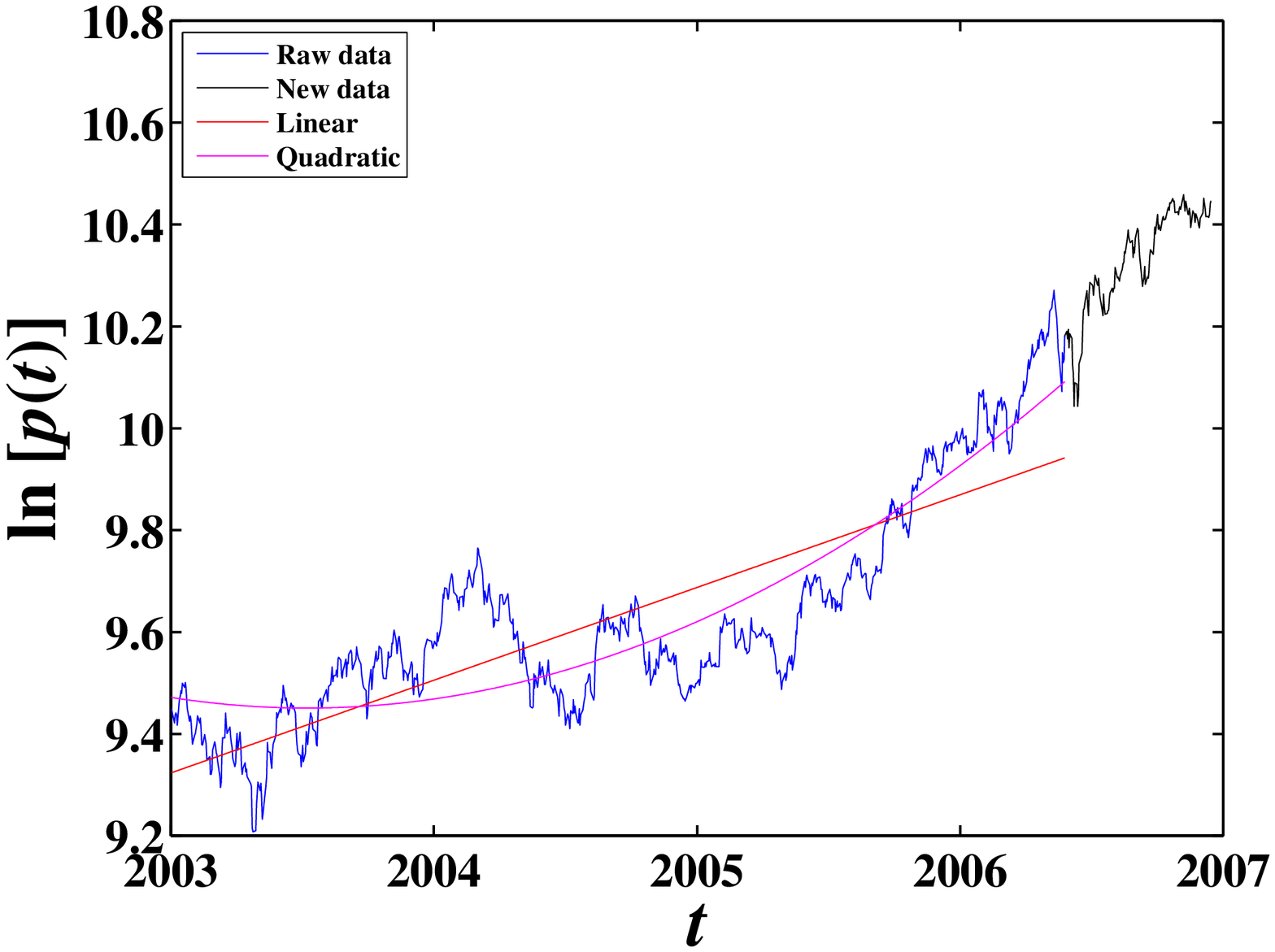}
  \end{minipage}}\\[10pt]
  \subfigure[Stock No. 7: AMS]{
  \label{Fig:SouthAfrica:LQ:AMS}
  \begin{minipage}[b]{0.31\textwidth}
    \includegraphics[width=4.5cm,height=4.5cm]{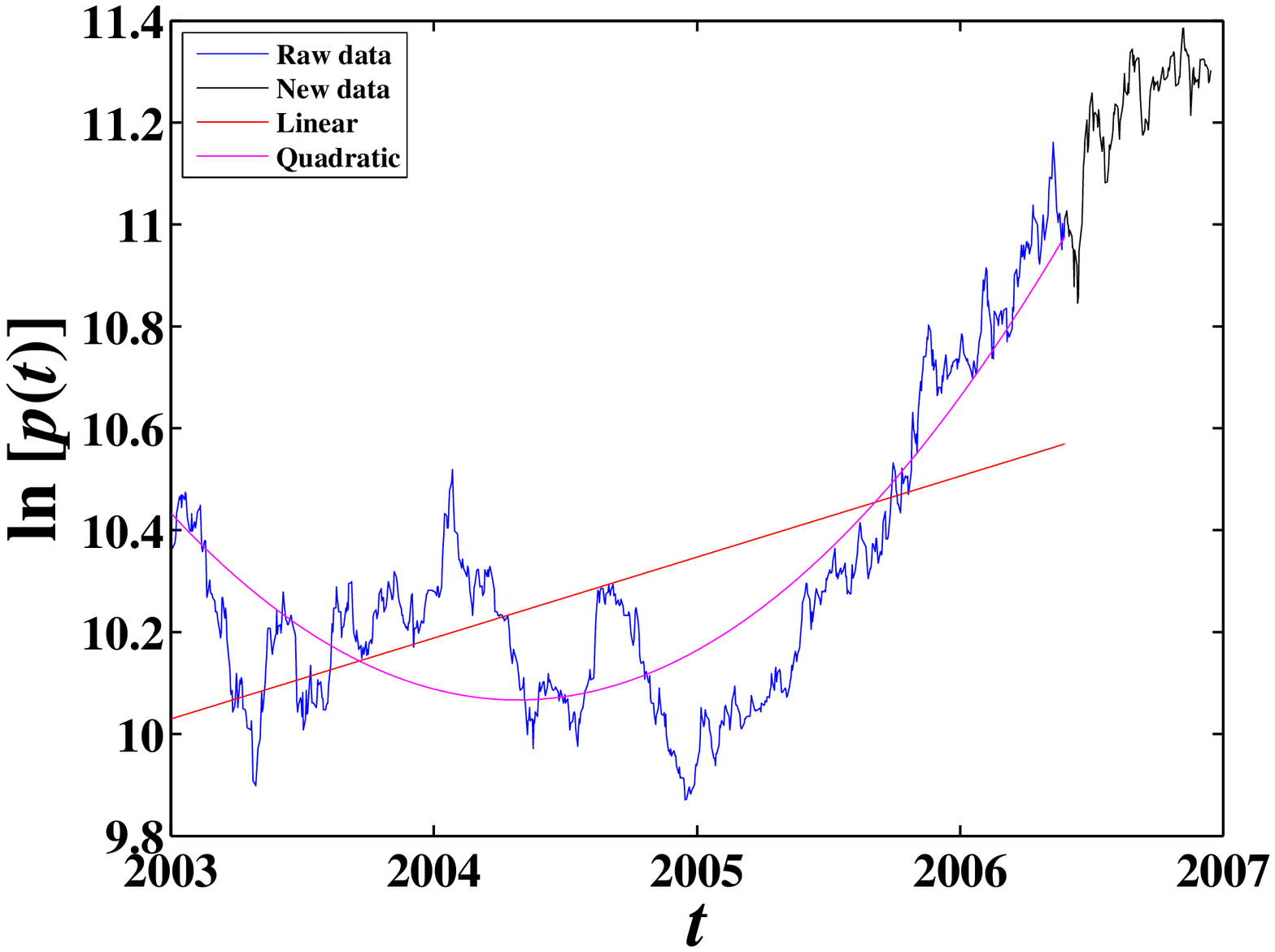}
  \end{minipage}}
  \hspace{0.1cm}
  \subfigure[Stock No. 8: ANG]{
  \label{Fig:SouthAfrica:LQ:ANG}
  \begin{minipage}[b]{0.31\textwidth}
    \includegraphics[width=4.5cm,height=4.5cm]{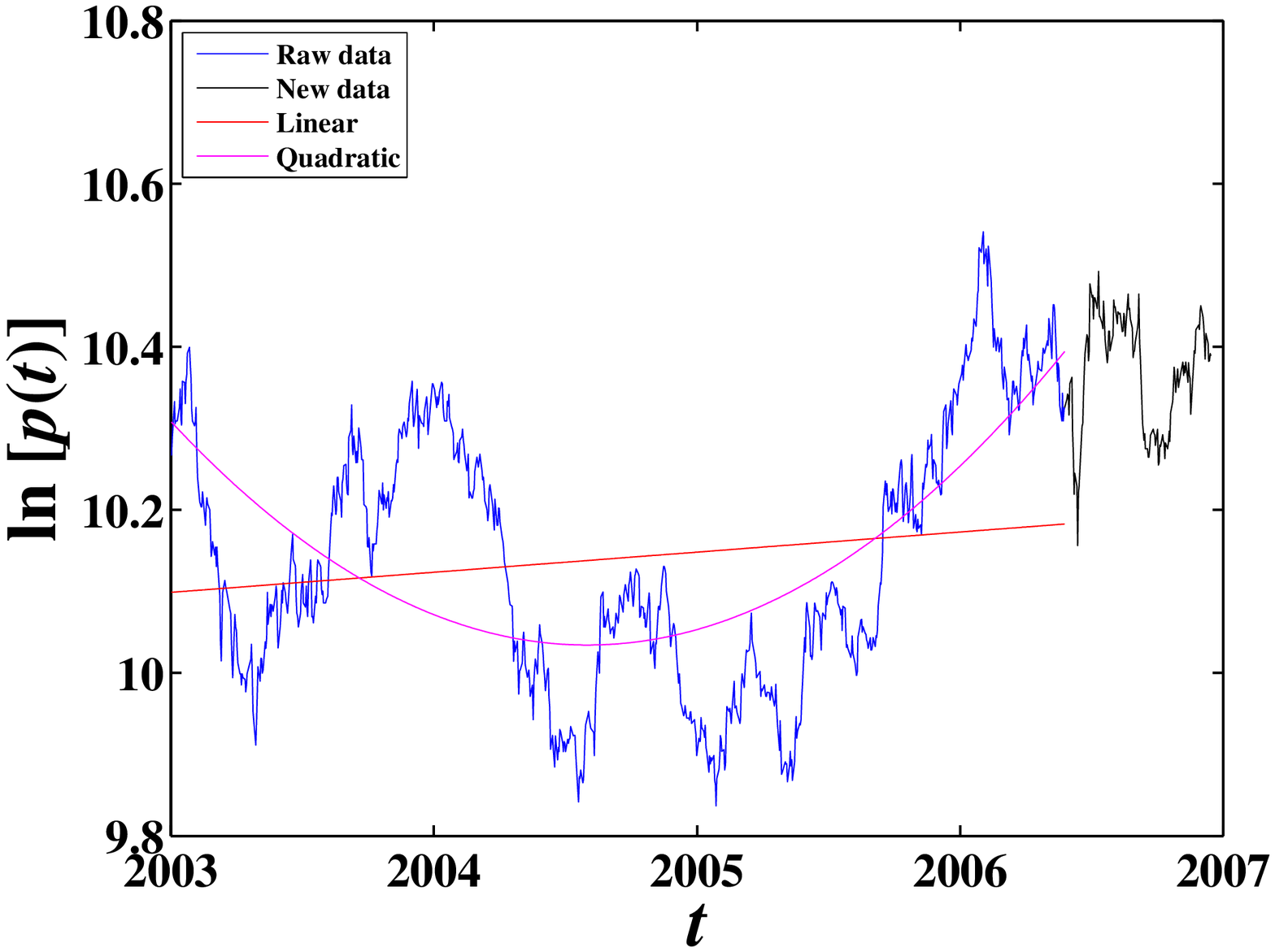}
  \end{minipage}}
  \hspace{0.1cm}
  \subfigure[Stock No. 9: APN]{
  \label{Fig:SouthAfrica:LQ:APN}
  \begin{minipage}[b]{0.31\textwidth}
    \includegraphics[width=4.5cm,height=4.5cm]{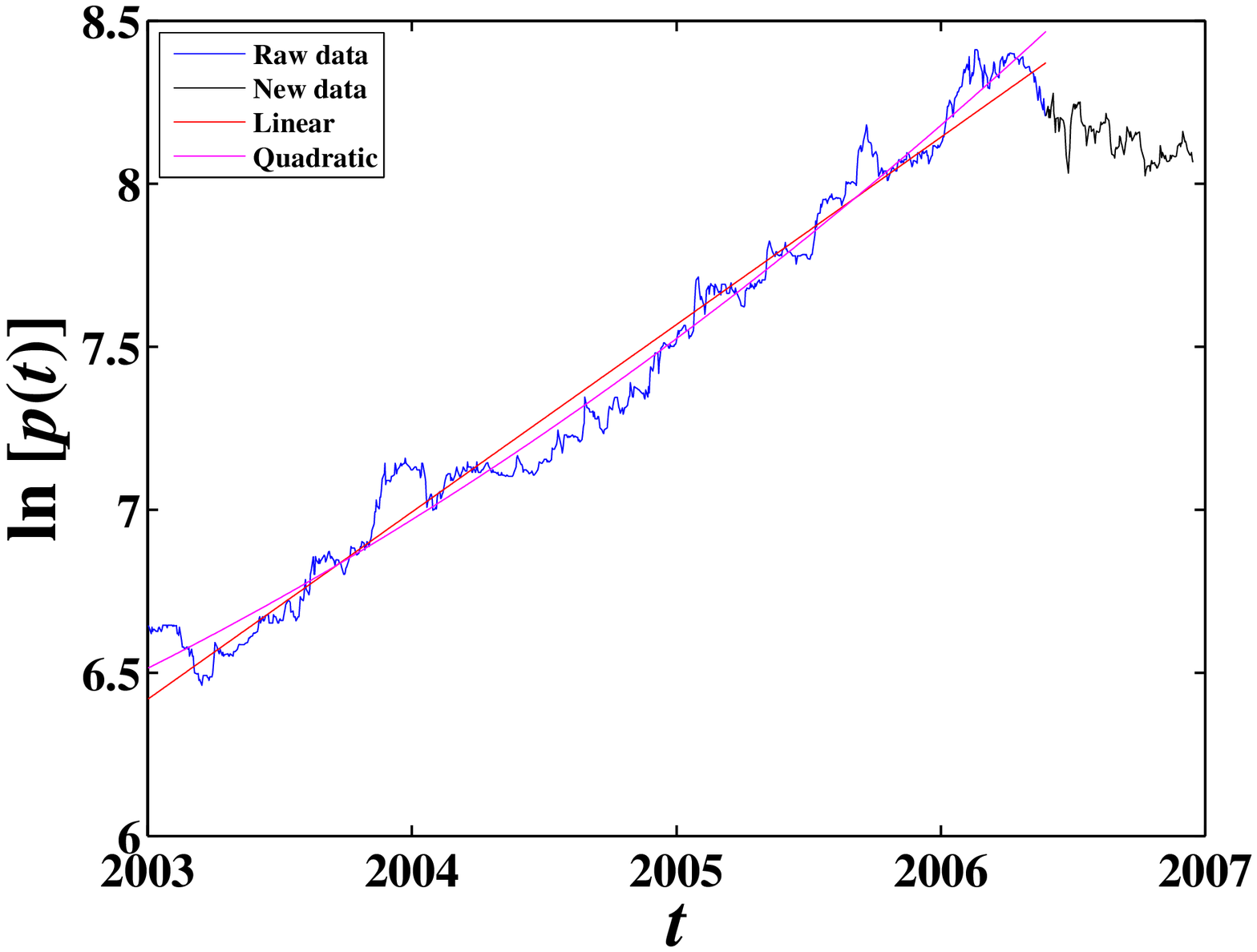}
  \end{minipage}}\\[10pt]
\end{center}
\caption{Linear fits and quadratic fits of the prices of stocks from
No. 1 to No. 9.} \label{Fig:SouthAfrica:LQ:1}
\end{figure}

\begin{figure}[htb]
\begin{center}
  \subfigure[Stock No.10: ASA]{
  \label{Fig:SouthAfrica:LQ:ASA}
  \begin{minipage}[b]{0.31\textwidth}
    \includegraphics[width=4.5cm,height=4.5cm]{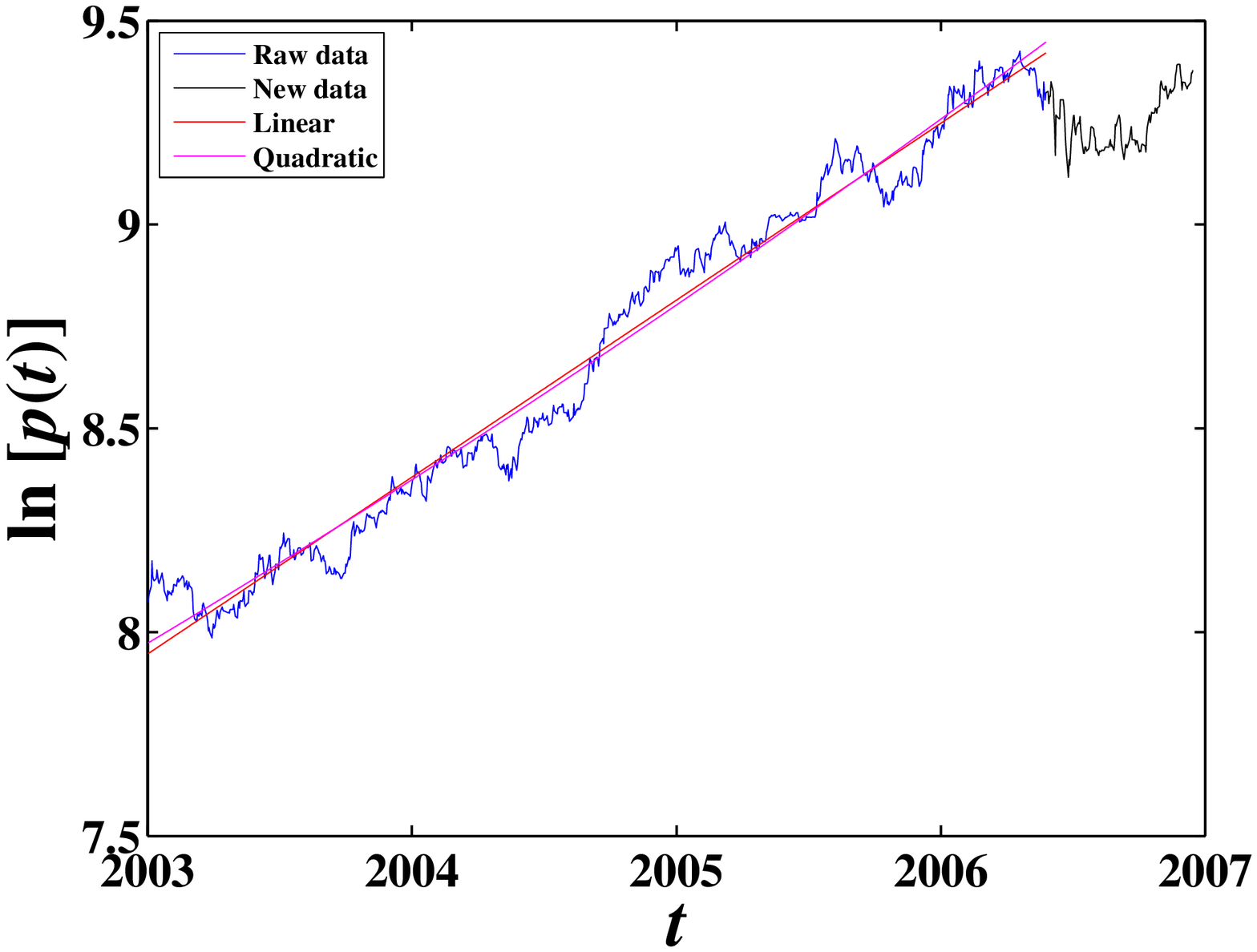}
  \end{minipage}}
  \hspace{0.1cm}
  \subfigure[Stock No.11: BAW]{
  \label{Fig:SouthAfrica:LQ:BAW}
  \begin{minipage}[b]{0.31\textwidth}
    \includegraphics[width=4.5cm,height=4.5cm]{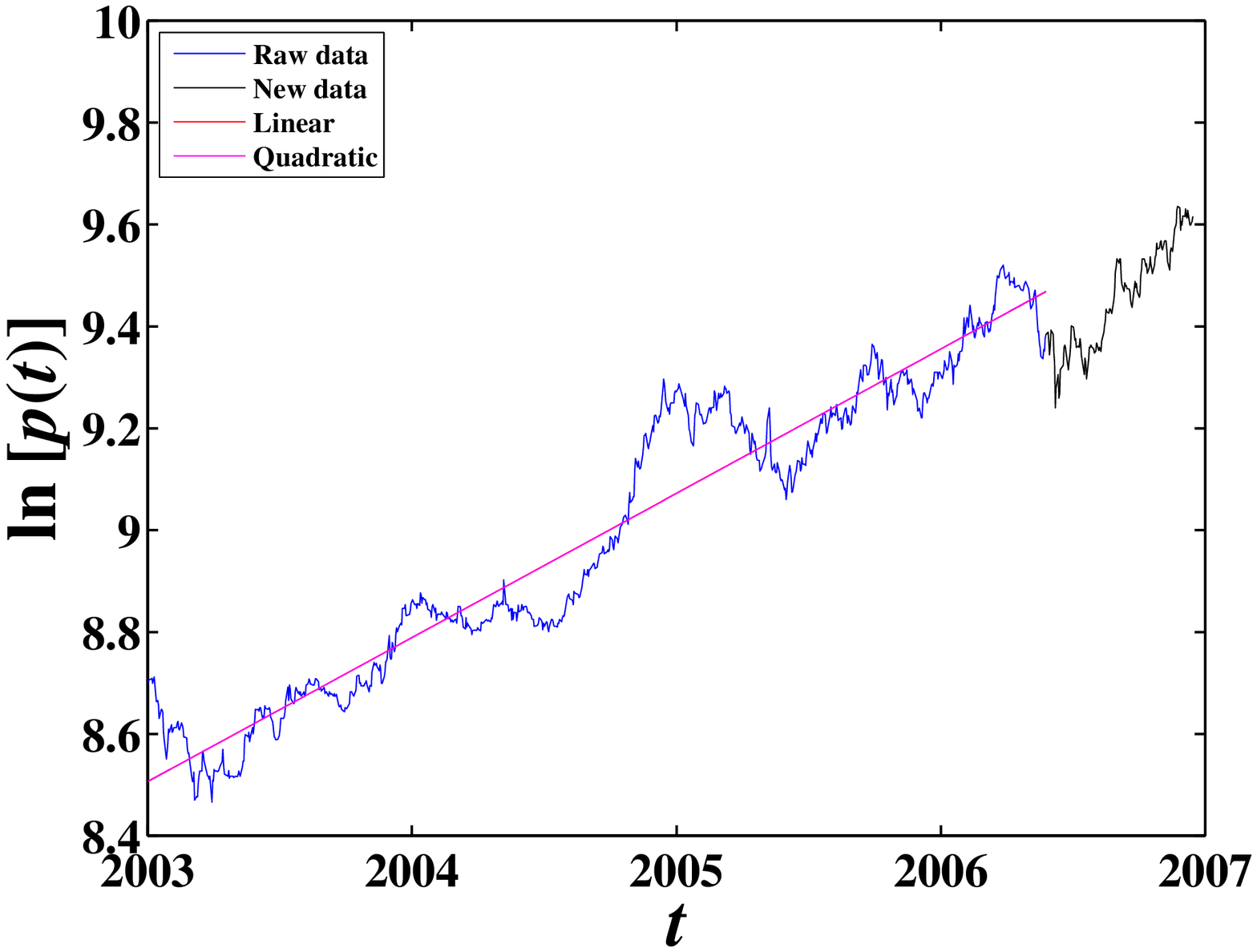}
  \end{minipage}}
  \hspace{0.1cm}
  \subfigure[Stock No.12: BIL]{
  \label{Fig:SouthAfrica:LQ:BIL}
  \begin{minipage}[b]{0.31\textwidth}
    \includegraphics[width=4.5cm,height=4.5cm]{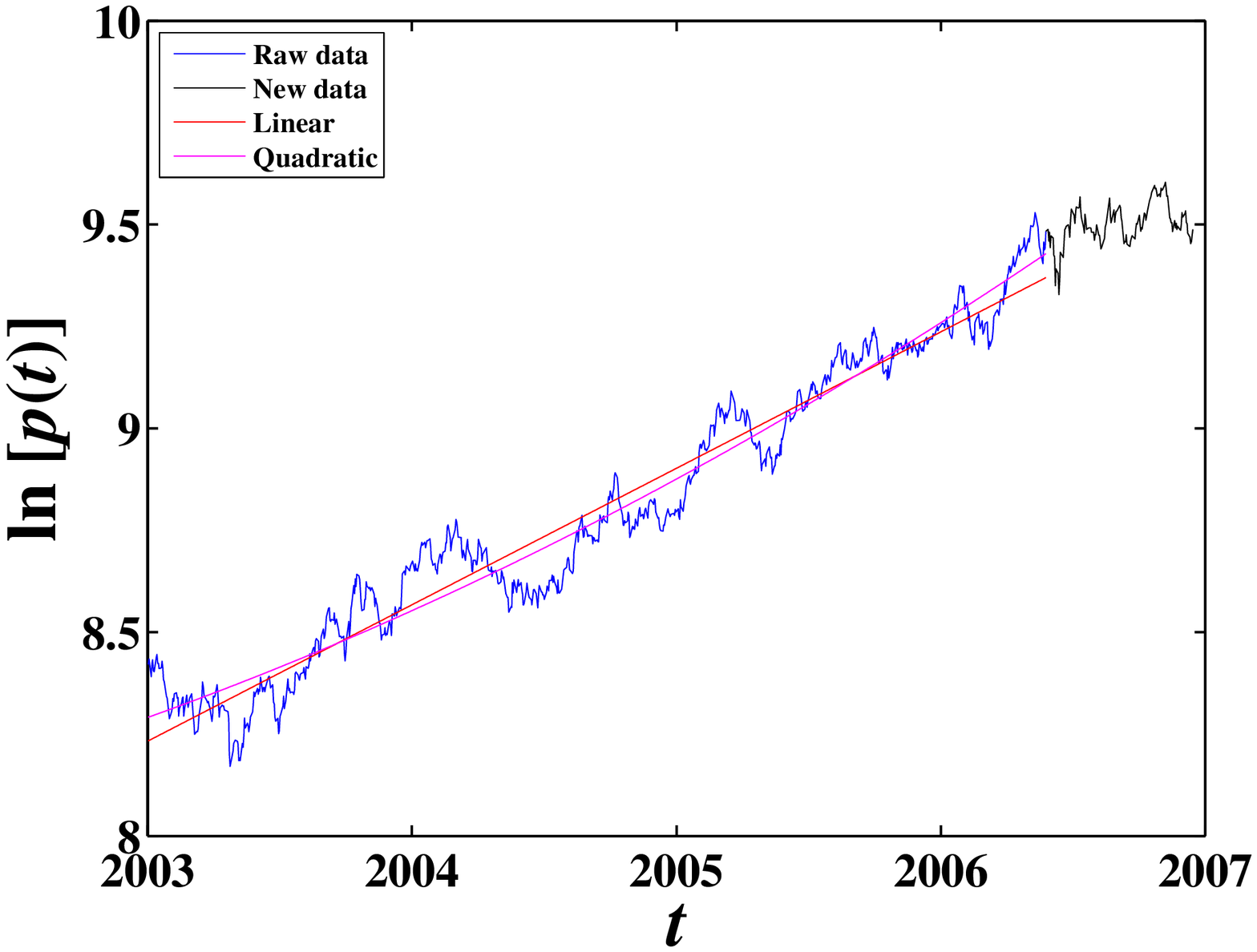}
  \end{minipage}}\\[10pt]
  \subfigure[Stock No.13: BVT]{
  \label{Fig:SouthAfrica:LQ:BVT}
  \begin{minipage}[b]{0.31\textwidth}
    \includegraphics[width=4.5cm,height=4.5cm]{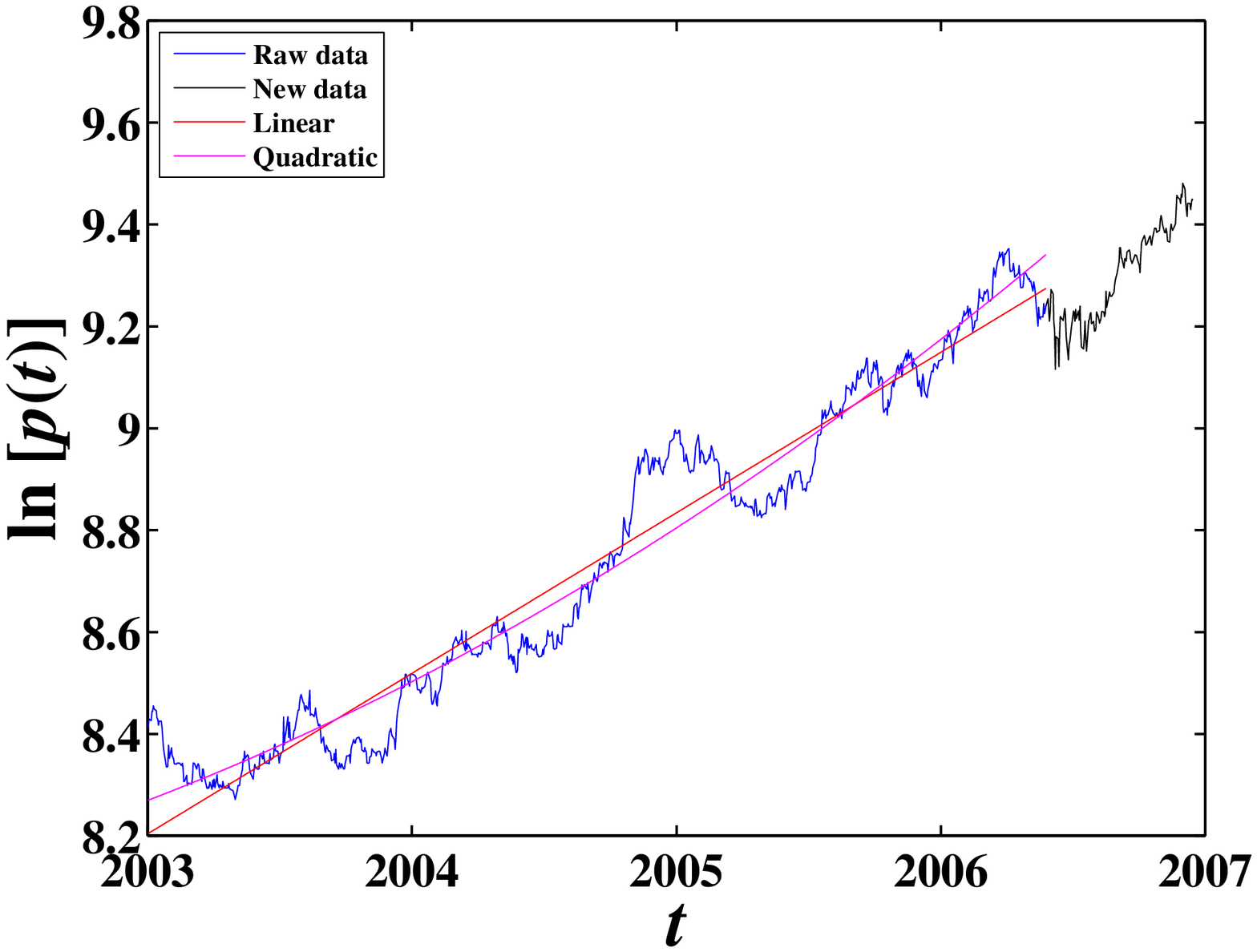}
  \end{minipage}}
  \hspace{0.1cm}
  \subfigure[Stock No.14: ECO]{
  \label{Fig:SouthAfrica:LQ:ECO}
  \begin{minipage}[b]{0.31\textwidth}
    \includegraphics[width=4.5cm,height=4.5cm]{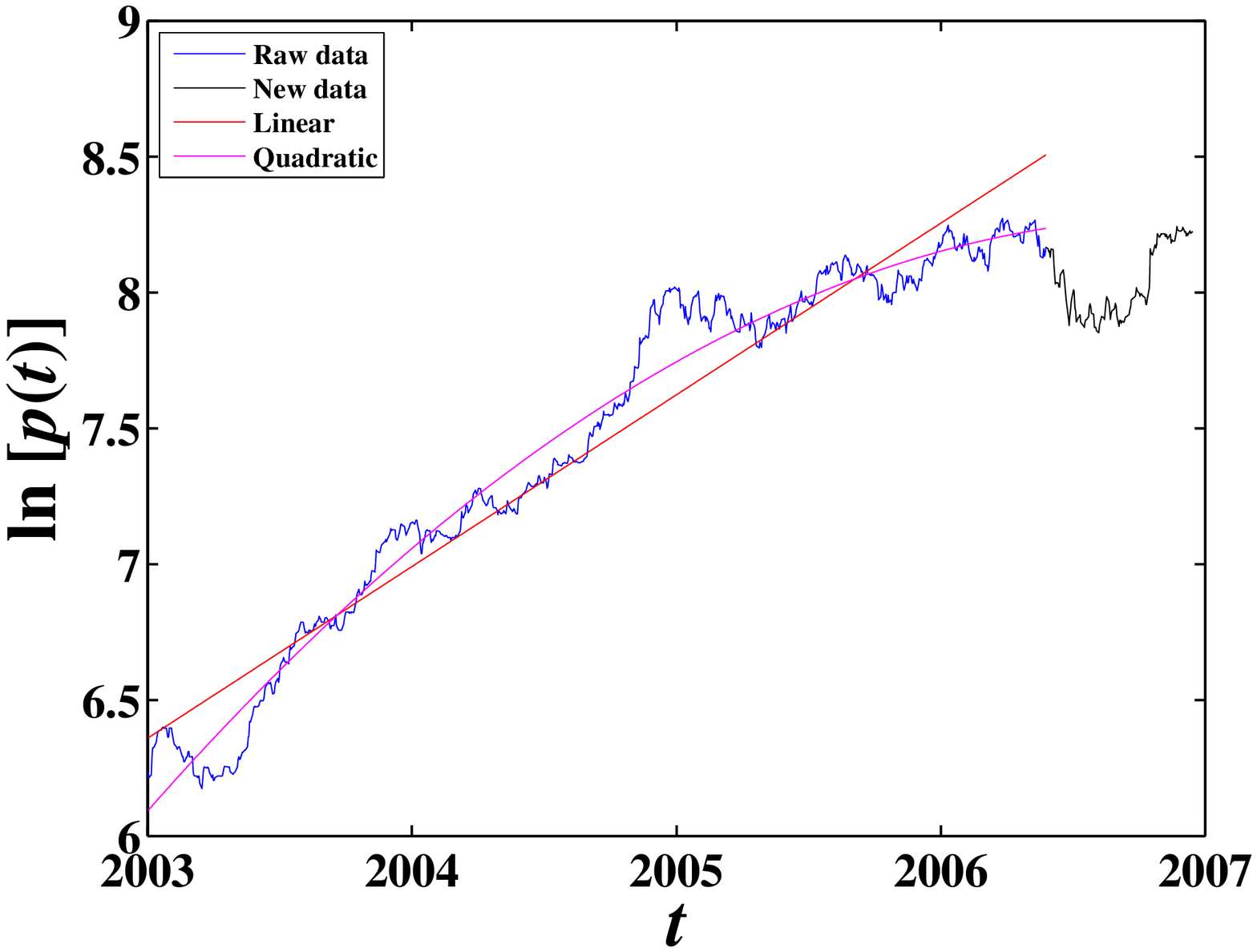}
  \end{minipage}}
  \hspace{0.1cm}
  \subfigure[Stock No.15: FSR]{
  \label{Fig:SouthAfrica:LQ:FSR}
  \begin{minipage}[b]{0.31\textwidth}
    \includegraphics[width=4.5cm,height=4.5cm]{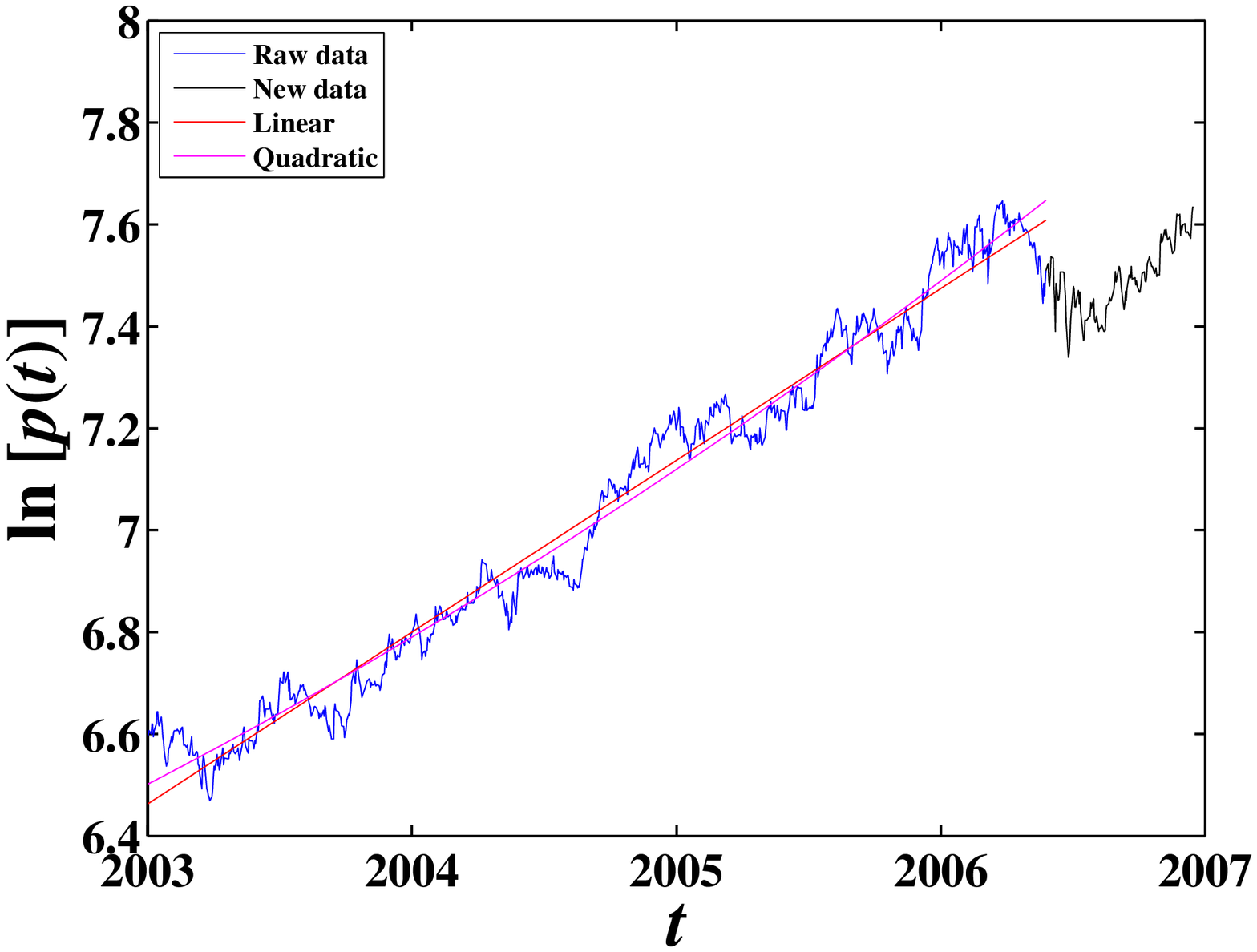}
  \end{minipage}}\\[10pt]
  \subfigure[Stock No.16: GFI]{
  \label{Fig:SouthAfrica:LQ:GFI}
  \begin{minipage}[b]{0.31\textwidth}
    \includegraphics[width=4.5cm,height=4.5cm]{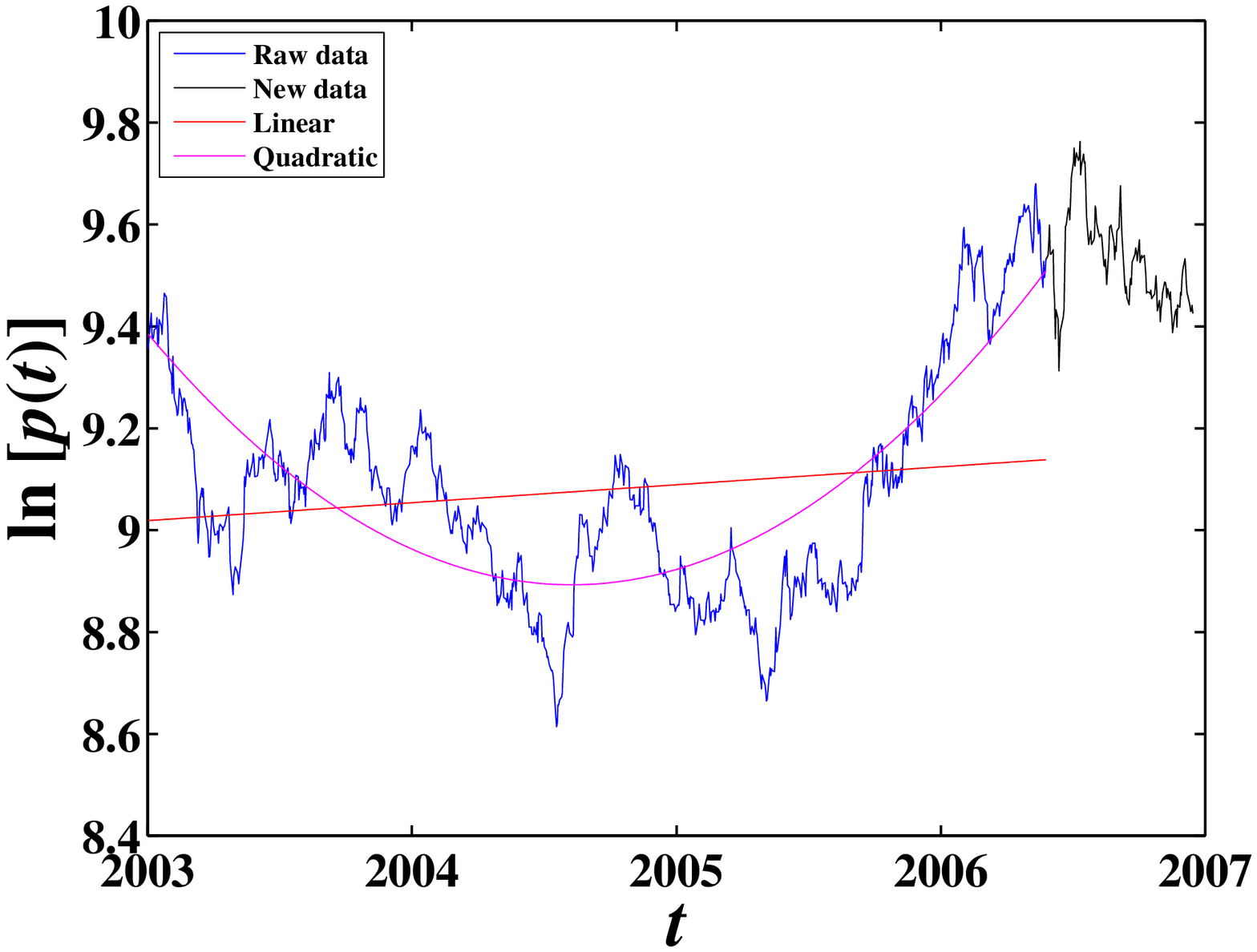}
  \end{minipage}}
  \hspace{0.1cm}
  \subfigure[Stock No.17: HAR]{
  \label{Fig:SouthAfrica:LQ:HAR}
  \begin{minipage}[b]{0.31\textwidth}
    \includegraphics[width=4.5cm,height=4.5cm]{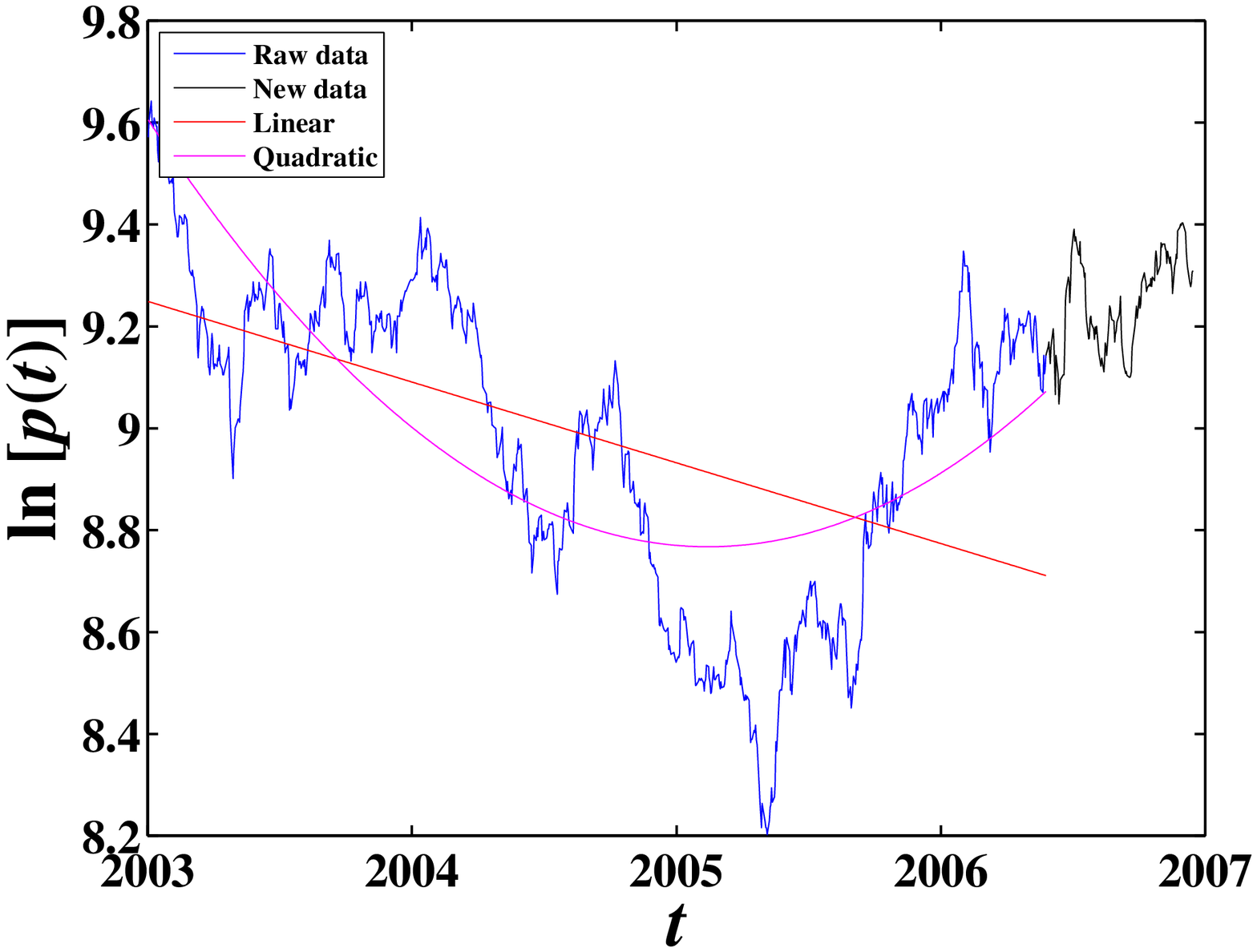}
  \end{minipage}}
  \hspace{0.1cm}
  \subfigure[Stock No.18: IMP]{
  \label{Fig:SouthAfrica:LQ:IMP}
  \begin{minipage}[b]{0.31\textwidth}
    \includegraphics[width=4.5cm,height=4.5cm]{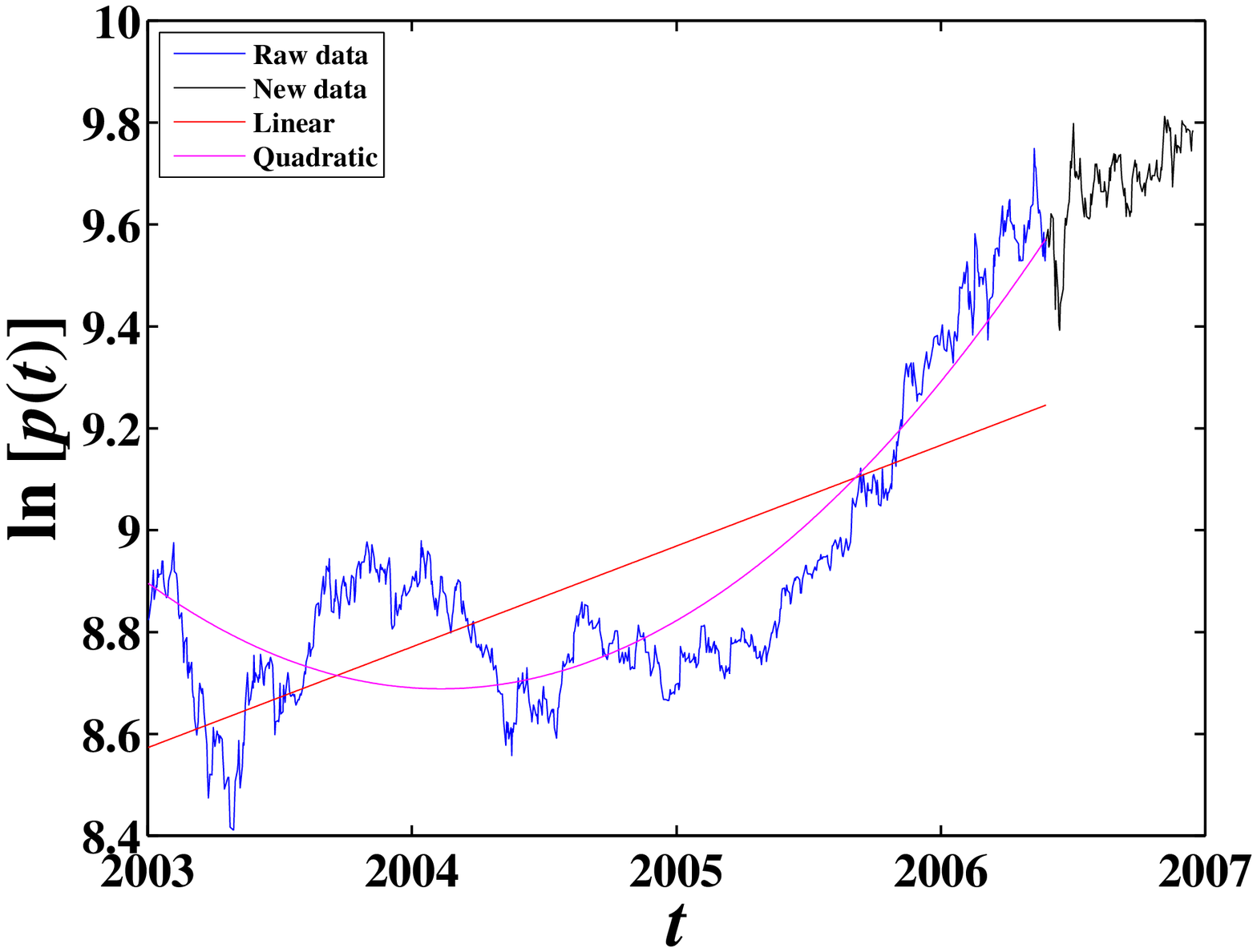}
  \end{minipage}}\\[10pt]
\end{center}
\caption{Linear fits and quadratic fits of the prices of stocks from
No.10 to No.18.} \label{Fig:SouthAfrica:LQ:2}
\end{figure}

\begin{figure}[htb]
\begin{center}
  \subfigure[Stock No.19: INL]{
  \label{Fig:SouthAfrica:LQ:INL}
  \begin{minipage}[b]{0.31\textwidth}
    \includegraphics[width=4.5cm,height=4.5cm]{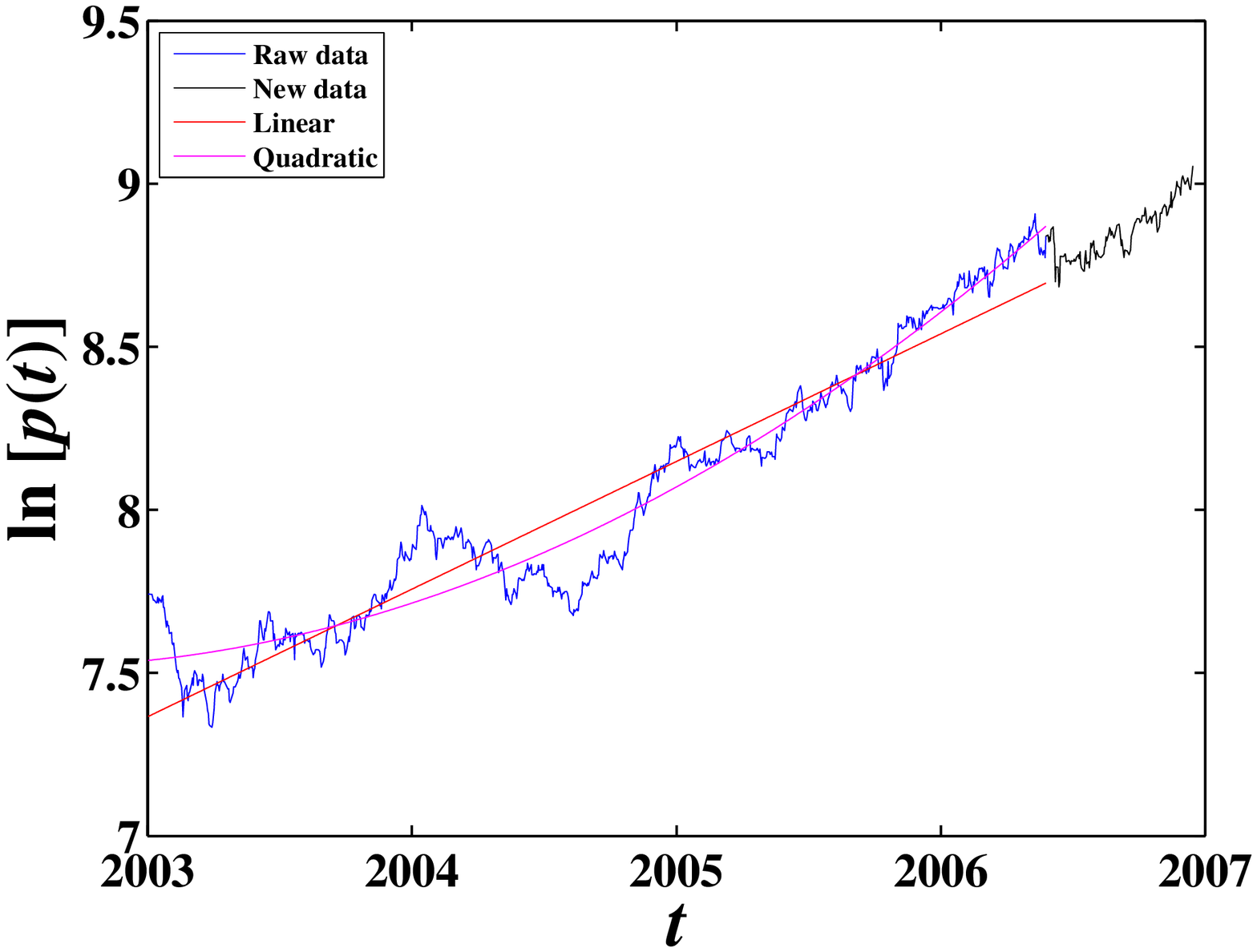}
  \end{minipage}}
  \hspace{0.1cm}
  \subfigure[Stock No.20: INP]{
  \label{Fig:SouthAfrica:LQ:INP}
  \begin{minipage}[b]{0.31\textwidth}
    \includegraphics[width=4.5cm,height=4.5cm]{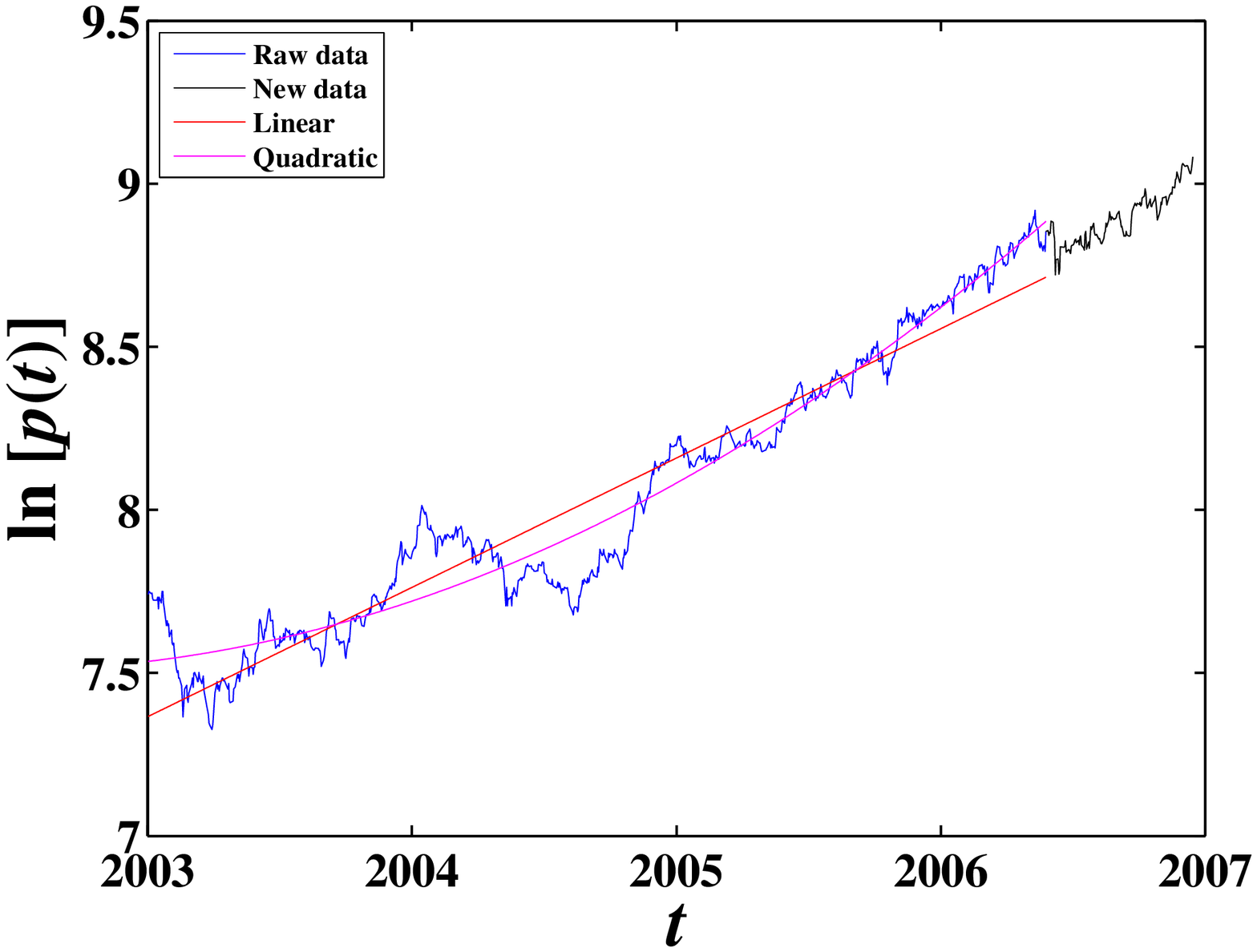}
  \end{minipage}}
  \hspace{0.1cm}
  \subfigure[Stock No.21: IPL]{
  \label{Fig:SouthAfrica:LQ:IPL}
  \begin{minipage}[b]{0.31\textwidth}
    \includegraphics[width=4.5cm,height=4.5cm]{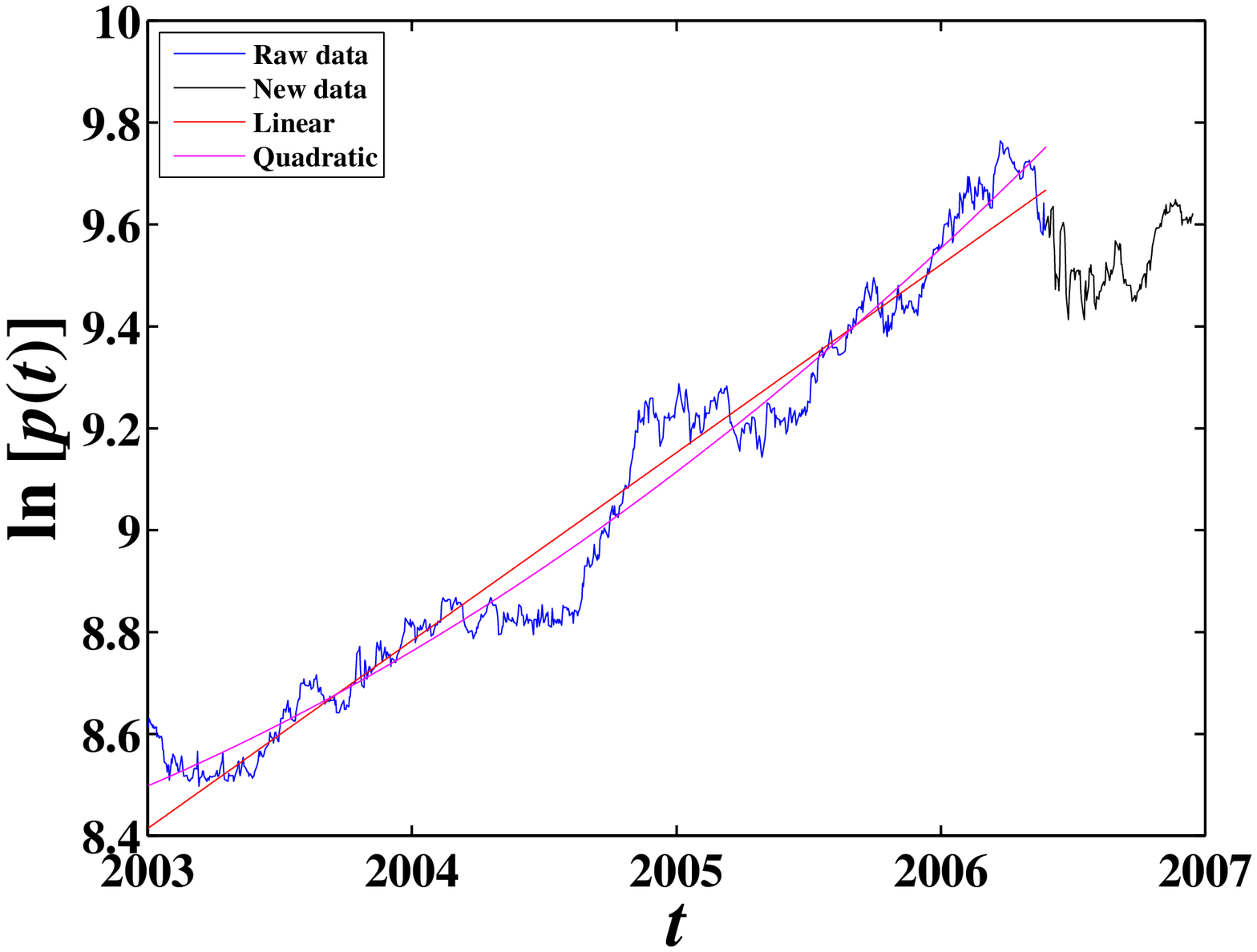}
  \end{minipage}}\\[10pt]
  \subfigure[Stock No.22: JDG]{
  \label{Fig:SouthAfrica:LQ:JDG}
  \begin{minipage}[b]{0.31\textwidth}
    \includegraphics[width=4.5cm,height=4.5cm]{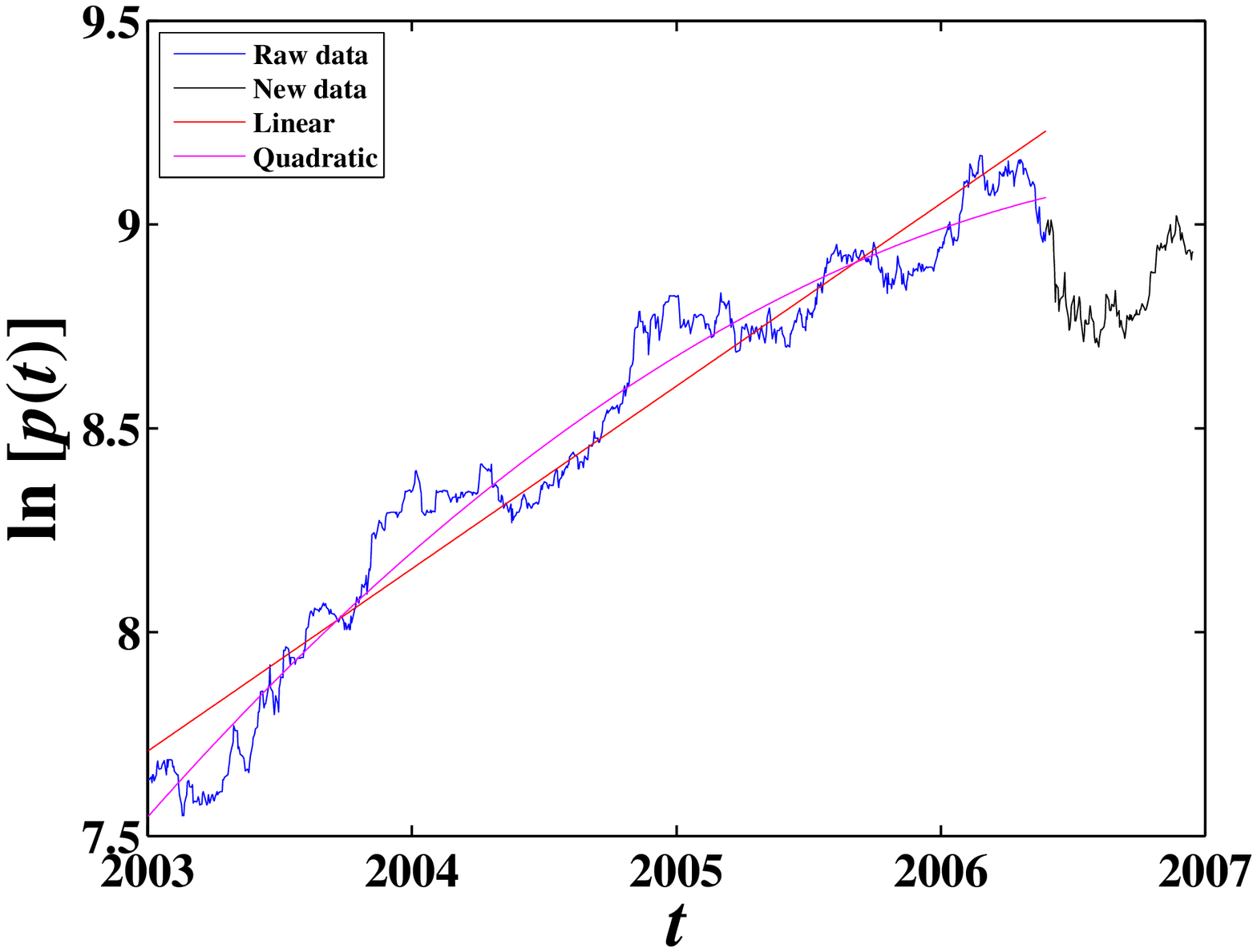}
  \end{minipage}}
  \hspace{0.1cm}
  \subfigure[Stock No.23: KMB]{
  \label{Fig:SouthAfrica:LQ:KMB}
  \begin{minipage}[b]{0.31\textwidth}
    \includegraphics[width=4.5cm,height=4.5cm]{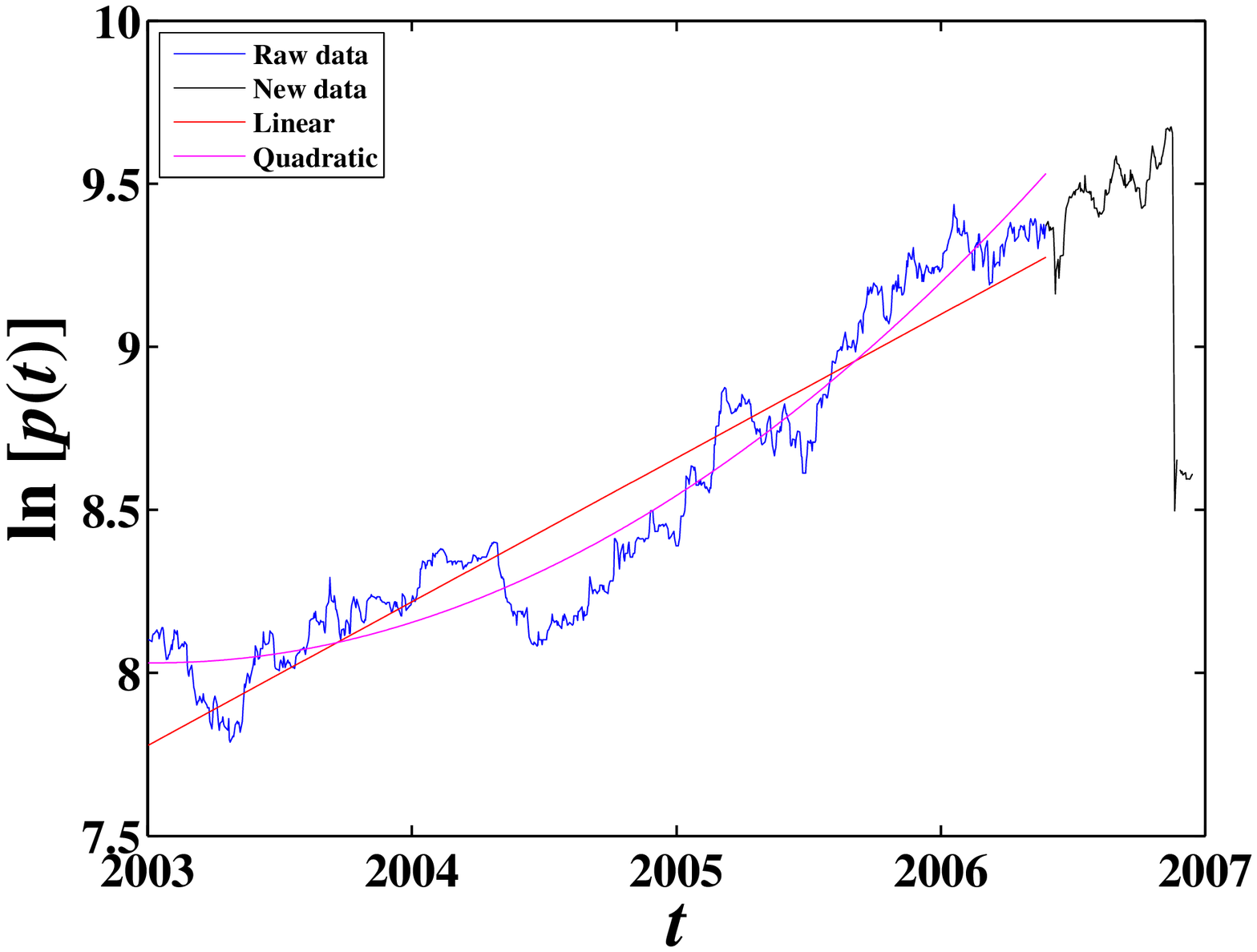}
  \end{minipage}}
  \hspace{0.1cm}
  \subfigure[Stock No.24: LBT]{
  \label{Fig:SouthAfrica:LQ:LBT}
  \begin{minipage}[b]{0.31\textwidth}
    \includegraphics[width=4.5cm,height=4.5cm]{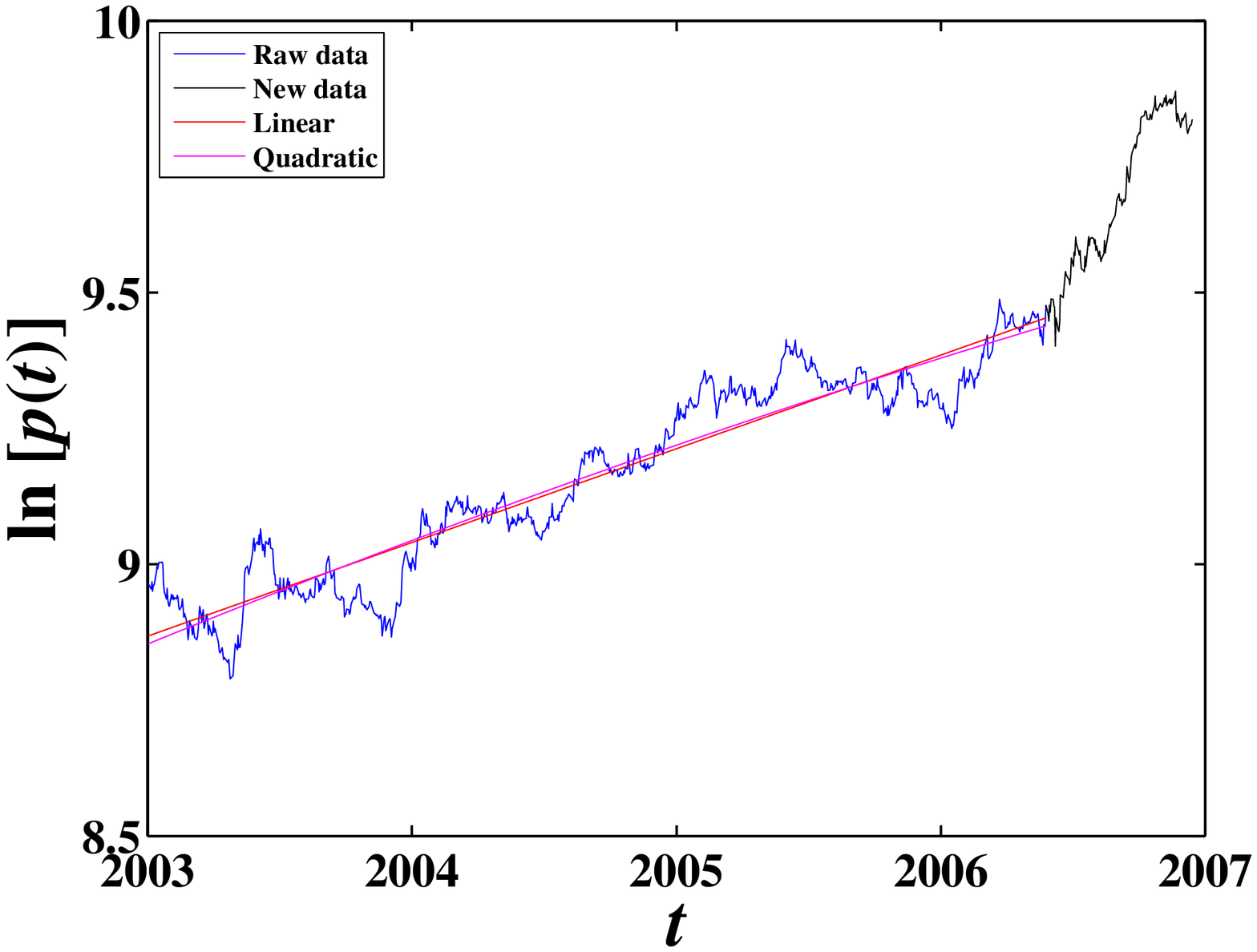}
  \end{minipage}}\\[10pt]
  \subfigure[Stock No.25: LGL]{
  \label{Fig:SouthAfrica:LQ:LGL}
  \begin{minipage}[b]{0.31\textwidth}
    \includegraphics[width=4.5cm,height=4.5cm]{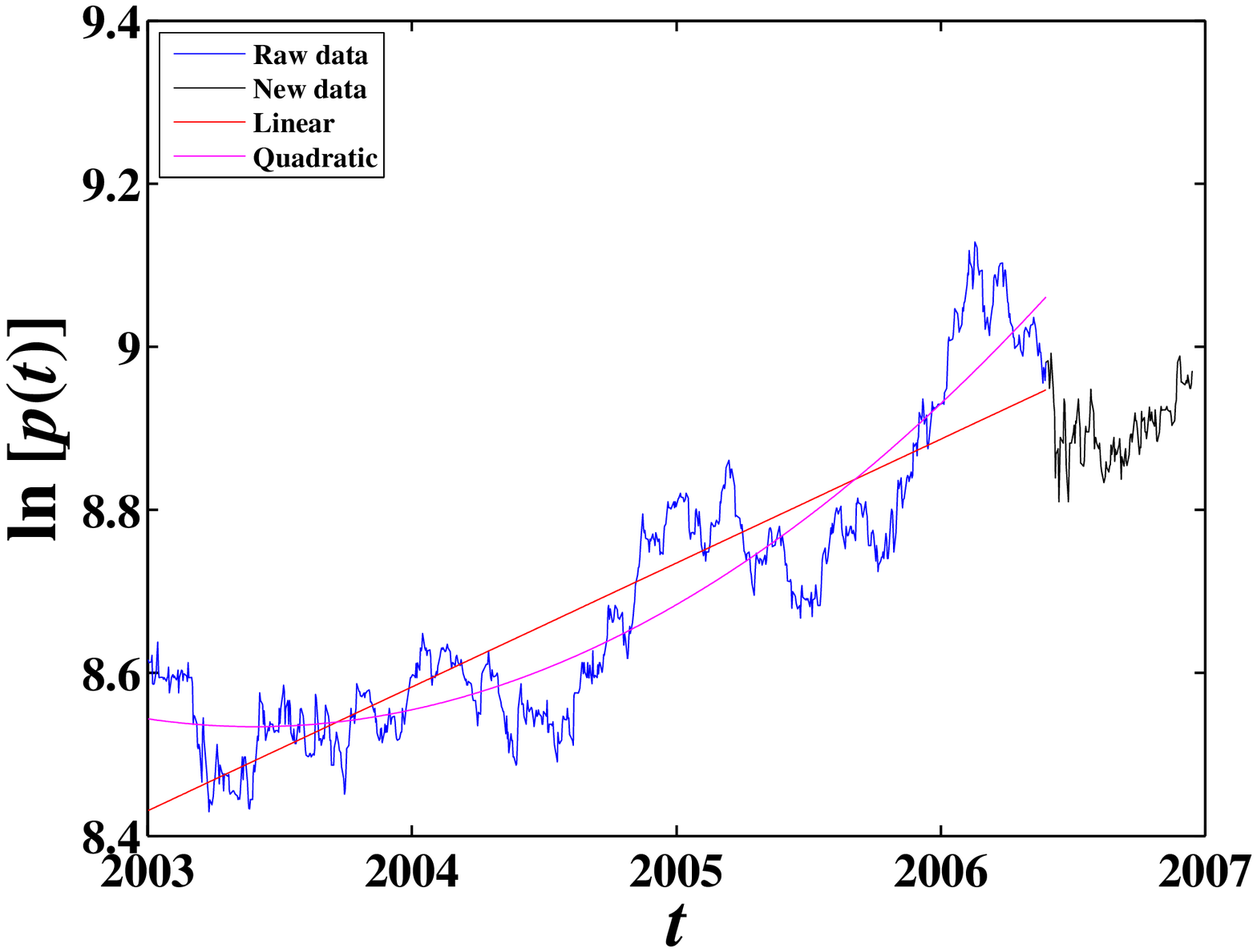}
  \end{minipage}}
  \hspace{0.1cm}
  \subfigure[Stock No.26: MLA]{
  \label{Fig:SouthAfrica:LQ:MLA}
  \begin{minipage}[b]{0.31\textwidth}
    \includegraphics[width=4.5cm,height=4.5cm]{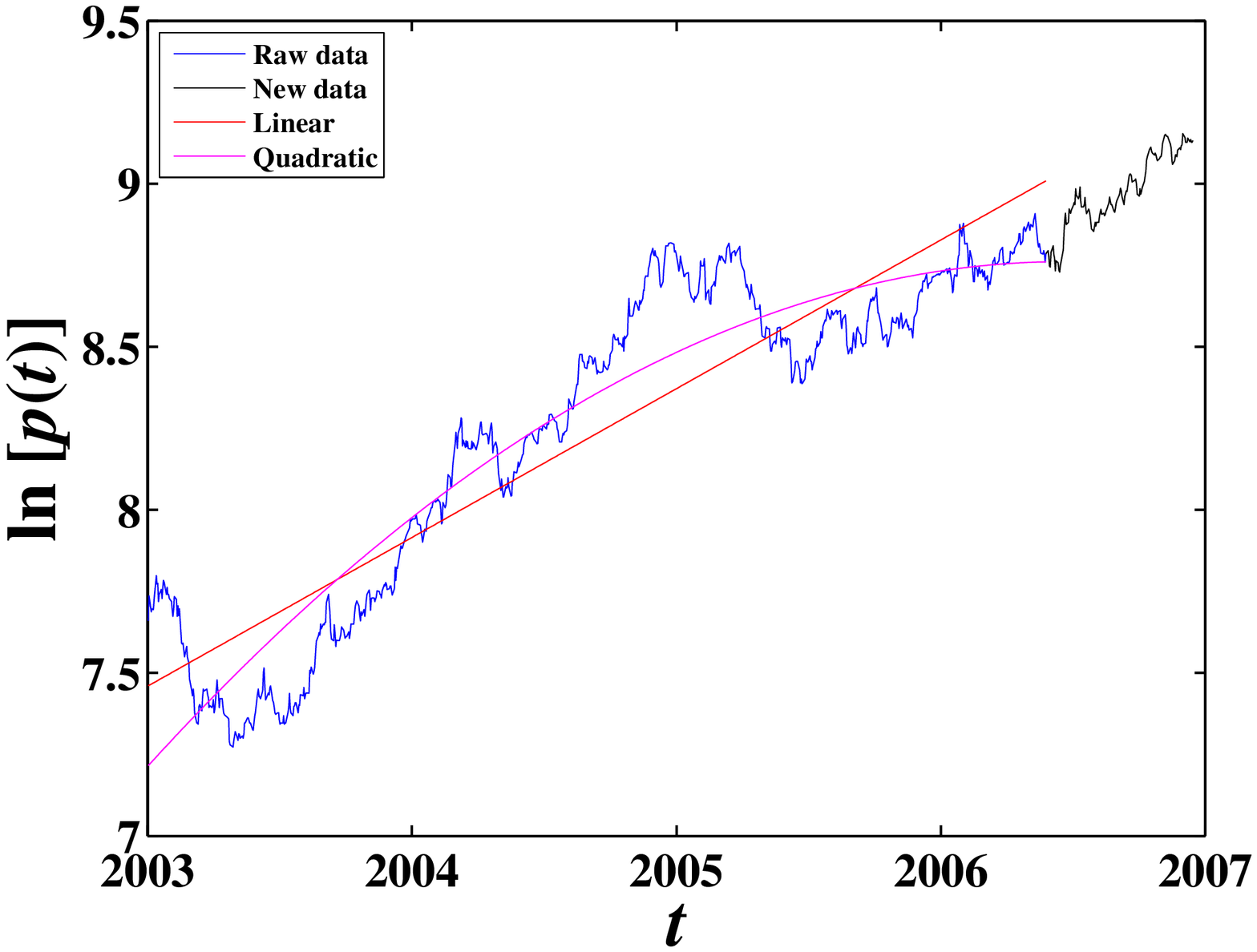}
  \end{minipage}}
  \hspace{0.1cm}
  \subfigure[Stock No.27: MTN]{
  \label{Fig:SouthAfrica:LQ:MTN}
  \begin{minipage}[b]{0.31\textwidth}
    \includegraphics[width=4.5cm,height=4.5cm]{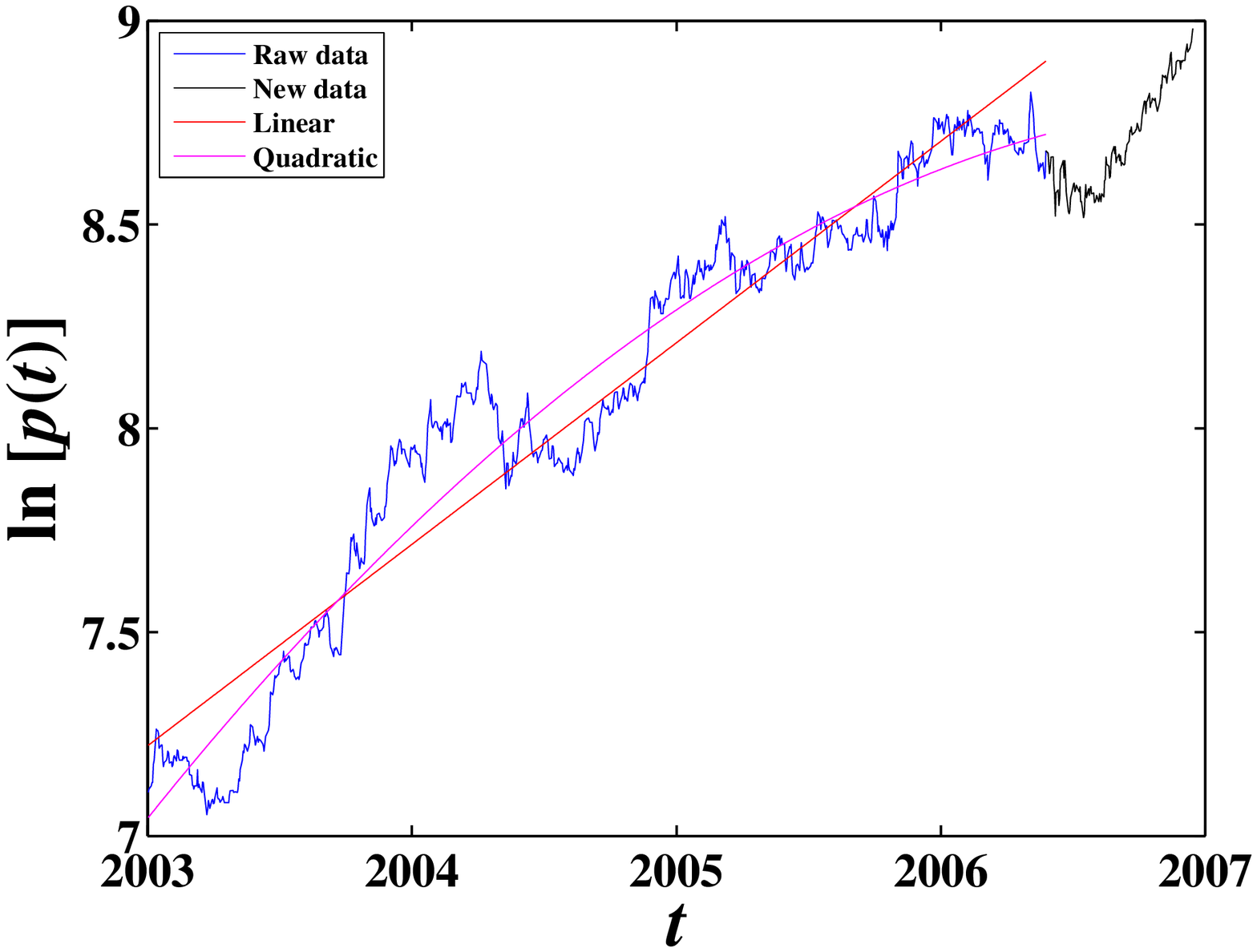}
  \end{minipage}}\\[10pt]
\end{center}
\caption{Linear fits and quadratic fits of the prices of stocks from
No.19 to No.27.} \label{Fig:SouthAfrica:LQ:3}
\end{figure}

\begin{figure}[htb]
\begin{center}
  \subfigure[Stock No.28: NED]{
  \label{Fig:SouthAfrica:LQ:NED}
  \begin{minipage}[b]{0.31\textwidth}
    \includegraphics[width=4.5cm,height=4.5cm]{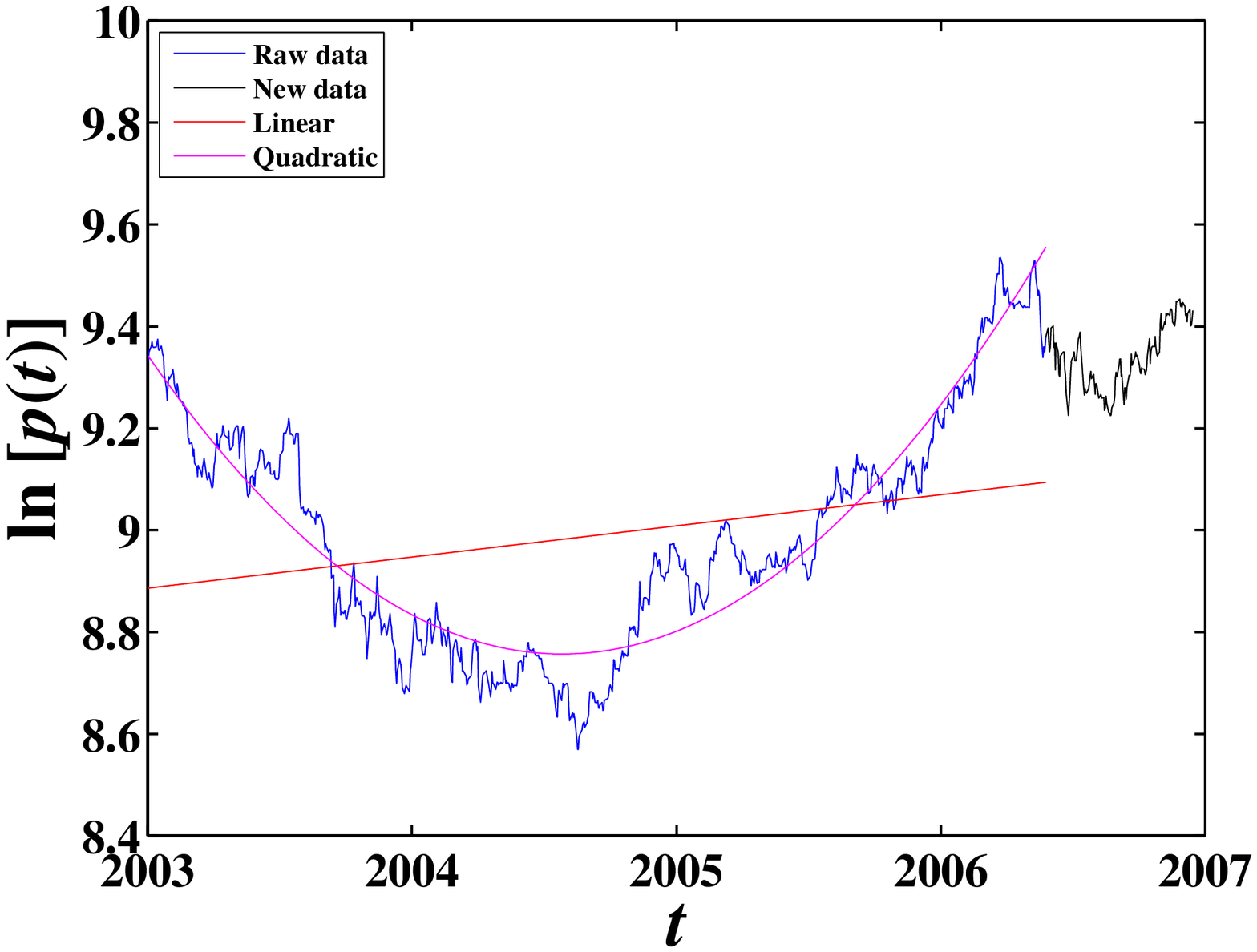}
  \end{minipage}}
  \hspace{0.1cm}
  \subfigure[Stock No.29: NPN]{
  \label{Fig:SouthAfrica:LQ:NPN}
  \begin{minipage}[b]{0.31\textwidth}
    \includegraphics[width=4.5cm,height=4.5cm]{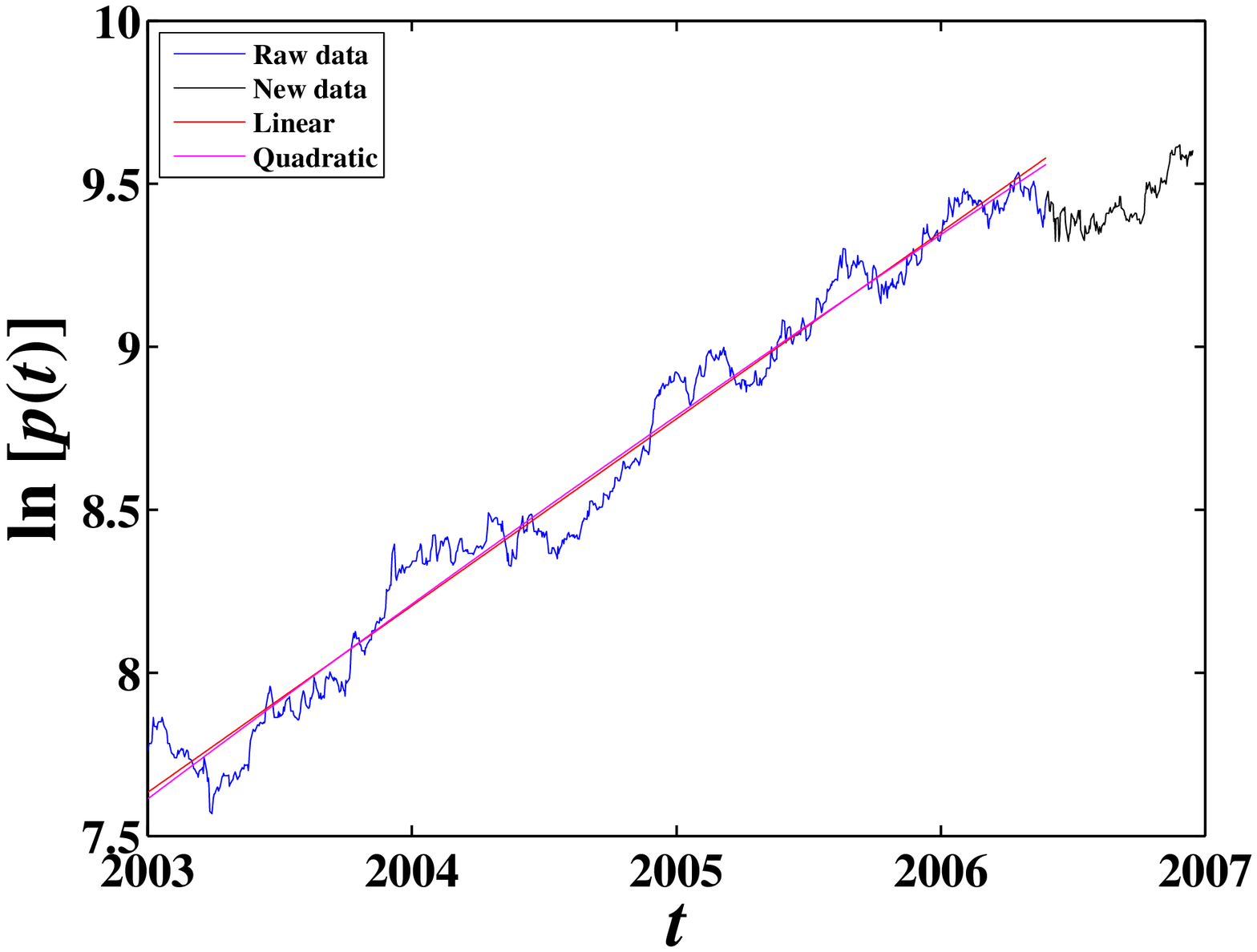}
  \end{minipage}}
  \hspace{0.1cm}
  \subfigure[Stock No.30: NTC]{
  \label{Fig:SouthAfrica:LQ:NTC}
  \begin{minipage}[b]{0.31\textwidth}
    \includegraphics[width=4.5cm,height=4.5cm]{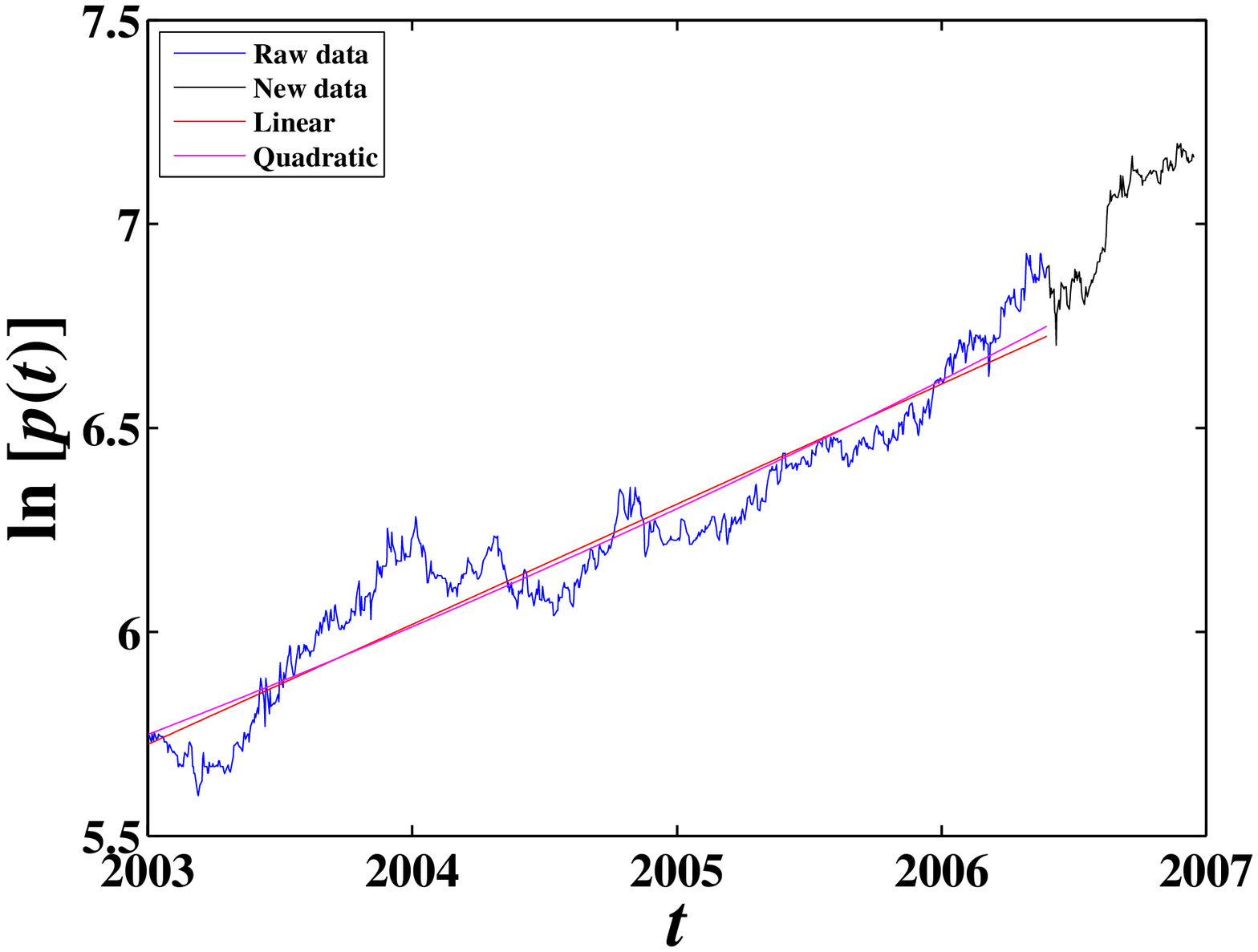}
  \end{minipage}}\\[10pt]
  \subfigure[Stock No.31: OML]{
  \label{Fig:SouthAfrica:LQ:OML}
  \begin{minipage}[b]{0.31\textwidth}
    \includegraphics[width=4.5cm,height=4.5cm]{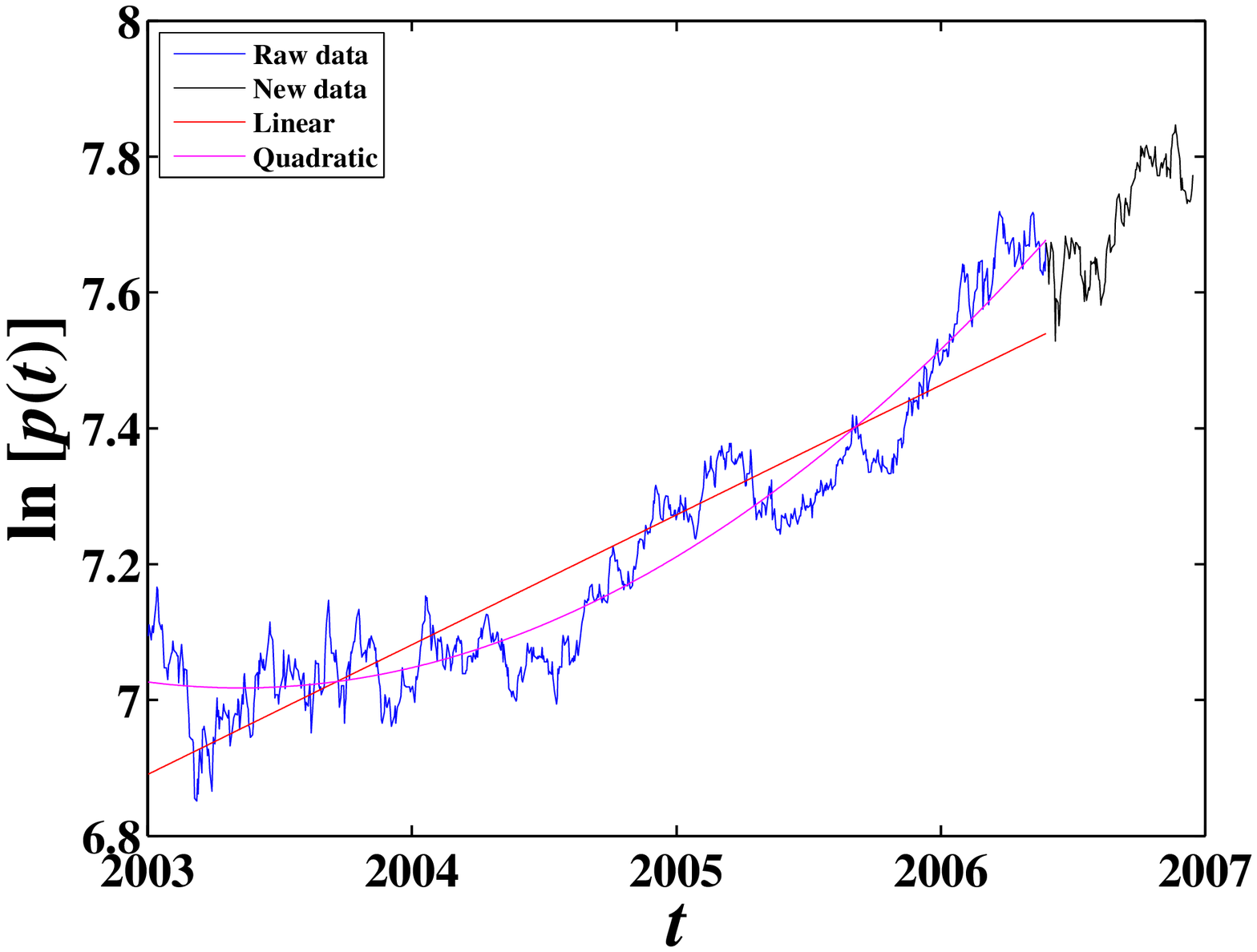}
  \end{minipage}}
  \hspace{0.1cm}
  \subfigure[Stock No.32: PIK]{
  \label{Fig:SouthAfrica:LQ:PIK}
  \begin{minipage}[b]{0.31\textwidth}
    \includegraphics[width=4.5cm,height=4.5cm]{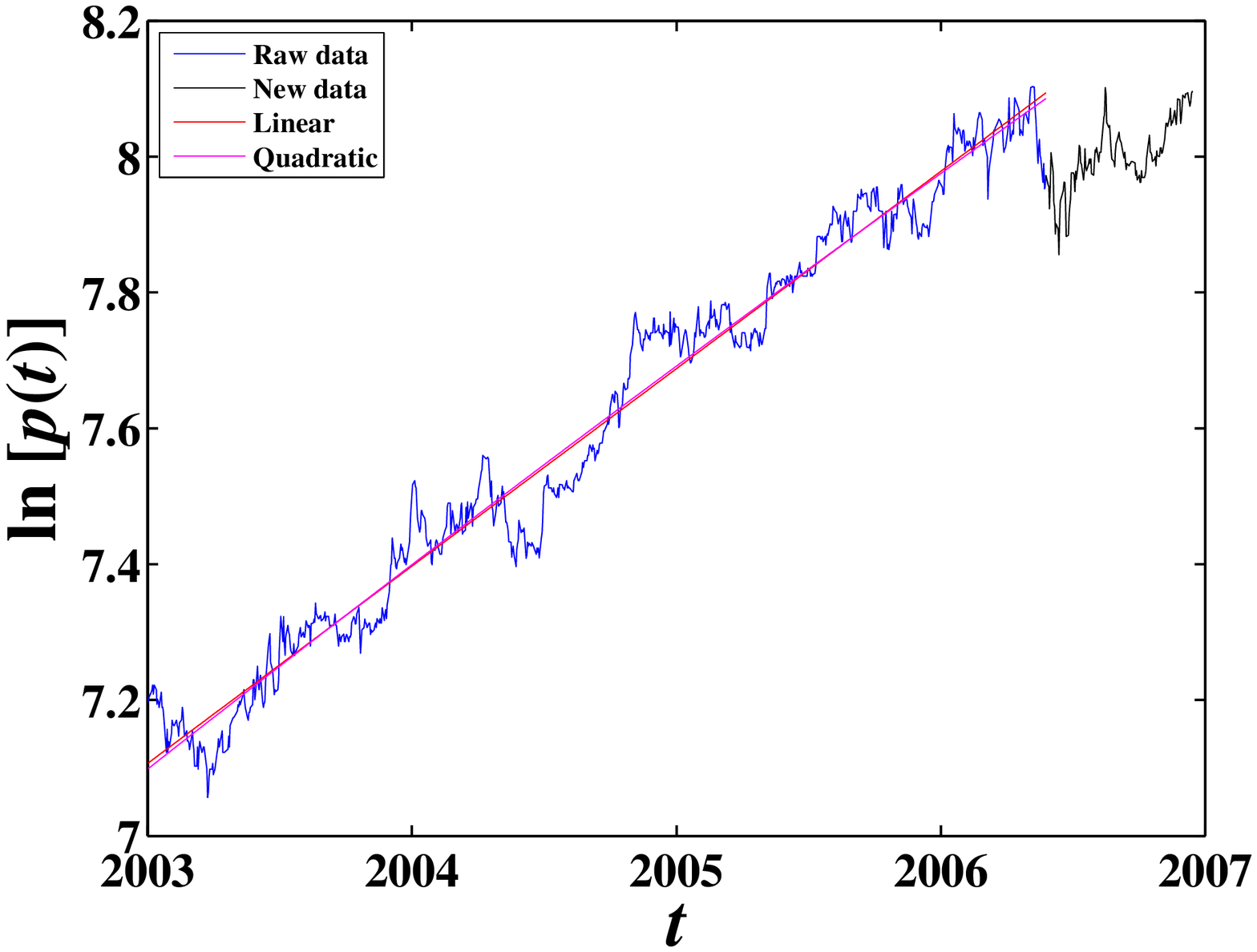}
  \end{minipage}}
  \hspace{0.1cm}
  \subfigure[Stock No.33: PPC]{
  \label{Fig:SouthAfrica:LQ:PPC}
  \begin{minipage}[b]{0.31\textwidth}
    \includegraphics[width=4.5cm,height=4.5cm]{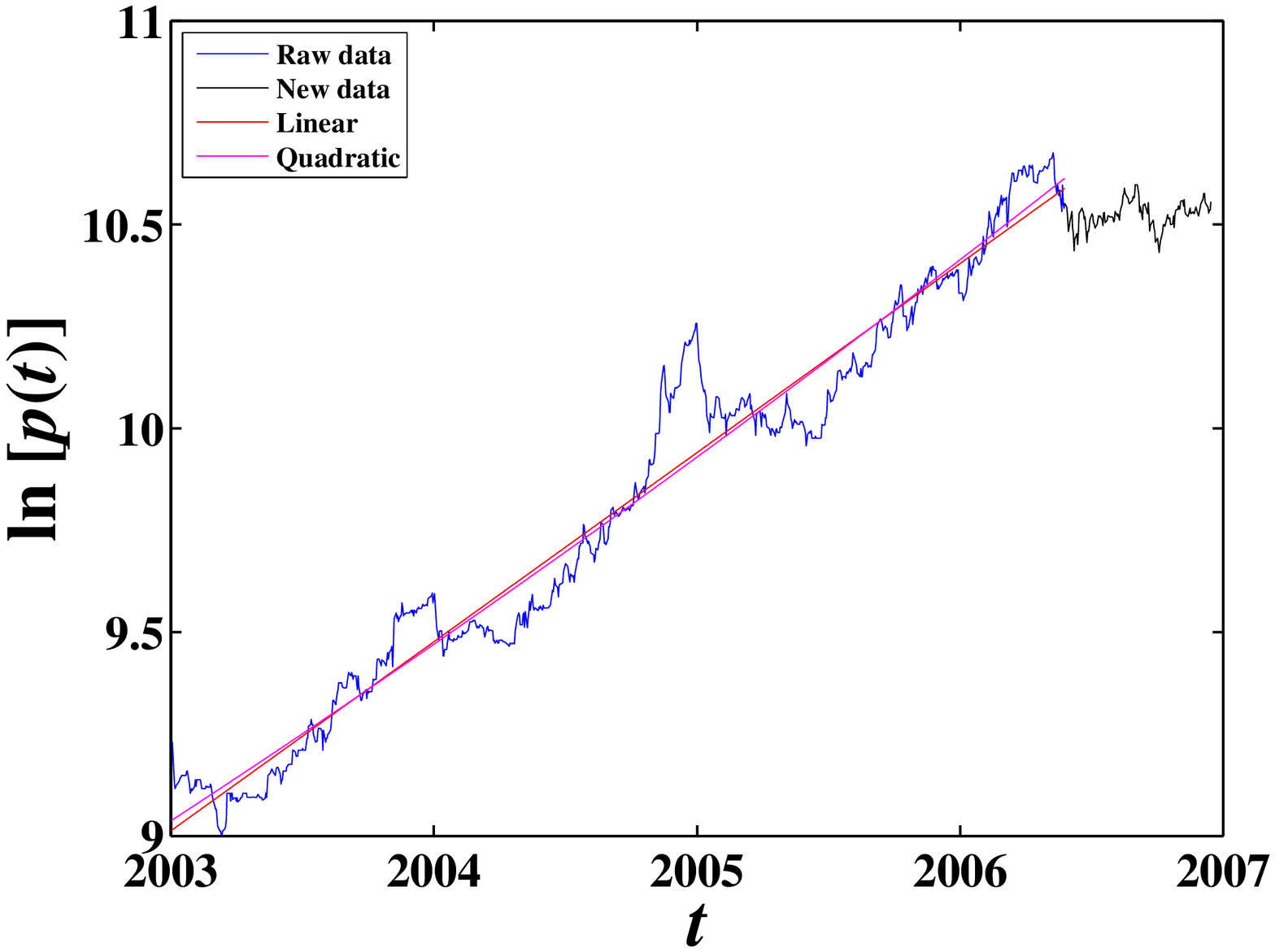}
  \end{minipage}}\\[10pt]
  \subfigure[Stock No.34: RCH]{
  \label{Fig:SouthAfrica:LQ:RCH}
  \begin{minipage}[b]{0.31\textwidth}
    \includegraphics[width=4.5cm,height=4.5cm]{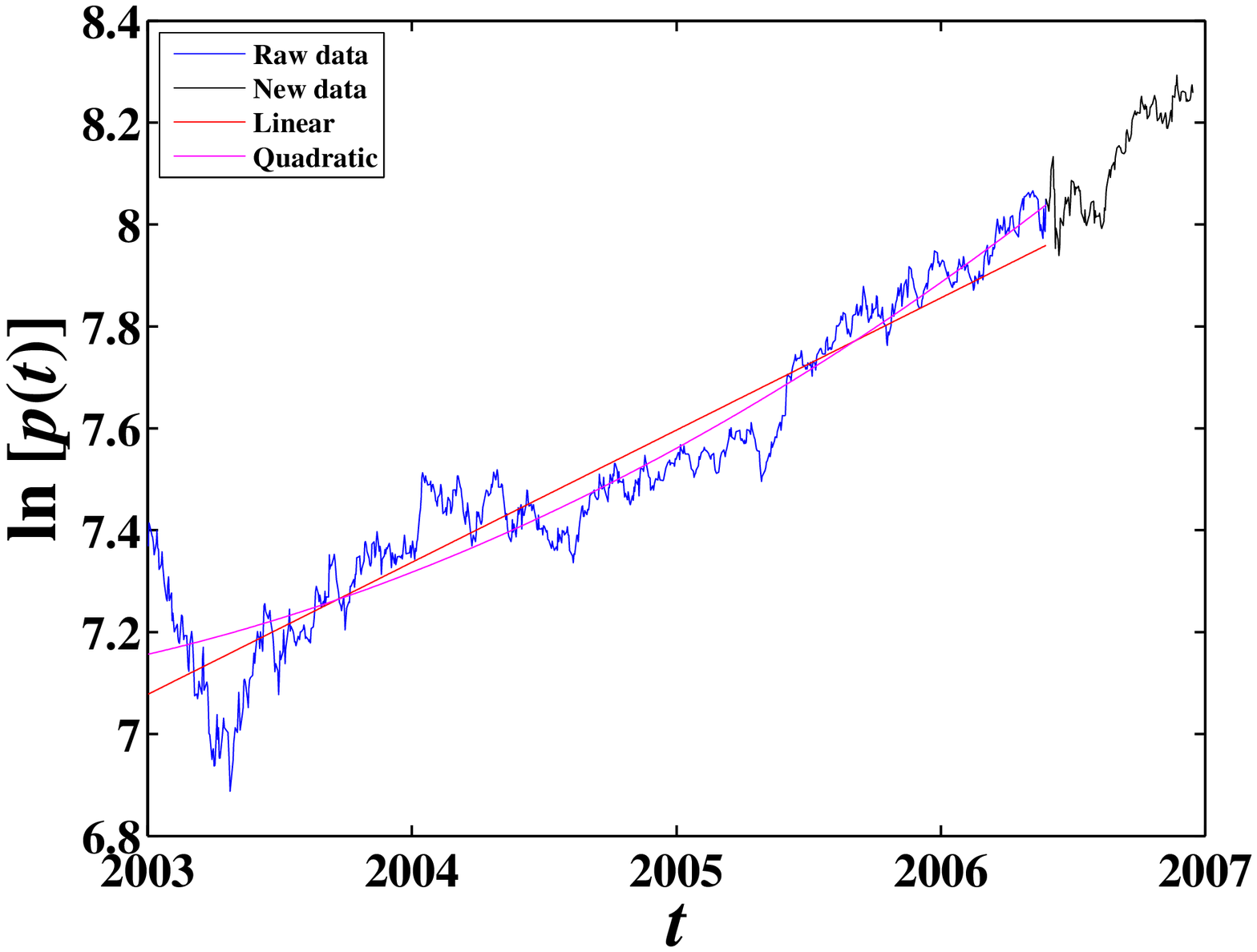}
  \end{minipage}}
  \hspace{0.1cm}
  \subfigure[Stock No.35: REM]{
  \label{Fig:SouthAfrica:LQ:REM}
  \begin{minipage}[b]{0.31\textwidth}
    \includegraphics[width=4.5cm,height=4.5cm]{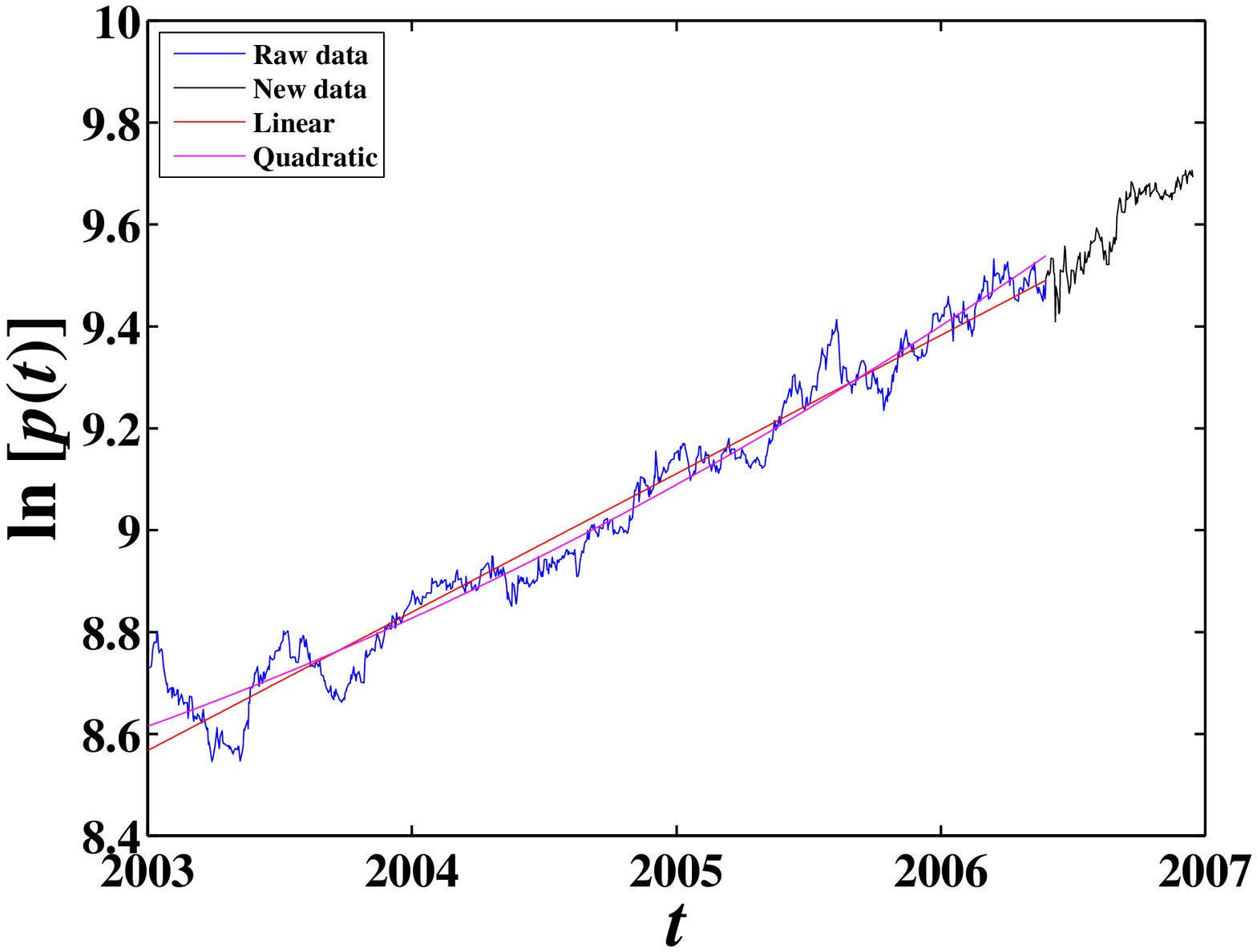}
  \end{minipage}}
  \hspace{0.1cm}
  \subfigure[Stock No.36: RMH]{
  \label{Fig:SouthAfrica:LQ:RMH}
  \begin{minipage}[b]{0.31\textwidth}
    \includegraphics[width=4.5cm,height=4.5cm]{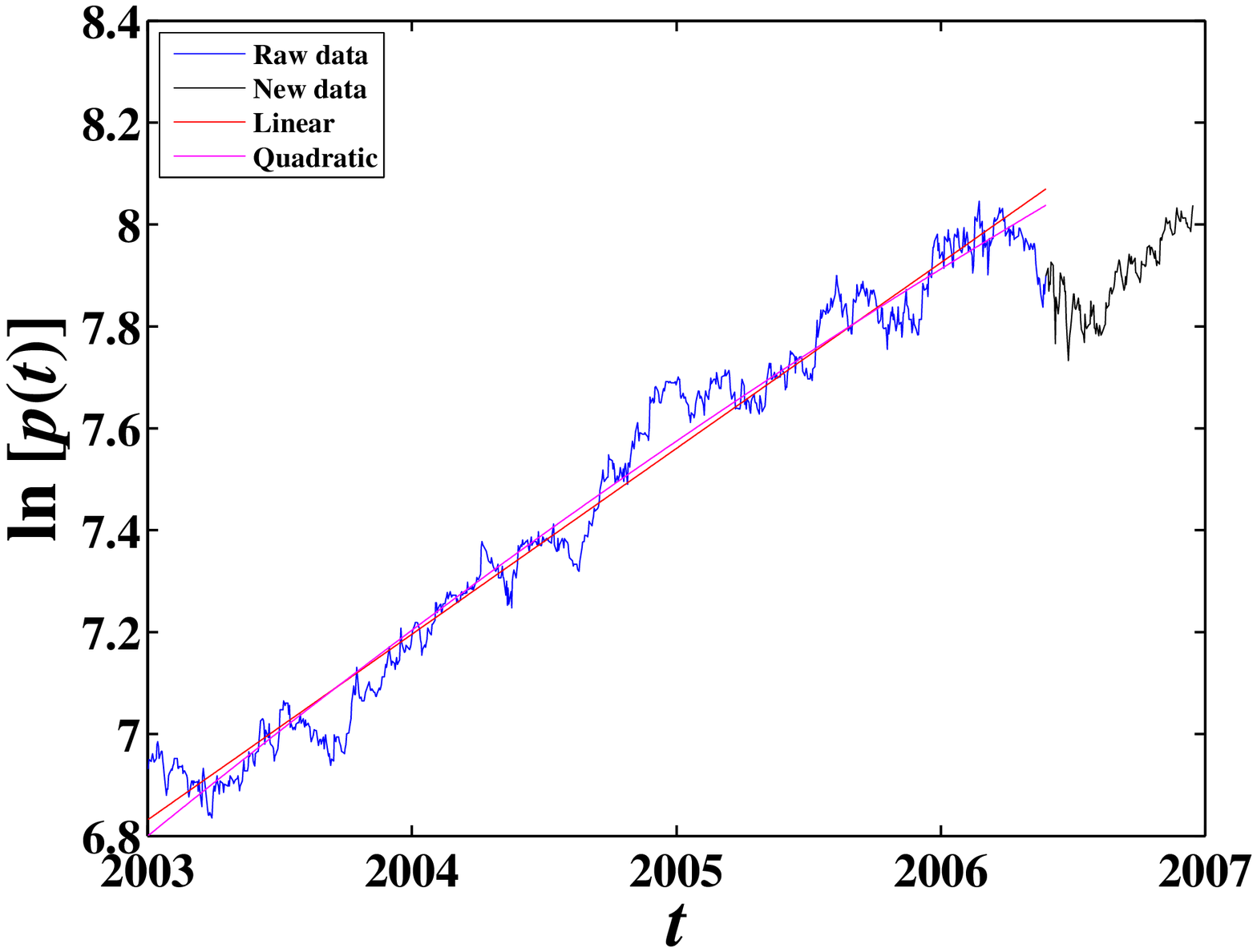}
  \end{minipage}}\\[10pt]
\end{center}
\caption{Linear fits and quadratic fits of the prices of stocks from
No.28 to No.36.} \label{Fig:SouthAfrica:LQ:4}
\end{figure}

\begin{figure}[htb]
\begin{center}
  \subfigure[Stock No.37: SAB]{
  \label{Fig:SouthAfrica:LQ:SAB}
  \begin{minipage}[b]{0.31\textwidth}
    \includegraphics[width=4.5cm,height=4.5cm]{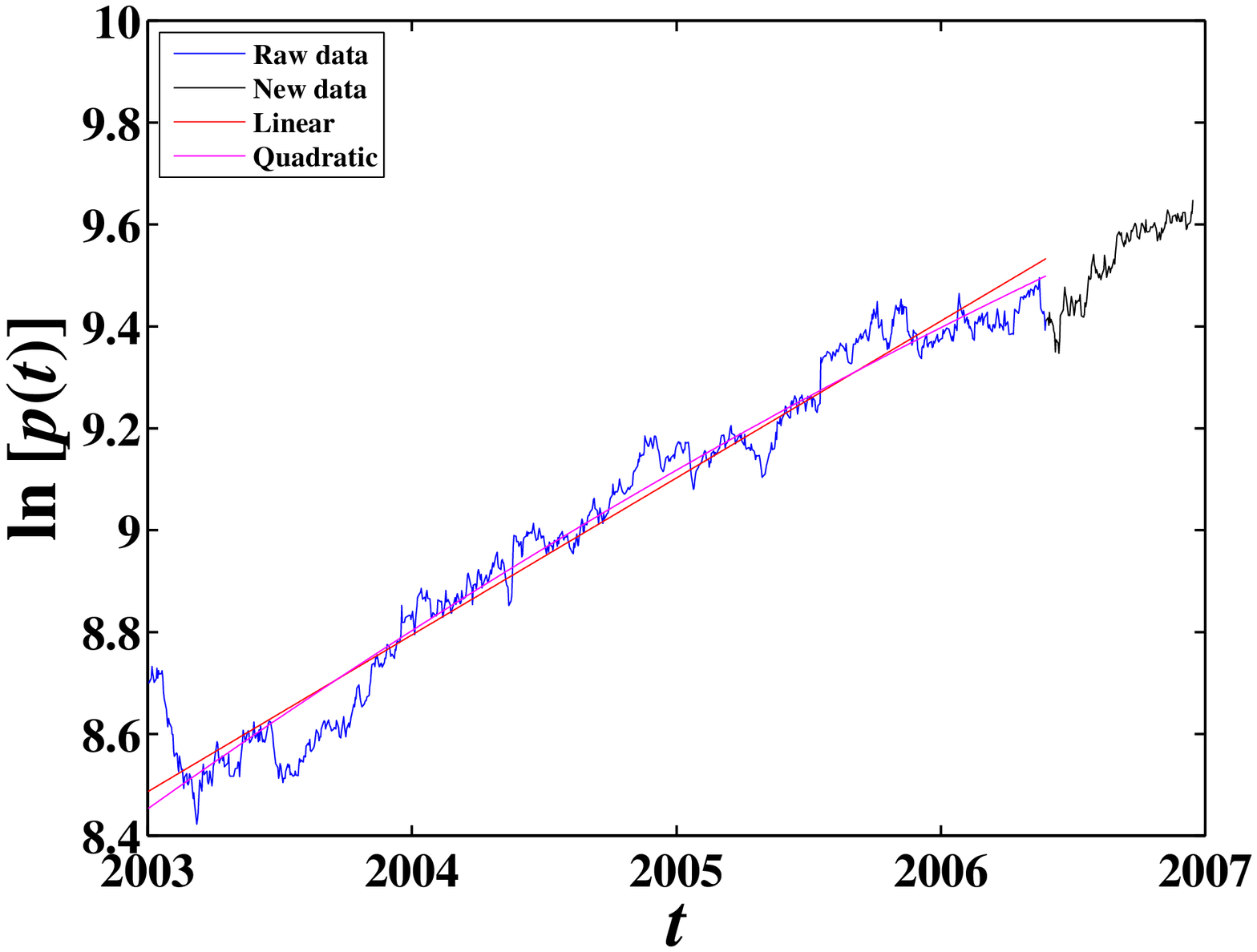}
  \end{minipage}}
  \hspace{0.1cm}
  \subfigure[Stock No.38: SAP]{
  \label{Fig:SouthAfrica:LQ:SAP}
  \begin{minipage}[b]{0.31\textwidth}
    \includegraphics[width=4.5cm,height=4.5cm]{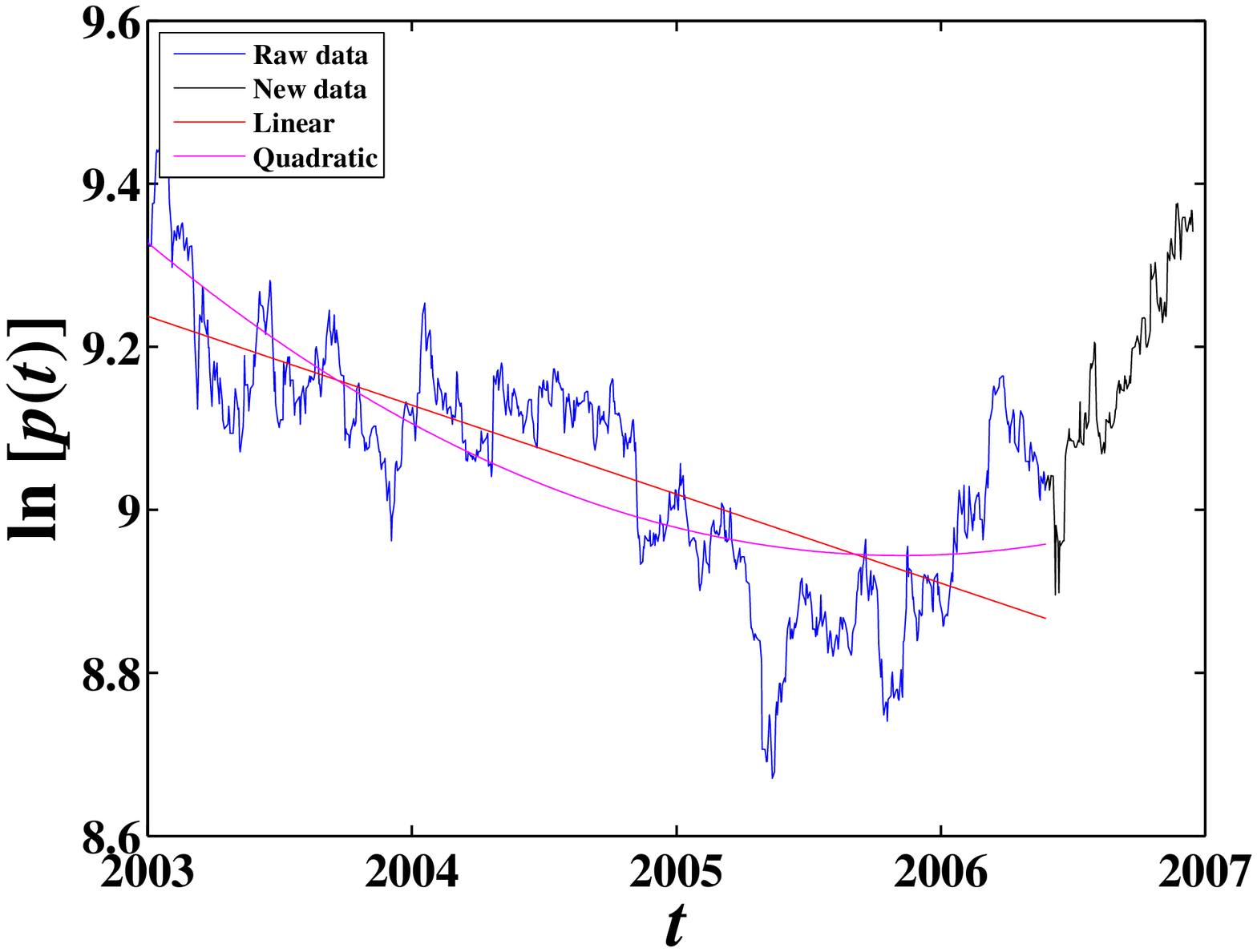}
  \end{minipage}}
  \hspace{0.1cm}
  \subfigure[Stock No.39: SBK]{
  \label{Fig:SouthAfrica:LQ:SBK}
  \begin{minipage}[b]{0.31\textwidth}
    \includegraphics[width=4.5cm,height=4.5cm]{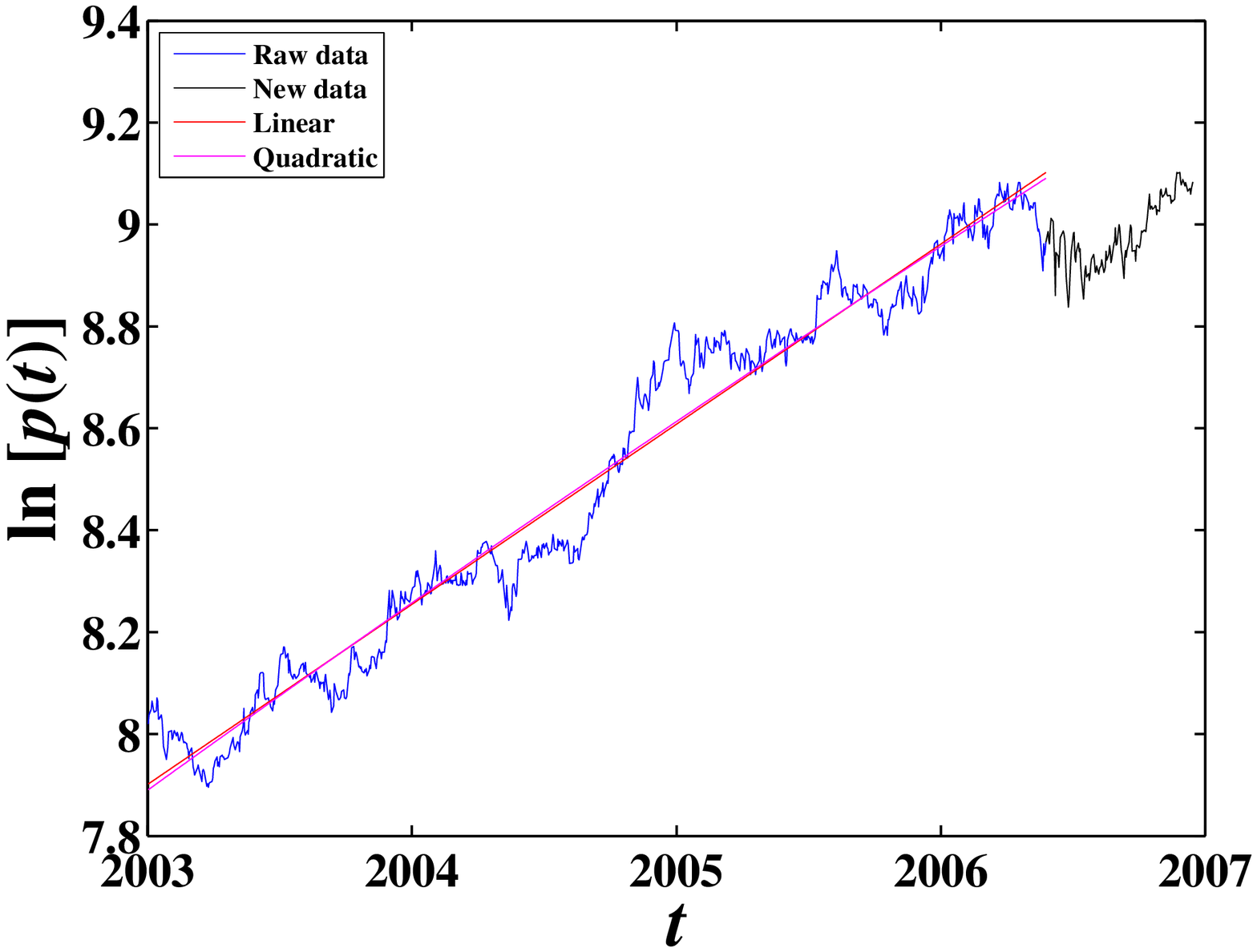}
  \end{minipage}}\\[10pt]
  \subfigure[Stock No.40: SHF]{
  \label{Fig:SouthAfrica:LQ:SHF}
  \begin{minipage}[b]{0.31\textwidth}
    \includegraphics[width=4.5cm,height=4.5cm]{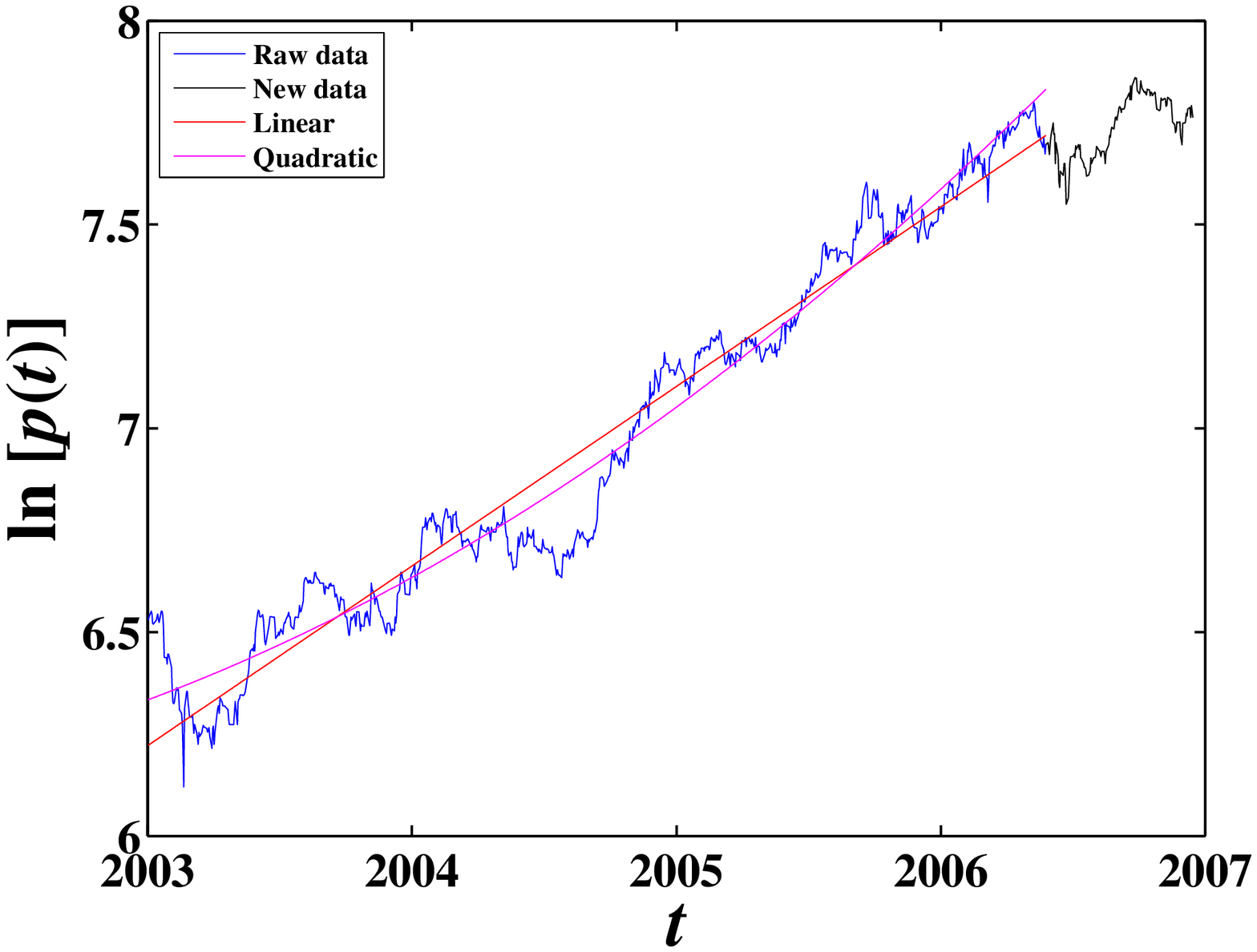}
  \end{minipage}}
  \hspace{0.1cm}
  \subfigure[Stock No.41: SLM]{
  \label{Fig:SouthAfrica:LQ:SLM}
  \begin{minipage}[b]{0.31\textwidth}
    \includegraphics[width=4.5cm,height=4.5cm]{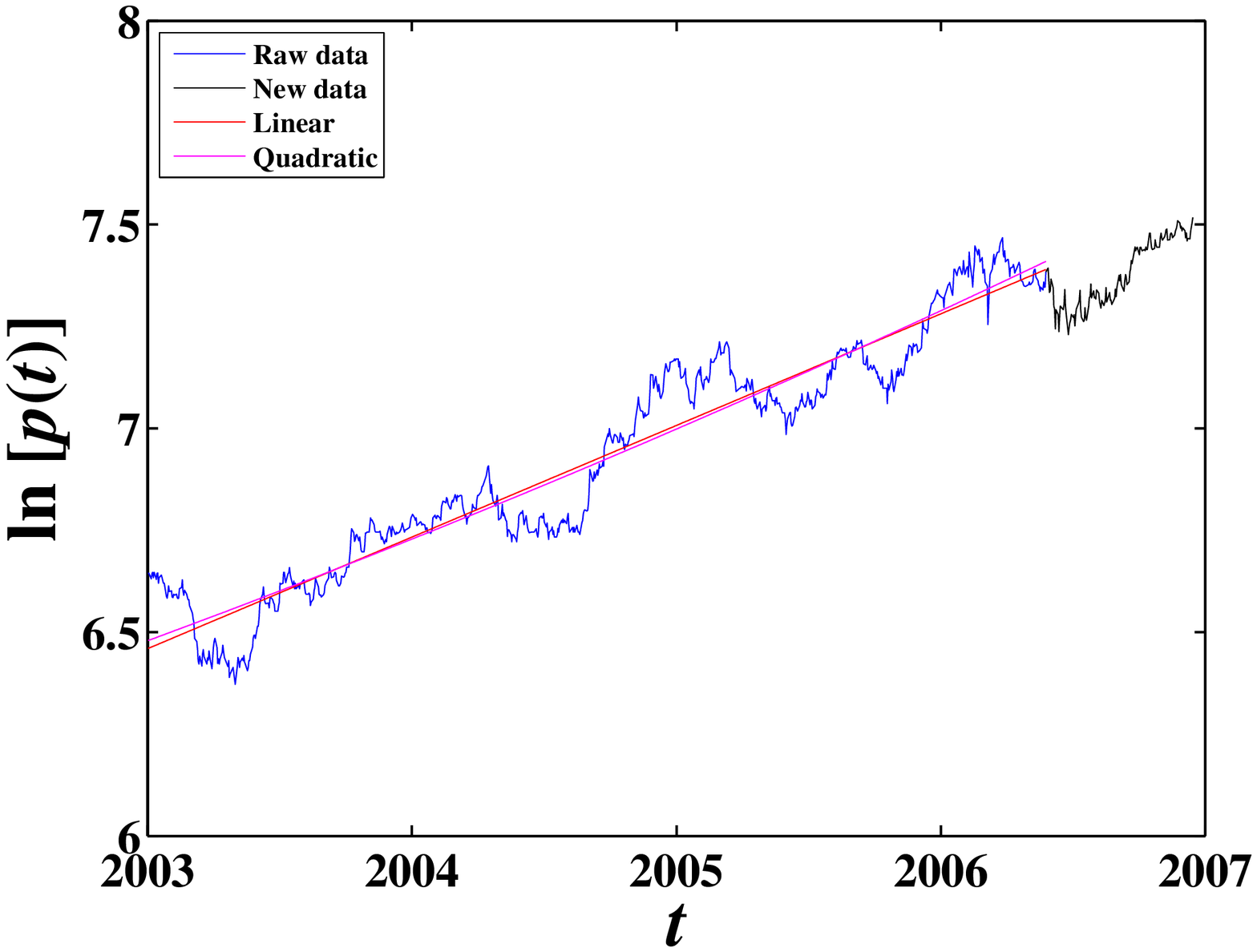}
  \end{minipage}}
  \hspace{0.1cm}
  \subfigure[Stock No.42: SOL]{
  \label{Fig:SouthAfrica:LQ:SOL}
  \begin{minipage}[b]{0.31\textwidth}
    \includegraphics[width=4.5cm,height=4.5cm]{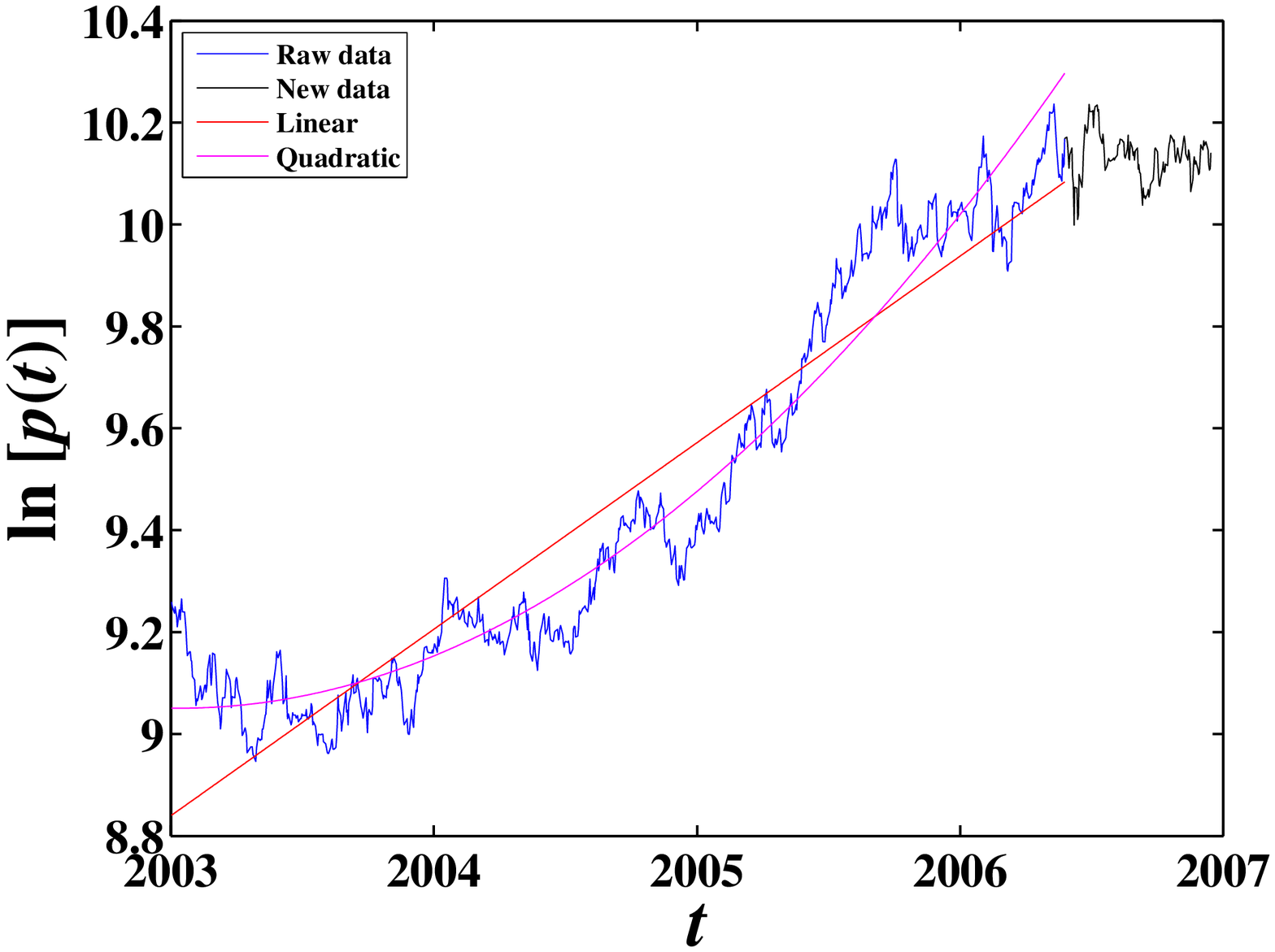}
  \end{minipage}}\\[10pt]
  \subfigure[Stock No.43: TBS]{
  \label{Fig:SouthAfrica:LQ:TBS}
  \begin{minipage}[b]{0.31\textwidth}
    \includegraphics[width=4.5cm,height=4.5cm]{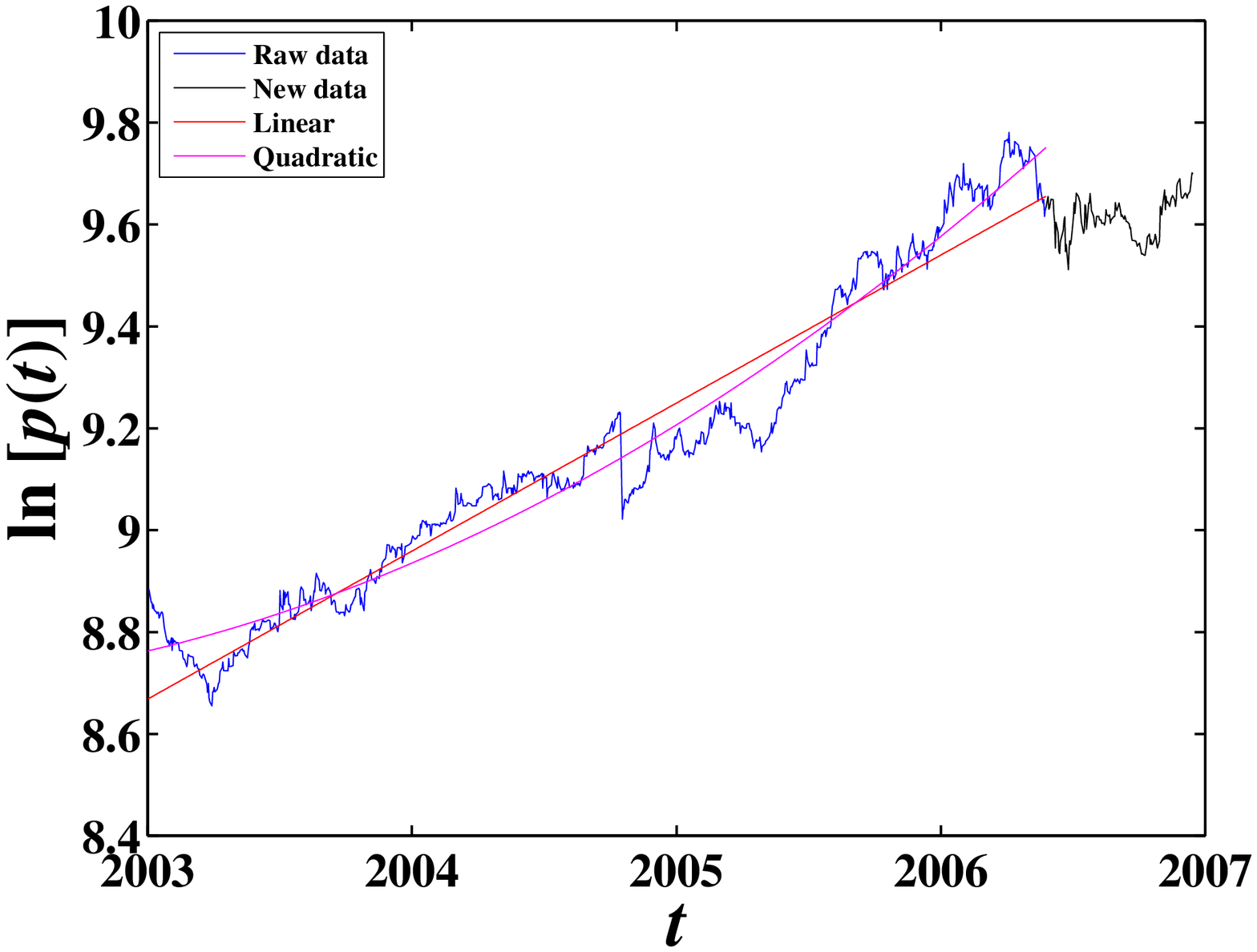}
  \end{minipage}}
  \hspace{0.1cm}
  \subfigure[Stock No.44: TKG]{
  \label{Fig:SouthAfrica:LQ:TKG}
  \begin{minipage}[b]{0.31\textwidth}
    \includegraphics[width=4.5cm,height=4.5cm]{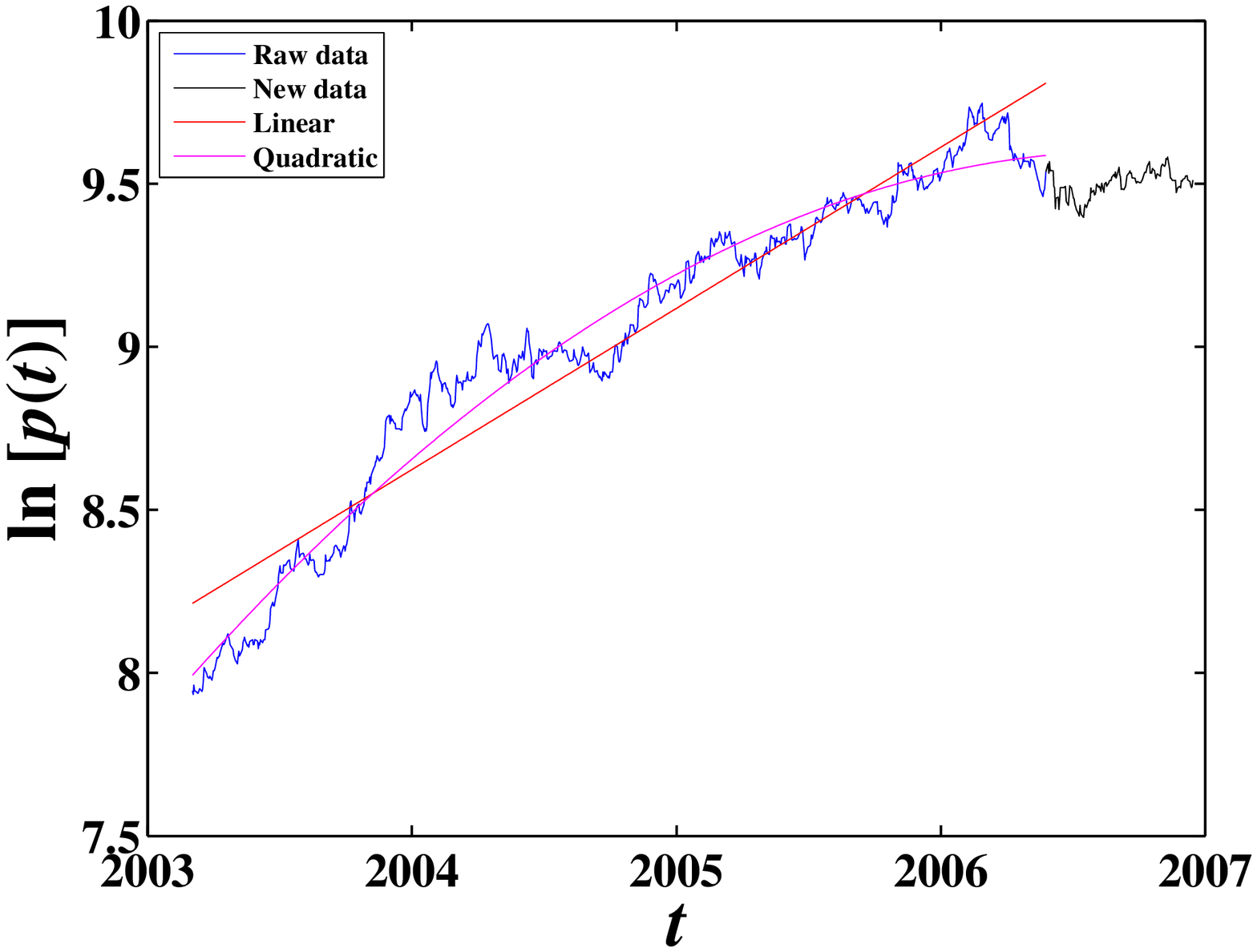}
  \end{minipage}}
  \hspace{0.1cm}
  \subfigure[Stock No.45: WHL]{
  \label{Fig:SouthAfrica:LQ:WHL}
  \begin{minipage}[b]{0.31\textwidth}
    \includegraphics[width=4.5cm,height=4.5cm]{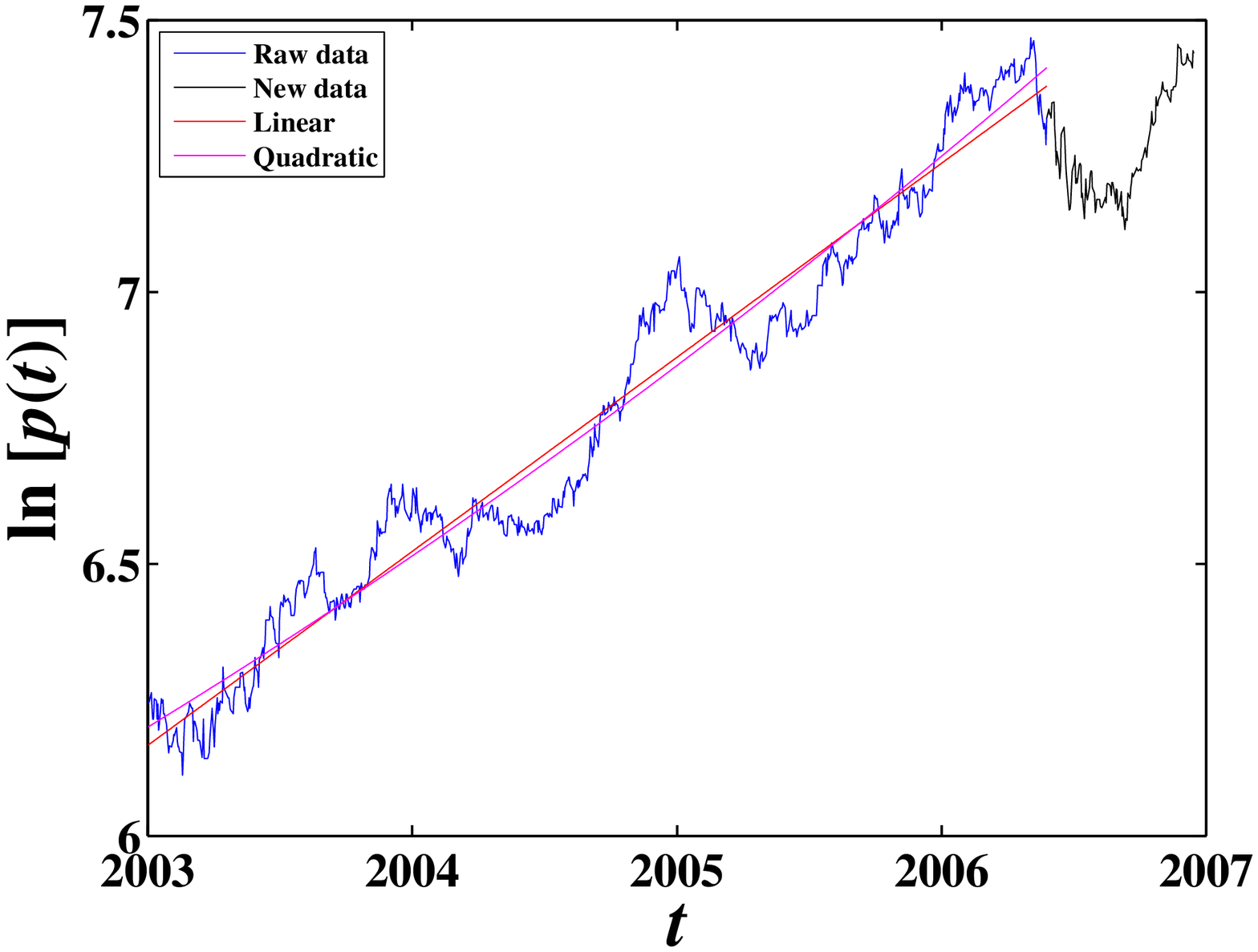}
  \end{minipage}}\\[10pt]
\end{center}
\caption{Linear fits and quadratic fits of the prices of stocks from
No.37 to No.45.} \label{Fig:SouthAfrica:LQ:5}
\end{figure}

\clearpage
\begin{figure}[htb]
\begin{center}
  \subfigure[Stock No. 1: J203]{
  \label{Fig:SouthAfrica:S0S1:J203}
  \begin{minipage}[b]{0.31\textwidth}
    \includegraphics[width=4.5cm,height=4.5cm]{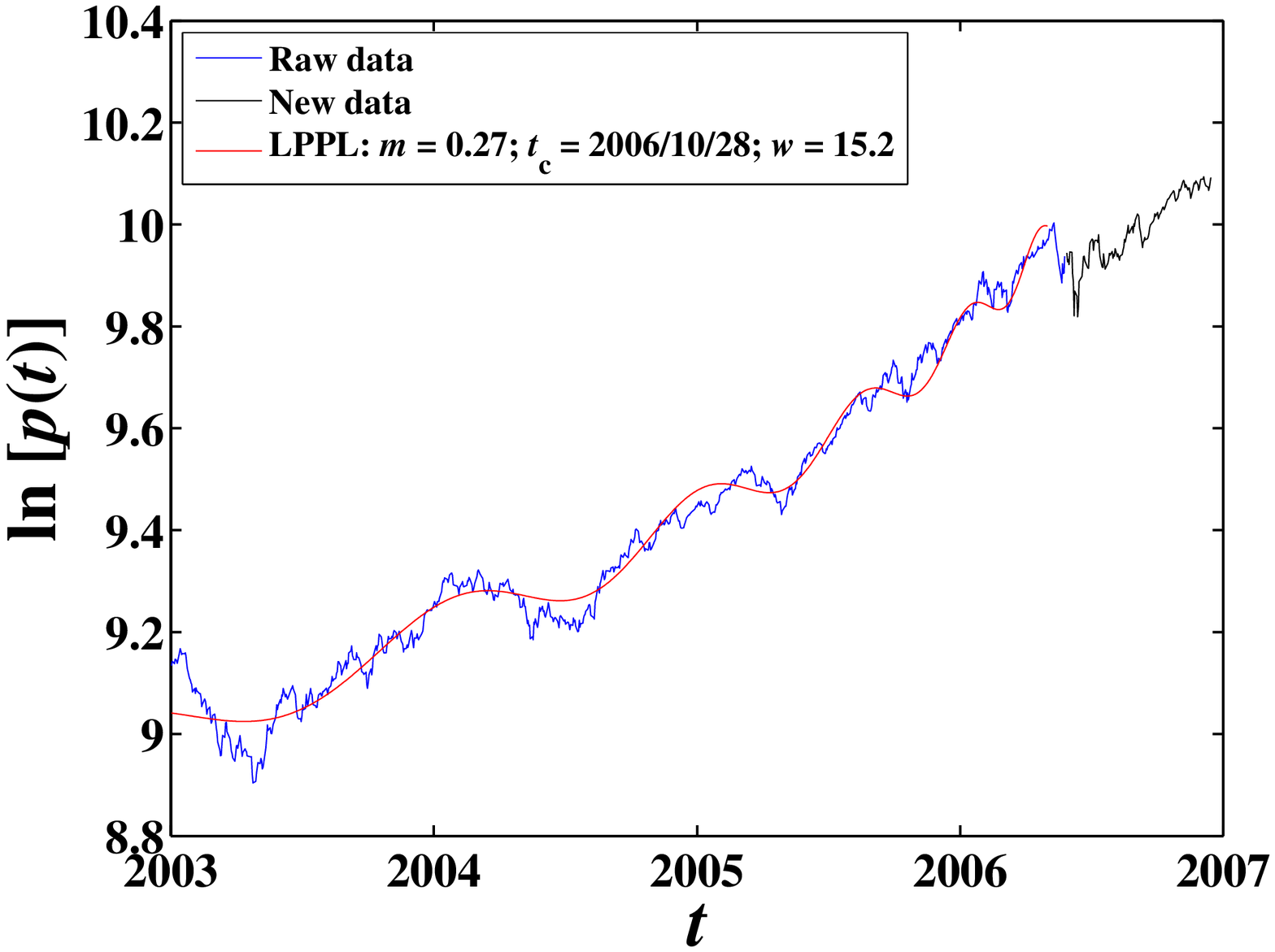}
  \end{minipage}}
  \hspace{0.1cm}
  \subfigure[Stock No. 2: J210]{
  \label{Fig:SouthAfrica:S0S1:J210}
  \begin{minipage}[b]{0.31\textwidth}
    \includegraphics[width=4.5cm,height=4.5cm]{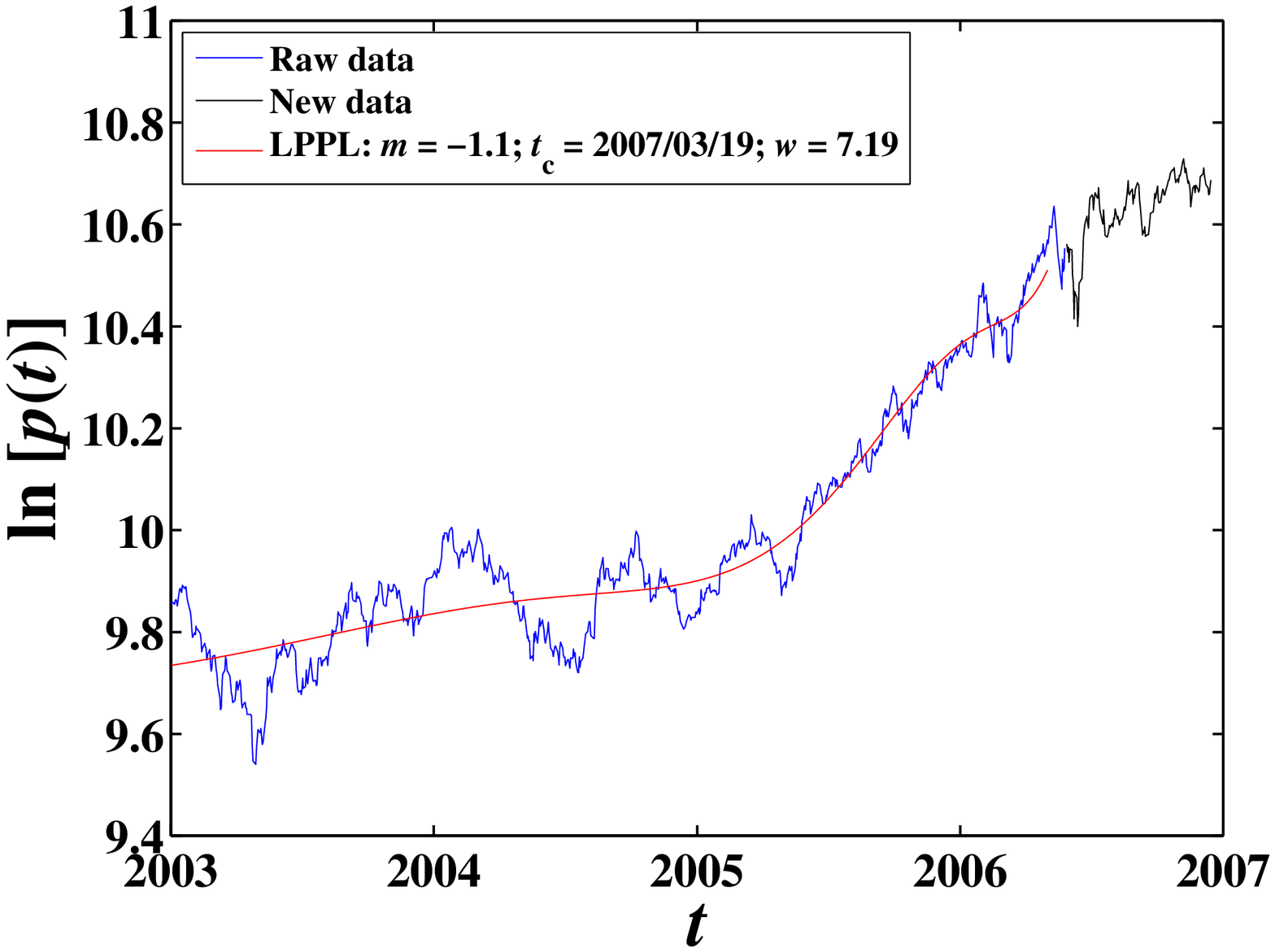}
  \end{minipage}}
  \hspace{0.1cm}
  \subfigure[Stock No. 3: J257]{
  \label{Fig:SouthAfrica:S0S1:J257}
  \begin{minipage}[b]{0.31\textwidth}
    \includegraphics[width=4.5cm,height=4.5cm]{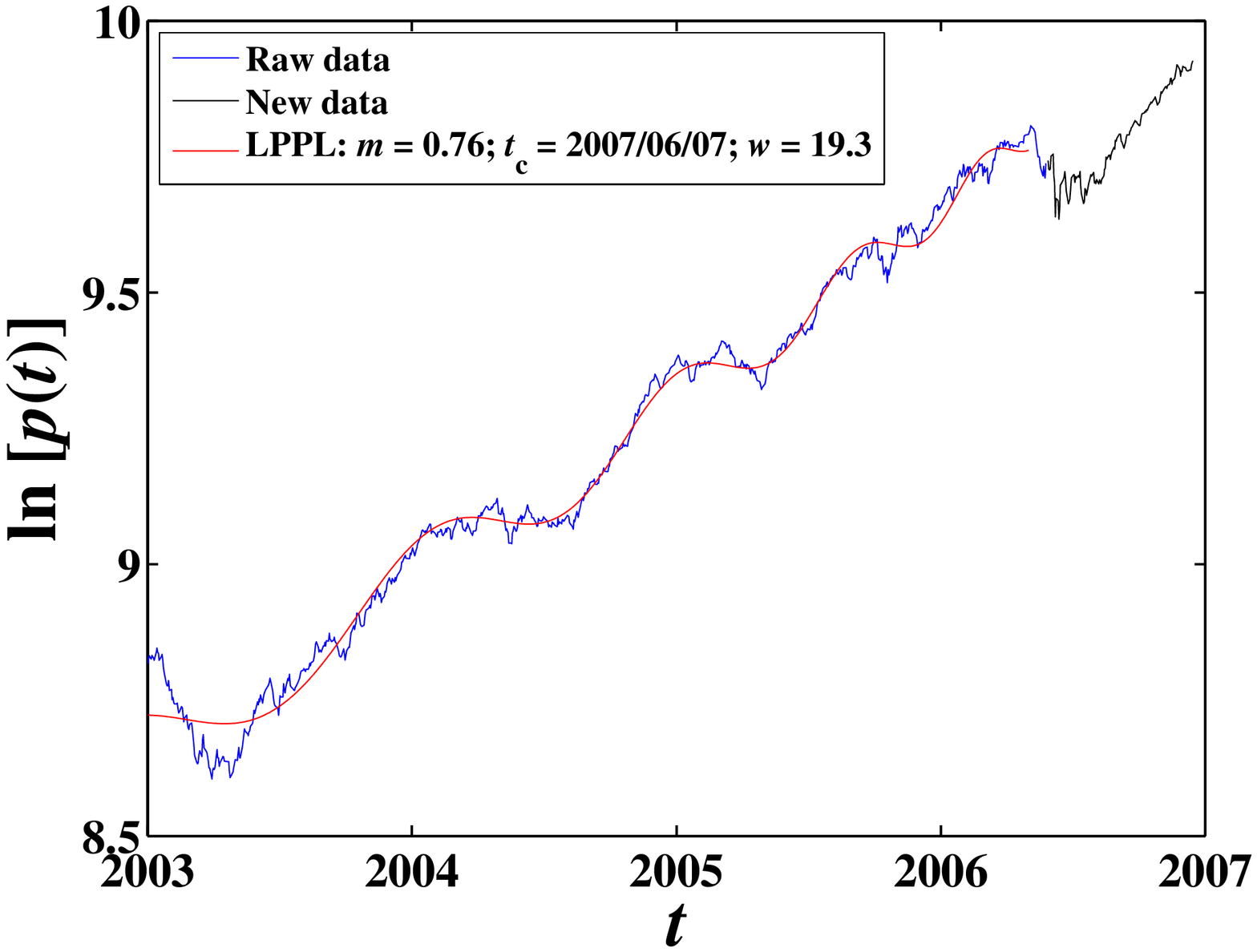}
  \end{minipage}}\\[10pt]
  \subfigure[Stock No. 4: J580]{
  \label{Fig:SouthAfrica:S0S1:J580}
  \begin{minipage}[b]{0.31\textwidth}
    \includegraphics[width=4.5cm,height=4.5cm]{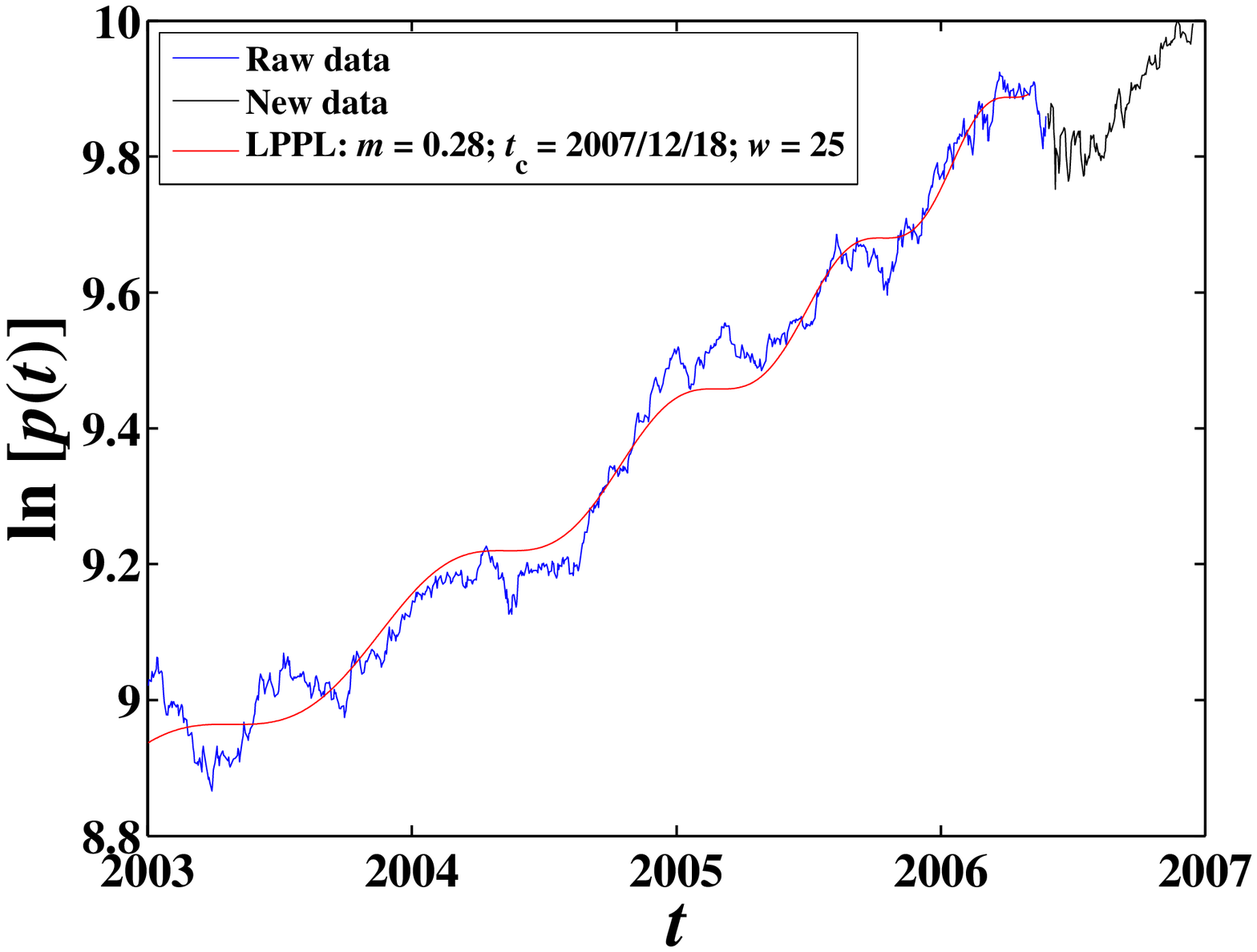}
  \end{minipage}}
  \hspace{0.1cm}
  \subfigure[Stock No. 5: ABL]{
  \label{Fig:SouthAfrica:S0S1:ABL}
  \begin{minipage}[b]{0.31\textwidth}
    \includegraphics[width=4.5cm,height=4.5cm]{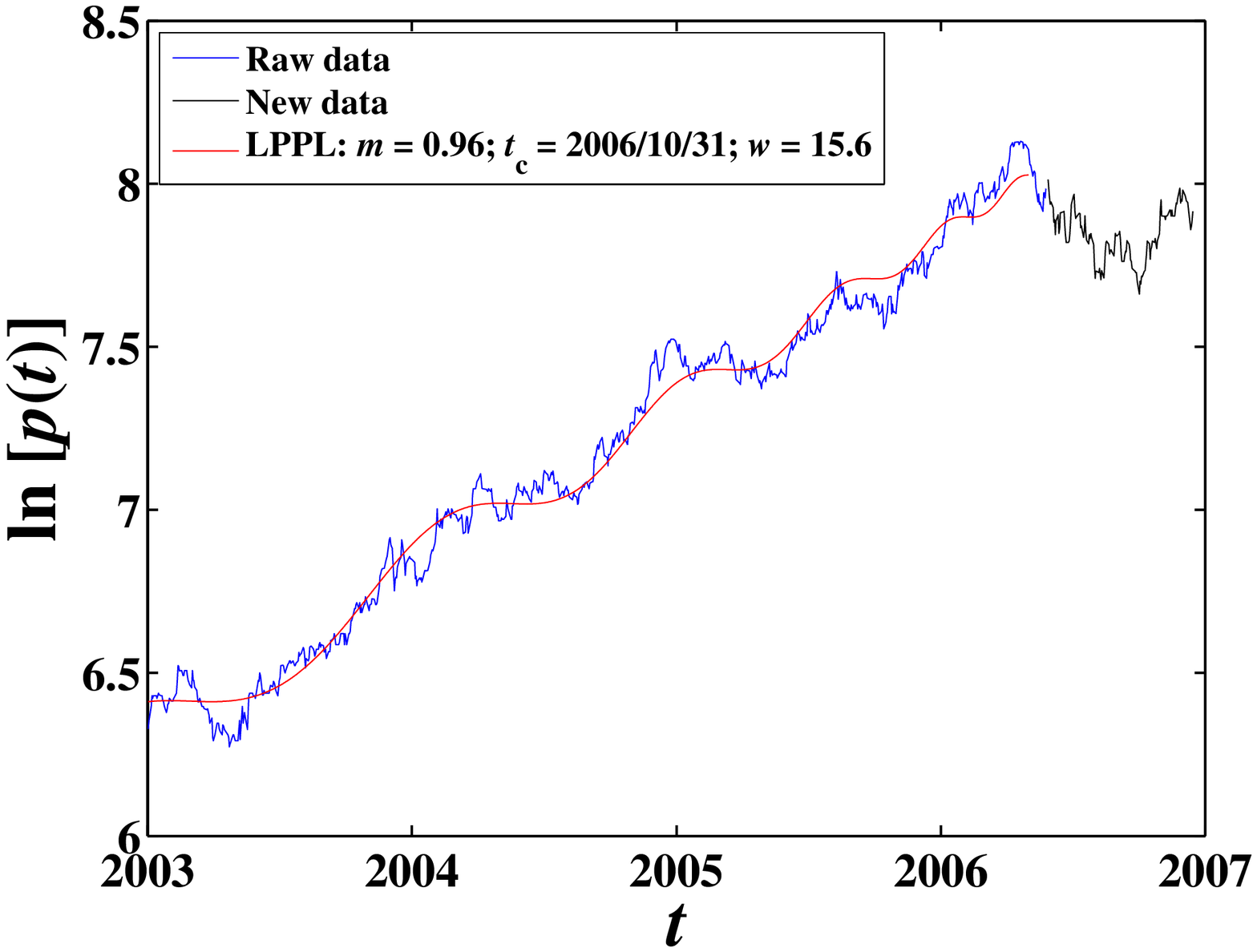}
  \end{minipage}}
  \hspace{0.1cm}
  \subfigure[Stock No. 6: AGL]{
  \label{Fig:SouthAfrica:S0S1:AGL}
  \begin{minipage}[b]{0.31\textwidth}
    \includegraphics[width=4.5cm,height=4.5cm]{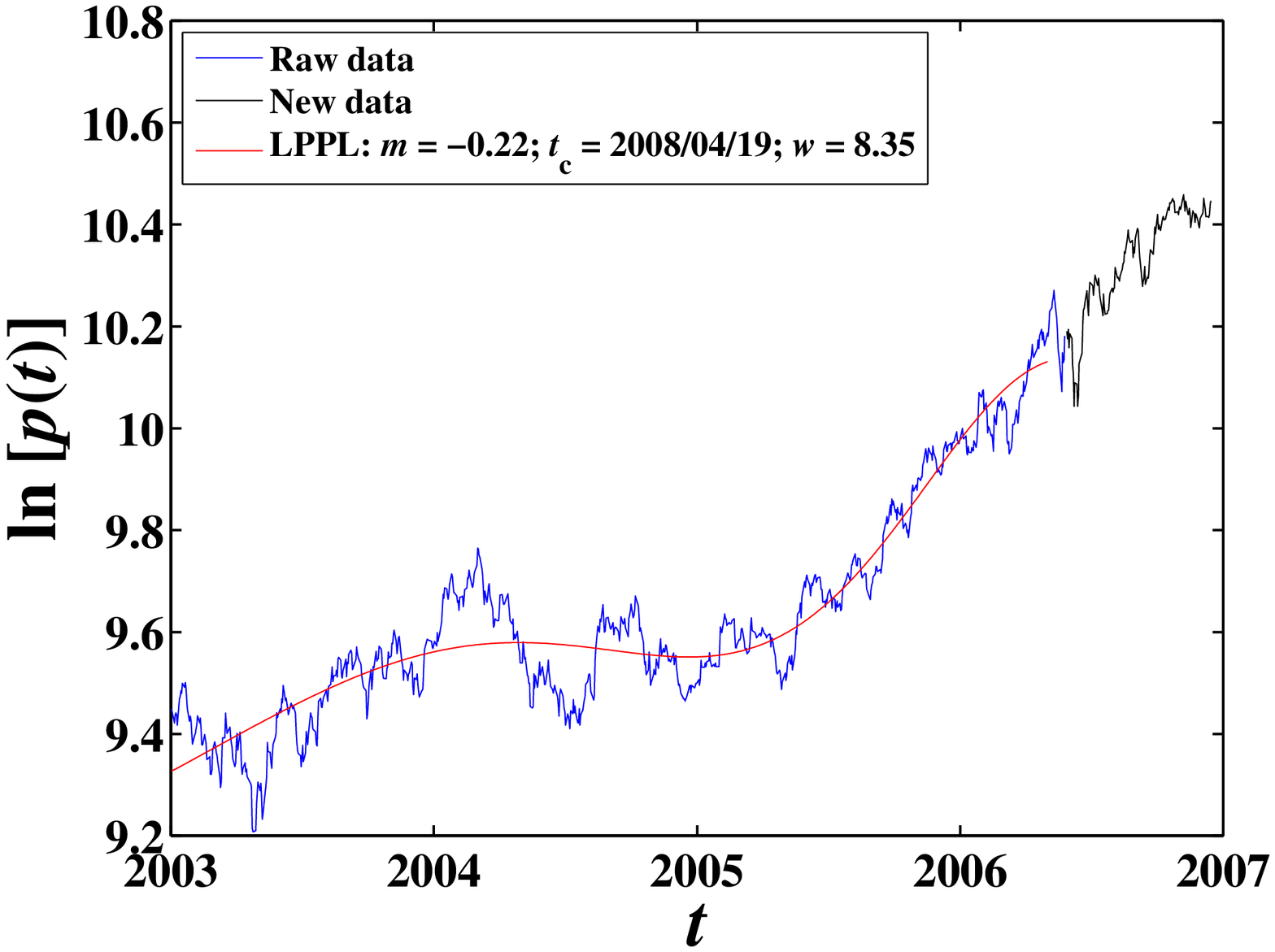}
  \end{minipage}}\\[10pt]
  \subfigure[Stock No. 7: AMS]{
  \label{Fig:SouthAfrica:S0S1:AMS}
  \begin{minipage}[b]{0.31\textwidth}
    \includegraphics[width=4.5cm,height=4.5cm]{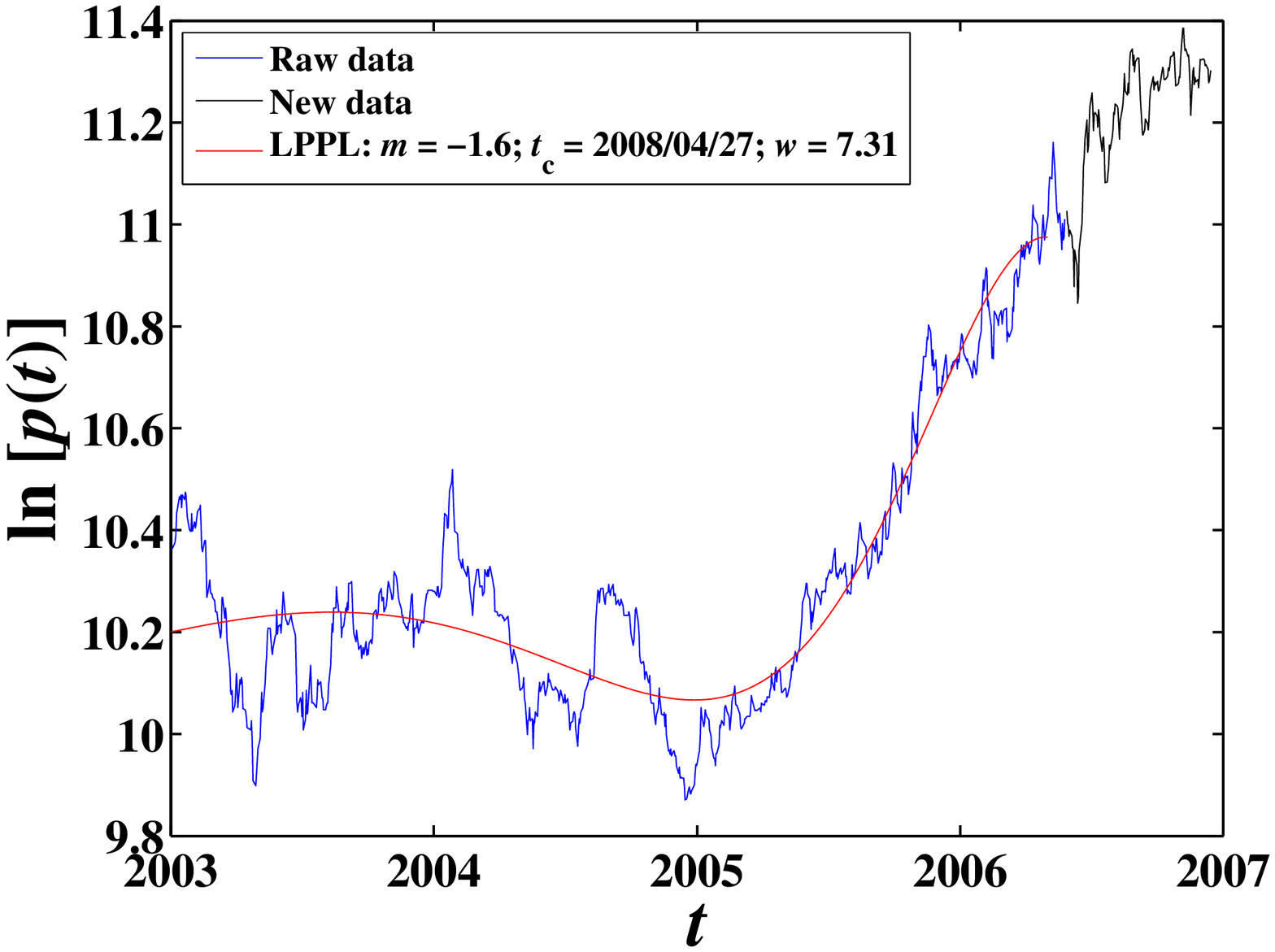}
  \end{minipage}}
  \hspace{0.1cm}
  \subfigure[Stock No. 8: ANG]{
  \label{Fig:SouthAfrica:S0S1:ANG}
  \begin{minipage}[b]{0.31\textwidth}
    \includegraphics[width=4.5cm,height=4.5cm]{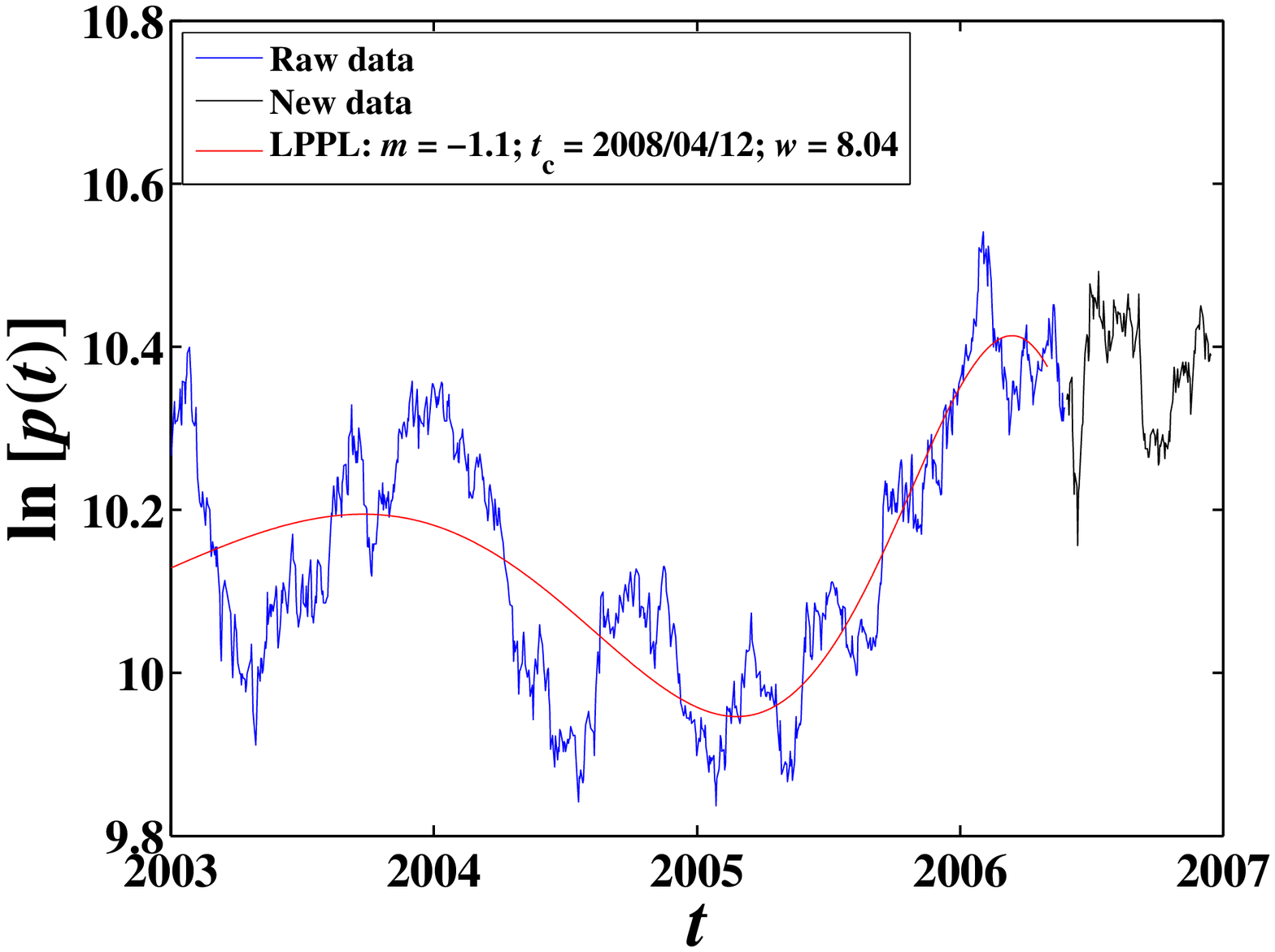}
  \end{minipage}}
  \hspace{0.1cm}
  \subfigure[Stock No. 9: APN]{
  \label{Fig:SouthAfrica:S0S1:APN}
  \begin{minipage}[b]{0.31\textwidth}
    \includegraphics[width=4.5cm,height=4.5cm]{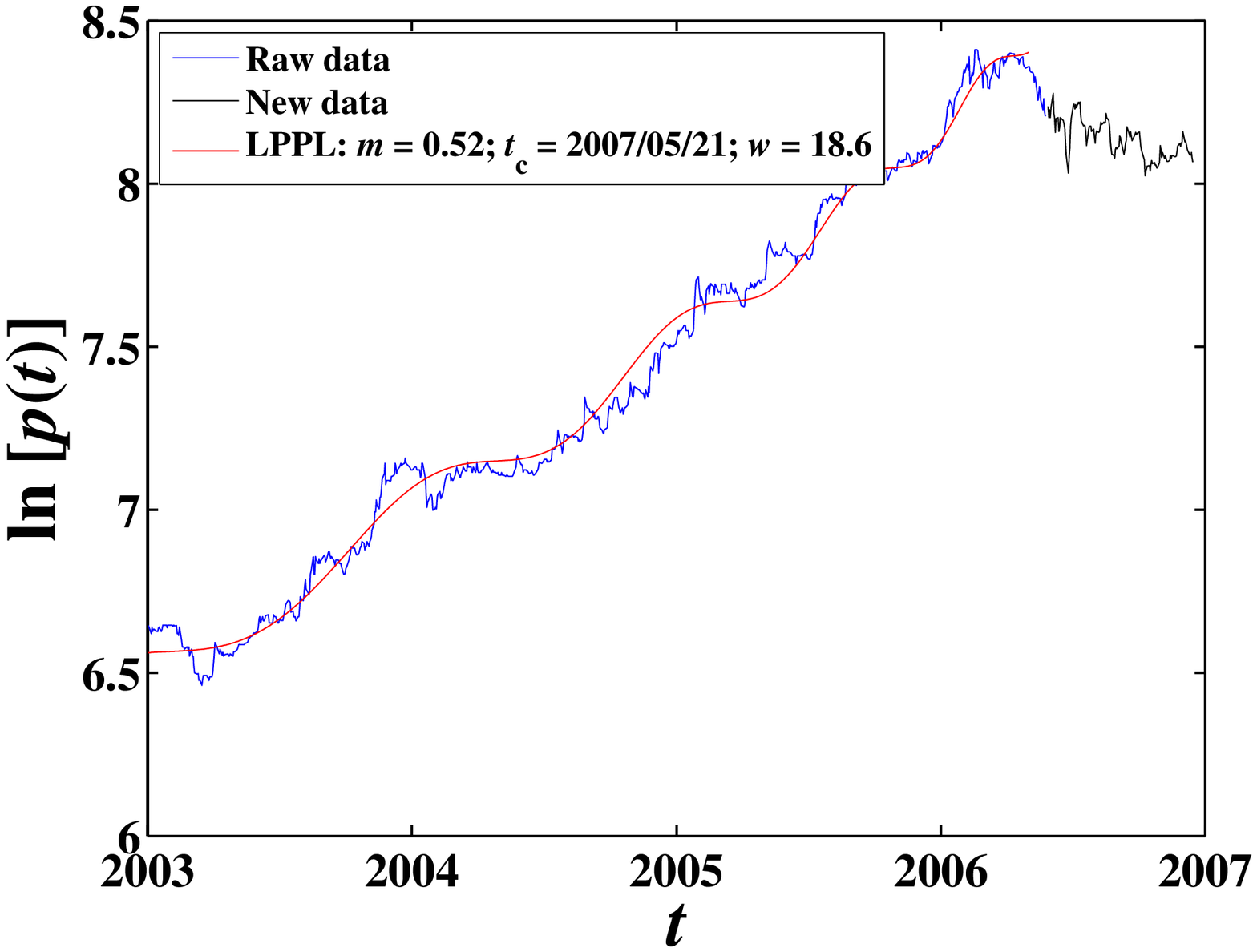}
  \end{minipage}}\\[10pt]
\end{center}
\caption{Log-periodic power-law fits of the prices of stocks from
No. 1 to No. 9.} \label{Fig:SouthAfrica:S0S1:1}
\end{figure}

\begin{figure}[htb]
\begin{center}
  \subfigure[Stock No.10: ASA]{
  \label{Fig:SouthAfrica:S0S1:ASA}
  \begin{minipage}[b]{0.31\textwidth}
    \includegraphics[width=4.5cm,height=4.5cm]{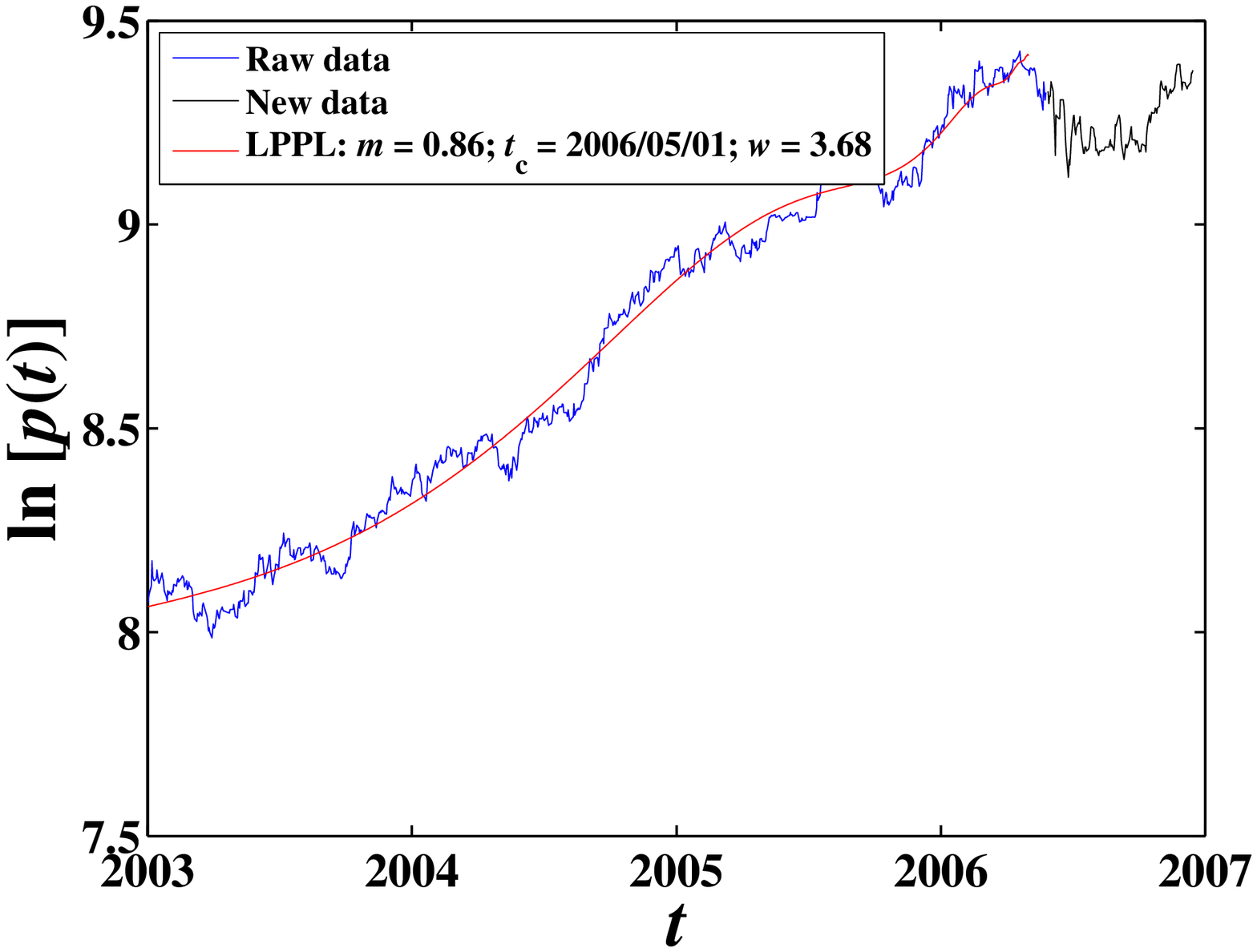}
  \end{minipage}}
  \hspace{0.1cm}
  \subfigure[Stock No.11: BAW]{
  \label{Fig:SouthAfrica:S0S1:BAW}
  \begin{minipage}[b]{0.31\textwidth}
    \includegraphics[width=4.5cm,height=4.5cm]{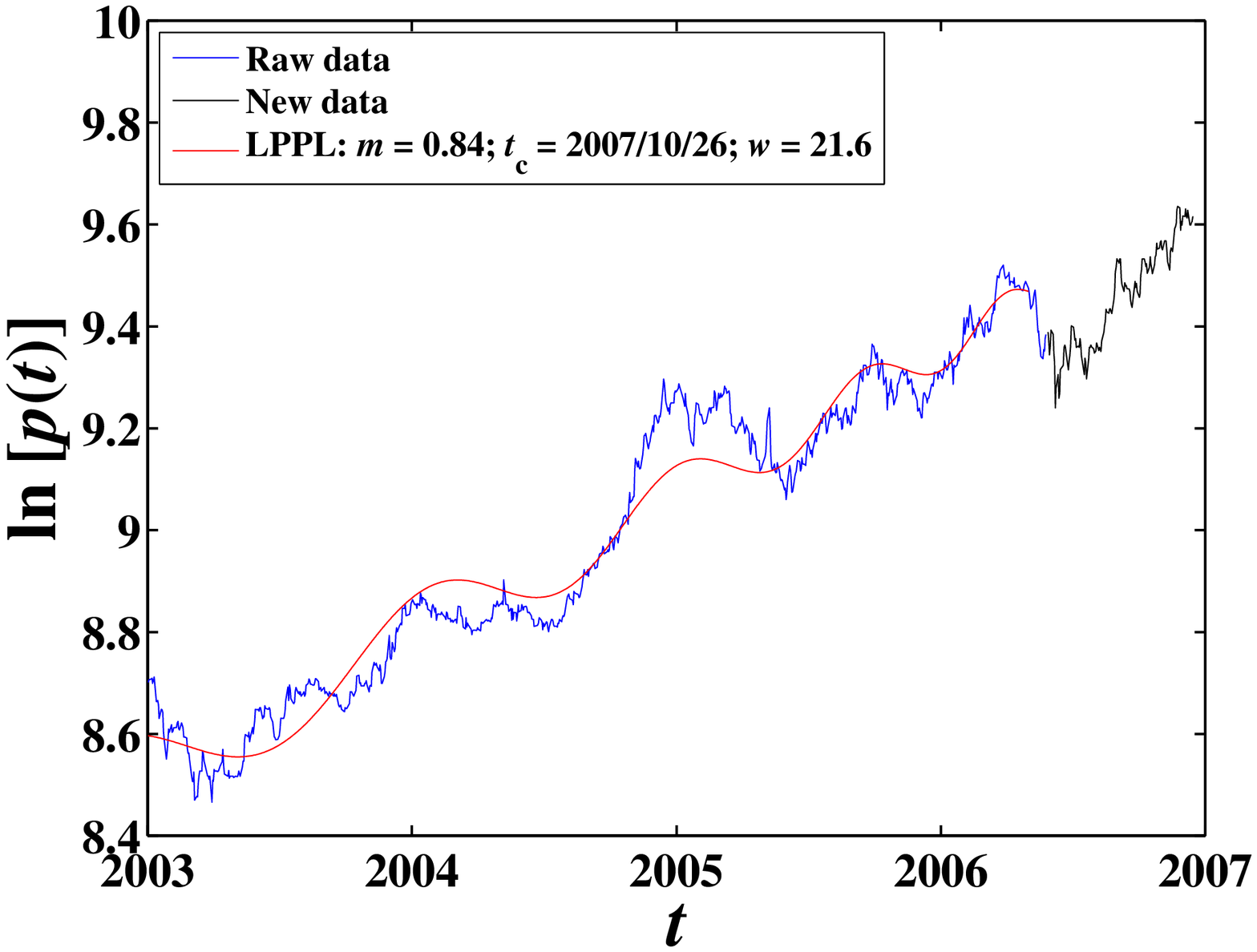}
  \end{minipage}}
  \hspace{0.1cm}
  \subfigure[Stock No.12: BIL]{
  \label{Fig:SouthAfrica:S0S1:BIL}
  \begin{minipage}[b]{0.31\textwidth}
    \includegraphics[width=4.5cm,height=4.5cm]{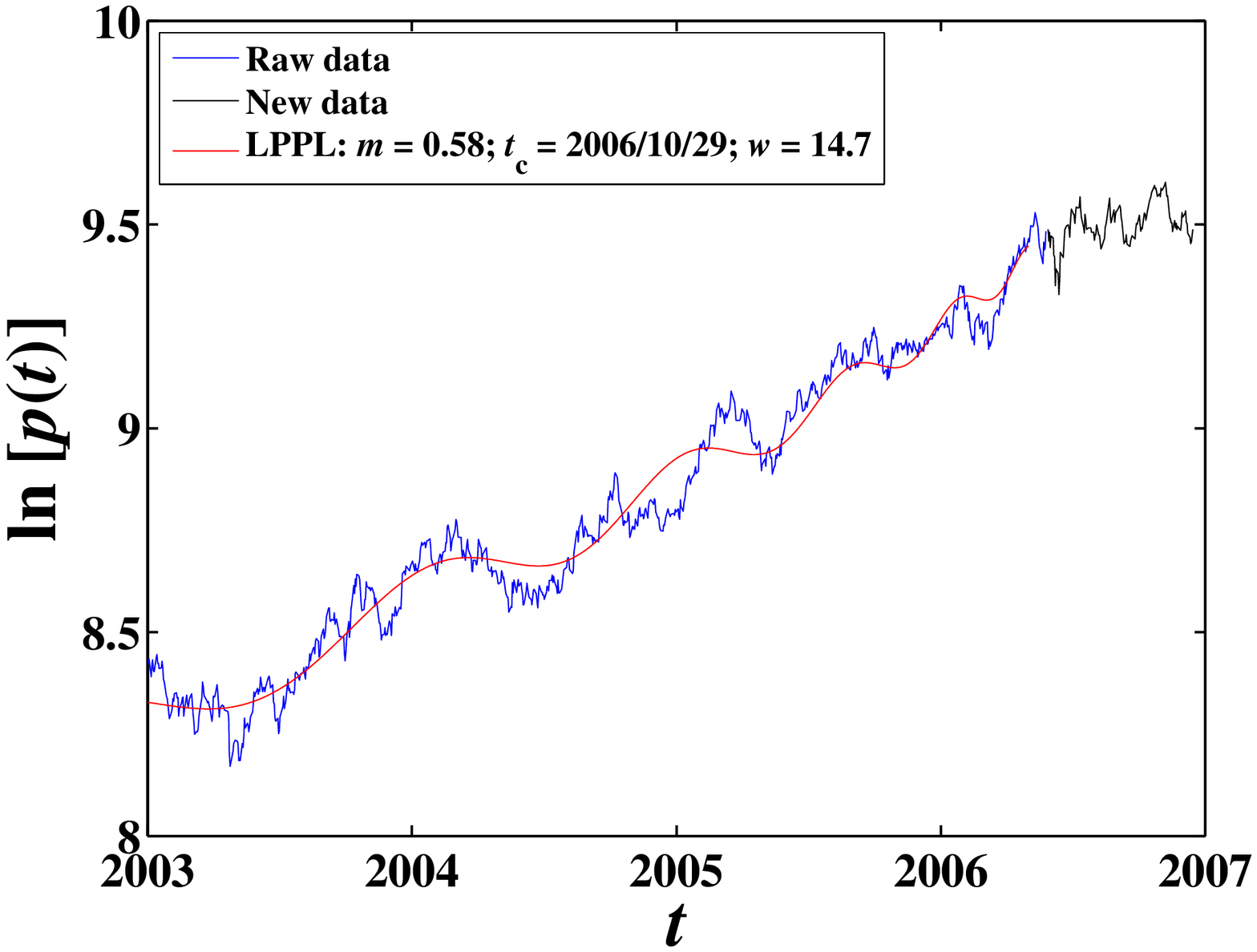}
  \end{minipage}}\\[10pt]
  \subfigure[Stock No.13: BVT]{
  \label{Fig:SouthAfrica:S0S1:BVT}
  \begin{minipage}[b]{0.31\textwidth}
    \includegraphics[width=4.5cm,height=4.5cm]{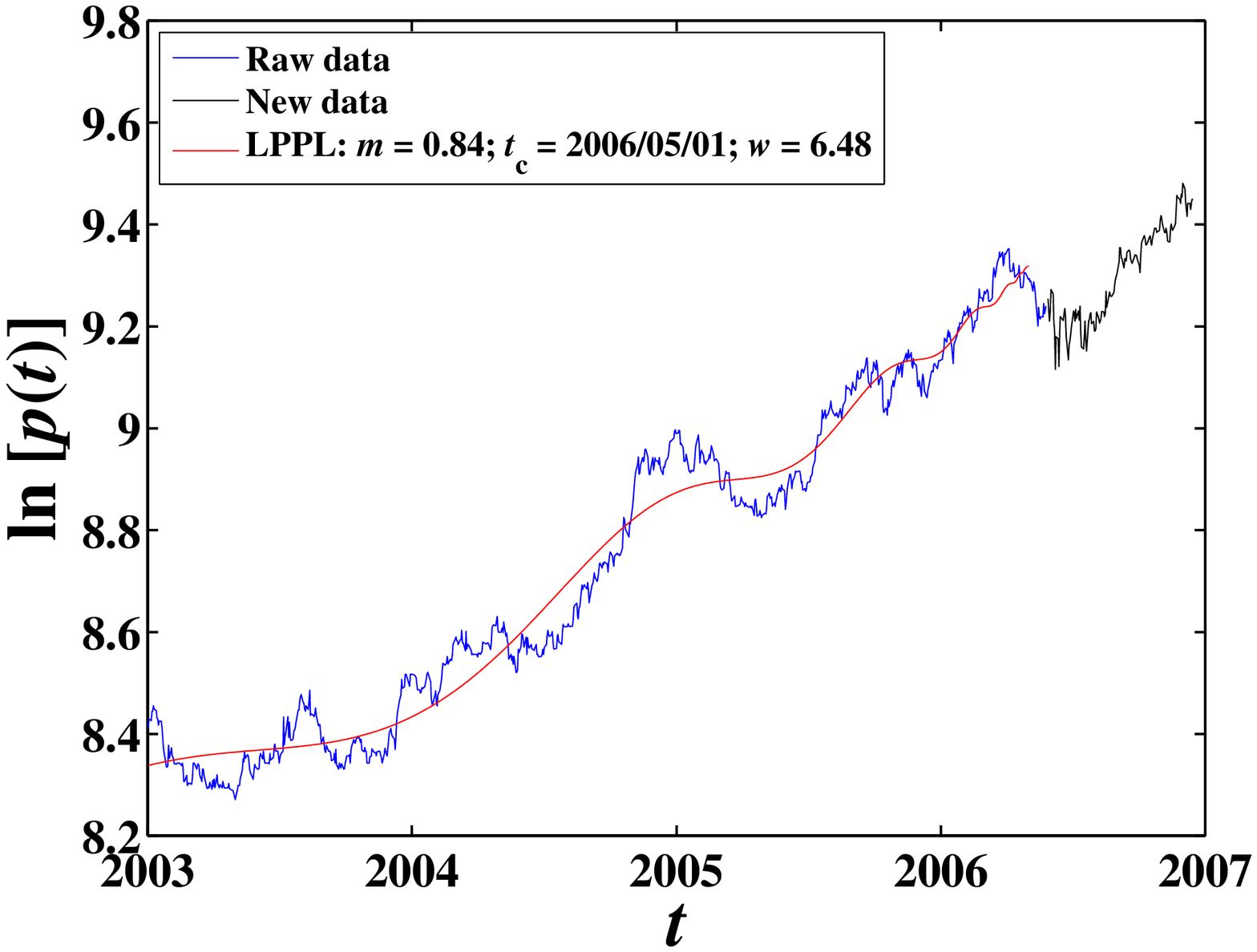}
  \end{minipage}}
  \hspace{0.1cm}
  \subfigure[Stock No.14: ECO]{
  \label{Fig:SouthAfrica:S0S1:ECO}
  \begin{minipage}[b]{0.31\textwidth}
    \includegraphics[width=4.5cm,height=4.5cm]{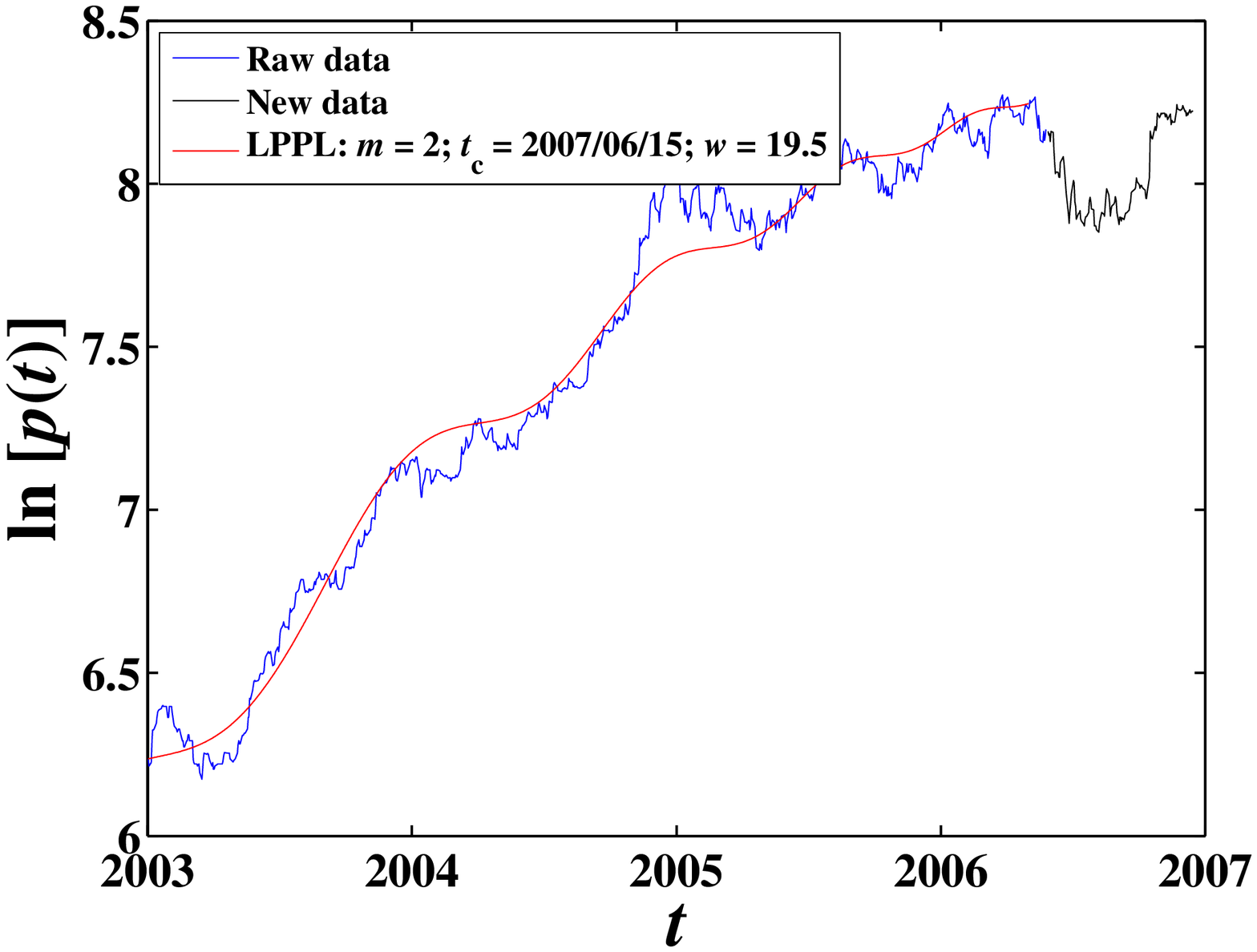}
  \end{minipage}}
  \hspace{0.1cm}
  \subfigure[Stock No.15: FSR]{
  \label{Fig:SouthAfrica:S0S1:FSR}
  \begin{minipage}[b]{0.31\textwidth}
    \includegraphics[width=4.5cm,height=4.5cm]{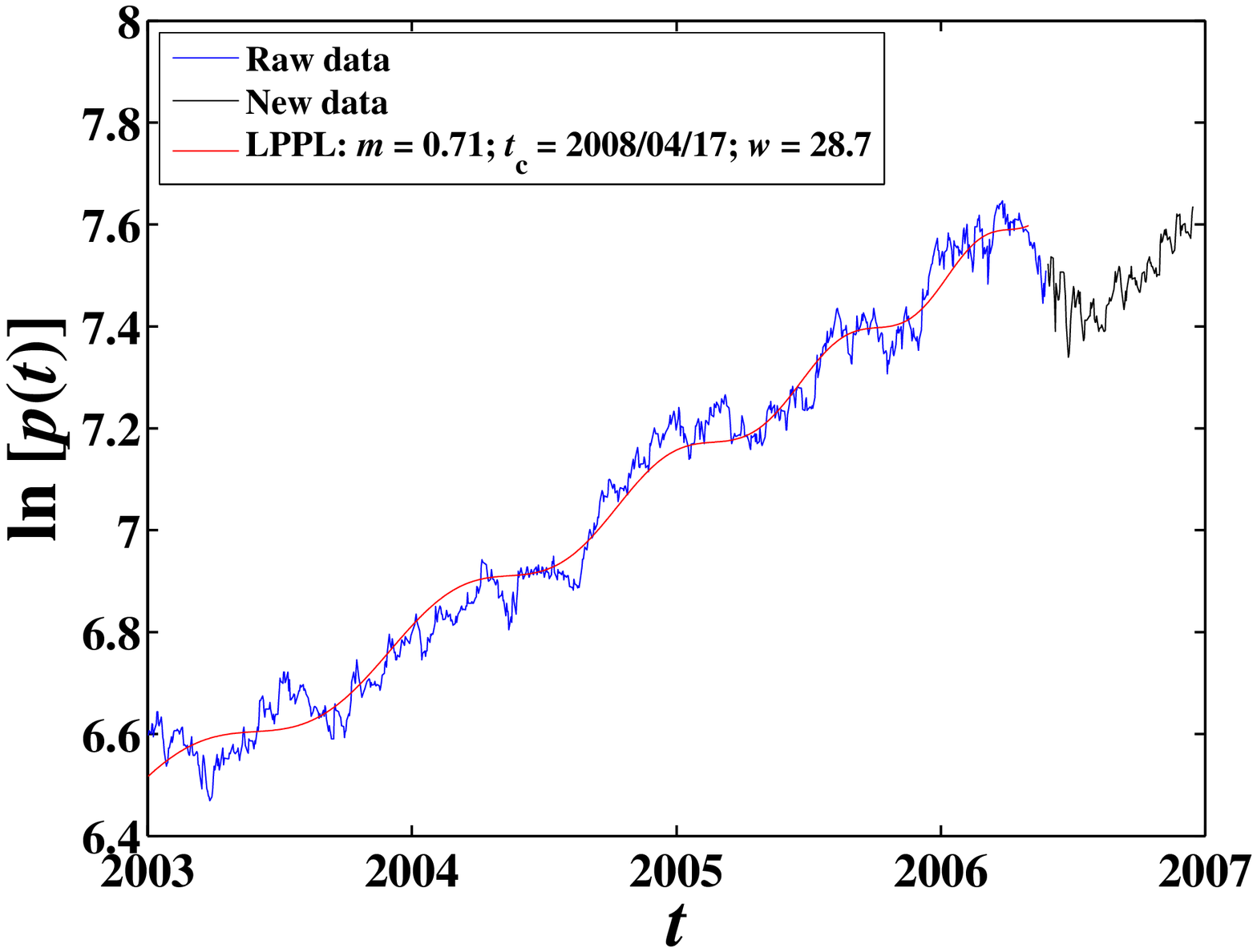}
  \end{minipage}}\\[10pt]
  \subfigure[Stock No.16: GFI]{
  \label{Fig:SouthAfrica:S0S1:GFI}
  \begin{minipage}[b]{0.31\textwidth}
    \includegraphics[width=4.5cm,height=4.5cm]{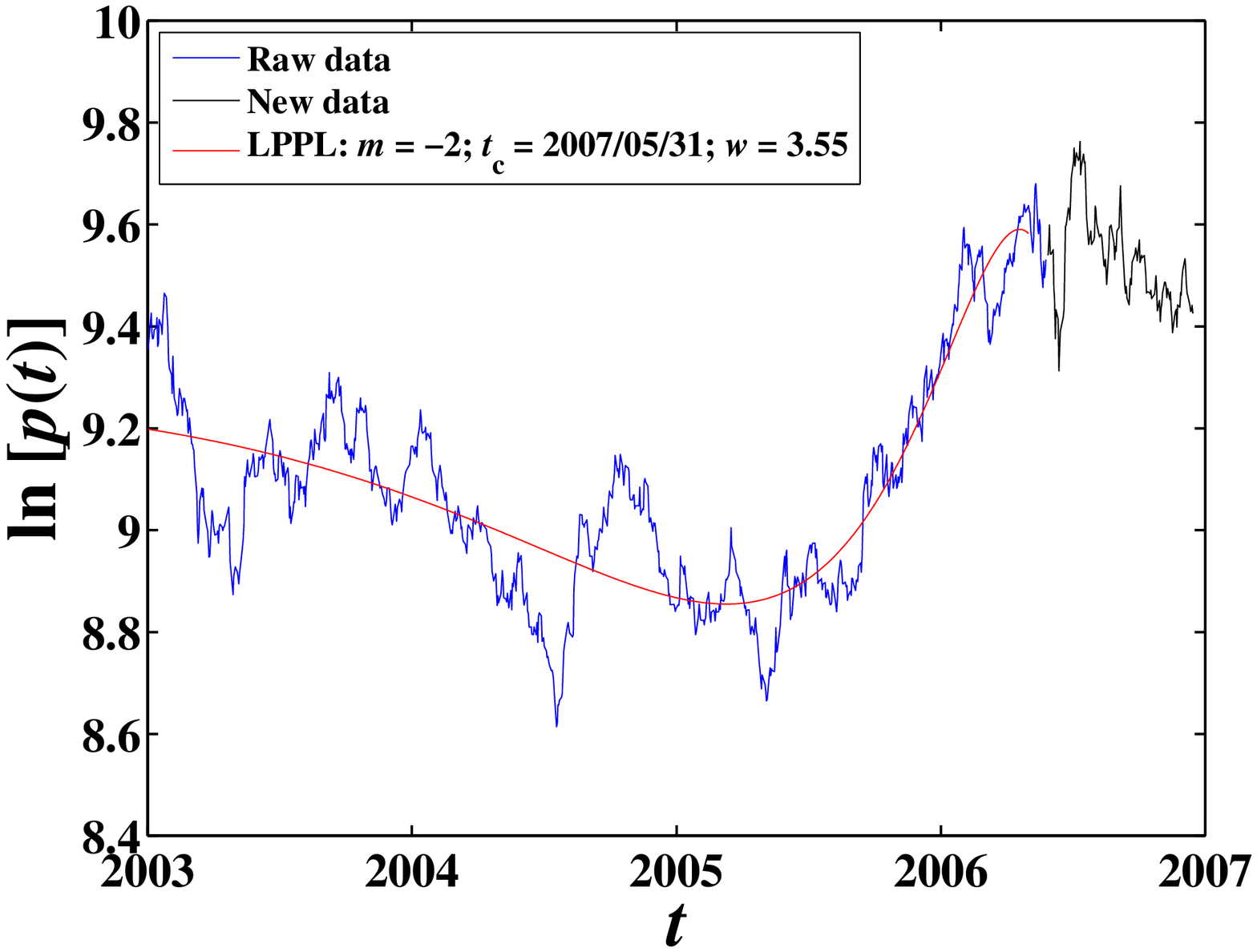}
  \end{minipage}}
  \hspace{0.1cm}
  \subfigure[Stock No.17: HAR]{
  \label{Fig:SouthAfrica:S0S1:HAR}
  \begin{minipage}[b]{0.31\textwidth}
    \includegraphics[width=4.5cm,height=4.5cm]{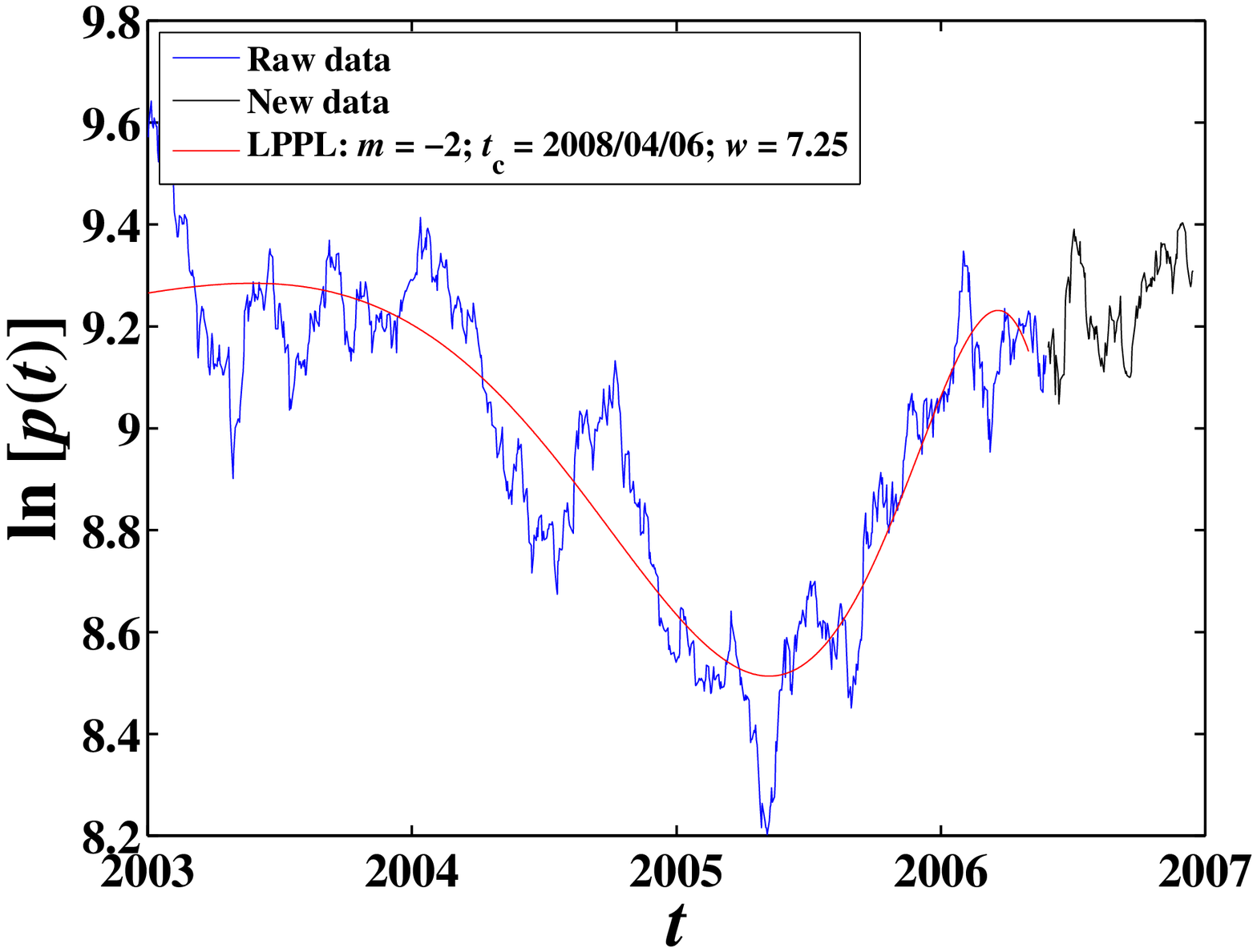}
  \end{minipage}}
  \hspace{0.1cm}
  \subfigure[Stock No.18: IMP]{
  \label{Fig:SouthAfrica:S0S1:IMP}
  \begin{minipage}[b]{0.31\textwidth}
    \includegraphics[width=4.5cm,height=4.5cm]{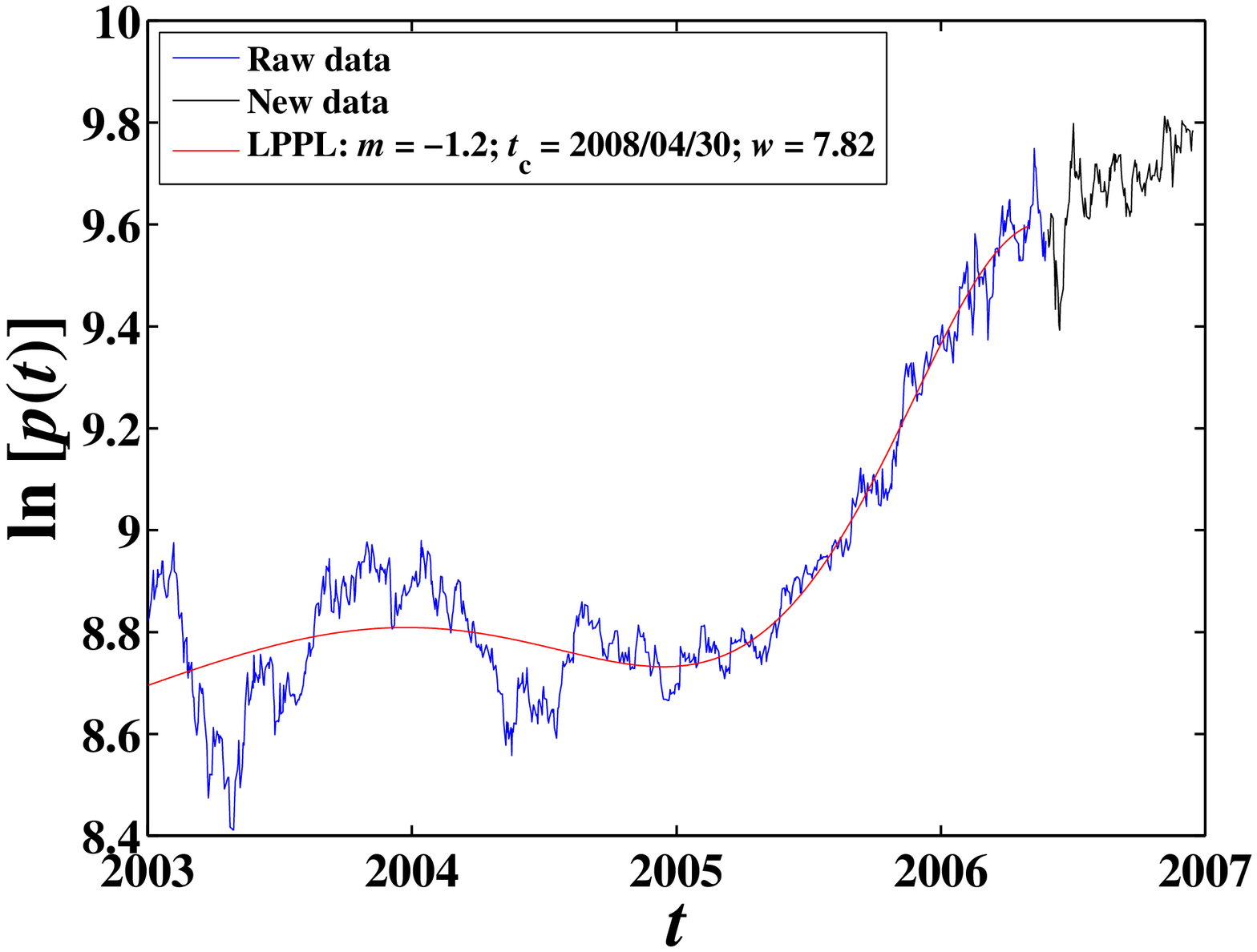}
  \end{minipage}}\\[10pt]
\end{center}
\caption{Log-periodic power-law fits of the prices of stocks from
No.10 to No.18.} \label{Fig:SouthAfrica:S0S1:2}
\end{figure}

\begin{figure}[htb]
\begin{center}
  \subfigure[Stock No.19: INL]{
  \label{Fig:SouthAfrica:S0S1:INL}
  \begin{minipage}[b]{0.31\textwidth}
    \includegraphics[width=4.5cm,height=4.5cm]{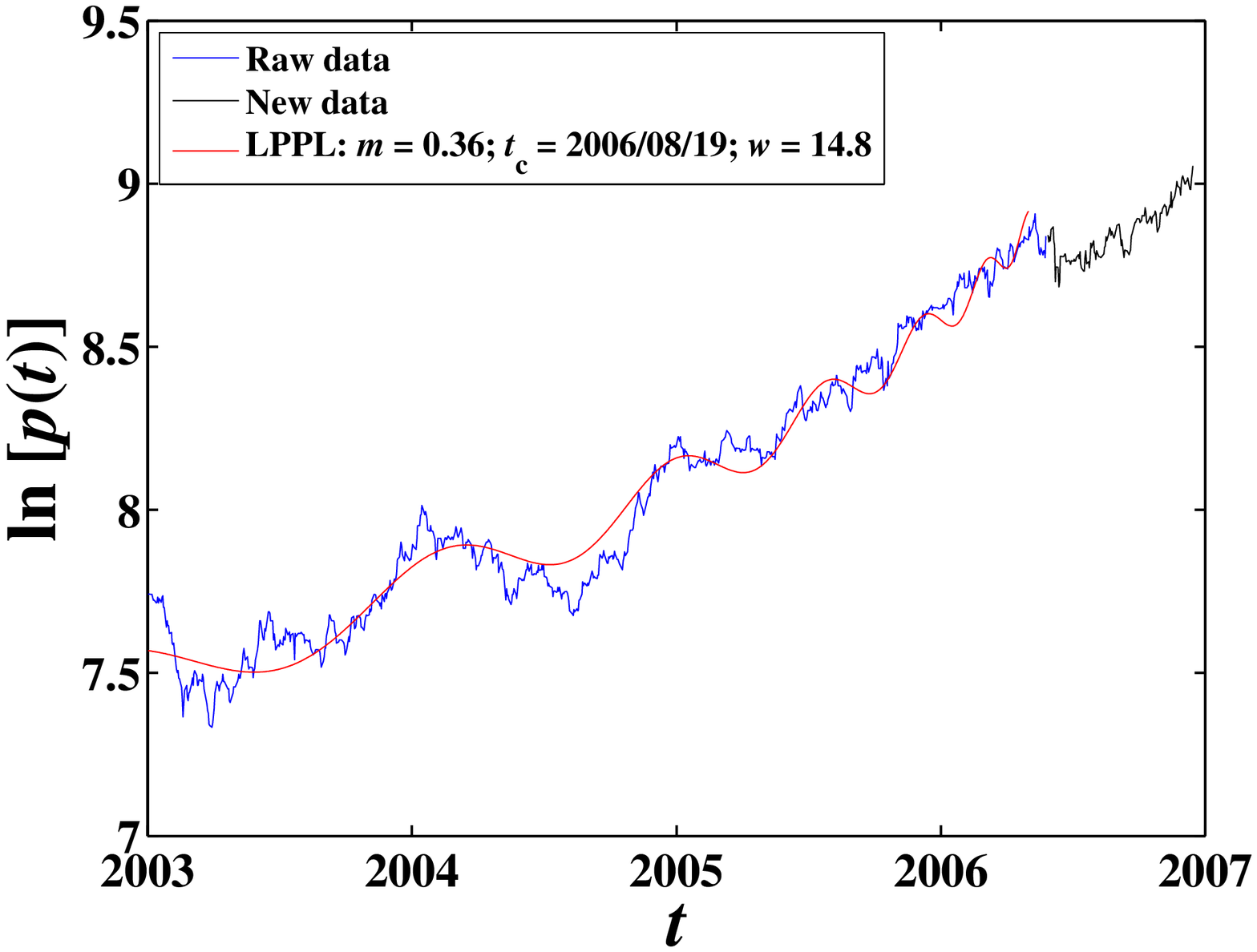}
  \end{minipage}}
  \hspace{0.1cm}
  \subfigure[Stock No.20: INP]{
  \label{Fig:SouthAfrica:S0S1:INP}
  \begin{minipage}[b]{0.31\textwidth}
    \includegraphics[width=4.5cm,height=4.5cm]{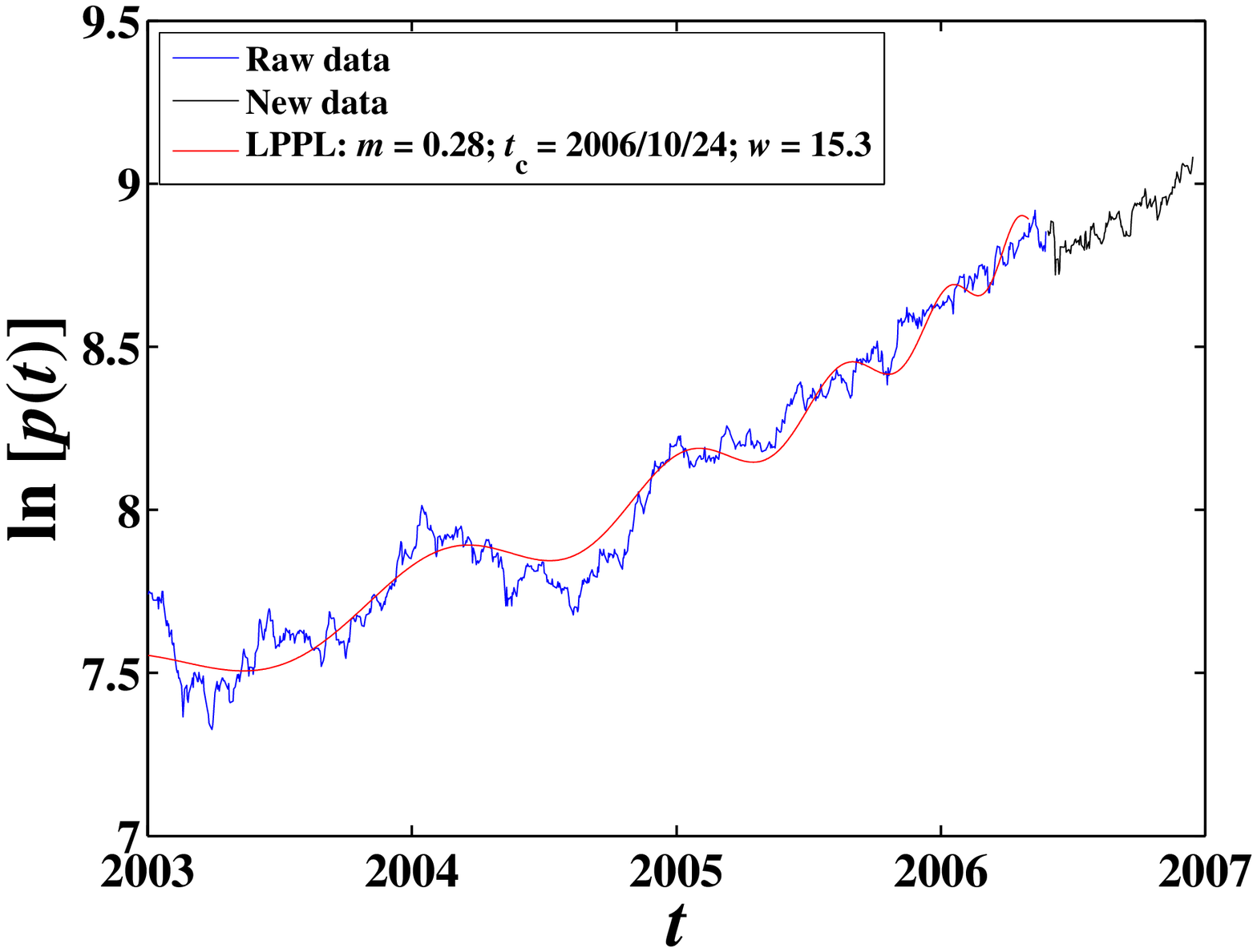}
  \end{minipage}}
  \hspace{0.1cm}
  \subfigure[Stock No.21: IPL]{
  \label{Fig:SouthAfrica:S0S1:IPL}
  \begin{minipage}[b]{0.31\textwidth}
    \includegraphics[width=4.5cm,height=4.5cm]{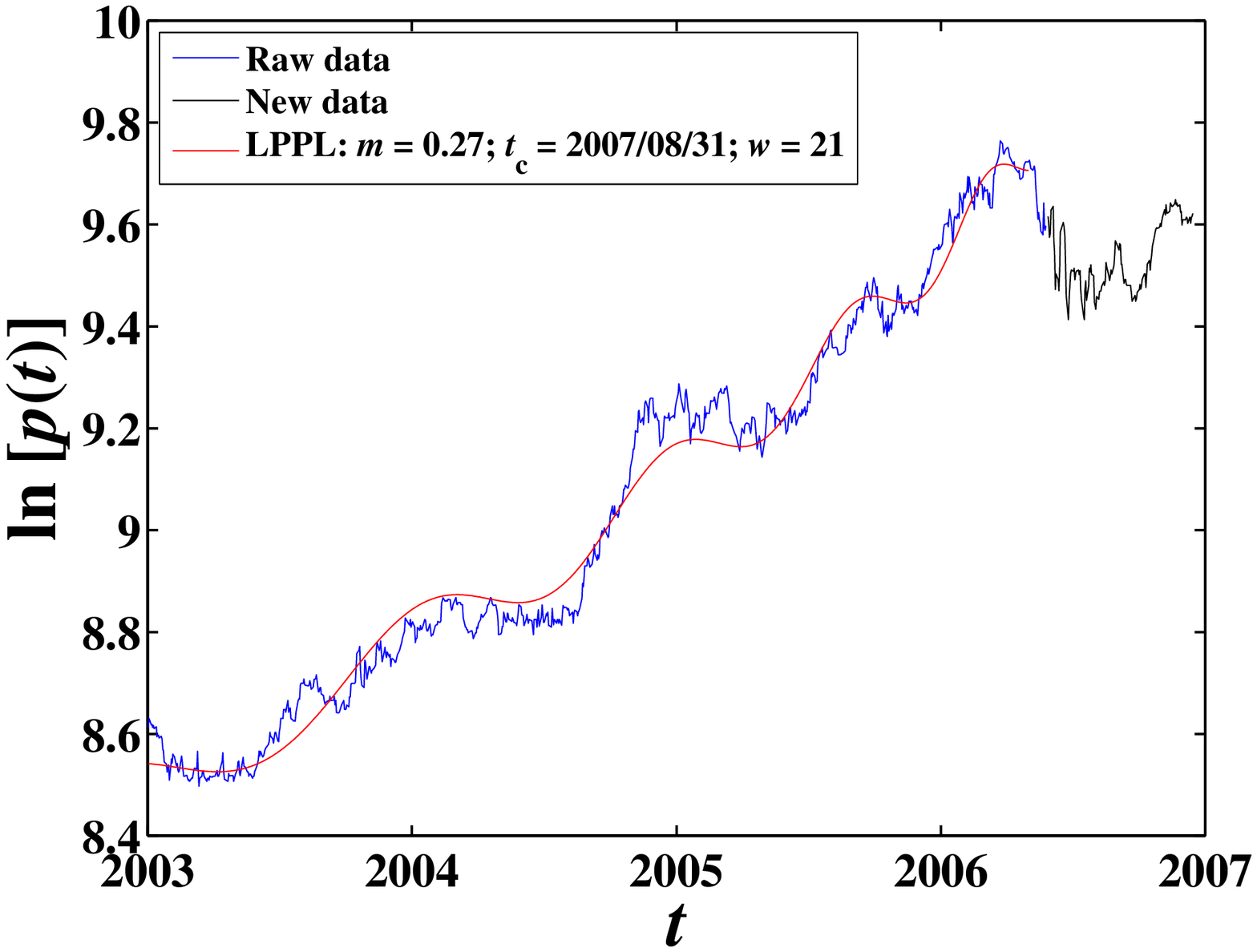}
  \end{minipage}}\\[10pt]
  \subfigure[Stock No.22: JDG]{
  \label{Fig:SouthAfrica:S0S1:JDG}
  \begin{minipage}[b]{0.31\textwidth}
    \includegraphics[width=4.5cm,height=4.5cm]{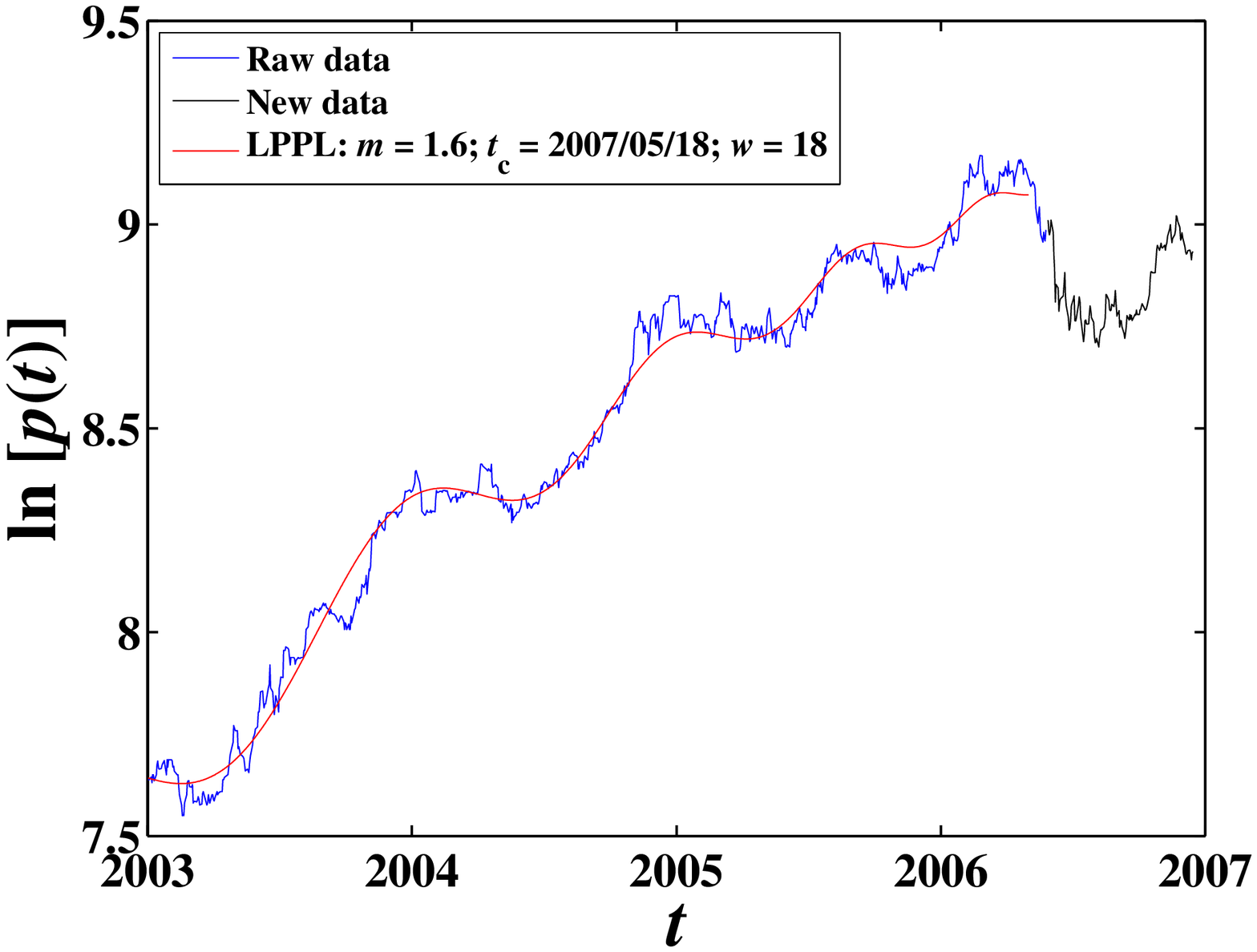}
  \end{minipage}}
  \hspace{0.1cm}
  \subfigure[Stock No.23: KMB]{
  \label{Fig:SouthAfrica:S0S1:KMB}
  \begin{minipage}[b]{0.31\textwidth}
    \includegraphics[width=4.5cm,height=4.5cm]{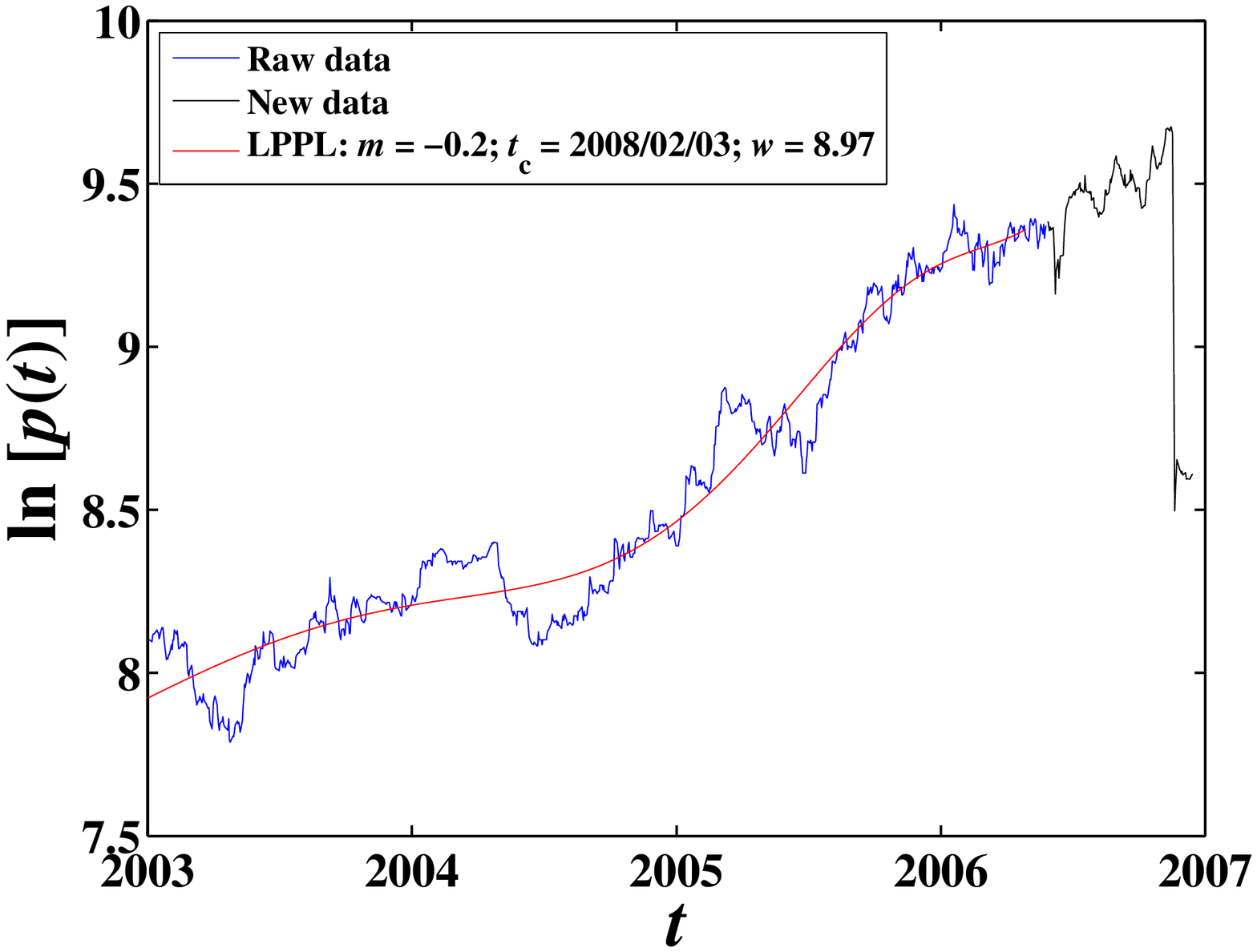}
  \end{minipage}}
  \hspace{0.1cm}
  \subfigure[Stock No.24: LBT]{
  \label{Fig:SouthAfrica:S0S1:LBT}
  \begin{minipage}[b]{0.31\textwidth}
    \includegraphics[width=4.5cm,height=4.5cm]{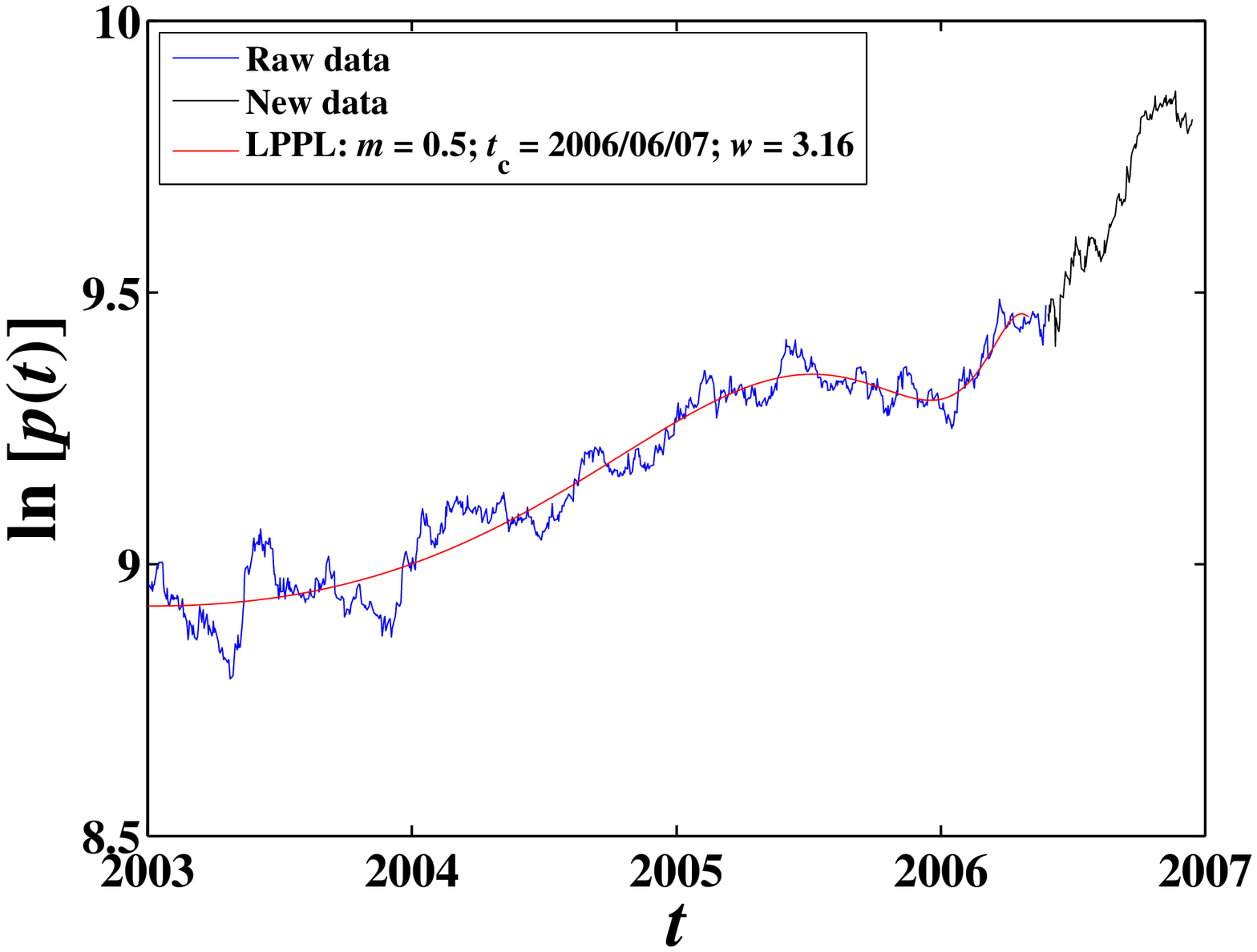}
  \end{minipage}}\\[10pt]
  \subfigure[Stock No.25: LGL]{
  \label{Fig:SouthAfrica:S0S1:LGL}
  \begin{minipage}[b]{0.31\textwidth}
    \includegraphics[width=4.5cm,height=4.5cm]{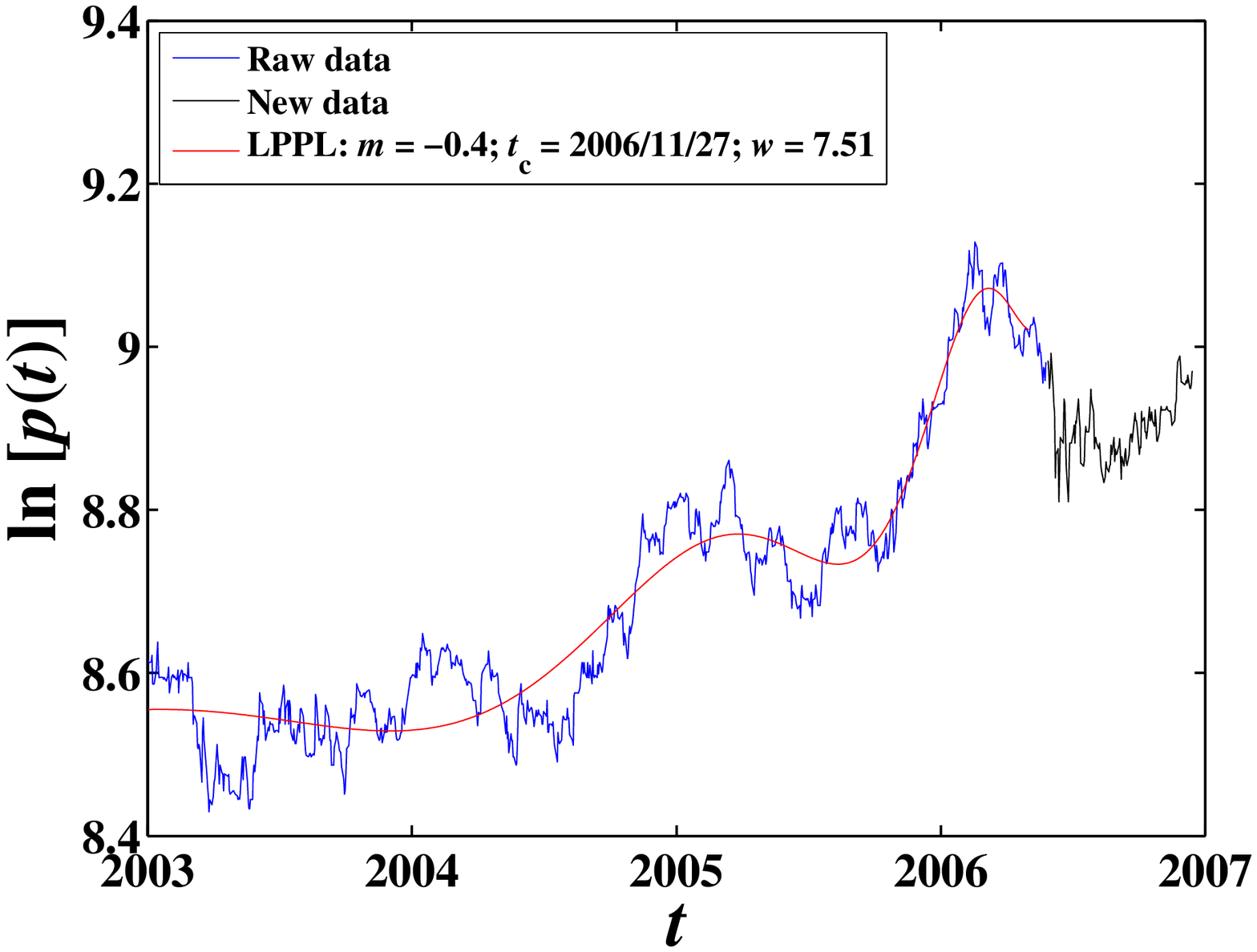}
  \end{minipage}}
  \hspace{0.1cm}
  \subfigure[Stock No.26: MLA]{
  \label{Fig:SouthAfrica:S0S1:MLA}
  \begin{minipage}[b]{0.31\textwidth}
    \includegraphics[width=4.5cm,height=4.5cm]{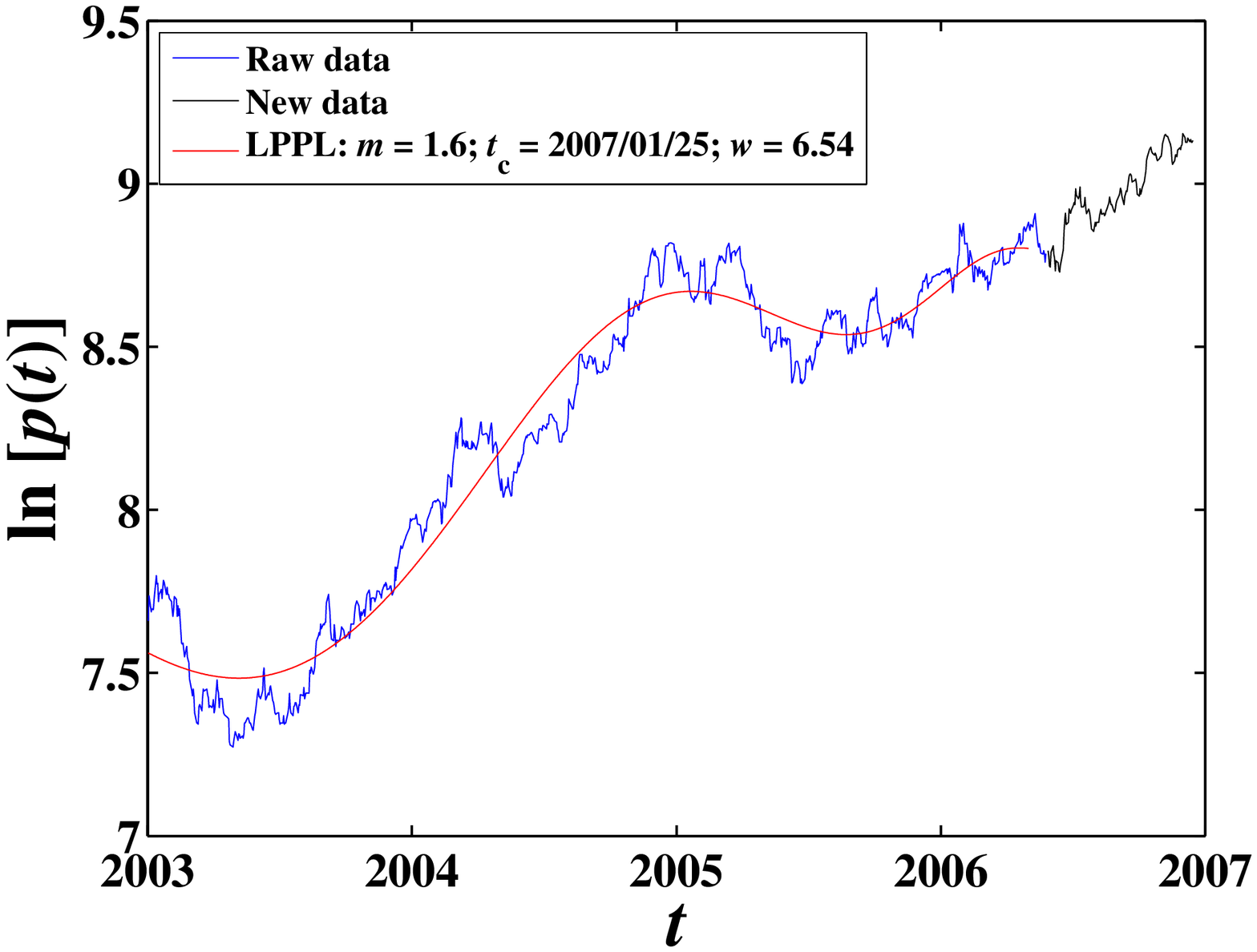}
  \end{minipage}}
  \hspace{0.1cm}
  \subfigure[Stock No.27: MTN]{
  \label{Fig:SouthAfrica:S0S1:MTN}
  \begin{minipage}[b]{0.31\textwidth}
    \includegraphics[width=4.5cm,height=4.5cm]{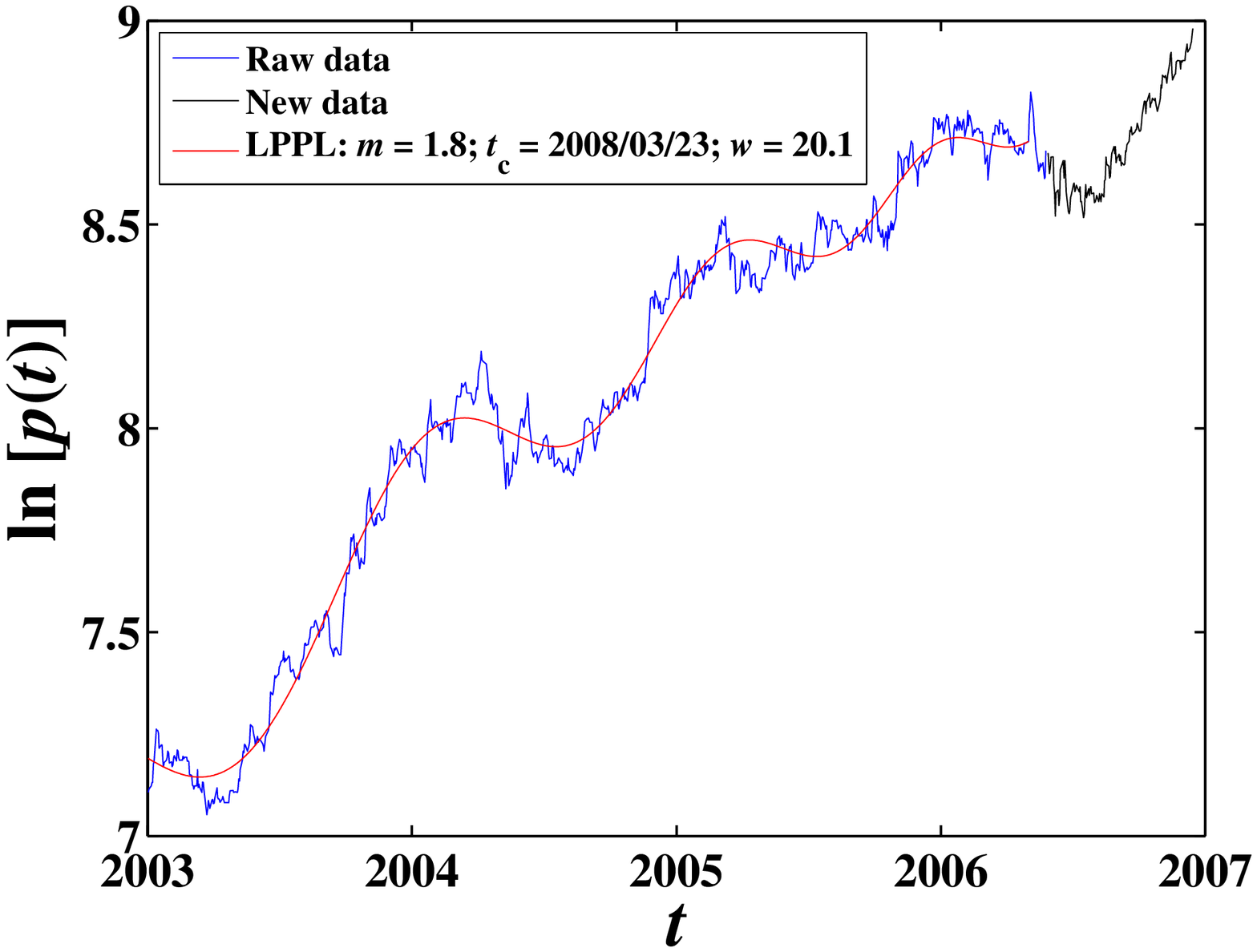}
  \end{minipage}}\\[10pt]
\end{center}
\caption{Log-periodic power-law fits of the prices of stocks from
No.19 to No.27.} \label{Fig:SouthAfrica:S0S1:3}
\end{figure}

\begin{figure}[htb]
\begin{center}
  \subfigure[Stock No.28: NED]{
  \label{Fig:SouthAfrica:S0S1:NED}
  \begin{minipage}[b]{0.31\textwidth}
    \includegraphics[width=4.5cm,height=4.5cm]{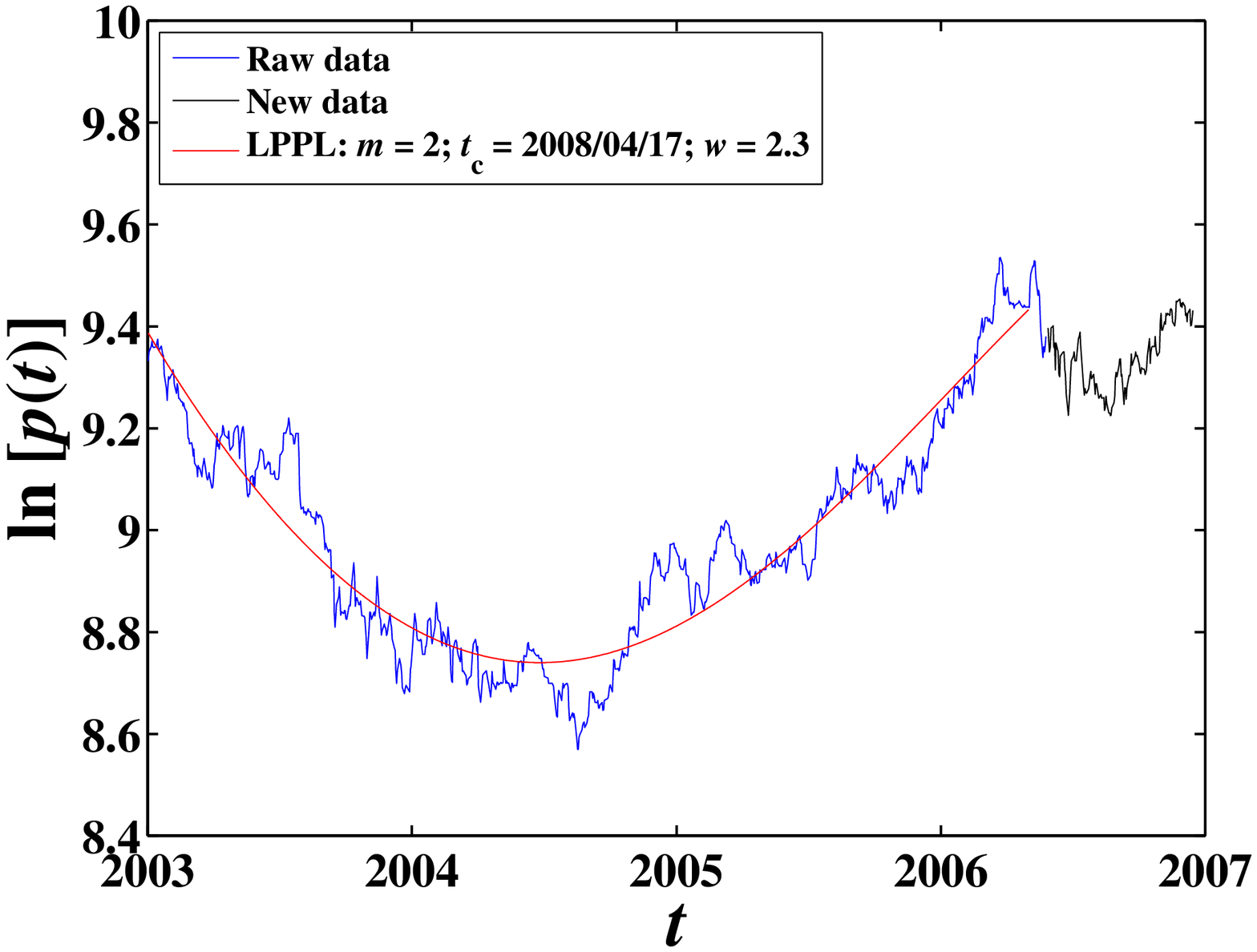}
  \end{minipage}}
  \hspace{0.1cm}
  \subfigure[Stock No.29: NPN]{
  \label{Fig:SouthAfrica:S0S1:NPN}
  \begin{minipage}[b]{0.31\textwidth}
    \includegraphics[width=4.5cm,height=4.5cm]{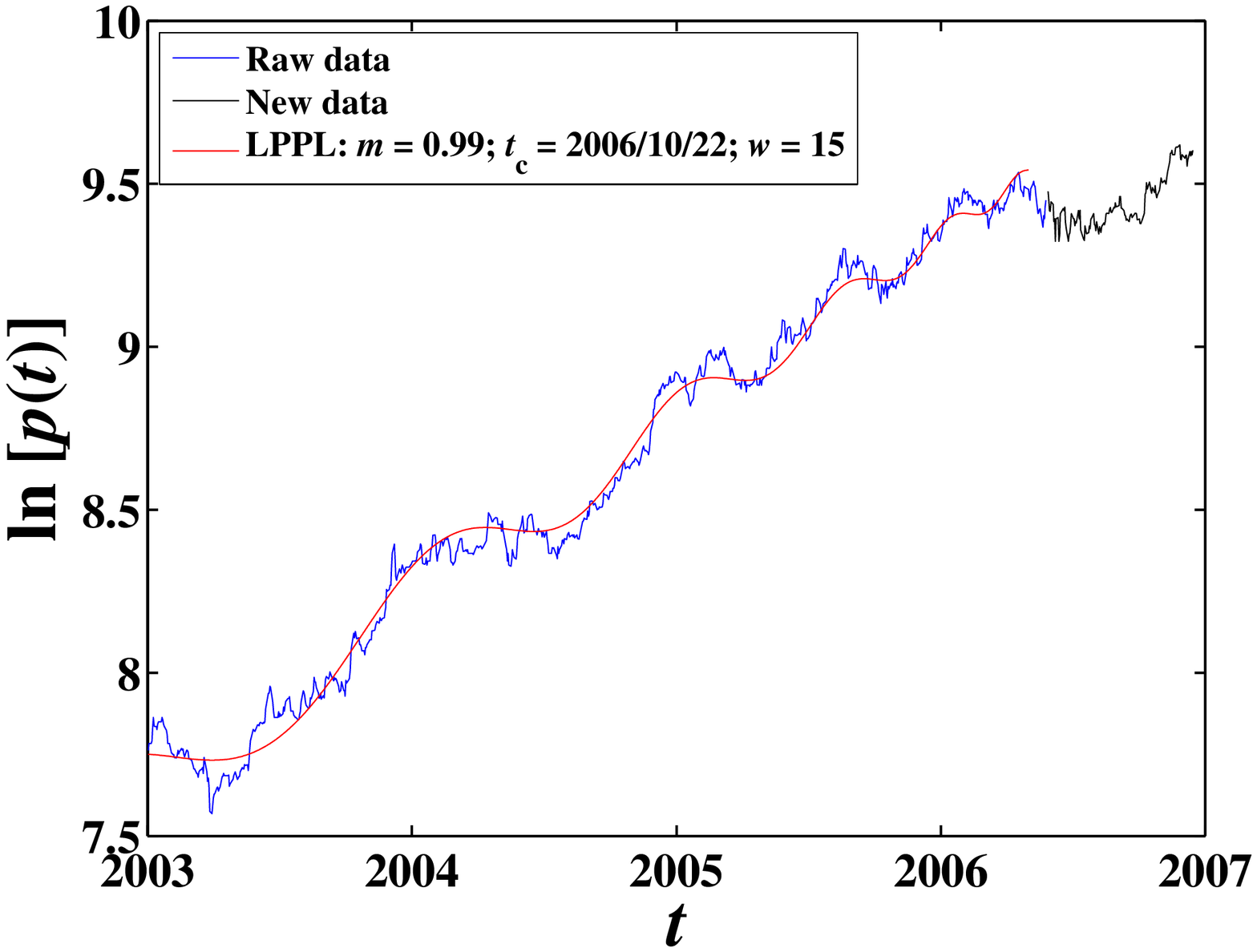}
  \end{minipage}}
  \hspace{0.1cm}
  \subfigure[Stock No.30: NTC]{
  \label{Fig:SouthAfrica:S0S1:NTC}
  \begin{minipage}[b]{0.31\textwidth}
    \includegraphics[width=4.5cm,height=4.5cm]{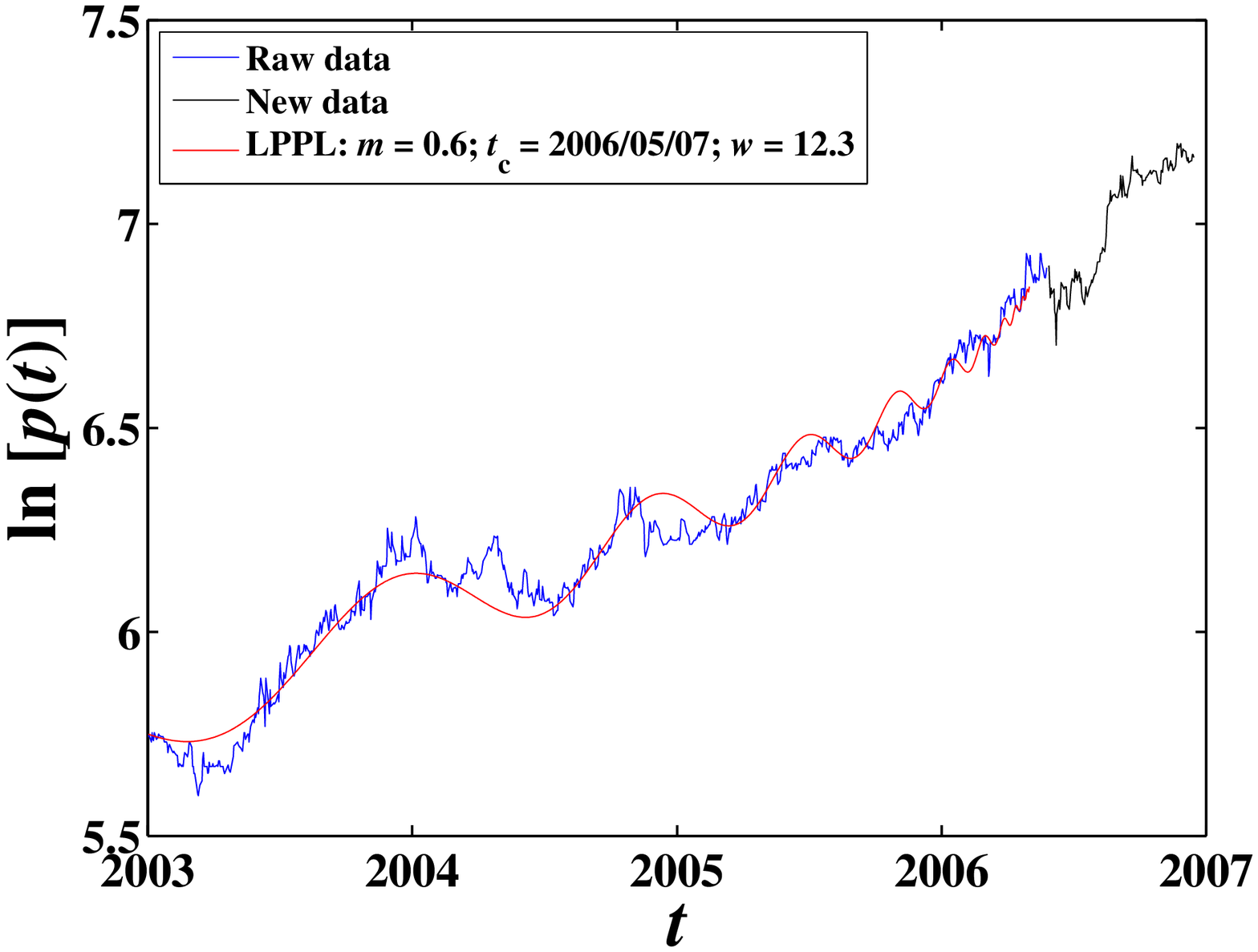}
  \end{minipage}}\\[10pt]
  \subfigure[Stock No.31: OML]{
  \label{Fig:SouthAfrica:S0S1:OML}
  \begin{minipage}[b]{0.31\textwidth}
    \includegraphics[width=4.5cm,height=4.5cm]{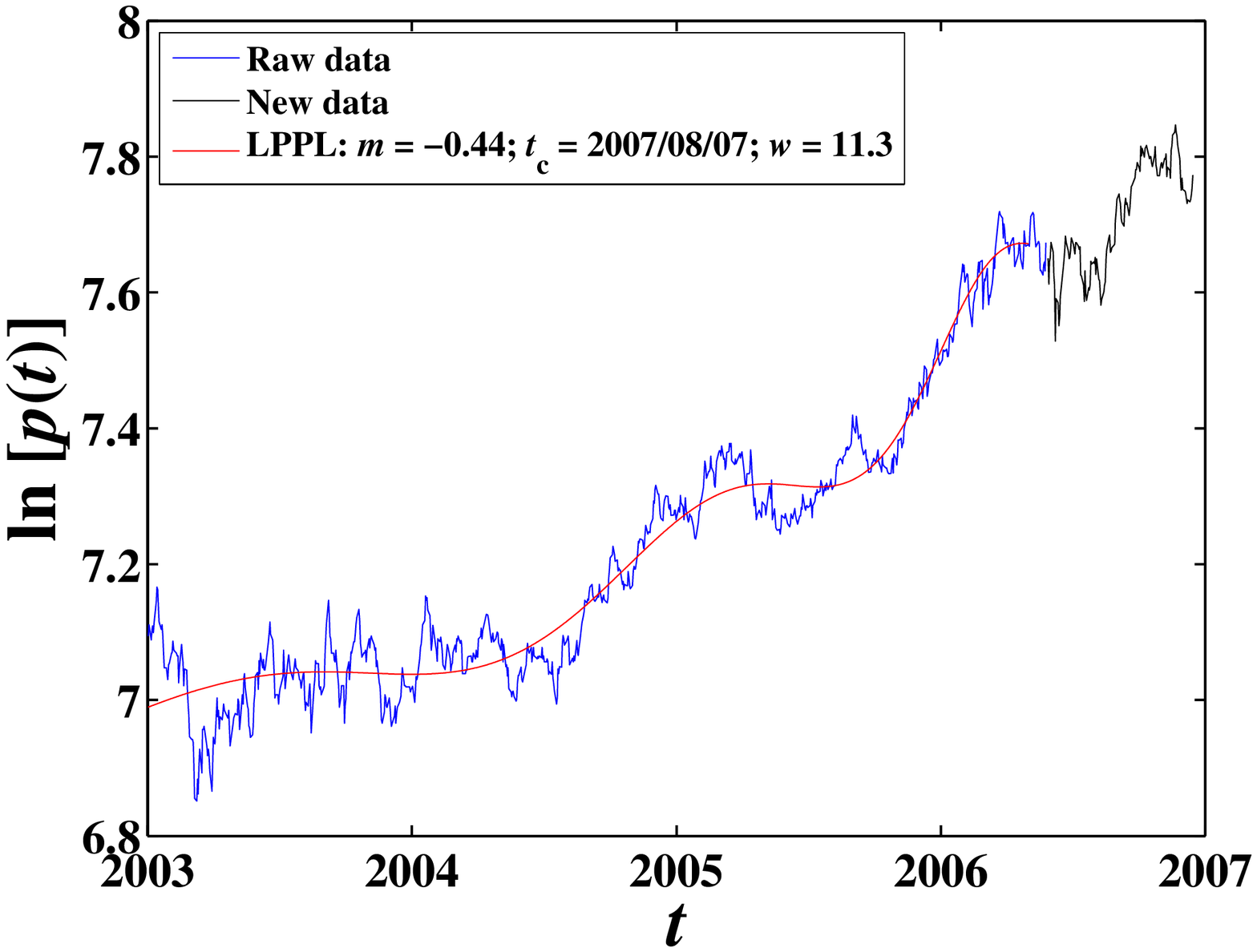}
  \end{minipage}}
  \hspace{0.1cm}
  \subfigure[Stock No.32: PIK]{
  \label{Fig:SouthAfrica:S0S1:PIK}
  \begin{minipage}[b]{0.31\textwidth}
    \includegraphics[width=4.5cm,height=4.5cm]{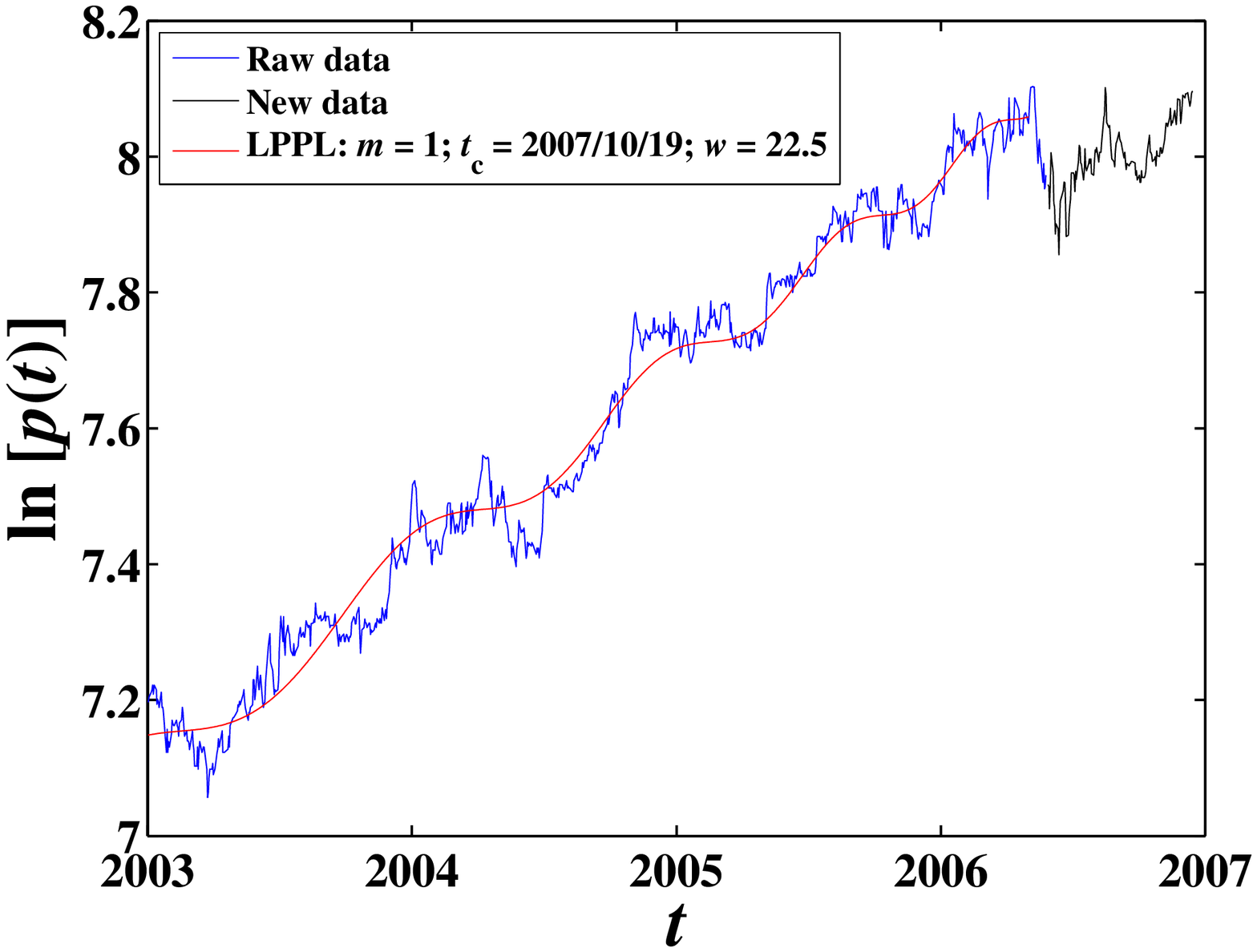}
  \end{minipage}}
  \hspace{0.1cm}
  \subfigure[Stock No.33: PPC]{
  \label{Fig:SouthAfrica:S0S1:PPC}
  \begin{minipage}[b]{0.31\textwidth}
    \includegraphics[width=4.5cm,height=4.5cm]{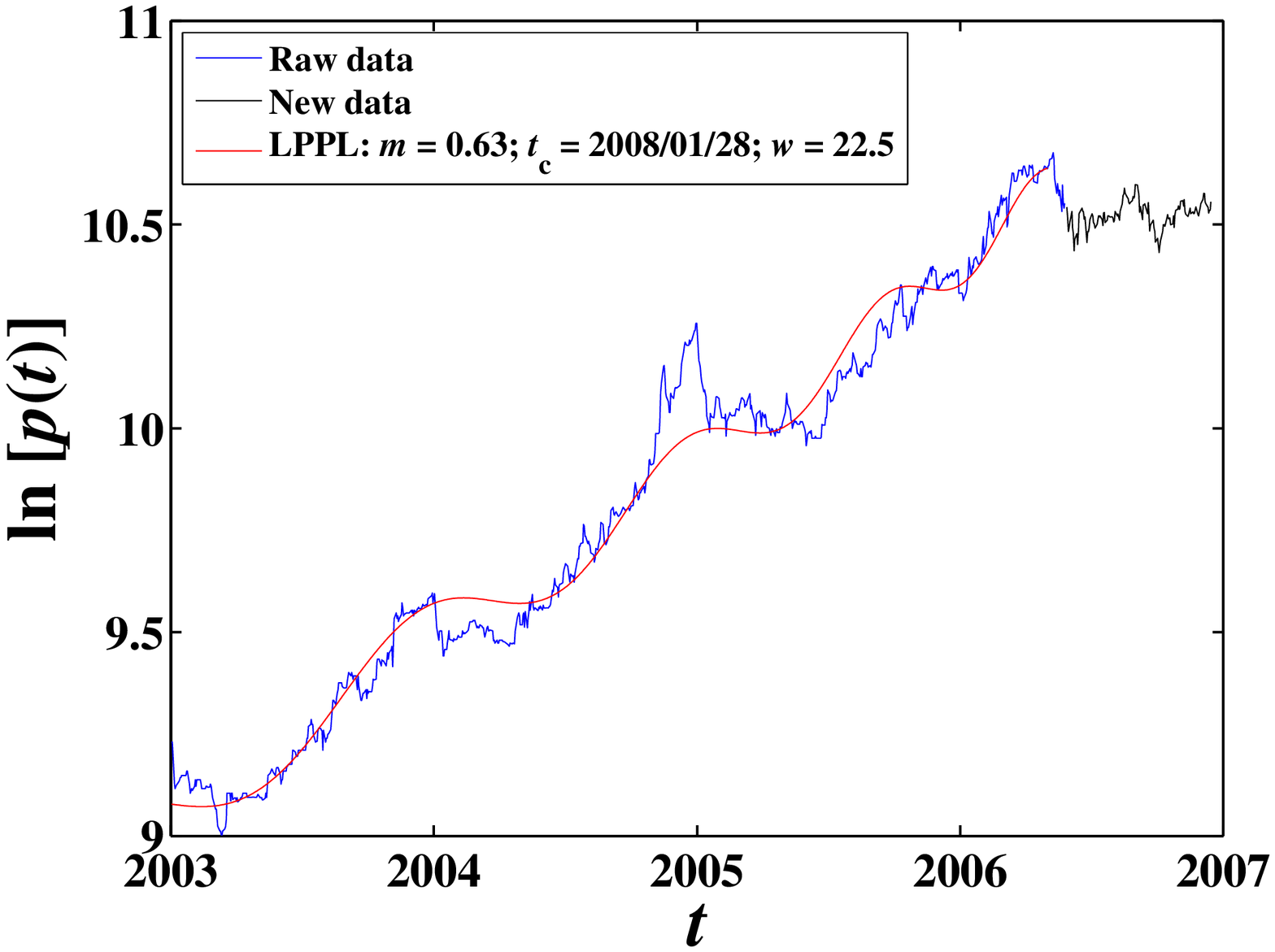}
  \end{minipage}}\\[10pt]
  \subfigure[Stock No.34: RCH]{
  \label{Fig:SouthAfrica:S0S1:RCH}
  \begin{minipage}[b]{0.31\textwidth}
    \includegraphics[width=4.5cm,height=4.5cm]{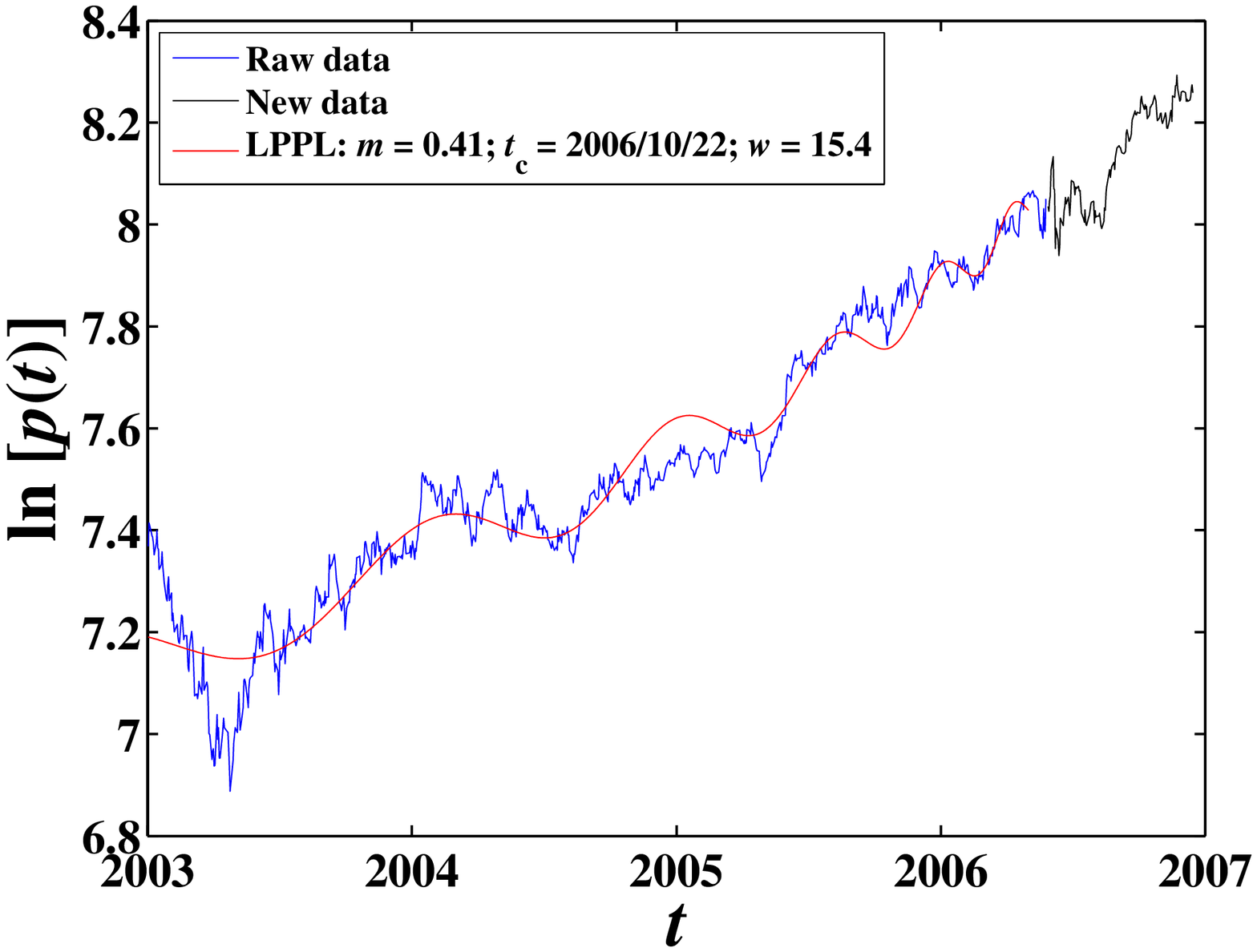}
  \end{minipage}}
  \hspace{0.1cm}
  \subfigure[Stock No.35: REM]{
  \label{Fig:SouthAfrica:S0S1:REM}
  \begin{minipage}[b]{0.31\textwidth}
    \includegraphics[width=4.5cm,height=4.5cm]{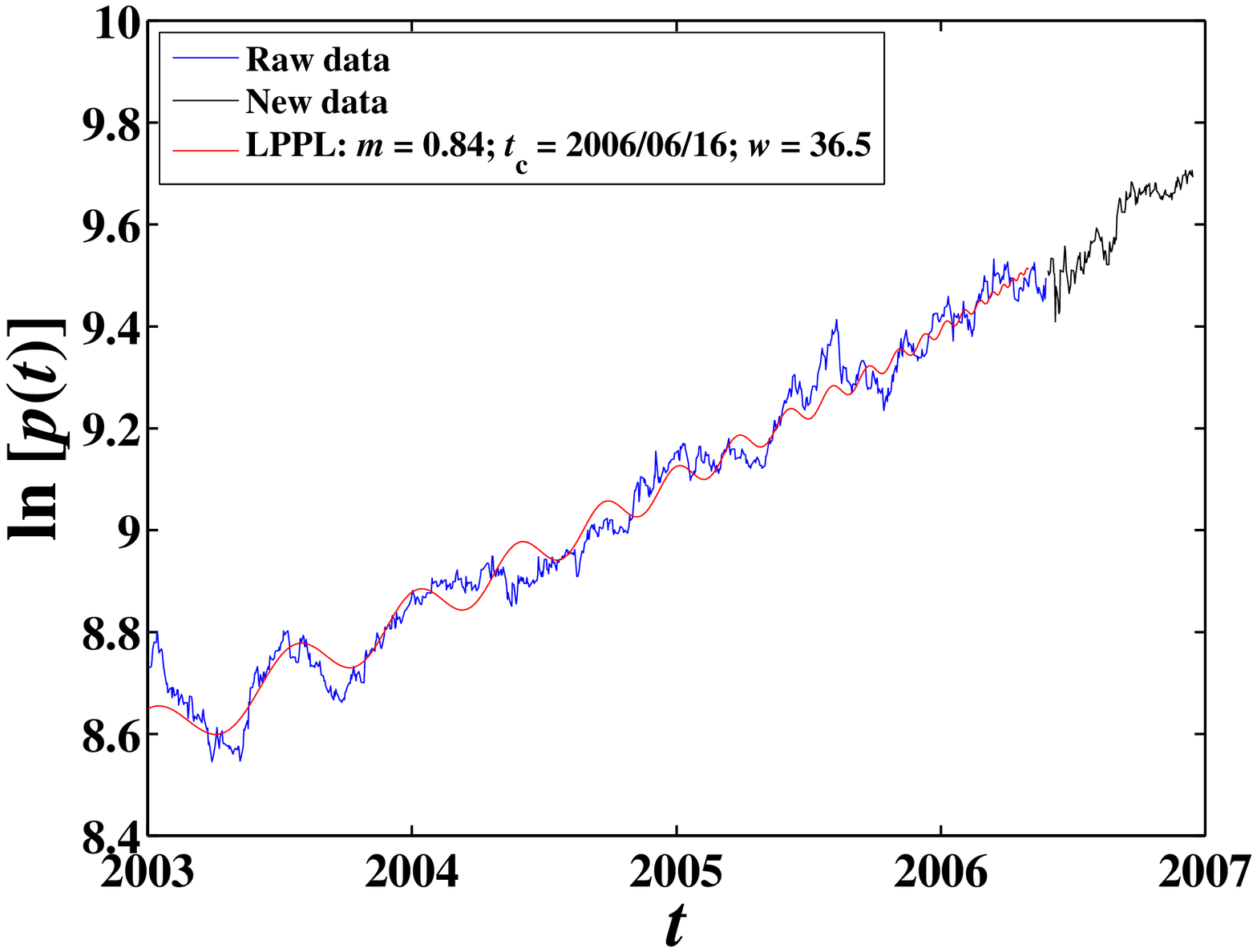}
  \end{minipage}}
  \hspace{0.1cm}
  \subfigure[Stock No.36: RMH]{
  \label{Fig:SouthAfrica:S0S1:RMH}
  \begin{minipage}[b]{0.31\textwidth}
    \includegraphics[width=4.5cm,height=4.5cm]{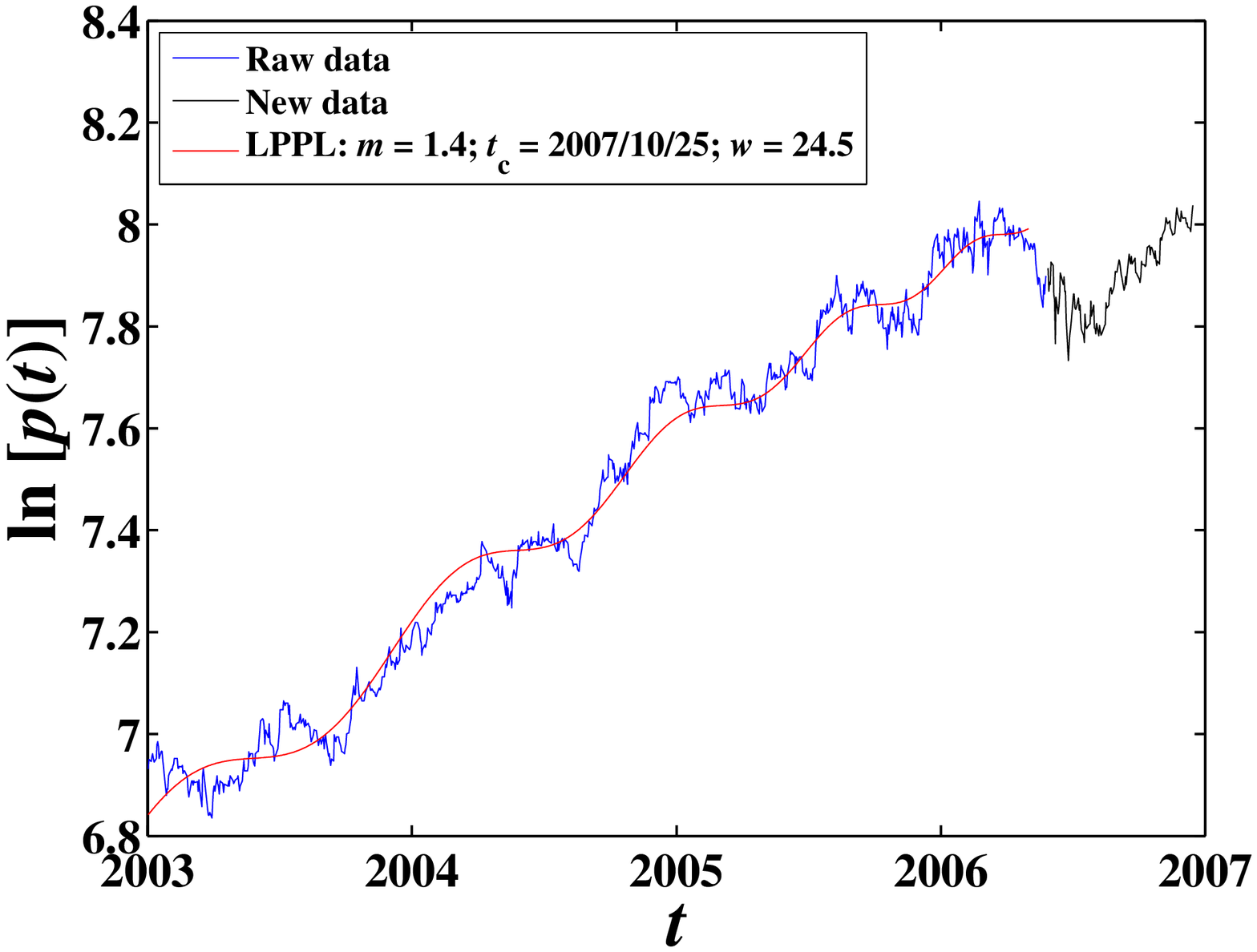}
  \end{minipage}}\\[10pt]
\end{center}
\caption{Log-periodic power-law fits of the prices of stocks from
No.28 to No.36.} \label{Fig:SouthAfrica:S0S1:4}
\end{figure}

\begin{figure}[htb]
\begin{center}
  \subfigure[Stock No.37: SAB]{
  \label{Fig:SouthAfrica:S0S1:SAB}
  \begin{minipage}[b]{0.31\textwidth}
    \includegraphics[width=4.5cm,height=4.5cm]{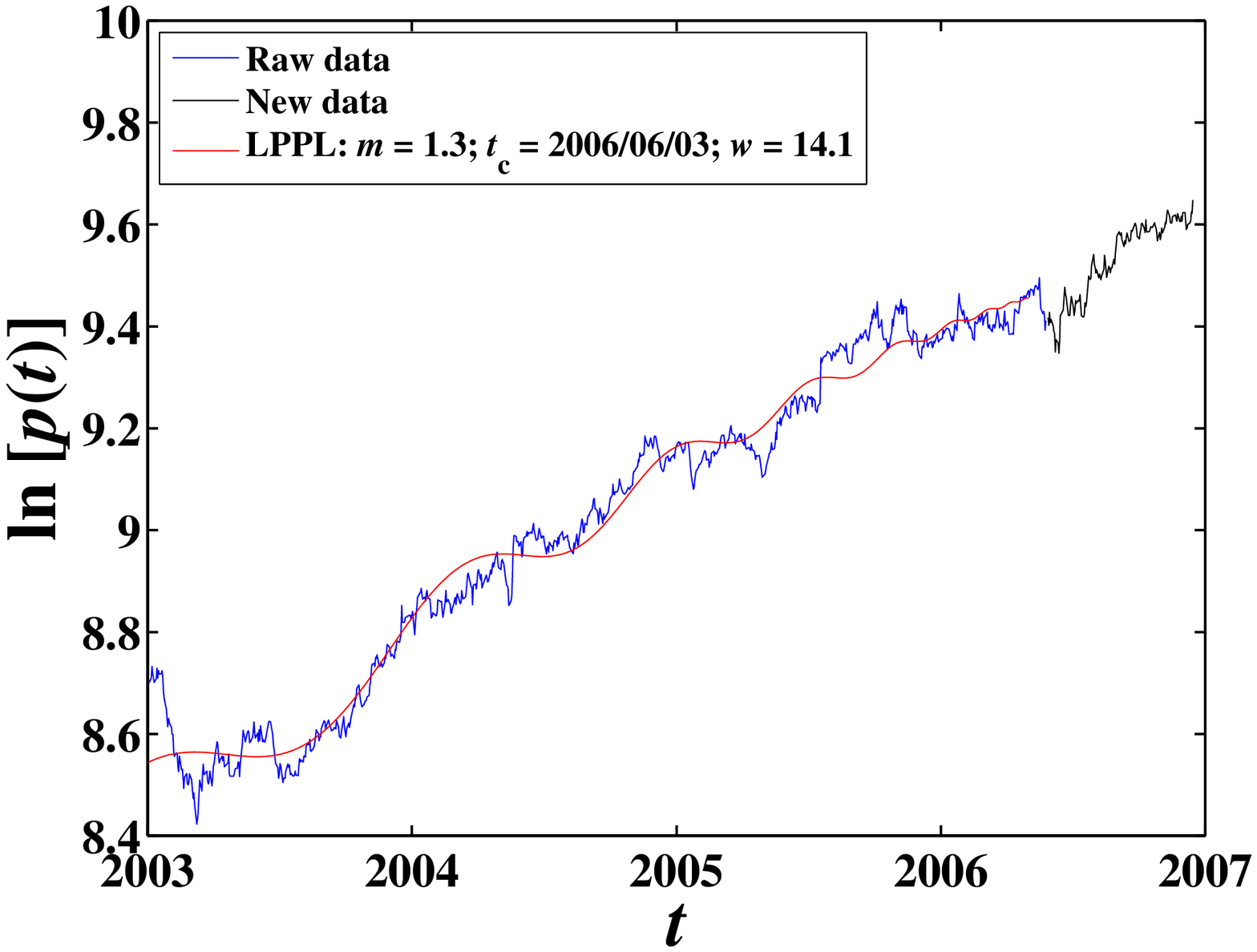}
  \end{minipage}}
  \hspace{0.1cm}
  \subfigure[Stock No.38: SAP]{
  \label{Fig:SouthAfrica:S0S1:SAP}
  \begin{minipage}[b]{0.31\textwidth}
    \includegraphics[width=4.5cm,height=4.5cm]{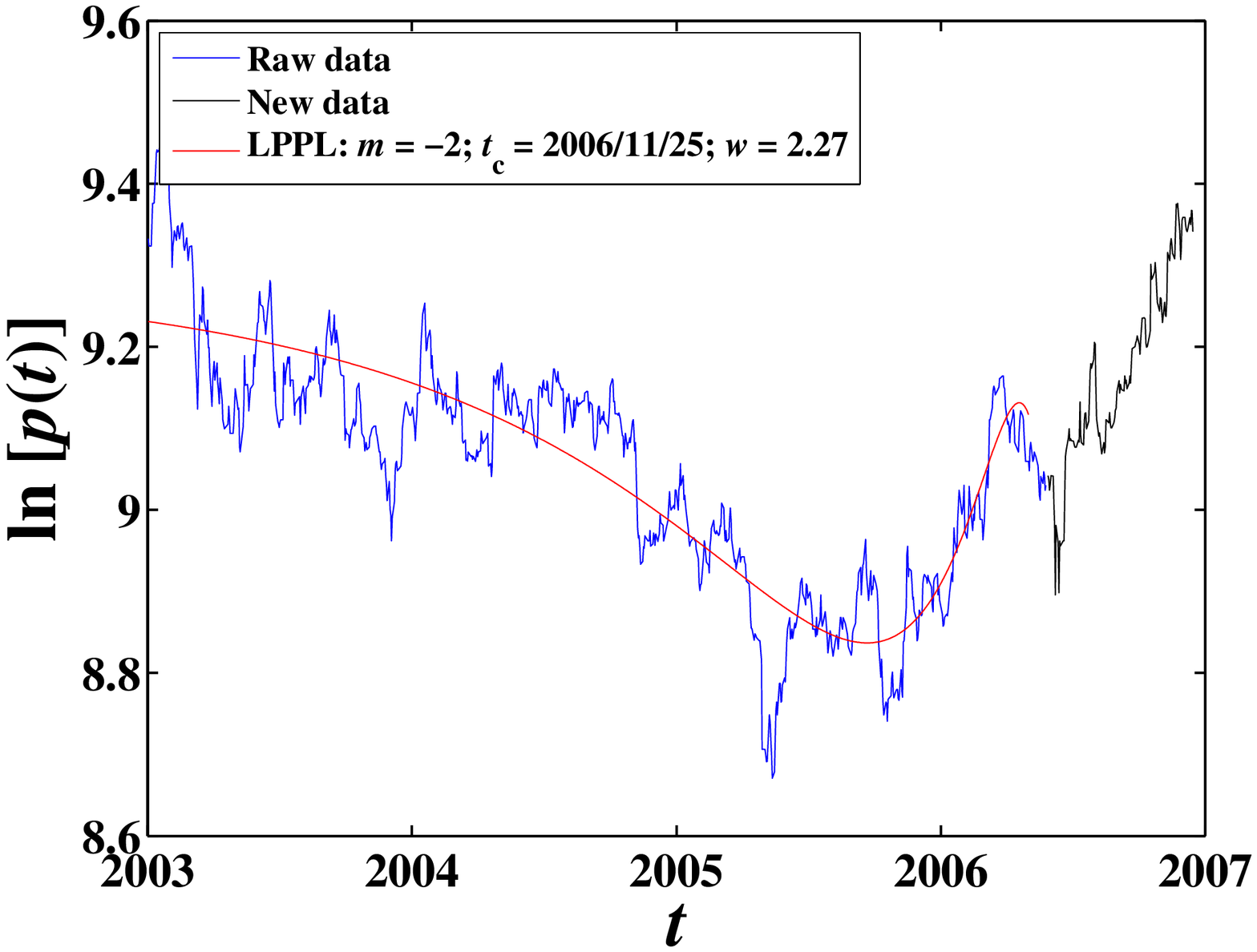}
  \end{minipage}}
  \hspace{0.1cm}
  \subfigure[Stock No.39: SBK]{
  \label{Fig:SouthAfrica:S0S1:SBK}
  \begin{minipage}[b]{0.31\textwidth}
    \includegraphics[width=4.5cm,height=4.5cm]{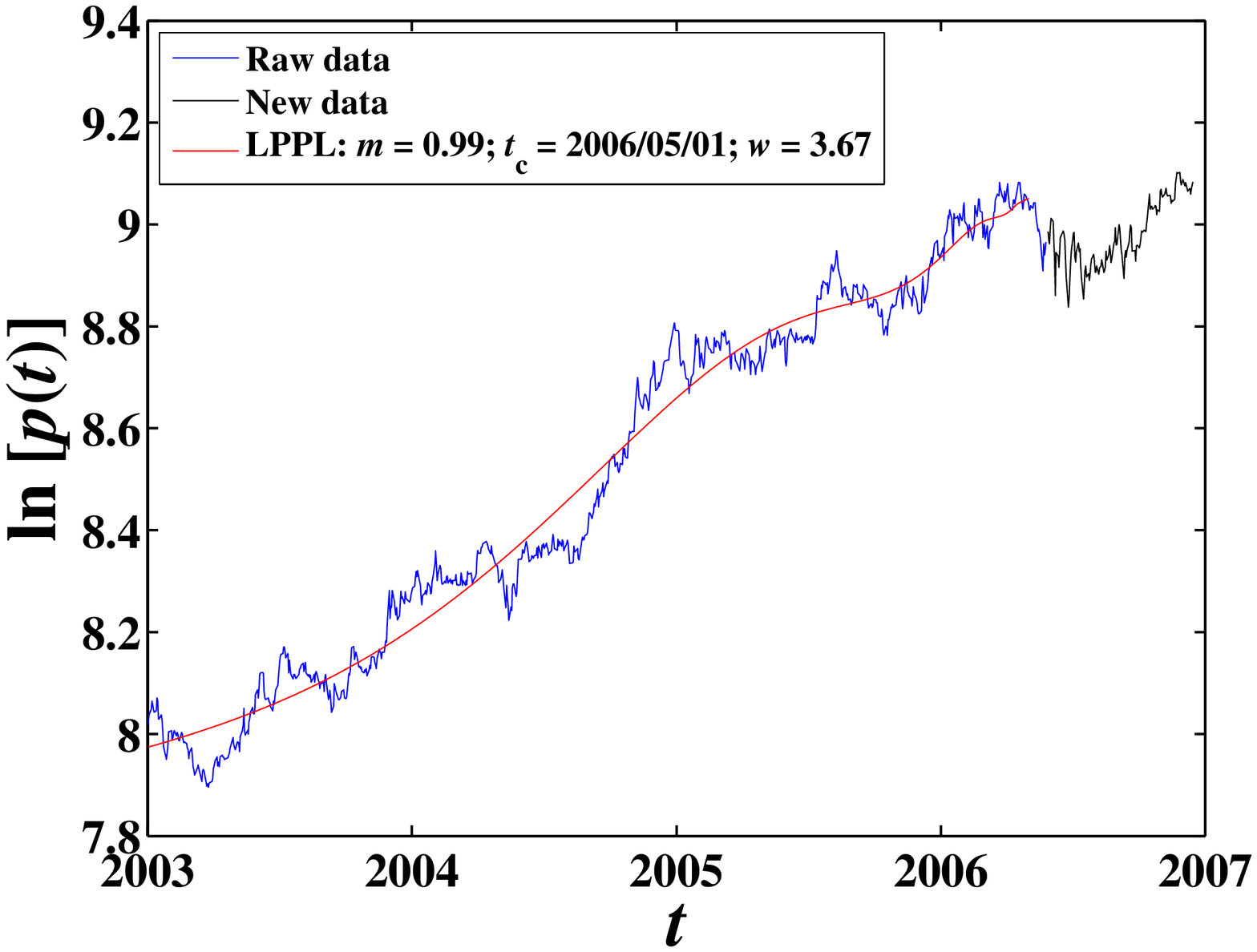}
  \end{minipage}}\\[10pt]
  \subfigure[Stock No.40: SHF]{
  \label{Fig:SouthAfrica:S0S1:SHF}
  \begin{minipage}[b]{0.31\textwidth}
    \includegraphics[width=4.5cm,height=4.5cm]{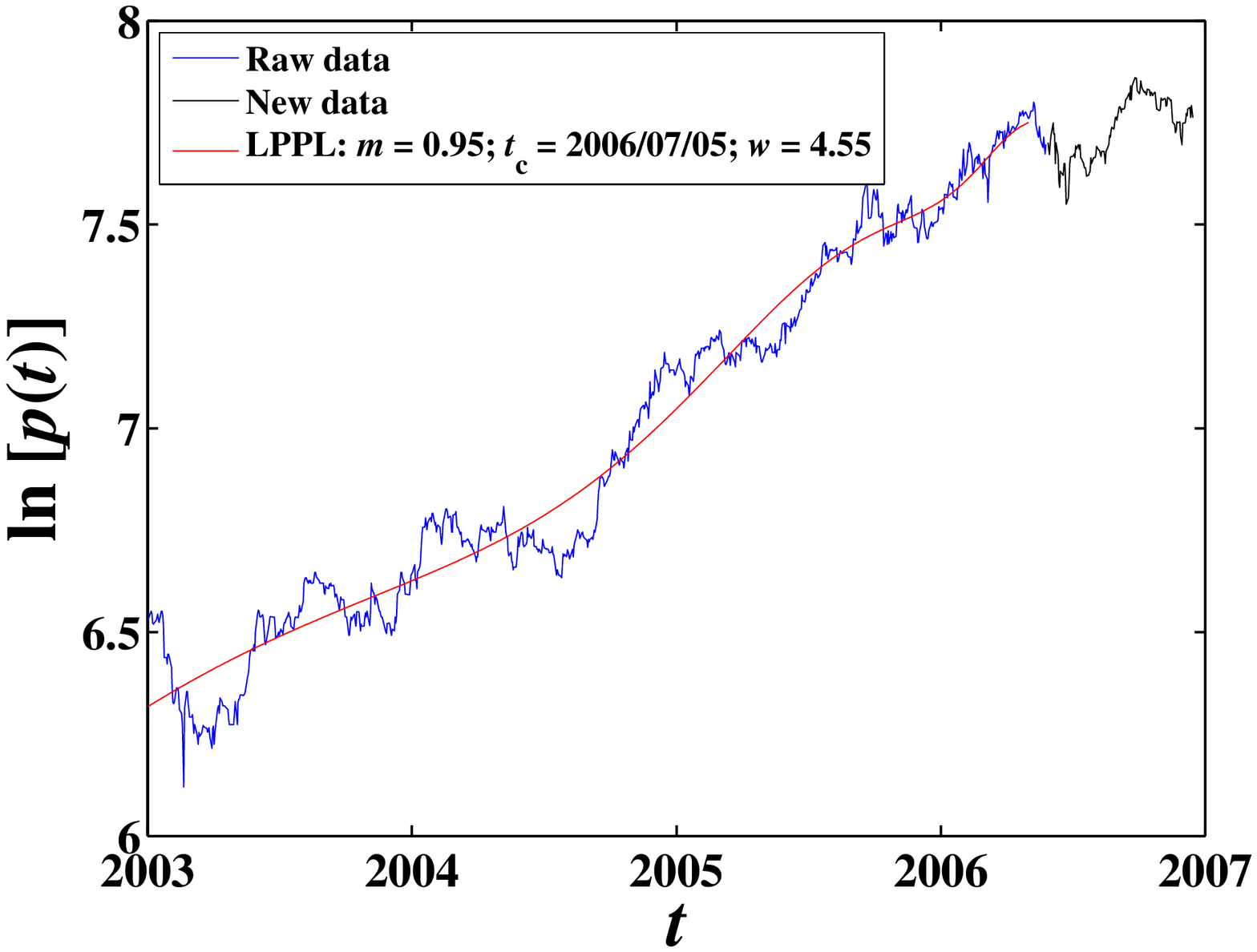}
  \end{minipage}}
  \hspace{0.1cm}
  \subfigure[Stock No.41: SLM]{
  \label{Fig:SouthAfrica:S0S1:SLM}
  \begin{minipage}[b]{0.31\textwidth}
    \includegraphics[width=4.5cm,height=4.5cm]{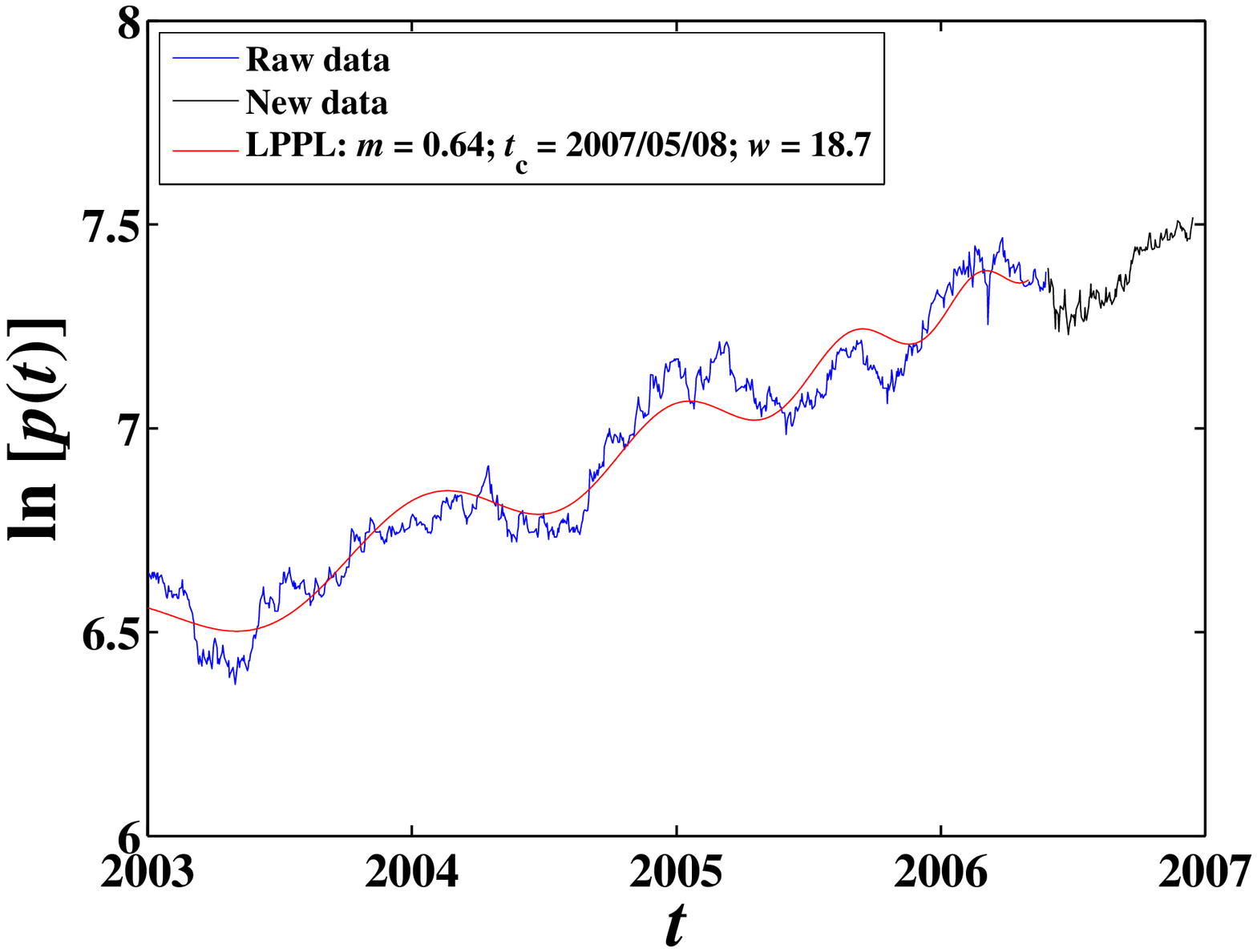}
  \end{minipage}}
  \hspace{0.1cm}
  \subfigure[Stock No.42: SOL]{
  \label{Fig:SouthAfrica:S0S1:SOL}
  \begin{minipage}[b]{0.31\textwidth}
    \includegraphics[width=4.5cm,height=4.5cm]{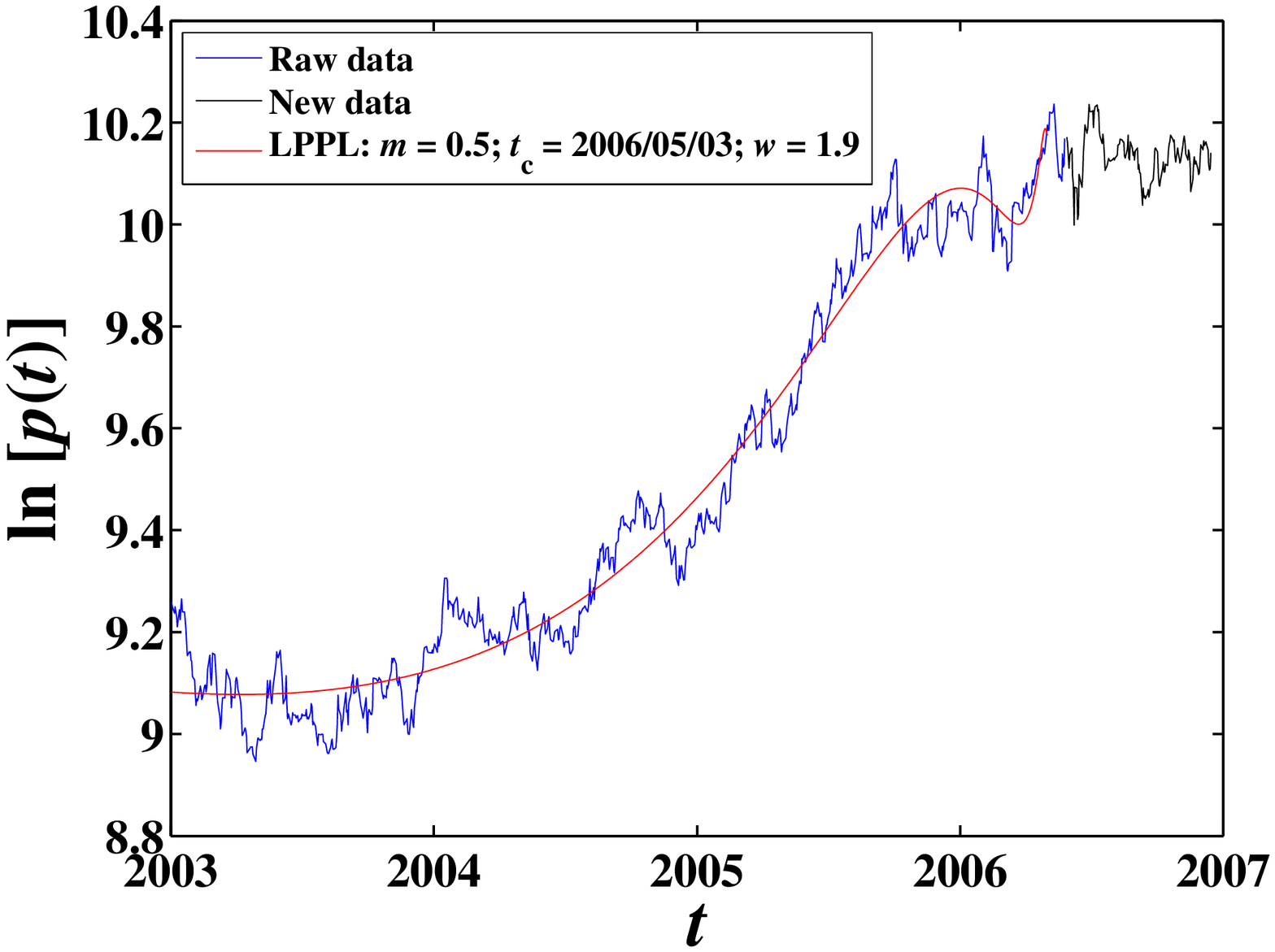}
  \end{minipage}}\\[10pt]
  \subfigure[Stock No.43: TBS]{
  \label{Fig:SouthAfrica:S0S1:TBS}
  \begin{minipage}[b]{0.31\textwidth}
    \includegraphics[width=4.5cm,height=4.5cm]{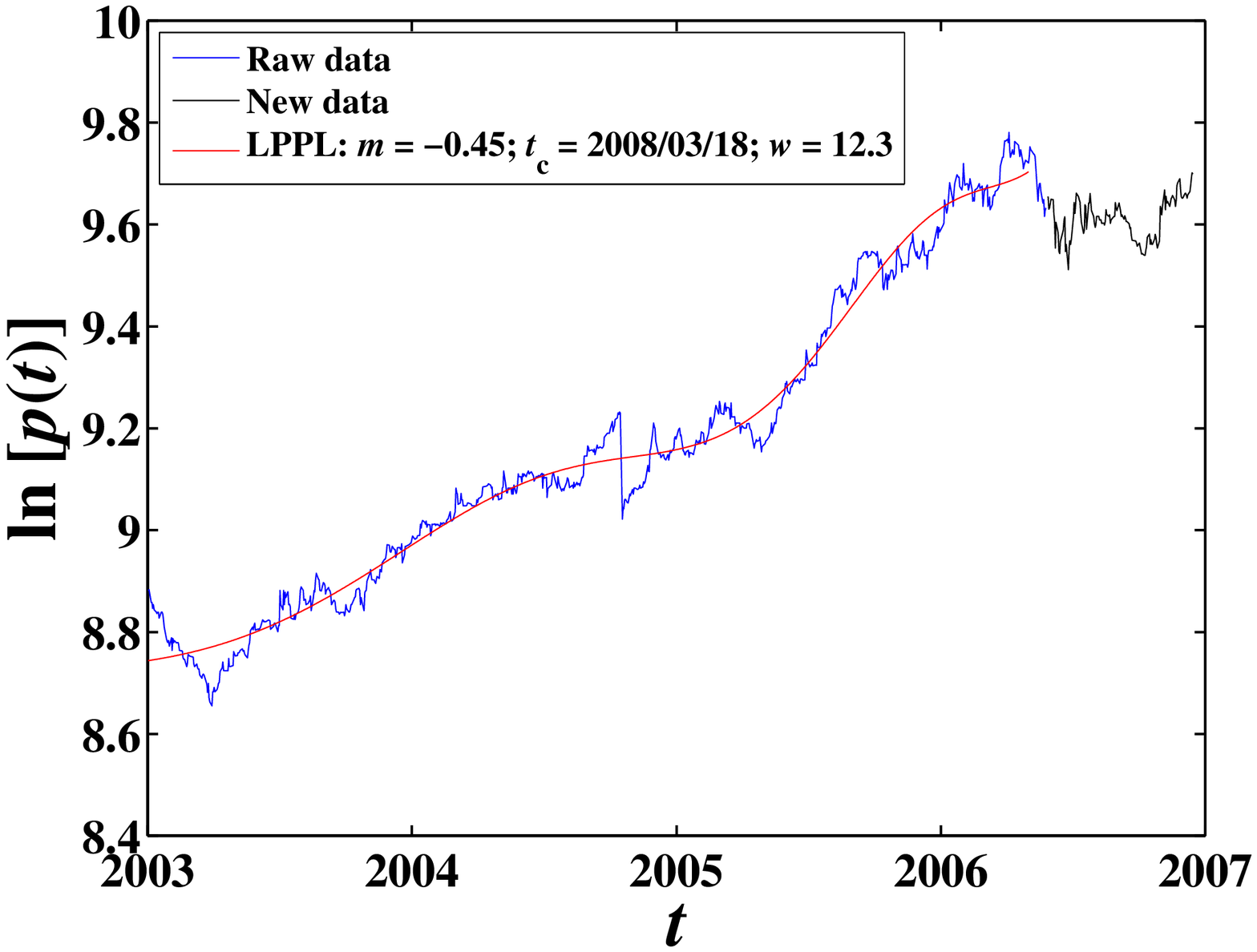}
  \end{minipage}}
  \hspace{0.1cm}
  \subfigure[Stock No.44: TKG]{
  \label{Fig:SouthAfrica:S0S1:TKG}
  \begin{minipage}[b]{0.31\textwidth}
    \includegraphics[width=4.5cm,height=4.5cm]{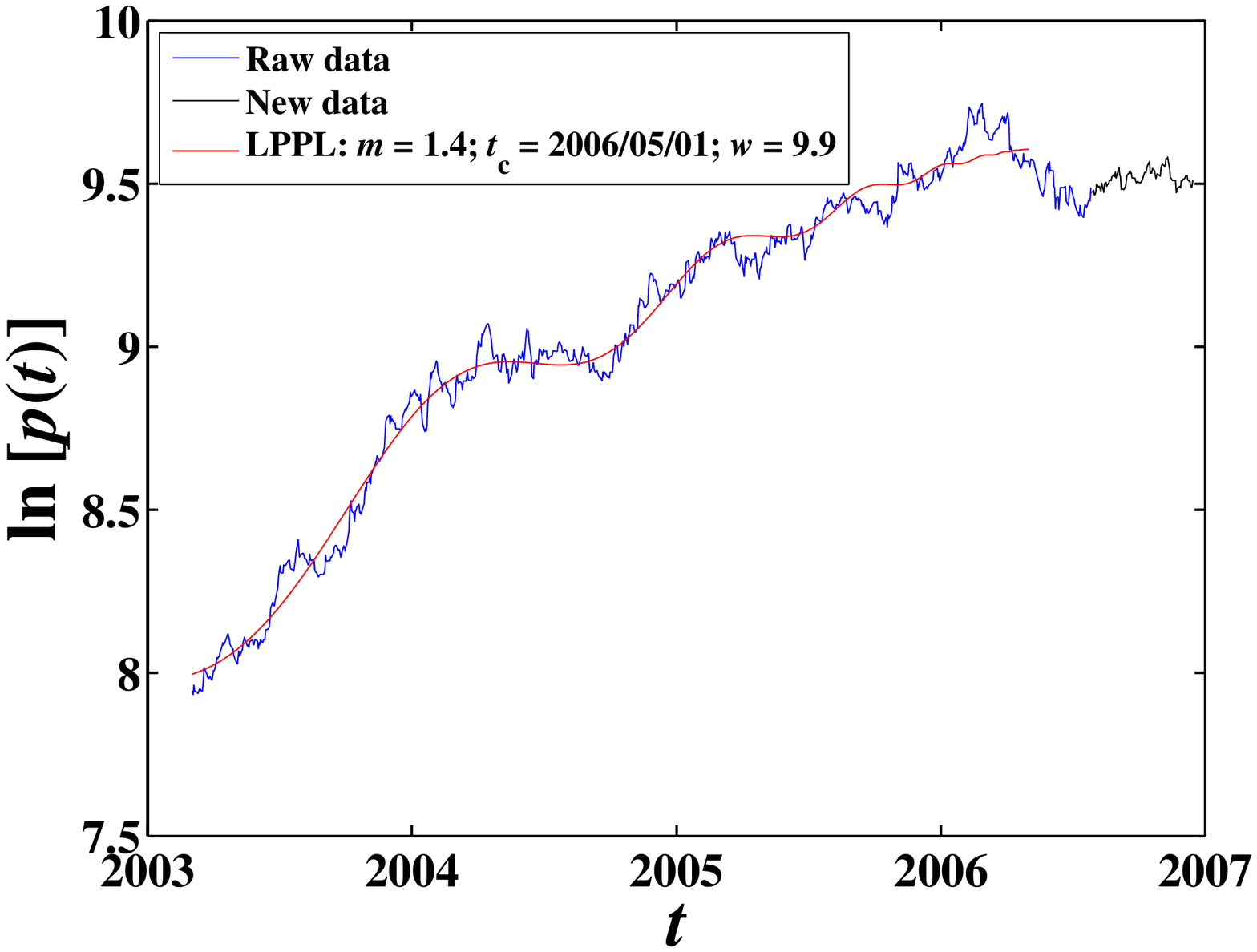}
  \end{minipage}}
  \hspace{0.1cm}
  \subfigure[Stock No.45: WHL]{
  \label{Fig:SouthAfrica:S0S1:WHL}
  \begin{minipage}[b]{0.31\textwidth}
    \includegraphics[width=4.5cm,height=4.5cm]{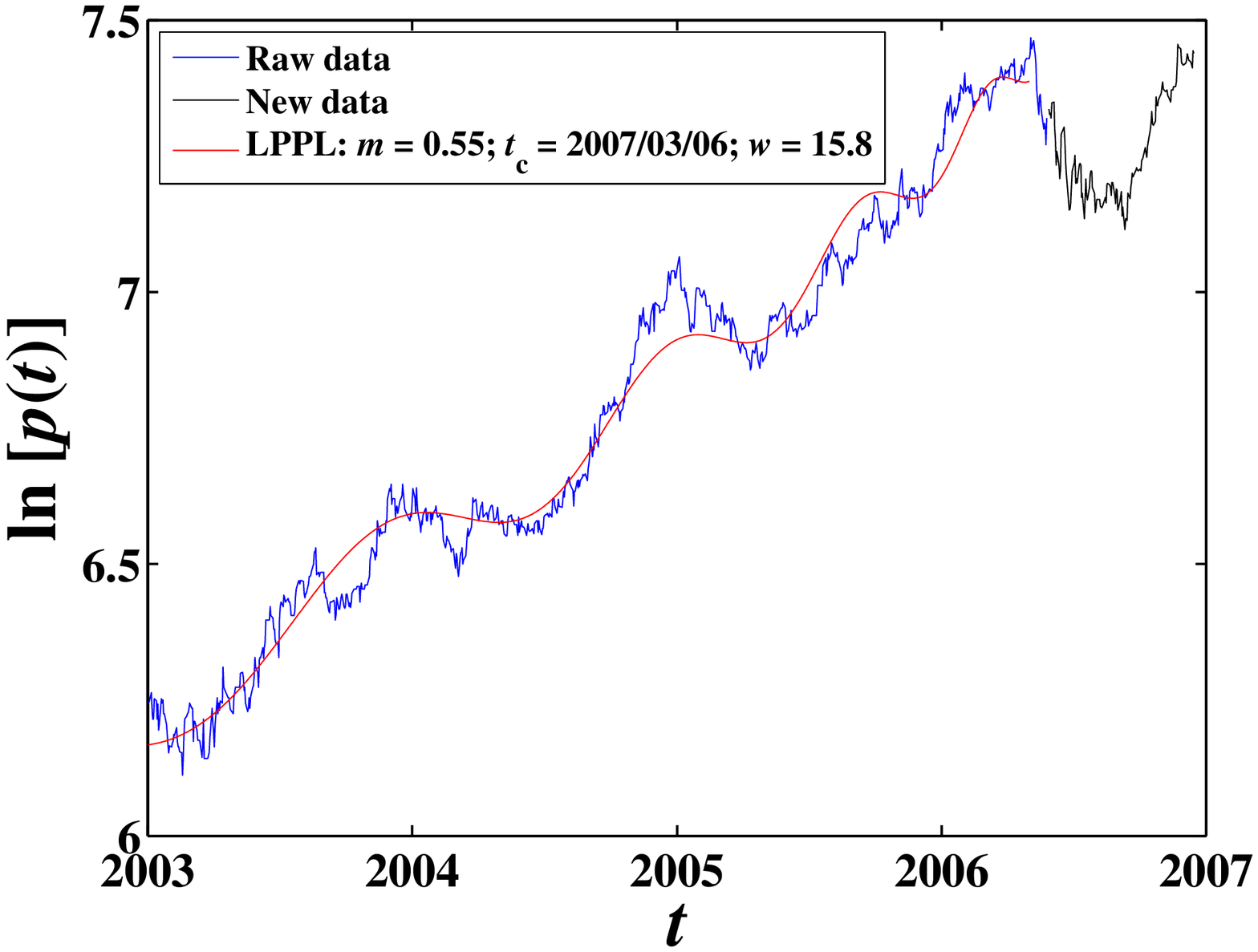}
  \end{minipage}}\\[10pt]
\end{center}
\caption{Log-periodic power-law fits of the prices of stocks from
No.37 to No.45.} \label{Fig:SouthAfrica:S0S1:5}
\end{figure}

\end{document}